\documentclass{aa}  

\usepackage{graphicx}
\usepackage{txfonts}
\usepackage[colorlinks=true,citecolor=blue,linkcolor=blue]{hyperref}
\usepackage{lscape}
\usepackage{xcolor}
\usepackage{siunitx}    
\usepackage{multirow}
\usepackage{lscape}
\usepackage{longtable}

\newcommand{\ditto}{$- \prime \prime -$}
\newcommand{\dg}{$^{\circ}$}
\newcommand{\Nobs}{1044\,}
\newcommand{\NobsRBPLR}{696\,}
\newcommand{\NobsRBPLsdssr}{296\,}
\newcommand{\Ntot}{107\,}
\newcommand{\Ngood}{65\,}
\newcommand{\Nvar}{18\,}
\newcommand{\Nunkn}{24\,}
\DeclareMathOperator\erf{erf}
\DeclareMathOperator\atantwo{atan2}
\begin{document}

   \title{The RoboPol sample of optical polarimetric standards}


   \author{D.\,Blinov\inst{1,2}
          \and
          S.\,Maharana\inst{3,1,4}
          \and
          F.\,Bouzelou\inst{2}
          \and
          C.\,Casadio\inst{1,2}
          \and
          E.~Gjerl{\o}w\inst{5}
          \and
          J.\,Jormanainen\inst{6,7}
          \and
          S.\,Kiehlmann\inst{1,2}
          \and
          J.\,A.\,Kypriotakis\inst{2,1}
          \and
          I.\,Liodakis\inst{6}
          \and
          N.\,Mandarakas\inst{2,1}
          \and
          L.\,Markopoulioti\inst{2}
          \and
          G.\,V.\,Panopoulou\inst{8}
          \and
          V.\,Pelgrims\inst{1,2}
          \and
          A.\,Pouliasi\inst{9}
          \and
          S.\,Romanopoulos\inst{2,1}
          \and
          R.\,Skalidis\inst{10,1}
          \and
          R.\,M.\,Anche\inst{11,3}
          \and
          E.\,Angelakis\inst{12}
          \and
          J.\,Antoniadis\inst{1,2}
          \and
          B.\,J.\,Medhi\inst{13}
          \and
          T.\,Hovatta\inst{6}
          \and
          A. Kus\inst{14}
          \and
          N.\,Kylafis\inst{1,2}
          \and
          A.\,Mahabal\inst{15}
          \and
          I.\,Myserlis\inst{16,17}
          \and
          E.\,Paleologou\inst{1,2}
          \and
          I.\,Papadakis\inst{1,2}
          \and
          V.\,Pavlidou\inst{1,2}
          \and
          I.\,Papamastorakis\inst{1,2}
          \and
          T.\,J.\,Pearson\inst{10}
          \and
          S.\,B.\,Potter\inst{3,18}
          \and
          A.\,N.\,Ramaprakash\inst{4,1,15}
          \and
          A.\,C.\,S.\,Readhead\inst{10}
          \and
          P.\,Reig\inst{1,2}
          \and
          A.\,S\l{}owikowska\inst{19,14}
          \and
          K.\,Tassis\inst{1,2}
          \and
          J.\,A.\,Zensus\inst{17}
          }

   \institute{Institute of Astrophysics, Foundation for Research and Technology-Hellas, N. Plastira 100, Vassilika Vouton, GR-71110 Heraklion, Greece\\
              \email{blinov@ia.forth.gr}
         \and
            Department of Physics, and Institute for Theoretical and Computational Physics, University of Crete, Voutes University campus, GR-70013 Heraklion, Greece
        \and
            South African Astronomical Observatory, PO Box 9, Observatory, 7935, Cape Town, South Africa
        \and
            Inter-University Centre for Astronomy and Astrophysics, Post Bag 4, Ganeshkhind, Pune - 411 007, India
        \and
            Institute of Theoretical Astrophysics, University of Oslo, P.O. Box 1029 Blindern, NO-0315 Oslo, Norway
        \and
            Finnish Centre for Astronomy with ESO, FINCA, University of Turku, Quantum, Vesilinnantie 5, FI-20014, Finland
        \and
            Department of Physics and Astronomy, University of Turku, FI-20014, Finland
        \and
            Department of Space, Earth and Environment, Chalmers University of Technology, Gothenburg, Sweden
        \and
            University of California, Los Angeles, Geophysics and Space Physics Department of Earth, Planetary, and Space Sciences, USA
        \and
            Owens Valley Radio Observatory, California Institute of Technology, MC 249-17, Pasadena, CA 91125, USA
        \and
            Department of Astronomy/Steward Observatory, Tucson, AZ, 85721-0065, USA
        \and
            Section of Astrophysics, Astronomy \& Mechanics, Department of Physics, National and Kapodistrian University of Athens,\\ Panepistimiopolis Zografos 15784, Greece
        \and
            Department of Physics, Gauhati University, Guwahati-781014, Assam, India
        \and
            Institute of Astronomy, Faculty of Physics, Astronomy and Informatics, Nicolaus Copernicus University in Toru\'n, Grudziadzka 5,\\PL-87-100 Toru\'n, Poland
        \and
            Cahill Center for Astronomy and Astrophysics, California Institute of Technology, 1200 E California Blvd, MC 249-17,\\Pasadena CA, 91125, USA
        \and
            Institut de Radioastronomie Millim\'{e}trique, Avenida Divina Pastora 7, Local 20, E-18012, Granada, Spain
        \and
            Max-Planck-Institut f\"{u}r Radioastronomie, Auf dem H\"{u}gel 69, 53121 Bonn, Germany
        \and
            Department of Physics, University of Johannesburg, PO Box 524, Auckland Park 2006, South Africa
        \and
            Joint Institute for VLBI ERIC, Oude Hoogevceensedijk 4, NL-7991 PD Dwingeloo, the Netherlands            
             }

   \date{Received X; accepted Y}

 
  \abstract
   {Optical polarimeters are typically calibrated using measurements of stars with known and stable polarization parameters. However, there is a lack of such stars available across the sky. Many of the currently available standards are not suitable for medium and large telescopes due to their high brightness. Moreover, as we find, some of the used polarimetric standards are in fact variable or have polarization parameters that differ from their cataloged values. }
   {Our goal is to establish a sample of stable standards suitable for calibrating linear optical polarimeters with an accuracy down to $10^{-3}$ in fractional polarization.}
   {For five years, we have been running a monitoring campaign of a sample of standard candidates  comprised of \Ntot stars distributed across the northern sky. We analyzed the variability of the linear polarization of these stars, taking into account the non-Gaussian nature of fractional polarization measurements. For a subsample of nine stars, we also performed multiband polarization measurements.}
   {We created a new catalog of \Ngood stars (see Table~\ref{tab:new_stand_data}) that are stable, have small uncertainties of measured polarimetric parameters, and can be used as calibrators of polarimeters at medium- and large-size telescopes.}
   {}

   \keywords{Polarization --
                Techniques: polarimetric --
                Standards
               }

   \maketitle
%

\section{Introduction} \label{sec:intro}
Polarimetry, on its own and in combination with other techniques, is a powerful tool for probing the physical conditions of astrophysical sources. As all experimental techniques, polarimetric observations require careful calibration and control of instrumental systematics. In the case of optical polarimetry, standard stars with known polarization properties are used for calibration purposes. Unfortunately, the number of reliable polarimetric standards is very limited. There are less than 30 stars in both hemispheres with  polarization degree (PD) known with an accuracy of 0.1\% or better, and proven to be stable in time \citep[e.g.][]{Schmidt1992,Hsu1982}. The lack of an appropriate unpolarized standard star in the night sky at a given moment is common.

The situation is particularly difficult for telescopes with aperture larger than 1-m. They often have a lower limit on the brightness of sources suitable for observations due to CCD saturation constraints. Meanwhile, most unpolarized standards are very bright ($<8^m$), making them unsuitable for calibration on such telescopes. This is because unpolarized standards are selected from nearby stars to ensure that their light does not pass through a significant column of dust in the interstellar medium.

Another problem is the lack of polarized standards with low, but not negligible, PD in the range between 0.1 and 2\%. Existing measurements of standards with PD > 2\% have been sufficient to calibrate conventional polarimeters, and there has been no need for covering a lower range of PD. It is because conventional polarimeters have (or are assumed to have) negligible crosstalk between the Stokes parameters, meaning that the parameters are independent and uncorrelated. In this case, one uses: unpolarized (zero- or negligibly-polarized) stars to find \textit{the offset} of the instrumental $Q/I$ - $U/I$ plane with respect to the standard one; (2) highly-polarized stars to find a \textit{rotation} of the instrumental relative Stokes parameters plane with respect to the standard one \citep[e.g.][]{Ramaprakash2019}. However, some new polarimeters have significant crosstalk \citep{Tinbergen2007,Wiersema2018,Maharana2022,Wiktorowicz2023}. This crosstalk must be modeled in the entire range of PDs of interest, including the 0.1 to 2\% range (the level of ISM-induced stellar polarization in the diffuse ISM). The lack of standards covering a range of polarization values hinders efficient calibration of modern polarimeters where crosstalk between the relative Stokes $Q/I$ and $U/I$ parameters is significant.

\begin{figure*}
 \centering
 \includegraphics[width=0.89\textwidth]{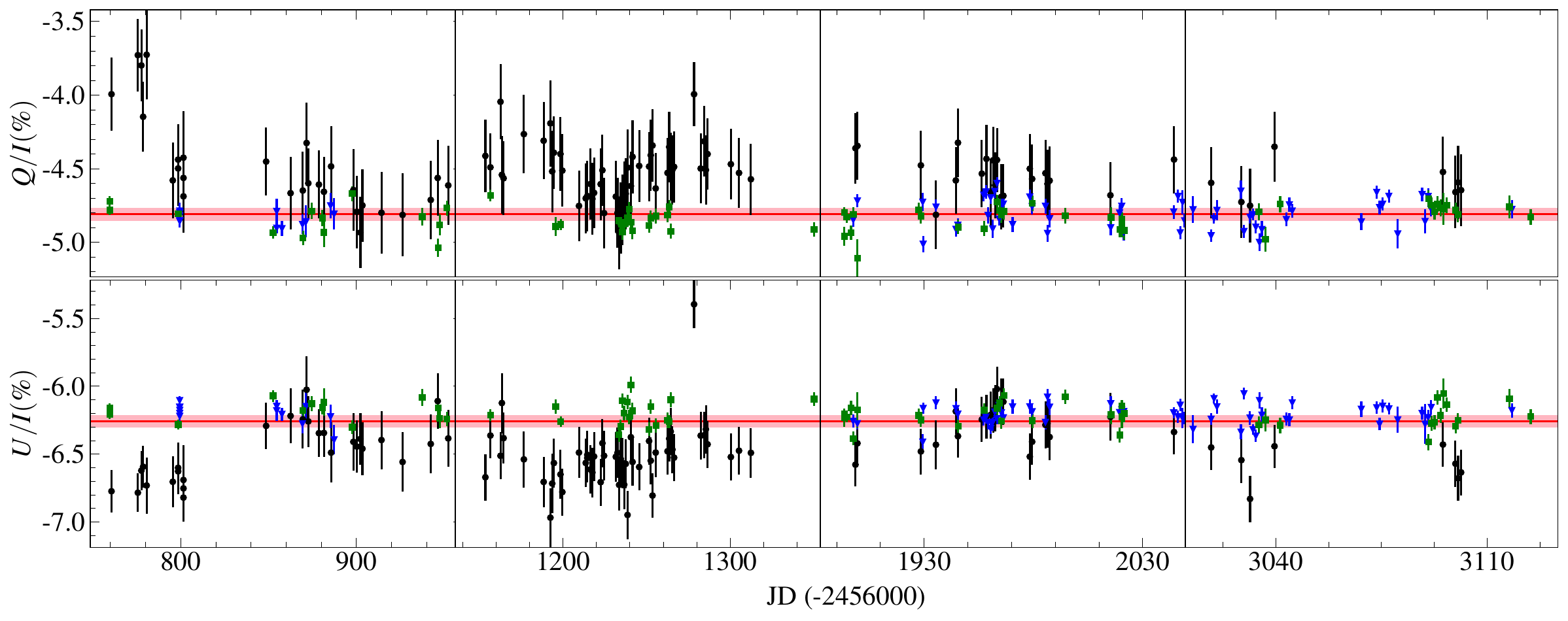}
 \caption{R-band relative Stokes parameters of VI Cyg 12 (black circles) in comparison with two other standards BD+32.3739 (blue trianlges) and HD212311 (green squares). The horizontal red lines and the pink areas represent $Q/I$ and $U/I$ values for VI Cyg 12 from \cite{Schmidt92} with corresponding uncertainties. Values for the two other stars are shifted in such a way that their average Stokes parameters match the red line for visualization purposes. In spite of larger photon noise uncertainties, it is clear that the Stokes parameters of VI Cyg 12 are significantly variable and systematically deviate from their catalogue values during long periods of time.}
 \label{fig:VICyg12}
\end{figure*}

Finally, a significant fraction of stars that are widely recognized as reliable standards exhibit inconsistent polarization parameters across different sources in the literature, and in some cases, they have been found to be variable (see, e.g. Table ~\ref{tab:pol_stnds}). A few examples of such studies follow. \cite{Hsu1982}, after monitoring 12 previously used standards, found that 3 of them are variable. \cite{Dolan1986} also questioned the stability of 3 standards. \cite{Bastien1988} monitored 13  previously known polarized standard stars and found 11 of them to be variable. Their methods were criticized by \cite{Clarke1994}. However, after considering this criticism and applying more rigorous statistical methods, \cite{Bastien2007} reached a very similar conclusion: out of these 13 standards, 7 show significant variability, while 4 others may also be variable. In a study by \cite{Clemens1990}, a single-epoch survey of 16 stars previously used as polarization standards was conducted. The study found that four of these stars had significantly different polarization parameters compared to the values previously published. \cite{Breus2021} found that 9 stars used as calibrators in previous studies show variability. As another example, while performing our polarimetric monitoring program, {\em RoboPol}\footnote{\url{http://robopol.org}}, we found that VI Cyg 12 (a.k.a. Cyg OB2 \#12 or Schulte 12), which is used as a highly-polarized standard in many observatories, is variable in polarization (see Fig.~\ref{fig:VICyg12}). Indeed, VI~Cyg~12 has been shown to be a Luminous Blue Variable with a circumstellar dust shell \citep{Chentsov2013}. The standard deviation of the Electric Vector Position Angle (EVPA) in our measurements of this star is $>0.8$\dg. Therefore, it should not be used for calibration if the desired accuracy of EVPA zero point calibration is more strict. Based on {\em RoboPol} monitoring data, BD$+$33.2642 is suspected to have different polarization values than previously reported \citep{Skalidis2018}.

There have been recent attempts to revise the parameters of polarization standards in use or to establish new samples of calibrators. \cite{Breus2021} report on their observations of a large sample $\sim$100 stars that had been considered as calibrators in various studies and offer revised/refined values of the polarization parameters of these stars. \cite{GilHutton2003} proposed a sample of nearby low-polarization stars in the southern hemisphere. Additionally, stars in the solar vicinity with measured polarization parameters obtained for the interstellar medium \citep[e.g.,][]{Piirola2020} and white dwarf physics \citep{Zejmo2017} studies can be used as unpolarized standards. Nevertheless, all candidate standard stars provided in these works are subject to one or several of the aforementioned deficiencies. They are either very bright or they are not proven to be stable, that is measured only a few times, or measured multiple times over a very short time interval.

In summary, there has been a long-standing need in the optical polarimetry community to establish a large homogeneous list of polarimetric standards that will facilitate easier characterization of instrument performance. The aim of this work is to contribute in this direction.

\begin{table*}
\centering
\caption{Polarization parameters of standard stars monitored by {\em RoboPol}, as reported in the literature.}
\label{tab:pol_stnds}
\begin{tabular}{lcccccc}
\hline
Source & RA & Dec & Band & PD (\%) &  EVPA (\degr) & Reference \\
\hline
\multicolumn{7}{c}{\textbf{polarized}} \\
BD$+$57.2615  & 22:47:49.6 & +58:08:50 & $R$ & 2.02  $\pm$ 0.05  & 41.0 $\pm$ 1.0  & \cite{Whittet1992}\\
BD$+$59.389${}^c$& 02:02:42.1 & +60:15:26 & $R$ & 6.430 $\pm$ 0.022 & 98.14 $\pm$ 0.10 & \cite{Schmidt1992}\\
BD$+$64.106   & 00:57:36.7 & +64:51:35 & $R$ & 5.150 $\pm$ 0.098 & 96.74 $\pm$ 0.54 & \cite{Schmidt1992}\\
CMaR1 24      & 07:04:47.4 &$-$10:56:18& $R$ & 3.18  $\pm$ 0.09  & 86.0 $\pm$ 1.0  & \cite{Whittet1992}\\
CygOB2 14     & 20:32:16.6 & +41:25:36 & $R$ & 3.13  $\pm$ 0.05  & 86.0 $\pm$ 1.0  & \cite{Whittet1992}\\
\multirow{2}{*}{HD147283 $\left\lbrace\begin{array}{l}{}\\{}\end{array}\right.$} & 16:21:57.7 &$-$24:29:44 & $R$ & 1.59  $\pm$ 0.03  & 174.0 $\pm$ 1.0*  & \cite{Whittet1992}\\
          & \ditto & \ditto & $R$  &   1.81      &     176.0   & \cite{Carrasco1973}\\
HD147343      & 16:22:19.9 & $-$24:21:48 & $R$ & 0.43  $\pm$ 0.05  & 151.0 $\pm$ 3.0  & \cite{Whittet1992}\\
HD150193      & 16:40:17.9 & $-$23:53:45 & $R$ & 5.19  $\pm$ 0.05  & 56.0  $\pm$ 1.0  & \cite{Whittet1992}\\
\multirow{2}{*}{HD154445${}^c$ $\left\lbrace\begin{array}{l}{}\\{}\end{array}\right.$} & 17:05:32.3 & $-$00:53:31 & $R$ & 3.683 $\pm$ 0.072 & 88.91 $\pm$ 0.56 & \cite{Schmidt1992}\\
\             & \ditto & \ditto & $R$ & 3.63  $\pm$ 0.01  & 90.0  $\pm$ 0.1*  & \cite{Hsu1982}\\
HD155197${}^c$& 17:10:15.8 & $-$04:50:04 & $R$ & 4.274 $\pm$ 0.027 & 102.88 $\pm$ 0.18 & \cite{Schmidt1992}\\
HD161056      & 17:43:47.0 & $-$07:04:47 & $R$ & 4.012 $\pm$ 0.032 & 67.33 $\pm$ 0.23 & \cite{Schmidt1992}\\
\multirow{2}{*}{HD183143${}^{b,c}$ $\left\lbrace\begin{array}{l}{}\\{}\end{array}\right.$} & 19:27:26.6 & +18:17:45 & $R$ & 5.90  $\pm$ 0.05  & 179.2 $\pm$ 0.2*  & \cite{Hsu1982}\\
              & \ditto & \ditto & $R$ & 5.7   $\pm$ 0.04  & 178.0 $\pm$ 1.0  & \cite{Bailey1982}\\
\multirow{3}{*}{HD204827${}^{b,c}$ $\left\lbrace\begin{array}{l}{}\\{}\\{}\end{array}\right.$} & 21:28:57.8 & +58:44:23 & $R$ & 4.893 $\pm$ 0.029 & 59.10 $\pm$ 0.17* & \cite{Schmidt1992}\\
               & \ditto & \ditto & $R$ & 4.86  $\pm$ 0.05  & 60.0  $\pm$ 1.0  & \cite{Bailey1982}\\
               & \ditto & \ditto & $R$ & 4.99  $\pm$ 0.05  & 59.9  $\pm$ 0.1  & \cite{Hsu1982}\\
HD215806       & 22:46:40.2 & +58:17:44 & $R$ & 1.83  $\pm$ 0.04  & 66.0  $\pm$ 1.0  & \cite{Whittet1992}\\
HD236633       & 01:09:12.3 & +60:37:41 & $R$ & 5.376 $\pm$ 0.028 & 93.04 $\pm$ 0.15 & \cite{Schmidt1992}\\
Hiltner960${}^a$ & 20:23:28.5 & +39:20:59 & $R$ & 5.210 $\pm$ 0.029 & 54.54 $\pm$ 0.16 & \cite{Schmidt1992}\\
\multirow{3}{*}{VICyg12${}^b$ $\left\lbrace\begin{array}{l}{}\\{}\\{}\end{array}\right.$} & 20:32:41.0 & +41:14:29 & $R$ & 7.97  $\pm$ 0.05  & 117.0 $\pm$ 1.0  & \cite{Whittet1992}\\
               & \ditto & \ditto & $R$  & 7.893 $\pm$ 0.037 & 116.23 $\pm$ 0.14* & \cite{Schmidt1992}\\
               & \ditto & \ditto & $R$  & 7.18  $\pm$ 0.04  & 117.0 $\pm$ 1.0  & \cite{Hsu1982}\\
\hline
\multicolumn{7}{c}{\textbf{unpolarized}} \\
BD$+$28.4211   & 21:51:11.0 & +28:51:50 & $V$ & 0.054 $\pm$ 0.027 &        54.22       & \cite{Schmidt1992}\\
BD$+$32.3739${}^c$& 20:12:02.1 & +32:47:44 & $V$ & 0.025 $\pm$ 0.017 &        35.79       & \cite{Schmidt1992}\\
BD$+$33.2642   & 15:51:59.9 & +32:56:54 & $R$ & 0.20  $\pm$ 0.15  &   78    $\pm$ 20   & \cite{Skalidis2018}\\
BD$+$40.2704   & 13:54:07.2 & +39:37:59 & ? & 0.07  $\pm$ 0.02  &   57    $\pm$ 9    & \cite{Berdyugin2002}\\
G191B2B        & 05:05:30.6 & +52:49:52 & $V$ & 0.061 $\pm$ 0.038 &       147.65       & \cite{Schmidt1992}\\
HD14069        & 02:16:45.2 & +07:41:11 & $V$ & 0.022 $\pm$ 0.019 &       156.57       & \cite{Schmidt1992}\\
HD154892       & 17:07:41.3 & +15:12:38 & $B$ & 0.05  $\pm$ 0.03  &         -          & \cite{Turnshek1990}\\
HD212311${}^c$ & 22:21:58.6 & +56:31:53 & $V$ & 0.034 $\pm$ 0.021 &       50.99        & \cite{Schmidt1992}\\
HD21447        & 03:30:00.2 & +55:27:07 & $V$ & 0.051 $\pm$ 0.020 &      171.49        & \cite{Schmidt1992}\\
HD94851        & 10:56:44.3 & $-$20:39:53 & $B$ & 0.057 $\pm$ 0.018 &         -          & \cite{Turnshek1990}\\
WD2149+021     & 21:52:25.4 & +02:23:20 & $R$ & 0.050 $\pm$ 0.006 & $-$63   $\pm$ 3    & \cite{Cikota2017}\\
 \hline
 \multicolumn{7}{l}{${}^a$ - possibly variable \cite{You2017}; ${}^b$ - variable in {\em RoboPol} data or/and in \cite{Hsu1982,Dolan1986};}\\
 \multicolumn{7}{l}{${}^c$ - possibly variable \cite{Breus2021}. For stars with multiple literature values, * marks the value used for the EVPA zero point}\\
 \multicolumn{7}{l}{calibration in Sect.~\ref{subsec:rbpl}.}\\
 \end{tabular}
\end{table*}

\section{Sample of polarization-standard candidates} \label{sec:samp}

To meet the challenges of establishing a large set of reliable polarization standards, we selected an initial sample of 121 candidate stars, which was comprised of four independent sub-samples:

\textbf{Sample B:}  35 polarized stars ($\mathrm{PD}/\sigma_\mathrm{PD} \ge 3$) in fields of blazars monitored within the {\em RoboPol}  
program, that did not show any significant variability between 2013 and 2016. {\em RoboPol} is a linear optical polarimeter designed for efficient monitoring of point sources such as blazars or stars \citep[see Sect.~\ref{subsec:rbpl} and ][]{Ramaprakash2019}. The point sources are placed in a central $22 \times 22$ arcsec masked area, where the sky background is reduced. However, the polarimeter also has a large unmasked field of view (FoV) of $13 \times 13$ arcmin, which allows linear polarimetry of all sources in the field, but with higher noise compared to the central target. High-cadence polarimetric monitoring of about one hundred blazars was performed between 2013 and 2016 \citep{Blinov2021}. Most of the sources were observed several tens to a few hundred times. This provided the same number of observation of stars in the corresponding fields. We analyzed the field stars data and selected 35 sources from these fields, which have shown stable polarization (see Sect.~\ref{sec:var}) throughout the monitoring period.

\textbf{Sample H:} Six stars were selected from \cite{Heiles2000} catalog, with brightness in the range $8^\mathrm{m} < R < 14^\mathrm{m}$. Three of them are highly-polarized stars. Additional three have low polarization and fill the range in Right Ascension (RA), where there is a lack of low-polarization stars in other samples.

\textbf{Sample L:} 54 photometric standard stars distributed along the celestial equator from \cite{Landolt1992}. For selecting these sources we used an atlas of Landolt standards compiled by P. S. Smith\footnote{\url{http://james.as.arizona.edu/~psmith/61inch/ATLAS/atlasinfo.html}}. The selection criteria of stars in this atlas were: 1) Declination $\delta > -20$\dg; 2) Observed by \cite{Landolt1992} on at least 5 nights; 3) Absence of confirmed or suspected variability.
The atlas stars are distributed in $6.8 \times 6.8$ arcmin fields every one hour in RA near $\rm Dec = 0$\dg. We selected 2 to 4 stars per such field, with brightness in the range $8^\mathrm{m} < R < 14^\mathrm{m}$. 

\textbf{Sample Z:} 26 unpolarized stars at high Galactic latitudes, from a single epoch survey by \cite{Berdyugin2014} that have fractional polarization ${\rm PD} < 0.1\%$, with uncertainties $\sigma_{\rm PD} < 0.05\%$;

\begin{figure*}
 \centering
 \includegraphics[width=0.86\textwidth]{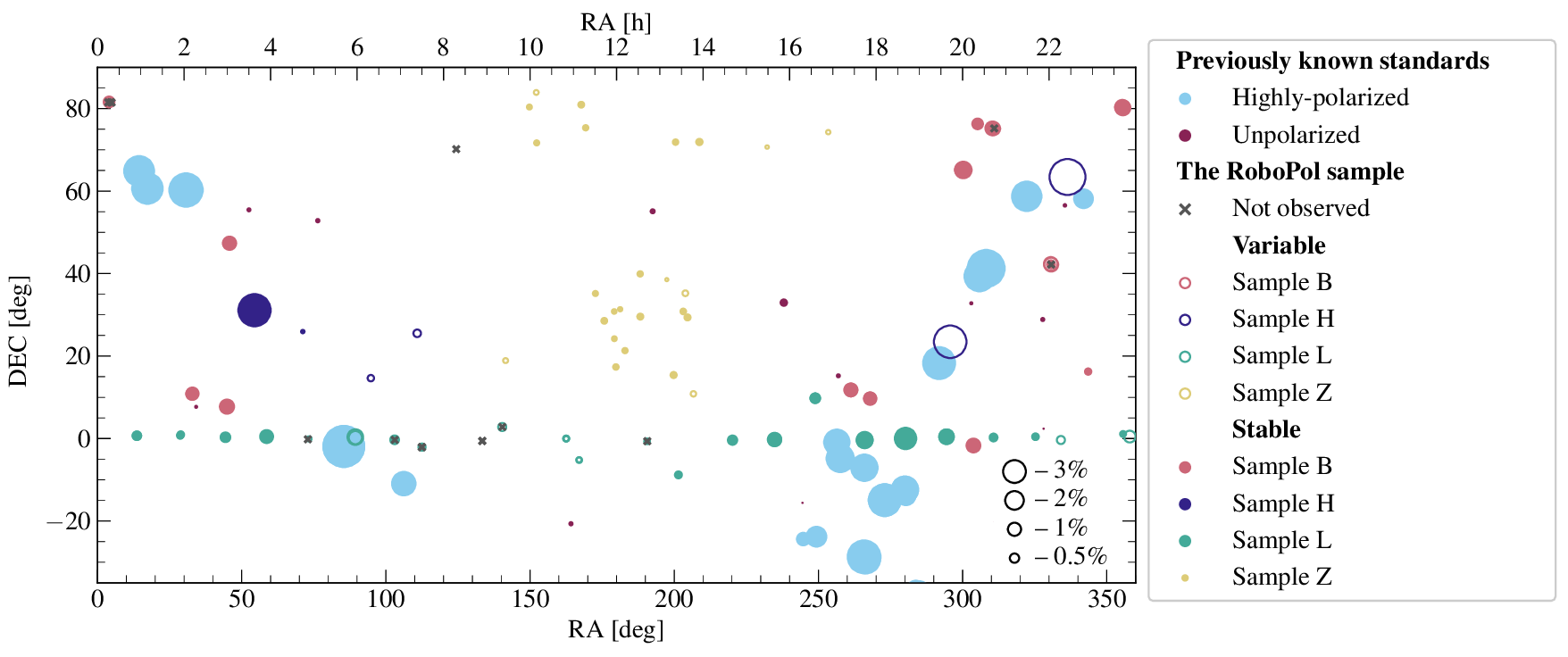}
 \caption{Distribution of the sample stars over the sky. Samples B, H, L and Z are described in Sect.~\ref{sec:samp}. 14 stars marked as ``not observed'' that have zero measurements are excluded from the final sample. Highly-polarized and Unpolarized stars are standards used in previous studies listed in Table~\ref{tab:pol_stnds}. The symbol size indicates polarization of individual stars from Table~\ref{tab:res}.}
 \label{fig:samp}
\end{figure*}

The properties of the selected standard-star candidates are summarized in Table~\ref{tab:sampinfo}, where \textit{Star ID} prefixes correspond to one of the four samples. The advantages of our sample are that it is widely distributed over the northern sky (see Fig.~\ref{fig:samp}) and partially available from the southern hemisphere. It contains relatively faint stars that are accessible to medium- and large-size telescopes. Moreover, a significant fraction of the sample are Landolt stars. Therefore, they can be used for simultaneous polarimetric and absolute photometric calibration of instruments (i.e., $I$, $Q$ and $U$ Stokes parameters can be calibrated together).

\section{Observations and data reduction} \label{Sec:obs}

We have been monitoring the linear polarization parameters of the sample of candidate stars for four consecutive years in order to confirm their stability. The monitoring was performed using the {\em RoboPol} polarimeter. Additionally, we performed single-epoch measurements of a small subsample using the Nordic Optical Telescope. These observations are described in the following subsections.

\subsection{{\em RoboPol} monitoring and data reduction} \label{subsec:rbpl}

We carried out our polarimetric monitoring of the selected sample in the Cousins $R$ and SDSS $r^\prime$ bands from May 2017 to June 2021 using the {\em RoboPol} polarimeter at the 1.3~m telescope of the Skinakas observatory. Every year observations were performed from May to November. Because the observatory does not operate during winter months, sources around $RA = 12\,\mathrm{h}$ were insufficiently sampled. Of the initial 121 stars selected for monitoring, 14 were never observed or have poor quality of measurements. For this reason, they have been dropped from the sample. However, most of them were located near other stars in the sample, and therefore, would not increase much the sky coverage of our final standards catalog.

The polarizing assembly of the polarimeter consists of two half-wave plates and two Wollaston prisms aligned in such a way that any incident ray is split into four rays/channels with the polarization state rotated by 45\dg with respect to each other. {\em RoboPol} has no moving parts except the filter wheel, which simplifies operations and instrumental polarization modeling. Three Stokes parameters $q=Q/I$, $u=U/I$ and $I$ (the latter only in the case when stars with known magnitude are present in the $13 \times 13$ arcmin FoV) can be measured simultaneously with a single exposure. The optical and mechanical design of {\em RoboPol} is described in \cite{Ramaprakash2019}. All data for this program were collected in the
central masked region of the FoV, where systematic uncertainties are < 0.1\% \citep{Ramaprakash2019}.

The data were processed using the standard {\em RoboPol} pipeline, which is described by \cite{King2014}, with modifications presented by \cite{Blinov2021}. Further corrections were introduced at the calibration stage, using known standard stars measurements. The details of this process are described below.

\begin{figure*}
 \centering
 \includegraphics[width=0.44\textwidth]{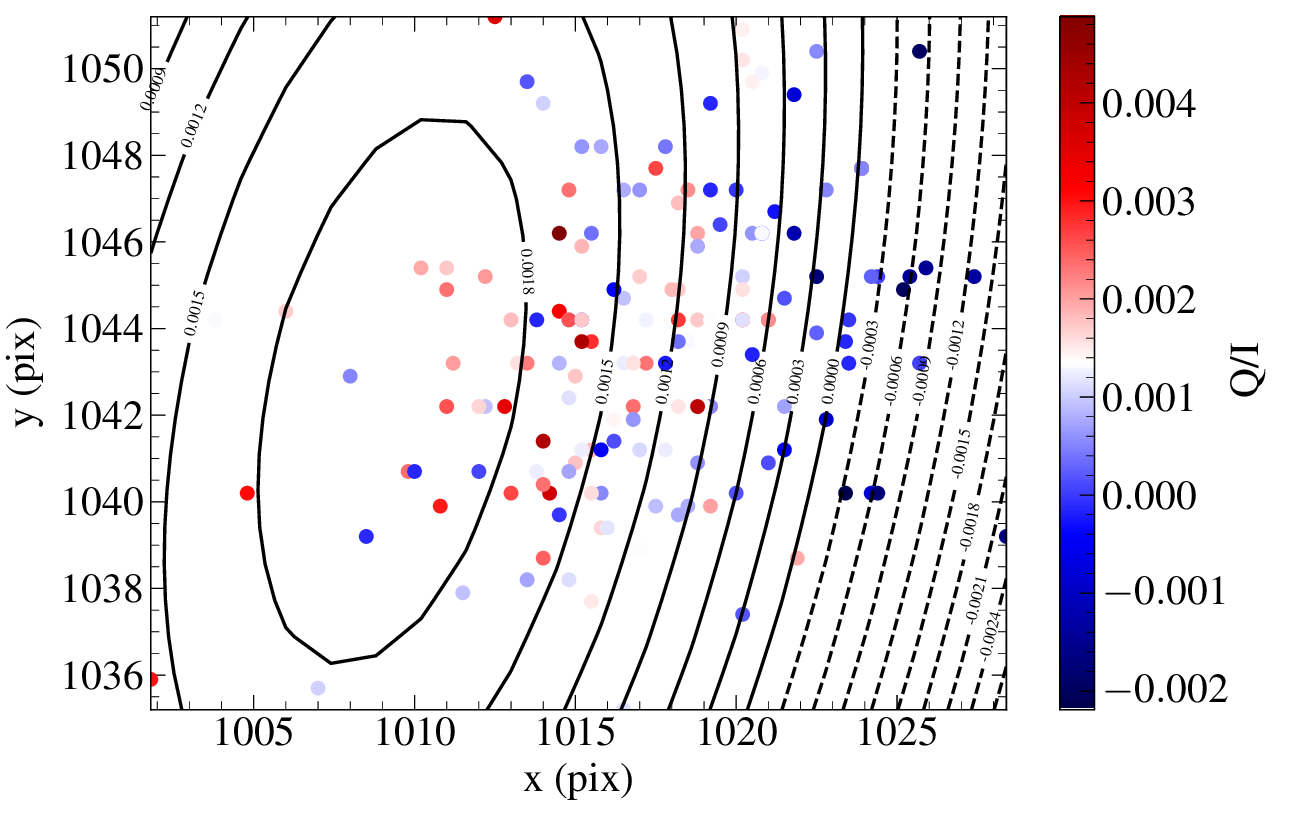}
 \includegraphics[width=0.44\textwidth]{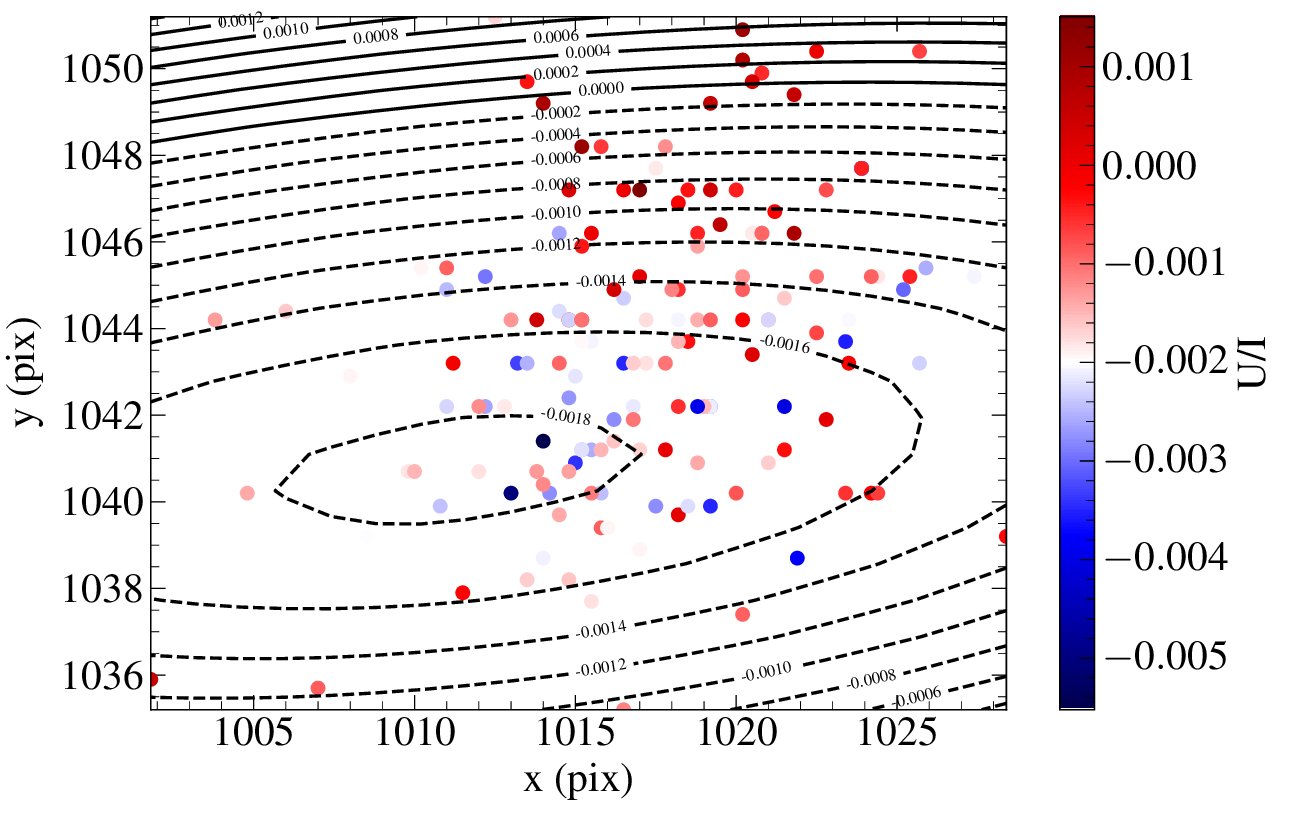}
 \caption{Relative Stokes parameters of zero-polarization standards from Table~\ref{tab:pol_stnds} observed in 2019 in the central masked area as a function of the position on the CCD. Color of points indicates the deviation of $Q/I$ or $U/I$ from zero. The planar contours show the fitted quadratic surface.}
 \label{fig:xyQUfit}
\end{figure*}

Standard processing of {\em RoboPol} data includes an instrumental polarization correction model. This model was created based on combined measurements obtained during multiple years of several unpolarized standard stars, in a grid of hundreds of positions uniformly covering the FoV \citep{King2014}. Therefore, it approximates well the large-scale instrumental polarization variation across the entire FoV. However, we discovered that for stars measured in the central masked area, there is a residual instrumental polarization, which is unaccounted for by the model, and depends on the $(x, y)$ source position on the CCD. In Fig.~\ref{fig:xyQUfit}, we show an example of such subtle instrumental polarization changes for unpolarized standards measured in 2019. A clear position-dependent trend with an amplitude of $\sim 0.5$\% in the measured values of relative Stokes parameters can be seen. Since all measurements discussed in this work were observed in the central masked area, we had to correct them for this trend. We approximated these $q(x,y) = \frac{Q}{I}(x,y)$ and $u(x,y) = \frac{U}{I}(x,y)$ dependencies with a quadratic surface for each observing season separately. Using these fits, we corrected all measurements of corresponding seasons. Then, we determined the standard deviations in the $q$ and $u$ estimates for the unpolarized standards, and propagated these values with the corresponding uncertainties of standard candidates measurements.

\begin{figure}
 \centering
 \includegraphics[width=0.40\textwidth]{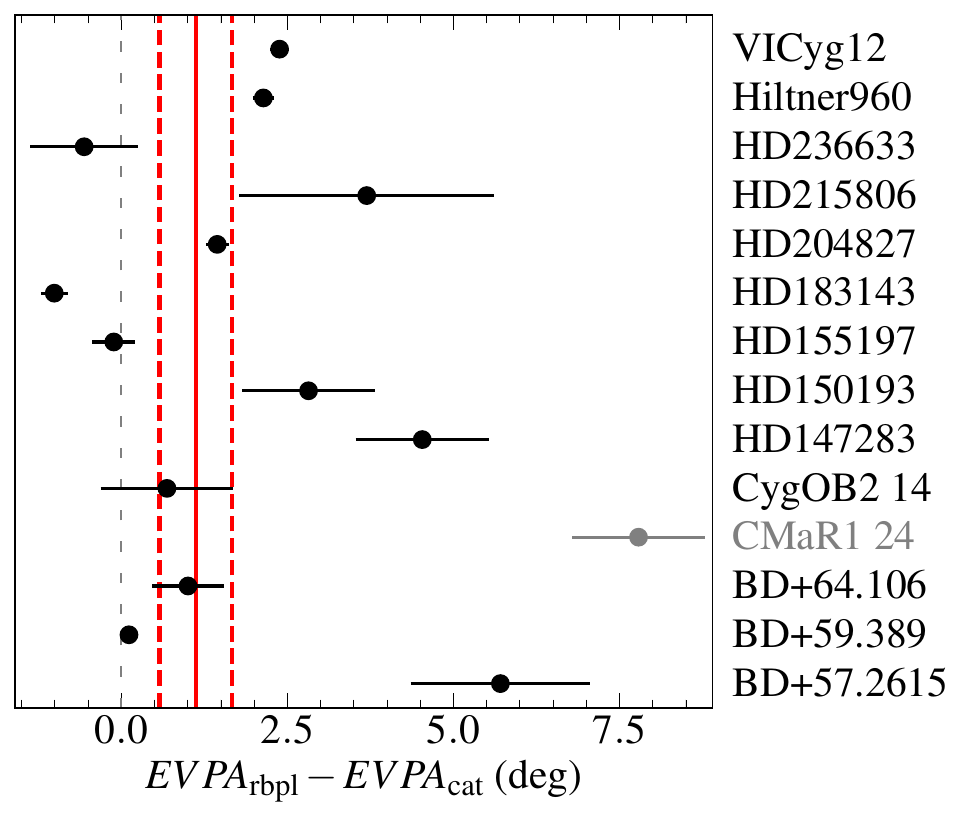}
 \caption{Differences between weighted average of observed EVPA and corresponding catalogue values for 14 most reliable highly-polarized standards. Weighted mean value for 13 stars (CMaR1 24 is excluded by $3\sigma$-clipping) is shown by the solid red line, while the standard error of the mean is shown by the red dashed lines.}
 \label{fig:pa_diff}
\end{figure}

We found the rotation of the instrumental $q$ - $u$ plane with respect to the standard reference frame using highly-polarized standards (Table~\ref{tab:pol_stnds}) that were monitored along with the standards candidate sample. Since individual catalog values for these standards are unreliable (see Sect.~\ref{sec:intro}), we used a statistical approach: the entire sample was considered, including stars that are known to be variable (e.g., VI Cyg 12). For each standard we computed the weighted mean of the relative Stokes parameters combining all measurements along the observing period. Then, using these $q$, $u$ estimates, we calculated the corresponding $EVPA_{\rm rbpl}$ values of each star in our measurements, and found the difference between this value and the one reported in the literature, $EVPA_{\rm cat}$. For stars with multiple values reported in the literature, we used either the value with the smallest uncertainty, or the most recent measurement if the uncertainties are comparable. The $EVPA_{\rm cat}$ values used are marked with asterisk symbols in Table~\ref{tab:pol_stnds}. The corresponding differences between the {\em RoboPol} and literature estimates are shown in Fig.~\ref{fig:pa_diff}.
We found the weighted mean of $EVPA_{\rm rbpl} - EVPA_{\rm cat}$ to be $1.1 \pm 0.5^{\circ}$, after applying $3 \sigma$-clipping, which excluded {\it CMaR1 24} from the averaging. This value was used as the instrumental EVPA zero-point correction, and all measurements were adjusted for it, while uncertainties were propagated accordingly.

We assessed the possibility that the polarimeter has a crosstalk between the relative Stokes parameters by measuring their covariance. We first corrected the measurements of unpolarized standards for the polarization and EVPA zero points as described before. Then, we calculated the correlation coefficient between $q$ and $u$ for each individual standard. There was no significant systematic correlation found among stars. We also calculated the correlation coefficient, $r$, for a set of standard star measurements for each season. In all cases, $|r|$ does not exceed 0.42, while the median value among seasons is $r=-0.17$. Therefore, we conclude that the crosstalk between channels of the polarimeter is negligible with respect to the noise level.

\begin{figure}
 \centering
 \includegraphics[width=0.5\textwidth]{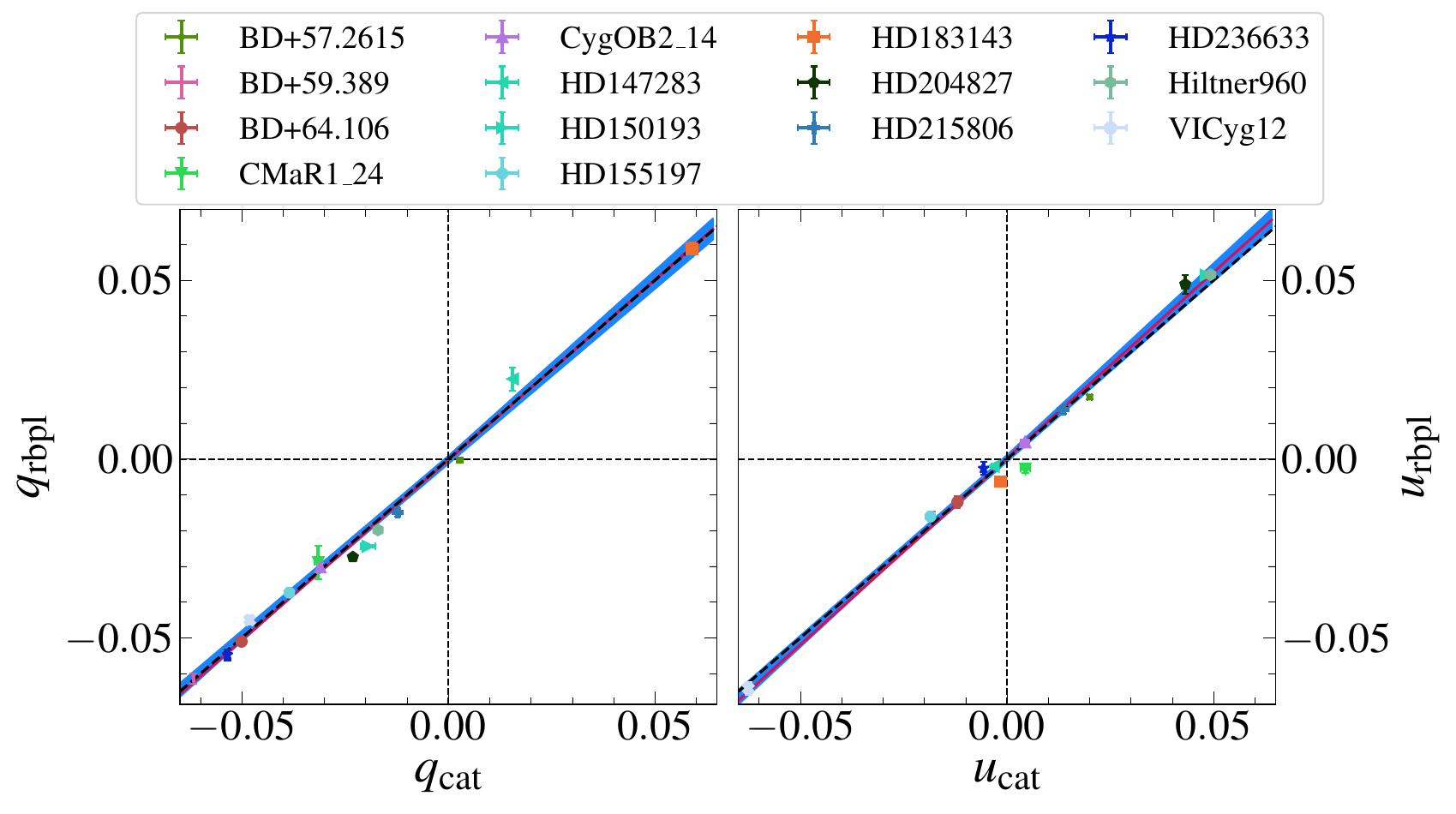}
 \caption{Weighted mean values of Stokes parameters of high-polarization standards as measured by {\em RoboPol} vs. catalogue values from Table~\ref{tab:pol_stnds}. The black dashed line is $y=x$ the red solid line is the ordinary least squares regression fit to the data. The light blue region is the $1 \sigma$ uncertainty region of the fit.}
 \label{fig:QQUU}
\end{figure}

We also verified that the polarimetric efficiency of the instrument is 100\% within measurement errors, i.e, the measured polarimetric accuracy is independent of the source polarization. To this end, we corrected the Stokes parameters of the highly-polarized stars for the zero points, and then compared them with the corresponding literature values. The orthogonal distance regression fits to the data are consistent with the expected $q_\mathrm{rbpl}=q_\mathrm{cat}$ and $u_\mathrm{rbpl}=u_\mathrm{cat}$ dependencies within $1 \sigma$, as shown in Fig.~\ref{fig:QQUU}.

\subsection{NOT observations and data analysis} \label{subsec:not}

We selected a subset of 10 standard candidates based on visibility, limited available observing time for this program, and preliminary stability observed in {\em RoboPol} data. Subsequently, we conducted observations of these candidates using the Nordic Optical Telescope (NOT) under proposal 61-608.
The Alhambra Faint Object Spectrograph and Camera (ALFOSC) instrument\footnote{\url{http://www.not.iac.es/instruments/alfosc/}} was used in its polarimetric mode. It is a two-channel polarimeter consisting of a rotating half-wave plate (HWP) and a calcite plate. For each object, two so-called $o$ and $e$ beam images are formed, corresponding to the $0^{\circ}$ and $90^{\circ}$ polarizations coming out of the WP, respectively.  Standard candidates were observed with sequences consisting of 8 exposures, corresponding to HWP positions of $0^{\circ}$ to $180^{\circ}$ in steps of $22.5^{\circ}$. This yields 4 $Q/I$ and $U/I$ measurements for each star.  Observations were performed during the nights of 8, 23 September and 13 December 2020 in multiple filters. Most of the stars were observed in SDSS $g$, $r$, $i$ and Johnson-Cousins $B$, $V$ and $R$ bands, while two stars were also observed in the $U$-band. Unpolarized standards stars BD+28.4211, HD 212311, and highly-polarized standards Hiltner 960, VI Cyg 12, BD+59.389, BD+64.106, HD 204827 were observed during the same nights as the program targets. These standards were observed with sequences consisting of 16 exposures corresponding to HWP positions of $0^{\circ}$ to $360^{\circ}$ in steps of $22.5^{\circ}$. This provided 8 $Q/I$ and $U/I$ measurements for each star.

For the analysis of the raw data, we developed a semi-automated data reduction pipeline in Python. Attention was paid to error estimation and propagation in each step of the analysis. Photometry was done using the  aperture photometry package of the \textit{Photutils} library\footnote{\url{https://photutils.readthedocs.io/en/stable/}}. To find the polarization parameters, we followed the procedure from \cite{Patat2006}. For each HWP position $\{\theta_i = i \times 22.5^{\circ} \mid i \in \{0, 1, ..., N\}\}$ we calculated the normalized flux differences between $o$ and $e$ star images:
\begin{equation}
 F_i = \frac{f_{o,i} - f_{e,i}}{f_{o,i} + f_{e,i}}.
\end{equation}
Then the relative Stokes parameters were expressed as:
\begin{equation}
 q \equiv \frac{Q}{I} = \frac{2}{N} \sum_{i=0}^{N-1} F_i \cos{\left(\frac{\pi}{2} i\right)}
\end{equation}
\begin{equation}
 u \equiv \frac{U}{I} = \frac{2}{N} \sum_{i=0}^{N-1} F_i \sin{\left(\frac{\pi}{2} i\right)}.
\end{equation}
Using these $q$ and $u$ estimates we inferred PD and EVPA, and their uncertainties as described in Sect.~\ref{subsec:calcPA}. Then, we examined dependencies of the PD and EVPA estimates on the photometry aperture radius, and selected optimal aperture and annulus radii, where both parameters reach a plateau and minimal uncertainties. The same procedure was performed for the observations of the standard stars. Using their polarization parameters, we found the instrumental polarization and EVPA zero points in each band individually. However, in the U-band, no standards were observed either during our observations or during adjacent nights. In the SDSS $g$ and $r$ bands, only a single standard star measurement (BD+28.4211) was available. Therefore, we fitted the dependencies of the instrumental $q$ and $u$ on the effective wavelength using all other bands with a linear function. By utilizing these fits, we determined the instrumental zero points of the relative Stokes parameters in each band and applied corrections to all measurements based on these values.

\subsection{Polarization parameter estimates and their uncertainties} \label{subsec:calcPA}

The polarization degree and its uncertainty were calculated assuming that the relative Stokes parameters $q=Q/I$ and $u=U/I$ follow a normal distribution,
\begin{equation}
 {\rm PD} = \sqrt{q^2 + u^2}, \,\,\, \sigma_{\rm PD} = \sqrt{ \frac{q^2 \sigma_{\rm q}^2 + u^2 \sigma_{\rm u}^2}{q^2 + u^2} }.
\end{equation}
Any linear polarization measurement is biased towards higher PD values \citep{Serkowski1958}. The PD follows a Rician distribution \citep{Rice1945} and significantly 
deviates from the normal distribution at low signal-to-noise ratios. There is a variety of methods suggested for correction of this bias \citep[e.g.][]{Simmons1985,Vaillancourt2006}. Our catalog provides the flexibility to select any debiasing method as the relative Stokes parameters constitute our ultimate data product. The data presented in this paper remain uncorrected for polarization bias. The relative Stokes parameters themselves, of course, are unbiased quantities.

The EVPA is defined as
\begin{equation}
 {\rm EVPA} = \frac{1}{2} \atantwo\left(\frac{u}{q}\right) \equiv
 \begin{cases}
  \arctan{\left(\frac{u}{q}\right)}       & q > 0 \\
  \arctan{\left(\frac{u}{q}\right)} + \pi & u \ge 0, \, q < 0 \\
  \arctan{\left(\frac{u}{q}\right)} - \pi & u < 0, \, q < 0 \\
  \frac{\pi}{2}                           & u > 0, \, q = 0 \\
  -\frac{\pi}{2}                          & u < 0, \, q = 0 \\
  {\rm undefined}                         & u = 0, \, q = 0
 \end{cases}
\end{equation}
while its measurements are also non-Gaussian and defined by the following probability density 
\citep{Naghizadeh1993}:
\begin{equation}
 G(\theta;\theta_o;{\rm PD}_o) = \frac{1}{\sqrt{\pi}} \left\{ \frac{1}{\sqrt{\pi}} + \eta_o e^{\eta_o^2}
 [ 1 + \erf(\eta_o) ] \right\} \exp\left(-\frac{{\rm PD}_o^2}{2 \sigma_{\rm PD}^2}\right),
\end{equation}
where $\eta_o = {\rm PD}_o \cos{2 (\theta - \theta_o)} /(\sigma_{\rm PD} \sqrt{2})$, $\erf$ is the 
Gaussian error function, ${\rm PD}_o$ and $\theta_o$ are the true values of PD and EVPA, and $\sigma_{\rm PD}$ is the uncertainty of PD\footnote{The equation for $\eta_o$ is missing the factor of 2 for the cosine argument in \cite{Clarke2009}, while the correct formula is provided in \cite{Naghizadeh1993}.}.

We determine the EVPA uncertainty $\sigma_\theta$ numerically, by solving the following integral:
\begin{equation}
 \int_{-1\sigma_\theta}^{1\sigma_\theta} G(\theta;{\rm PD}_o) d\theta = 68.27\%.
\end{equation}
The true value of PD in this procedure was estimated following \cite{Vaillancourt2006} as:
\begin{equation}
  {\rm PD}_o =
    \begin{cases}
      0                        & \text{for ${\rm PD}/\sigma_{\rm PD} < \sqrt{2}$}\\
      \sqrt{{\rm PD}^2 - \sigma_{\rm PD}^2} & \text{for ${\rm PD}/\sigma_{\rm PD} \ge \sqrt{2}$}.\\
    \end{cases}       
\end{equation}

For high SNR values, ${\rm PD}/\sigma_{\rm PD} \ge 20$, the uncertainty of EVPA was approximated as
$\sigma_\theta = {\rm PD}/(2\sigma_{\rm PD})$.

\section{Analysis of variability} \label{sec:var}

We assessed variability of the sample stars following \cite{Clarke1993} and \cite{Bastien2007}. The method can be summarized as follows. If measurements of the relative Stokes parameters $q$ and $u$ are independent and follow normal distributions with means $q_0$ and $u_0$, then the statistic
\begin{equation}
    \wp = \sqrt{\left(\frac{q}{\sigma_q}\right)^2 + \left(\frac{u}{\sigma_u}\right)^2},
\end{equation}
as demonstrated by \cite{Simmons1985}, follows the Rician distribution \citep{Rice1945}:
\begin{equation}
    \label{rice}
    f(\wp,\wp_0) = \wp \exp\left(-\frac{\wp^2+\wp_0^2}{2}\right) J_0(i \wp \wp_0),
\end{equation}
where $i$ is the unit imaginary number, $J_0$ is the zeroth order Bessel function, and $\wp_0 = \sqrt{(q_0/\sigma_{\rm q0})^2 + (u_0/\sigma_{\rm u0})^2}$.
In the case of an unpolarized source ($\wp_0 = 0$), Eq.~\ref{rice} reduces to the Rayleigh distribution:
\begin{equation}
    \label{eq:rayleigh}
    f(\wp,0) = \wp \exp{\left(-\frac{\wp^2}{2}\right)}.
\end{equation}
Then, the cumulative distribution function (CDF) of $\wp$ is expressed as:
\begin{equation}
   \label{eq:cdf}
   {\rm CDF}(\wp) = \frac{\int_0^\wp{f(\wp,0) d\wp}}{\int_0^{\infty}{f(\wp,0) d\wp}} = 1 - \exp{\left(-\frac{\wp^2}{2}\right)}.
\end{equation}
In practice, the CDF is approximated by the empirical cumulative distribution function
\begin{equation}
  {\rm EDF}(\wp) = \frac{\textrm{ number of observations} < \wp}{\textrm{total number of observations}}.
\end{equation}
The EDF can deviate significantly from the CDF in two cases: (1) the source has variable polarization; (2) the uncertainties $\sigma_q$ and $\sigma_u$ are incorrectly estimated. Since in either of these cases the measurements of a star cannot be considered for establishing it as a standard, we do not distinguish between them. 

In the case of a polarized source, its weighted means of the measured normalized Stokes parameters $\overline{q}$ and $\overline{u}$ were used as estimates for $q_0$ and $u_0$ in order to reduce the polarization to zero:
\begin{equation}
    \wp_{\rm reduced} = \sqrt{\left(\frac{q-\overline{q}}{\sigma_{q}}\right)^2 + \left(\frac{u-\overline{u}}{\sigma_{u}}\right)^2}.
\label{eq:zeroed_pol}
\end{equation}

In order to assess whether the EDF significantly deviates from the CDF of a constant source given by Eq.~\ref{eq:cdf}, we used a two-sided Kolmogorov-Smirnov (KS) test. If the p-value of the KS test exceeds a given threshold, we consider the star as non-variable and suitable for use as a standard. Otherwise, we consider it unsuitable. As mentioned earlier, if the p-value of the KS test is below the threshold, the star may indeed be variable, or our uncertainty estimates of its measurements may be incorrect. We do not discriminate between these two cases. We use a threshold of $p=0.0455$, corresponding to a $2\sigma$ confidence level, to assess the variability of stars. However, we also provide the p-values for all stars in the sample with $\ge 5$ measurements in Table~\ref{tab:res}. This allows one to select a more or less robust sample of standards by filtering stars based on the p-value and choosing a different confidence level if desired. We do not perform the test for stars with fewer than 5 measurements and mark them as uncertain.

We note that in the procedure described above the variability evaluation is based purely on the fractional polarization behaviour, while information about the EVPA is completely ignored. However, there is a possibility that in peculiar cases the polarization vector can produce nearly perfect loops on the $Q/I$ - $U/I$ plane. In such a situation, the $\wp_{\rm reduced}$ remains constant, while the EVPA changes with time. For instance, binary stars with envelopes symmetric about their orbital plane can produce such polarization variability \citep{Brown1978}. In order to avoid identification of stars with this variability pattern as stable, we visually inspected distributions of measurements on the relative Stokes parameters plane for each source. We did not find any false stable stars during this inspection.

\section{Results} \label{subsec:res}

\begin{figure*}
  \centering
  \includegraphics[width=0.95\textwidth]{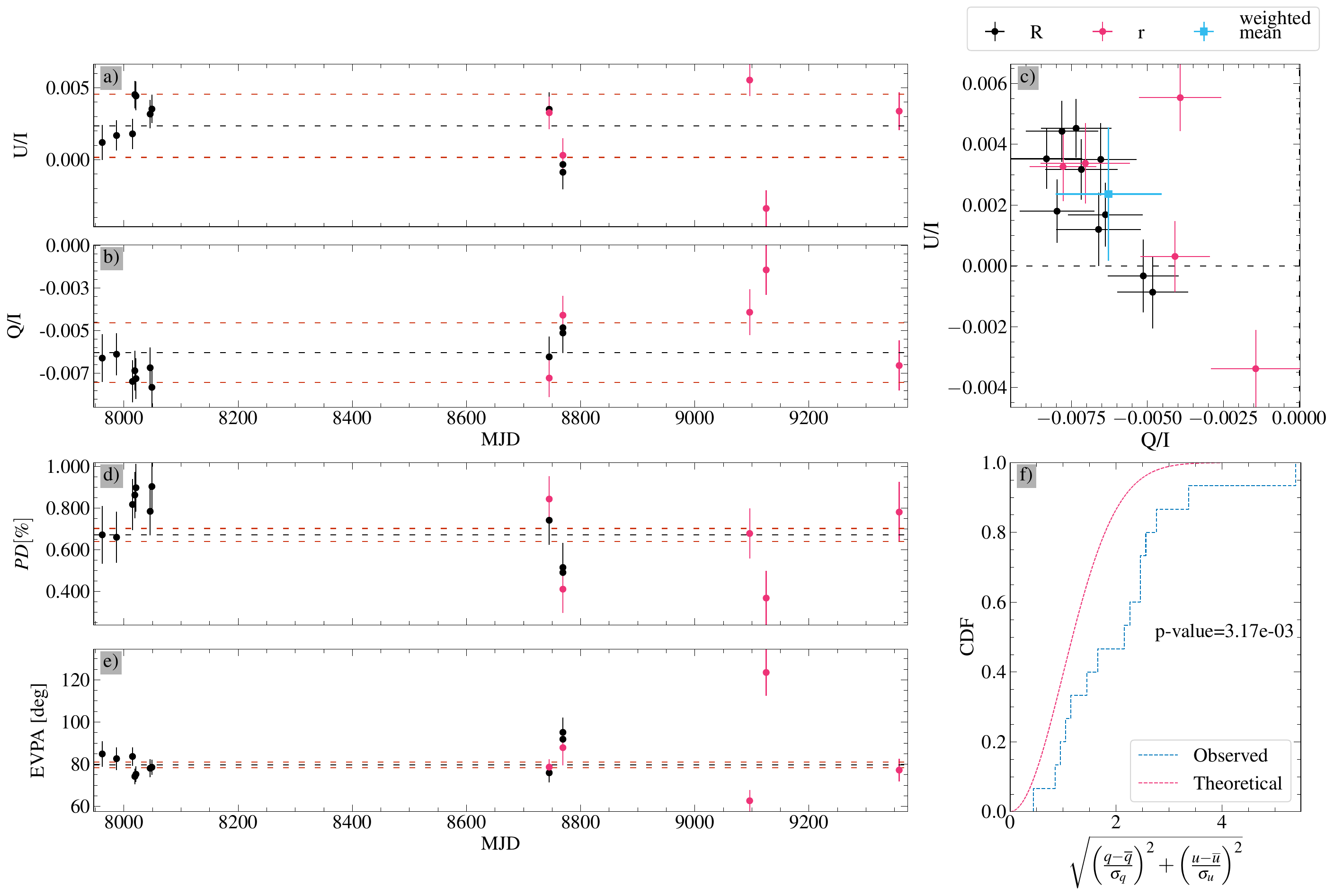}
  \caption{Evolution of polarization parameters of B\_0017+8135\_82, which is found to be variable. (a, b) - Evolution of the relative Stokes parameters. The dashed black line shows the weighted average, the red dashed lines show the corresponding 1$\sigma$ uncertainty. (c) - Distribution of measurements on the relative Stokes parameters plane. (d, e) - Evolution of the polarization degree and the electric vector position angle. The dashed black line shows the weighted average, the red dashed lines show the corresponding 1$\sigma$ uncertainty. (f) - EDF of measured polarization in both bands together with expected CDF of polarization measurements for a constant source with similar uncertainties.}
  \label{fig:B_0017+8135_82}
\end{figure*}

We obtained \NobsRBPLR $R$-band and \NobsRBPLsdssr SDSS $r^\prime$-band measurements of \Ntot stars with {\em RoboPol} that are listed in Table~\ref{tab:mon_data}. Additionally, for nine stars we obtained multi-band polarization measurements with ALFOSC that are presented in Table~\ref{tab:NOT}. We did not find any significant systematic difference in relative Stokes parameters between the $R$ and $r^\prime$-bands. Therefore, we combine all measurements in these two bands and consider them together. For each star in the sample, we constructed plots of the time series data showing the evolution of the fractional polarization PD, the EVPA, and the relative Stokes parameters $Q/I$ and $U/I$. These monitoring data were analyzed using the method described in Sect.~\ref{sec:var}, so that the EDF of the normalized fractional polarization given by Eq.~\ref{eq:zeroed_pol} was computed for each star. Then we compared it with the distribution given by Eq.~\ref{eq:cdf}, which is the expected cumulative distribution of the same quantity for a stable source with the same noise level. The time series, CDF, EDF and the distribution of measurements on the relative Stokes parameters plane for B\_0017+8135\_82, as an example, are shown in Figure~\ref{fig:B_0017+8135_82}.
Similar plots for all other sources in the sample are available only in the electronic version in Appendix~\ref{ap:b}. As the result, we found the average polarimetric parameters for each star in the sample and classified them as stable or variable with the $2\sigma$ confidence level. These parameters and classes are listed in Table~\ref{tab:res}. For reader's convenience, we list only the stable stars in a separate Table~\ref{tab:new_stand_data}, together with their average relative Stokes parameters and Gaia's $G$-band magnitude. We arbitrarily place the limit between high- and low-polarization stars at ${\rm PD}=0.5$\% in Table~\ref{tab:new_stand_data}. This information, along with finding charts for all sample stars, can also be accessed online at \url{https://robopol.physics.uoc.gr/standards}.

\begin{table*}
\scriptsize
\caption{Average polarization parameters of stable ($2\sigma$ S.L.) stars in the established standards sample: (1) - Star identifier; (2,3,4,5) - Equatorial coordinates; (6,7) - Weighted average $Q/I$ and $U/I$ relative Stokes parameters corrected for the instrumental polarization and EVPA zero-point; (8) - Gaia DR3 G-band magnitude. This table is available in a machine-readable format at \url{https://doi.org/10.7910/DVN/IV9TXX}.}
\label{tab:new_stand_data}      
\centering
\begin{tabular}{lccSSS[table-format=1.4(4)]S[table-format=1.4(4)]S}     
\hline\hline       
Star ID & RA (J2000) & DEC (J2000)            & \si{RA (J2000)} & \si{DEC (J2000)} & \si{Q/I} & \si{U/I} & \si{G mag}\\
        & h:m:s      & \degr:\arcmin:\arcsec  & \si{\degr}      & \si{\degr}       &          &          &    \\
\hline                    
\multicolumn{8}{c}{\textbf{Low-polarization (${\rm PD} \le 0.5$\%) stars}}\\
L\_92\_245& 00:54:16.16 & +00:39:55.0 & 13.56734 & 0.66527 & 0.0014(11)& -0.0010(8)& 13.53 \\ 
L\_92\_250& 00:54:37.16 & +00:38:57.6 & 13.65482 & 0.64933 & -0.0000(11)& -0.0006(9)& 12.99 \\ 
L\_93\_317& 01:54:37.73 & +00:43:00.6 & 28.65719 & 0.71682 & 0.0011(9)& -0.0020(9)& 11.42 \\ 
L\_93\_333& 01:55:05.21 & +00:45:42.8 & 28.77172 & 0.76188 & 0.0010(26)& -0.0013(10)& 12.42 \\ 
L\_93\_424& 01:55:26.36 & +00:56:42.6 & 28.85984 & 0.94517 & 0.0008(10)& -0.0024(11)& 11.34 \\ 
L\_94\_251& 02:57:46.98 & +00:16:02.7 & 44.44575 & 0.26742 & 0.0039(10)& 0.0017(10)& 10.83 \\ 
H\_HD283807& 04:44:42.60 & +25:56:09.6 & 71.17525 & 25.93600 & -0.0003(11)& 0.0006(8)&  9.96 \\ 
Z\_HD85471& 09:59:04.94 & +80:22:47.6 & 149.77059 & 80.37989 & 0.0005(13)& 0.0013(14)&  7.83 \\ 
Z\_HD87582& 10:09:10.40 & +71:41:15.5 & 152.29185 & 71.68763 & 0.0011(13)& 0.0010(9)&  7.07 \\ 
Z\_HD96589& 11:11:00.40 & +80:56:55.9 & 167.75166 & 80.94887 & -0.0008(13)& 0.0018(11)&  7.54 \\ 
Z\_HD97853& 11:17:09.82 & +75:21:11.1 & 169.29090 & 75.35309 & 0.0001(7)& 0.0016(7)&  7.16 \\ 
Z\_BD+35.2256& 11:30:39.95 & +35:09:50.3 & 172.66646 & 35.16397 & 0.0006(23)& 0.0013(9)& 10.40 \\ 
Z\_BD+29.2198& 11:42:56.23 & +28:32:39.2 & 175.73429 & 28.54423 & -0.0003(14)& 0.0019(9)& 10.19 \\ 
Z\_BD+31.2314& 11:56:43.33 & +30:48:45.5 & 179.18056 & 30.81264 & -0.0003(11)& 0.0012(10)& 10.83 \\ 
Z\_BD+25.2439& 11:56:55.10 & +24:12:09.6 & 179.22922 & 24.20266 & 0.0000(13)& 0.0013(7)& 10.23 \\ 
Z\_BD+18.2549& 11:59:04.54 & +17:21:07.4 & 179.76893 & 17.35205 & -0.0002(11)& 0.0017(9)& 10.34 \\ 
Z\_BD+32.2217& 12:04:58.34 & +31:21:06.1 & 181.24309 & 31.35169 & -0.0002(8)& 0.0010(9)& 10.46 \\ 
Z\_BD+22.2446& 12:11:40.13 & +21:19:08.7 & 182.91720 & 21.31909 & 0.0001(11)& 0.0016(10)& 10.75 \\ 
Z\_BD+40.2546& 12:32:44.89 & +39:55:20.2 & 188.18703 & 39.92229 & -0.0001(11)& 0.0016(10)& 10.85 \\ 
Z\_BD+30.2290& 12:32:55.79 & +29:33:52.9 & 188.23247 & 29.56469 & 0.0001(10)& 0.0020(9)& 11.39 \\ 
L\_104\_334& 12:42:20.43 & $-$00:40:28.7 & 190.58513 & -0.67454 & -0.0001(15)& 0.0019(16)& 13.37 \\ 
Z\_BD+16.2491& 13:19:08.90 & +15:23:53.6 & 199.78371 & 15.39821 & -0.0004(15)& 0.0021(9)& 11.09 \\ 
Z\_HD116513& 13:21:49.84 & +71:52:18.8 & 200.45765 & 71.87189 & -0.0011(13)& 0.0012(9)&  7.24 \\ 
L\_PG1323$-$085B& 13:25:50.65 & $-$08:50:55.9 & 201.46104 & -8.84863 & 0.0005(10)& 0.0016(5)& 13.21 \\ 
Z\_BD+31.2505& 13:32:32.16 & +30:49:09.5 & 203.13398 & 30.81931 & -0.0005(14)& 0.0017(8)& 10.14 \\ 
Z\_BD+30.2431& 13:38:24.77 & +29:21:56.0 & 204.60319 & 29.36556 & -0.0005(13)& 0.0020(10)& 10.03 \\ 
Z\_HD121859& 13:54:53.28 & +71:54:11.6 & 208.72206 & 71.90322 & -0.0018(10)& 0.0014(8)&  7.99 \\ 
L\_106\_700& 14:40:50.94 & $-$00:23:36.2 & 220.21226 & -0.39356 & -0.0040(9)& 0.0029(7)&  9.33 \\ 
L\_112\_805& 20:42:46.75 & +00:16:08.1 & 310.69480 & 0.26891 & -0.0018(15)& 0.0008(19)& 12.07 \\ 
L\_112\_822& 20:42:54.91 & +00:15:01.9 & 310.72881 & 0.25053 & -0.0033(15)& 0.0010(11)& 11.26 \\ 
L\_113\_339& 21:40:55.68 & +00:27:58.1 & 325.23198 & 0.46613 & -0.0018(13)& -0.0009(10)& 12.10 \\ 
L\_113\_241& 21:41:09.19 & +00:25:48.2 & 325.28828 & 0.43005 & 0.0006(21)& -0.0024(18)& 13.73 \\ 
B\_2202+4216\_129& 22:02:33.40 & +42:14:25.3 & 330.63915 & 42.24037 & -0.0028(19)& 0.0022(13)& 14.10 \\ 
B\_2253+1608\_23& 22:54:11.22 & +16:14:04.8 & 343.54675 & 16.23468 & -0.0003(10)& -0.0022(11)& 13.48 \\ 
L\_115\_420& 23:42:36.48 & +01:05:58.8 & 355.65202 & 1.09968 & -0.0017(10)& -0.0011(10)& 11.05 \\ 
\multicolumn{8}{c}{\textbf{High-polarization (${\rm PD} > 0.5$\%) stars}}\\
B\_0211+1051\_37& 02:10:59.84 & +10:48:05.7 & 32.74935 & 10.80158 & 0.0074(11)& -0.0016(11)& 13.05 \\ 
B\_0211+1051\_18& 02:11:39.60 & +10:53:15.4 & 32.91501 & 10.88762 & 0.0086(13)& 0.0004(15)& 13.50 \\ 
L\_94\_242& 02:57:21.21 & +00:18:38.7 & 44.33837 & 0.31074 & 0.0050(13)& 0.0016(10)& 11.67 \\ 
B\_0259+0747\_30& 02:59:31.71 & +07:44:09.4 & 44.88213 & 7.73594 & 0.0007(31)& -0.0115(15)& 14.68 \\ 
B\_0303+4716\_121& 03:03:04.56 & +47:21:25.9 & 45.76902 & 47.35720 & -0.0089(37)& -0.0048(10)& 14.55 \\ 
H\_GSC02355& 03:37:45.11 & +31:06:58.2 & 54.43797 & 31.11618 & 0.0348(25)& -0.0488(22)& 12.74 \\ 
L\_95\_330& 03:54:30.75 & +00:29:05.4 & 58.62814 & 0.48482 & 0.0094(11)& -0.0018(9)& 11.16 \\ 
L\_95\_275& 03:54:44.24 & +00:27:20.3 & 58.68435 & 0.45564 & -0.0003(11)& 0.0075(10)& 12.65 \\ 
L\_107\_599& 15:39:09.46 & $-$00:14:28.3 & 234.78940 & -0.24131 & -0.0067(11)& 0.0078(10)& 14.48 \\ 
L\_107\_602& 15:39:18.88 & $-$00:15:29.0 & 234.82865 & -0.25832 & -0.0048(37)& 0.0083(15)& 11.82 \\ 
L\_PG1633+099B& 16:35:33.30 & +09:46:20.7 & 248.88877 & 9.77241 & -0.0036(14)& 0.0039(10)& 12.65 \\ 
L\_PG1633+099D& 16:35:40.90 & +09:46:41.5 & 248.91704 & 9.77820 & -0.0036(20)& 0.0038(13)& 13.54 \\ 
B\_1725+1152\_11& 17:24:47.28 & +11:46:50.2 & 261.19704 & 11.78062 & -0.0080(17)& 0.0050(9)& 13.81 \\ 
B\_1725+1152\_24& 17:24:56.26 & +11:55:41.3 & 261.23440 & 11.92815 & -0.0040(20)& 0.0046(8)& 13.86 \\ 
B\_1725+1152\_35& 17:25:15.72 & +11:46:34.6 & 261.31550 & 11.77627 & -0.0044(18)& 0.0052(11)& 12.71 \\ 
B\_1725+1152\_113& 17:25:16.51 & +11:47:06.2 & 261.31881 & 11.78506 & -0.0077(22)& 0.0055(8)& 12.79 \\ 
L\_109\_381& 17:44:12.27 & $-$00:20:32.2 & 266.05112 & -0.34245 & -0.0139(13)& 0.0047(9)& 11.51 \\ 
B\_1751+0939\_376& 17:51:43.60 & +09:43:51.6 & 267.93168 & 9.73101 & -0.0077(30)& 0.0019(16)& 10.13 \\ 
B\_1751+0939\_129& 17:51:45.73 & +09:40:45.3 & 267.94055 & 9.67924 & -0.0083(20)& 0.0032(15)& 14.60 \\ 
L\_110\_229& 18:40:45.66 & +00:01:49.8 & 280.19026 & 0.03049 & 0.0189(13)& 0.0150(18)& 12.53 \\ 
L\_110\_233& 18:40:52.71 & +00:00:50.8 & 280.21962 & 0.01412 & 0.0240(17)& 0.0106(7)& 12.19 \\ 
L\_111\_1965& 19:37:41.56 & +00:26:51.0 & 294.42315 & 0.44749 & -0.0104(9)& 0.0064(10)& 10.70 \\ 
B\_1959+6508\_179& 19:59:12.96 & +65:12:13.7 & 299.80399 & 65.20380 & -0.0039(13)& -0.0101(9)& 13.38 \\ 
B\_1959+6508\_73& 19:59:34.24 & +65:06:19.1 & 299.89267 & 65.10531 & -0.0007(18)& -0.0062(7)& 12.56 \\ 
B\_1959+6508\_104& 20:00:37.89 & +65:13:59.0 & 300.15787 & 65.23306 & -0.0005(18)& -0.0095(11)& 12.73 \\ 
B\_1959+6508\_38& 20:00:58.14 & +65:07:11.9 & 300.24225 & 65.11996 & -0.0033(11)& -0.0155(8)& 12.50 \\ 
B\_2015$-$0137\_102& 20:15:04.91 & $-$01:41:26.1 & 303.77046 & -1.69081 & -0.0105(22)& -0.0030(16)& 12.84 \\ 
B\_2022+7611\_1& 20:20:52.70 & +76:17:42.8 & 305.21694 & 76.29522 & 0.0062(10)& -0.0021(10)& 12.04 \\ 
B\_2042+7508\_28& 20:41:12.89 & +75:12:30.6 & 310.30369 & 75.20850 & 0.0037(10)& 0.0102(11)& 11.78 \\ 
B\_2340+8015\_34& 23:41:38.22 & +80:20:16.9 & 355.40926 & 80.33802 & 0.0030(22)& 0.0119(16)& 13.73 \\ 
\hline
\end{tabular}
\end{table*}

\section{Notes on individual stars} \label{sec:indiv}

Several stars in the sample exhibited unexpected polarization or variability. We list such sources below with a brief description of their peculiar properties.

L\_PG2349+002, L\_92\_249, L\_92\_248, L\_111\_1969 and L\_PG2213$-$006A were selected among Landolt photometric standards, that is, they are expected to have stable total flux density. However, these stars exhibit significant polarization variability.

H\_GSC02355 was selected as a high-polarization star from \cite{Heiles2000}, where it has $\mathrm{PD}=5.058\%$. However, in our measurements, this star is variable and has a higher average polarization of 6.0\%. H\_HD57702 was selected as a low-polarization star from \cite{Heiles2000}, where it has a PD of $0.040\pm0.069\%$. However, in our measurements, this star is $0.33\%$ polarized.

For stars L\_PG1323$-$085D, Z\_HD153752, H\_HD344776 and L\_111\_1969, the EDF of the reduced PD (see Sect.~\ref{sec:var}) is located entirely to the left of the theoretical CDF. It means that these stars are more stable than one would expect from the uncertainties of their PD measurements. Since we used the two-sided KS test, these stars are classified as variable. However, it is likely that the uncertainties in the relative Stokes parameters for these four sources are overestimated, and the stars are in fact stable.

\section{Conclusions}

We obtained \Nobs polarization measurements of \Ntot stars using two different polarimeters. Most  observations were performed in the Cousins $R$ and the SDSS $r^\prime$ bands with the {\em RoboPol} polarimeter along a four-year time interval. After applying a variability analysis to these monitoring data, we have selected \Ngood stars that have $\ge 5$ measurements and do not demonstrate significant variability in linear polarization in the red bands. These stars are listed in Table~\ref{tab:new_stand_data} and they can be used as optical polarimetric standards for calibration of instrumental polarization. For \Nunkn stars, we did not have enough data to conclude whether they are variable or stable, while the remaining \Nvar stars were found to exhibit significant variability in polarization. 

\section*{Data availability}

All data discussed in this paper are available in Harvard Dataverse at \url{https://doi.org/10.7910/DVN/IV9TXX}.

\section*{Acknowledgements}
We thank T. Pursimo and S. Armas P\'{e}rez for assistance with the NOT observations.
D.B., S.K, N.M., V.P., R.S., and K.T. acknowledge support from the European Research Council (ERC) under the European Union Horizon 2020 research and innovation program under the grant agreement No 771282. A.S. acknowledges the Polish National Science Centre grant 2017/25/B/ST9/02805. This work was supported by the NSF grant AST-2109127. The data presented here were obtained in part with ALFOSC, which is provided by the Instituto de Astrofisica de Andalucia (IAA) under a joint agreement with the University of Copenhagen and NOT.

%
%

\bibliographystyle{aa}
\bibliography{bibliography}

\appendix

\section{Sample stars information and monitoring data} \label{ap:a}

{\onecolumn
\begin{landscape}
\begin{longtable}{lcccccccccc}
 \caption[]{\label{tab:sampinfo}Standards candidates sample information: (1) - Star identifier; (2) - Alternative identifier; (3, 4) - Equatorial coordinates; (5) - Gaia EDR3 id; (6) - synthetic SDSS r\textquotesingle\, magnitude from \cite{Montegriffo2022}; (7) - Gaia EDR3 G-band magnitude; (8) - Spectral type; (9) - Sample; (10) - Number of observations in the SDSS-r band; (11) - Number of observations in the Cousins-R band.}\\
 \hline
  Star ID &
  Alt. ID &
  RA (J2000) &
 DEC (J2000) &
  Gaia ID &
  r\textquotesingle &
  G &
  Spec. typ. &
  Sample &
  $N^{r}_{\rm obs}$ &
  $N^{R}_{\rm obs}$ \\
  &
  &
  h:m:s  &
  \degr:\arcmin:\arcsec &
  &
  mag &
  mag &
  &
  &
  &
  \\ \hline
B\_0017+8135\_82 & - & 00:15:57.61 & +81:36:56.6 & 566773475044000896 & 13.40 & 13.42 & - &  B  &  5  &  10 \\
B\_0017+8135\_79 & - & 00:16:33.82 & +81:36:33.9 & 566773406324526592 & 14.05 & 14.07 & - &  B  &  0  &  1 \\
L\_92\_245 & SA 92-245 & 00:54:16.16 & +00:39:54.10 & 2537310861359601024 & - & 13.53 & - &  L  &  2  &  11 \\
L\_92\_248 & SA 92-248 & 00:54:30.78 & +00:40:17.0 & 2537313644497864064 & 14.91 & 14.97 & - &  L  &  2  &  8 \\
L\_92\_249 & SA 92-249 & 00:54:33.59 & +00:41:05.4 & 2537313678857608448 & 14.11 & 14.15 & - &  L  &  1  &  9 \\
L\_92\_250 & SA 92-250 & 00:54:37.16 & +00:38:57.6 & 2537313369619964416 & 12.93 & 12.99 & - &  L  &  2  &  11 \\
L\_93\_317 & SA 93-317 & 01:54:37.73 & +00:43:00.6 & 2510809435673401472 & 11.42 & 11.42 & F5 &  L  &  2  &  10 \\
L\_93\_333 & SA 93-333 & 01:55:05.21 & +00:45:42.8 & 2510902962881171968 &  & 12.42 & G5 &  L  &  4  &  12 \\
L\_93\_424 & SA 93-424 & 01:55:26.36 & +00:56:42.6 & 2510924536502154112 & 11.29 & 11.34 & G8III &  L  &  2  &  10 \\
B\_0211+1051\_37 & Pul -3 170384 & 02:10:59.84 & +10:48:05.7 & 72423941863975552 & - & 13.05 & - &  B  &  1  &  8 \\
B\_0211+1051\_18 & - & 02:11:39.60 & +10:53:15.4 & 72425797289881344 & 13.51 & 13.50 & - &  B  &  1  &  7 \\
L\_94\_242 & SA 94-242 & 02:57:21.21 & +00:18:38.7 & 2498069944198626560 & 11.69 & 11.67 & A2 &  L  &  3  &  9 \\
L\_94\_251 & SA 94-251 & 02:57:46.98 & +00:16:02.7 & 2498066061548190592 & - & 10.83 & K1III &  L  &  3  &  8 \\
B\_0259+0747\_30 & - & 02:59:31.71 & +07:44:09.4 & 8525892335265920 & 14.67 & 14.68 & - &  B  &  1  &  8 \\
B\_0259+0747\_22 & - & 02:59:40.87 & +07:41:59.9 & 8477204585999872 & 14.66 & 14.65 & - &  B  &  2  &  6 \\
B\_0303+4716\_121 & - & 03:03:04.56 & +47:21:25.9 & 434503531895117440 & 14.52 & 14.55 & - &  B  &  0  &  5 \\
H\_GSC02355 & GSC 02355-00137 & 03:37:45.11 & +31:06:58.2 & 121211368733067392 & 13.02 & 12.74 & - &  H  &  3  &  7 \\
L\_95\_330 & SA 95-330 & 03:54:30.75 & +00:29:05.4 & 3257868006962104320 & 11.45 & 11.16 & - &  L  &  3  &  9 \\
L\_95\_275 & SA 95-275 & 03:54:44.24 & +00:27:20.3 & 3257866941810214400 & 12.81 & 12.65 & - &  L  &  1  &  5 \\
L\_95\_276 & SA 95-276 & 03:54:45.88 & +00:25:54.1 & 3257866838731000192 & 13.63 & 13.64 & - &  L  &  0  &  4 \\
H\_HD283807 & BD+25.728 & 04:44:42.6 & +25:56:09.6 & 148254441333464576 &  9.95 &  9.96 & G0 &  H  &  1  &  6 \\
L\_96\_235 & SA 96-235 & 04:53:18.87 & $-$00:05:01.5 & 3228325297755515008 & 10.81 & 10.85 & - &  L  &  0  &  3 \\
L\_97\_345 & SA 97-345 & 05:57:33.18 & +00:21:16.6 & 3218646468694578048 & 11.01 & 10.91 & G8III &  L  &  1  &  2 \\
L\_97\_351 & SA 97-351 & 05:57:37.28 & +00:13:44.0 & 3218641447876998784 &  9.80 &  9.74 & A0 &  L  &  0  &  1 \\
H\_HD255017 & HD 255017 & 06:19:08.20 & +14:38:31.0 & 3344737878054909312 &  9.36 &  9.35 & A5Ib &  H  &  0  &  2 \\
L\_98\_653 & HD 50188 & 06:52:04.95 & $-$00:18:18.7 & 3113277417552594560 & - &  9.52 & B8/9IV &  L  &  0  &  1 \\
L\_98\_685 & SA 98-685 & 06:52:18.47 & $-$00:20:19.5 & 3113276249321476992 & - & 11.84 & F8 &  L  &  0  &  1 \\
H\_HD57702 & HD 57702 & 07:23:24.60 & +25:30:58.5 & 870304410193552128 & - &  8.87 & B9 &  H  &  0  &  1 \\
L\_RU\_152D & RU 152 D & 07:30:06.7 & $-$02:04:37.5 & 3061899472568519808 & 10.82 & 10.85 & - &  L  &  0  &  1 \\
L\_PG0918+029D & PG 0918+029 D & 09:21:21.93 & +02:47:28.3 & 3845583974466346752 & 11.93 & 11.96 & - &  L  &  1  &  0 \\
Z\_HD81418 & BD+19.2212 & 09:26:02.70 & +18:54:02.2 & 633976652229184640 &  8.84 &  8.82 & A0 &  Z  &  1  &  1 \\
Z\_HD85471 & BD+81.319 & 09:59:04.94 & +80:22:47.6 & 1132568002485076352 &  7.78 &  7.83 & G5 &  Z  &  4  &  2 \\
Z\_HD86321 & BD+84.225 & 10:08:34.32 & +83:55:06.2 & 1147461505958427264 &  5.78 &  5.62 & K0 &  Z  &  0  &  1 \\
Z\_HD87582 & BD+72.479 & 10:09:10.4 & +71:41:15.5 & 1077571702173596928 &  7.06 &  7.07 & K0 &  Z  &  3  &  2 \\
L\_PG1047+003 & V* UY Sex & 10:50:02.83 & $-$00:00:36.1 & 3806303066866089216 & - & 13.42 & sdO9VIIHe6 &  L  &  1  &  0 \\
L\_PG1047+003B & PG 1047+003 B & 10:50:07.92 & $-$00:02:04.6 & 3806302787692664832 & 14.55 & 14.58 & - &  L  &  1  &  0 \\
L\_PG1047+003C & PG 1047+003 C & 10:50:13.68 & $-$00:00:32.7 & 3806303200009524480 & - & 12.31 & - &  L  &  1  &  0 \\
L\_G163\_50 & WD 1105-048 & 11:07:59.95 & $-$05:09:26.10 & 3788194488314248832 & 13.24 & 13.09 & DA3 &  L  &  2  &  0 \\
L\_G163\_51 & L 970-27 & 11:08:06.54 & $-$05:13:47.9 & 3788190605663811840 & - & 11.47 & M3V &  L  &  2  &  1 \\
Z\_HD96589 & BD+81.362 & 11:11:00.40 & +80:56:55.9 & 1133527773057098752 &  7.51 &  7.54 & G0 &  Z  &  4  &  1 \\
Z\_HD97853 & BD+76.421 & 11:17:09.82 & +75:21:11.1 & 1079936717325036160 &  7.14 &  7.16 & K0 &  Z  &  3  &  3 \\
Z\_BD+35.2256 & BD+35.2256 & 11:30:39.95 & +35:09:50.3 & 759519993695325696 & 10.41 & 10.40 & F2 &  Z  &  3  &  6 \\
Z\_BD+29.2198 & BD+29.2198 & 11:42:56.23 & +28:32:39.2 & 4018335981542938112 & - & 10.19 & A2 &  Z  &  4  &  6 \\
Z\_BD+31.2314 & BD+31.2314 & 11:56:43.33 & +30:48:45.5 & 4026504219066025600 & - & 10.83 & kA3hA7:m &  Z  &  5  &  3 \\
Z\_BD+25.2439 & BD+25.2439 & 11:56:55.1 & +24:12:09.6 & 4004411731929387904 & - & 10.23 & F5 &  Z  &  3  &  4 \\
Z\_BD+18.2549 & BD+18.2549 & 11:59:04.54 & +17:21:07.4 & 3926601527414968704 & 10.33 & 10.34 & G0 &  Z  &  3  &  4 \\
Z\_BD+32.2217 & BD+32.2217 & 12:04:58.34 & +31:21:06.1 & 4026404644544171520 & - & 10.46 & kA4hA7VmF2 &  Z  &  4  &  4 \\
Z\_BD+22.2446 & BD+22.2446 & 12:11:40.13 & +21:19:08.7 & 3951825148789721472 & - & 10.75 & - &  Z  &  3  &  4 \\
Z\_BD+40.2546 & BD+40.2546 & 12:32:44.89 & +39:55:20.2 & 1533606794177007104 & 10.85 & 10.85 & F6 &  Z  &  3  &  8 \\
Z\_BD+30.2290 & BD+30.2290 & 12:32:55.79 & +29:33:52.9 & 4011002135906720768 & 11.41 & 11.39 & A2 &  Z  &  4  &  4 \\
L\_104\_334 & SA 104-334 & 12:42:20.43 & $-$00:40:28.7 & 3683802417670914688 & 13.36 & 13.37 & - &  L  &  5  &  3 \\
Z\_BD+39.2611 & * 15 CVn & 13:09:42.3 & +38:32:01.10 & 1523140405554059008 &  6.44 &  6.29 & B7III &  Z  &  0  &  1 \\
Z\_BD+16.2491 & BD+16.2491 & 13:19:08.9 & +15:23:53.6 & 3744299780814144896 & 11.06 & 11.09 & - &  Z  &  5  &  3 \\
Z\_HD116513 & BD+72.613 & 13:21:49.84 & +71:52:18.8 & 1687792271812816384 &  7.20 &  7.24 & K2 &  Z  &  5  &  8 \\
L\_PG1323$-$085B & PG 1323-086 B & 13:25:50.65 & $-$08:50:55.9 & 3624196212997860992 & 13.18 & 13.21 & - &  L  &  3  &  3 \\
L\_PG1323$-$085C & PG 1323-086 C & 13:25:50.22 & $-$08:48:38.0 & 3624199374093791616 & 13.79 & 13.83 & - &  L  &  2  &  0 \\
L\_PG1323$-$085D & PG 1323-086 D & 13:26:05.25 & $-$08:50:36.8 & 3624196109918646016 & 11.91 & 11.93 & - &  L  &  3  &  3 \\
Z\_BD+31.2505 & BD+31.2505 & 13:32:32.16 & +30:49:09.5 & 1468318824514203264 & 10.20 & 10.14 & - &  Z  &  5  &  6 \\
Z\_BD+35.2465 & BD+35.2465 & 13:35:23.76 & +35:13:39.1 & 1471664741475242112 & 10.86 & 10.82 & - &  Z  &  0  &  4 \\
Z\_BD+30.2431 & Feige 86 & 13:38:24.77 & +29:21:56.0 & 1455929733649174528 & 10.20 & 10.03 & B5Vp &  Z  &  3  &  4 \\
Z\_HD120010 & BD+11.2600 & 13:46:37.63 & +10:50:26.2 & 3726354449674413696 &  8.26 &  8.30 & K2 &  Z  &  0  &  2 \\
Z\_HD121859 & BD+72.631 & 13:54:53.28 & +71:54:11.6 & 1687308933373463808 &  7.96 &  7.99 & K2 &  Z  &  4  &  9 \\
L\_106\_700 & BD+00.3222 & 14:40:50.94 & $-$00:23:36.2 & 3651584807127464960 &  9.33 &  9.33 & K0III &  L  &  2  &  9 \\
Z\_HD138733 & BD+71.733 & 15:28:51.17 & +70:40:52.5 & 1695728233208611328 &  8.31 &  8.35 & K0 &  Z  &  5  &  10 \\
L\_107\_599 & SA 107-599 & 15:39:09.46 & $-$00:14:28.3 & 4416475498513533952 & 14.45 & 14.48 & - &  L  &  3  &  5 \\
L\_107\_602 & SA 107-602 & 15:39:18.88 & $-$00:15:29.0 & 4416474639520070144 & - & 11.82 & F8 &  L  &  3  &  7 \\
L\_PG1633+099B & PG 1633+099 B & 16:35:33.30 & +09:46:20.7 & 4446607820836297344 & 12.60 & 12.65 & - &  L  &  3  &  10 \\
L\_PG1633+099D & PG 1633+099 D & 16:35:40.9 & +09:46:41.5 & 4446607752116815232 & 13.53 & 13.54 & - &  L  &  5  &  9 \\
Z\_HD153752 & BD+74.690 & 16:53:38.18 & +74:17:30.1 & 1656269376524601472 &  7.66 &  7.64 & F0 &  Z  &  5  &  11 \\
B\_1725+1152\_11 & - & 17:24:47.28 & +11:46:50.2 & 4540367777239708416 & 13.80 & 13.81 & - &  B  &  2  &  7 \\
B\_1725+1152\_24 & - & 17:24:56.26 & +11:55:41.3 & 4540374516046715648 & 13.82 & 13.86 & - &  B  &  3  &  8 \\
B\_1725+1152\_35 & - & 17:25:15.72 & +11:46:34.6 & 4492329534405398016 & 12.67 & 12.71 & - &  B  &  5  &  10 \\
B\_1725+1152\_113 & - & 17:25:16.51 & +11:47:06.2 & 4492331011874148480 & 12.78 & 12.79 & - &  B  &  3  &  9 \\
L\_109\_71 & SA 109-71 & 17:44:06.79 & $-$00:24:58.9 & 4371937237413220096 & 11.46 & 11.43 & A0 &  L  &  0  &  3 \\
L\_109\_381 & SA 109-381 & 17:44:12.27 & $-$00:20:32.2 & 4371985993881977856 & 11.49 & 11.51 & F2 &  L  &  5  &  10 \\
B\_1751+0939\_376 & - & 17:51:43.60 & +09:43:51.6 & 4488790412635906176 & 11.84 & 10.13 & - &  B  &  4  &  10 \\
B\_1751+0939\_129 & - & 17:51:45.73 & +09:40:45.3 & 4488787973094456448 & 14.58 & 14.60 & - &  B  &  3  &  10 \\
B\_1751+0939\_204 & - & 17:51:48.7 & +09:41:44.8 & 4488789484922948352 & 14.93 & 14.24 & - &  B  &  2  &  9 \\
L\_110\_229 & SA 110-229 & 18:40:45.66 & +00:01:49.8 & 4272464794108661760 & 12.89 & 12.53 & - &  L  &  6  &  13 \\
L\_110\_233 & SA 110-233 & 18:40:52.71 & +00:00:50.8 & 4272452974358663680 & 12.31 & 12.19 & - &  L  &  4  &  10 \\
L\_111\_1965 & SA 111-1965 & 19:37:41.56 & +00:26:50.10 & 4239473470987190016 & 10.83 & 10.70 & K0 &  L  &  6  &  11 \\
L\_111\_1969 & BD+00.4260 & 19:37:43.28 & +00:25:48.5 & 4239285763738341632 &  9.69 &  9.28 & M5III &  L  &  6  &  13 \\
H\_HD344776 & BD+23.3745 & 19:42:49.62 & +23:27:47.9 & 2020105333313222144 &  8.54 &  8.53 & B0.5Ib &  H  &  4  &  12 \\
B\_1959+6508\_179 & - & 19:59:12.96 & +65:12:13.7 & 2247538007835538176 & 13.35 & 13.38 & - &  B  &  3  &  7 \\
B\_1959+6508\_73 & - & 19:59:34.24 & +65:06:19.1 & 2247536152409628032 & 12.68 & 12.56 & - &  B  &  5  &  9 \\
B\_1959+6508\_108 & - & 20:00:21.39 & +65:13:59.5 & 2247560822701907200 & 14.36 & 14.39 & - &  B  &  3  &  9 \\
B\_1959+6508\_104 & - & 20:00:37.89 & +65:13:59.0 & 2247550583499874304 & 12.76 & 12.73 & - &  B  &  4  &  10 \\
B\_1959+6508\_38 & - & 20:00:58.14 & +65:07:11.9 & 2247547525483073792 & 12.51 & 12.50 & - &  B  &  4  &  9 \\
B\_2015$-$0137\_102 & - & 20:15:04.91 & $-$01:41:26.1 & 4223920844642460032 & 12.83 & 12.84 & - &  B  &  4  &  14 \\
B\_2022+7611\_1 & - & 20:20:52.7 & +76:17:42.8 & 2290191431129629952 & 12.05 & 12.04 & - &  B  &  5  &  9 \\
B\_2042+7508\_28 & - & 20:41:12.89 & +75:12:30.6 & 2277833706411214464 & 11.83 & 11.78 & - &  B  &  5  &  8 \\
L\_112\_805 & SA 112-805 & 20:42:46.75 & +00:16:08.1 & 4228196566186425216 & 12.15 & 12.07 & A0 &  L  &  5  &  13 \\
L\_112\_822 & SA 112-822 & 20:42:54.91 & +00:15:01.9 & 4228184578933327872 & 11.22 & 11.26 & G8III &  L  &  3  &  13 \\
B\_2042+7508\_17 & - & 20:42:55.65 & +75:09:42.5 & 2277830961927902336 & 14.89 & 14.89 & - &  B  &  4  &  7 \\
L\_113\_339 & SA 113-339 & 21:40:55.68 & +00:27:58.1 & 2687421308383972992 & 12.09 & 12.10 & F8 &  L  &  0  &  8 \\
L\_113\_241 & SA 113-241 & 21:41:09.19 & +00:25:48.2 & 2687420170216933888 & 13.77 & 13.73 & - &  L  &  3  &  10 \\
B\_2202+4216\_25 & - & 22:02:17.6 & +42:21:06.0 & 1960081928389196800 & 12.82 & 12.69 & - &  B  &  4  &  10 \\
B\_2202+4216\_129 & - & 22:02:33.40 & +42:14:25.3 & 1960065779311967488 & 14.08 & 14.10 & - &  B  &  5  &  13 \\
B\_2202+4216\_239 & - & 22:02:54.47 & +42:10:24.9 & 1960061823644122112 & 15.05 & 15.03 & - &  B  &  4  &  9 \\
H\_BD+62.2078 & BD+62.2078 & 22:25:33.57 & +63:25:02.6 & 2205414029452444800 &  9.34 &  9.25 & O7V((f))z &  H  &  5  &  12 \\
L\_PG2213$-$006A & PG 2213-006 A & 22:16:23.21 & $-$00:21:27.10 & 2678666889428745088 & 13.98 & 13.99 & - &  L  &  4  &  9 \\
B\_2253+1608\_23 & - & 22:54:11.22 & +16:14:04.8 & 2828722296448918144 & - & 13.48 & - &  B  &  3  &  8 \\
B\_2340+8015\_34 & - & 23:41:38.22 & +80:20:16.9 & 2283216988357240064 & 13.74 & 13.73 & - &  B  &  4  &  7 \\
B\_2340+8015\_109 & - & 23:42:21.63 & +80:11:05.0 & 2283199499250447488 & 13.65 & 13.68 & - &  B  &  0  &  4 \\
B\_2340+8015\_99 & - & 23:42:26.8 & +80:12:06.3 & 2283211353360181376 & 13.34 & 13.36 & - &  B  &  5  &  10 \\
L\_115\_420 & SA 115-420 & 23:42:36.48 & +01:05:58.8 & 2646050087444966272 & - & 11.05 & F5 &  L  &  2  &  13 \\
L\_PG2349+002 & - & 23:51:53.23 & +00:28:17.7 & 2642341469085008000 & - & 13.26 & sdB2VIIHe0 &  L  &  3  &  3 \\
\hline
\end{longtable}
\end{landscape}
}\twocolumn

\begin{table*}
\caption{Polarimetric monitoring data: (1) - Star identifier; (2) - Julian Date; (3) - band; (4,5) - $Q/I$ and $U/I$ relative Stokes parameters corrected for the instrumental polarization and EVPA zero-point; (6,7) - Fractional polarization and the polarization vector position angle corrected for the instrumental polarization and EVPA zero-point. This table gives only first 3 rows of the entire dataset, which can be accessed at \url{https://doi.org/10.7910/DVN/IV9TXX}.}
\label{tab:mon_data}      
\centering          
\begin{tabular}{cccccS[table-format=2.1(1)]S[table-format=2.0(1)]}     
\hline\hline       
Star ID & JD & band & $q=Q/I$ & $u=U/I$ & \si{PD} & \si{EVPA}\\
        &    &      &         &         & \si{\%} & \si{\degr}\\
\hline                    
B\_0017+8135\_82 & 2457962.42898 & R & $-0.0066\pm0.0014$  &  $0.0012\pm0.0012$  & 0.7(1) & 85(6)\\
B\_0017+8135\_82 & 2457987.52889 & R & $-0.0064\pm0.0012$  &  $0.0017\pm0.0011$  & 0.7(1) & 83(5)\\
B\_0017+8135\_82 & 2458015.46020 & R & $-0.0080\pm0.0012$  &  $0.0018\pm0.0010$  & 0.8(1) & 84(4)\\  
\multicolumn{7}{c}{\dots} \\
\hline
\end{tabular}
\end{table*}

\begin{table*}[]
\normalsize
\centering
\caption{Multiband single epoch measurements with ALFOSC/NOT: (1) - Star identifier; (2) - Alternative identifier; (3) - observing bandpass name; (4,5) - relative Stokes parameter $\frac{Q}{I}$ and its uncertainty; (6,7) - relative Stokes parameter $\frac{U}{I}$ and its uncertainty; (8, 9) - fractional polarization and its uncertainty; (10, 11) - the electric vector position angle.}
\begin{tabular}{lcccccccccc}
\hline
Star ID & Alt. ID & band & q & $\sigma_q$ & u & $\sigma_u$ & PD & $\sigma_{\rm PD}$ & EVPA & $\sigma_{\rm EVPA}$ \\
        &         &      &               &                      &               &                      & \% &      \%     & \dg  &    \dg        \\
\hline
B\_1959+6508\_38 &  -  & B & $-$0.0009 & 0.0016 & $-$0.0181 & 0.0017 & 1.8 & 0.2 & $-$47 & 3\\
                 &     & V & $-$0.0016 & 0.0009 & $-$0.0186 & 0.0011 & 1.9 & 0.1 & $-$48 & 2\\
RA=20h00m58.14s  &     & R & $-$0.0061 & 0.0009 & $-$0.0172 & 0.0012 & 1.8 & 0.1 & $-$55 & 2\\
DEC=+65d07m11.9s &     & g & $-$0.0016 & 0.0013 & $-$0.0196 & 0.0014 & 2.0 & 0.1 & $-$47 & 2\\
                 &     & r & $-$0.0039 & 0.0009 & $-$0.0175 & 0.0011 & 1.8 & 0.1 & $-$51 & 2\\
                 &     & i & $-$0.0048 & 0.0008 & $-$0.0160 & 0.0011 & 1.7 & 0.1 & $-$53 & 2\\
\hline
B\_2015$-$0137\_102 &  -  & B & $-$0.0108 & 0.0009 & $-$0.0067 & 0.0011 & 1.27 & 0.10 & $-$74 & 2 \\
                    &     & V & $-$0.0119 & 0.0010 & $-$0.0061 & 0.0011 & 1.34 & 0.10 & $-$77 & 2 \\
RA=20h15m04.91s     &     & R & $-$0.0114 & 0.0009 & $-$0.0049 & 0.0011 & 1.23 & 0.09 & $-$78 & 2 \\
DEC=$-$01d41m26.1s  &     & g & $-$0.0096 & 0.0007 & $-$0.0087 & 0.0009 & 1.29 & 0.08 & $-$69 & 2 \\
                    &     & r & $-$0.0116 & 0.0008 & $-$0.0062 & 0.0011 & 1.31 & 0.09 & $-$76 & 2 \\
                    &     & i & $-$0.0105 & 0.0009 & $-$0.0047 & 0.0012 & 1.15 & 0.10 & $-$78 & 2 \\
\hline
B\_2042+7508\_17    &  -  & B & $-$0.0024 & 0.0016 &  0.0097 & 0.0018 & 1.0 & 0.2 & 52 & 5  \\
                    &     & V & $-$0.0017 & 0.0012 &  0.0102 & 0.0013 & 1.0 & 0.1 & 50 & 4  \\
RA=20h42m55.65s     &     & R & $-$0.0024 & 0.0011 &  0.0101 & 0.0013 & 1.0 & 0.1 & 51 & 4  \\
DEC=+75d09m42.5s    &     & g &    0.0001 & 0.0042 &  0.0070 & 0.0038 & 0.7 & 0.4 & 45 & 19 \\
                    &     & r & $-$0.0013 & 0.0013 &  0.0093 & 0.0013 & 0.9 & 0.1 & 49 & 4  \\
                    &     & i & $-$0.0011 & 0.0009 &  0.0078 & 0.0012 & 0.8 & 0.1 & 49 & 4  \\
\hline
H\_HD344776 & BD+23.3745  & U & 0.0287 & 0.0014 & 0.0426 & 0.0015 & 5.14 & 0.15 & 28.0 & 0.8 \\
                    &     & B & 0.0292 & 0.0006 & 0.0531 & 0.0008 & 6.06 & 0.07 & 30.6 & 0.3 \\
RA=19h42m49.62s     &     & V & 0.0374 & 0.0006 & 0.0527 & 0.0008 & 6.46 & 0.07 & 27.3 & 0.3 \\
DEC=+23d27m47.9s    &     & R & 0.0424 & 0.0006 & 0.0455 & 0.0009 & 6.22 & 0.07 & 23.5 & 0.3 \\
                    &     & g & 0.0321 & 0.0005 & 0.0547 & 0.0008 & 6.34 & 0.07 & 29.8 & 0.3 \\
\hline
L\_93\_333   & SA 93$-$333& B & $-$0.0001 & 0.0011 & $-$0.0046 & 0.0013 & 0.46 & 0.13 & $-$46 & 8 \\
                    &     & V & $-$0.0001 & 0.0010 & $-$0.0026 & 0.0011 & 0.26 & 0.11 & $-$45 & 14 \\
RA=01h55m05.21s     &     & R & $-$0.0002 & 0.0007 & $-$0.0036 & 0.0010 & 0.36 & 0.10 & $-$47 & 8 \\
DEC=+00d45m42.8s    &     & g &    0.0012 & 0.0007 & $-$0.0027 & 0.0009 & 0.30 & 0.09 & $-$33 & 9 \\
                    &     & r &    0.0008 & 0.0007 & $-$0.0030 & 0.0010 & 0.31 & 0.09 & $-$38 & 10 \\
                    &     & i & $-$0.0014 & 0.0007 & $-$0.0040 & 0.0010 & 0.43 & 0.10 & $-$55 & 7 \\
\hline
L\_94\_242   & SA 94$-$242& U &  0.0044 & 0.0031 &    0.0024 & 0.0031 & 0.50 & 0.31 & 14    & 23 \\
                    &     & B &  0.0036 & 0.0007 &    0.0017 & 0.0010 & 0.40 & 0.08 & 13    & 6 \\
RA=02h57m21.21s     &     & V &  0.0043 & 0.0007 &    0.0006 & 0.0009 & 0.43 & 0.07 &  4    & 5 \\
DEC=+00d18m38.7s    &     & R &  0.0044 & 0.0007 & $-$0.0009 & 0.0010 & 0.44 & 0.07 & $-$6  & 5 \\
                    &     & r &  0.0047 & 0.0006 &    0.0004 & 0.0009 & 0.47 & 0.06 &  3    & 4  \\
                    &     & i &  0.0040 & 0.0007 & $-$0.0021 & 0.0010 & 0.45 & 0.08 & $-$14 & 5 \\
\hline
L\_110\_229 & SA 110$-$229& B &  0.0133 & 0.0010 &  0.0172 & 0.0012 & 2.17 & 0.11 & 26.1 & 1.4 \\
                    &     & V &  0.0188 & 0.0011 &  0.0160 & 0.0012 & 2.47 & 0.11 & 20.2 & 1.3 \\
RA=18h40m45.66s     &     & R &  0.0192 & 0.0010 &  0.0156 & 0.0012 & 2.47 & 0.11 & 19.5 & 1.2 \\
DEC=+00d01m49.8s    &     & g &  0.0164 & 0.0007 &  0.0169 & 0.0009 & 2.35 & 0.08 & 22.9 & 1.0 \\
                    &     & r &  0.0183 & 0.0010 &  0.0153 & 0.0012 & 2.39 & 0.11 & 20.0 & 1.3 \\
                    &     & i &  0.0174 & 0.0008 &  0.0120 & 0.0011 & 2.11 & 0.09 & 17.3 & 1.2 \\
\hline
L\_111\_1965 &SA 111$-$1965& B & $-$0.0105 & 0.0010 & 0.0008 & 0.0012 & 1.06 & 0.10 & 87.9 & 3 \\
                     &     & V & $-$0.0113 & 0.0008 & 0.0040 & 0.0010 & 1.20 & 0.08 & 80.2 & 2 \\
RA=19h37m41.56s      &     & R & $-$0.0119 & 0.0007 & 0.0046 & 0.0010 & 1.27 & 0.08 & 79.5 & 2 \\
DEC=+00d26m50.10s    &     & g & $-$0.0106 & 0.0007 & 0.0026 & 0.0010 & 1.09 & 0.08 & 83.2 & 2 \\
                     &     & r & $-$0.0098 & 0.0007 & 0.0041 & 0.0010 & 1.06 & 0.08 & 78.6 & 2 \\
                     &     & i & $-$0.0087 & 0.0006 & 0.0038 & 0.0009 & 0.95 & 0.07 & 78.1 & 2 \\
\hline
L\_112\_805  &SA 112$-$805& B & $-$0.0016 & 0.0009 & $-$0.0004 & 0.0011 & 0.16 & 0.09 & $-$83 & 20 \\
                    &     & V & $-$0.0026 & 0.0010 & $-$0.0005 & 0.0012 & 0.27 & 0.10 & $-$85 & 12 \\
RA=20h42m46.75s     &     & R & $-$0.0026 & 0.0014 & $-$0.0005 & 0.0015 & 0.27 & 0.14 & $-$84 & 18 \\
DEC=+00d16m08.1s    &     & g & $-$0.0019 & 0.0007 & $-$0.0003 & 0.0010 & 0.19 & 0.07 & $-$85 & 12 \\
                    &     & r & $-$0.0035 & 0.0013 &    0.0010 & 0.0014 & 0.36 & 0.13 &    82 & 11 \\
                    &     & i & $-$0.0027 & 0.0014 & $-$0.0006 & 0.0016 & 0.28 & 0.14 & $-$84 & 18 \\
\end{tabular}
\label{tab:NOT}
\end{table*}

{\onecolumn
\small
\LTcapwidth=\textwidth
\begin{longtable}{lccccccccccc} 
\caption[]{Average polarization parameters of the sample stars: (1) - Star identifier; (2) - Number of measurements; (3, 4) - Mean of the fractional polarization and its standard error; (5, 6) - Mean of the EVPA and its uncertainty ; (7, 8) - $Q/I$ relative Stokes parameter and its standard error; (9, 10) - $U/I$ relative Stokes parameter and its standard error; (11) - Variability flag for $2 \sigma$ S.L. (``NV'' stands for non-variable, ``V'' -- variable, ``UN'' -- uncertain). This table is available in a machine-readable format at \url{https://doi.org/10.7910/DVN/IV9TXX}.}\\
\hline
Star ID & N meas. & PD & $\sigma$PD & EVPA & $\sigma$ EVPA & q & $\sigma_q$ & u & $\sigma_u$  & Var. flag & p-value\\
        &         & \% &     \%     &  \dg &  \dg          &   &            &   &             &           & \\
\hline                    
B\_0017+8135\_82& 15 & 0.671 & 0.031 & 79.7 &  1.3 & -0.00628 & 0.00174 & 0.00236 & 0.00220 & \textcolor{red}{V} & 0.0032 \\ 
B\_0017+8135\_79& 1 & 0.461 & 0.119 & 84.8 &  7.8 & -0.00454 & 0.00120 & 0.00083 & 0.00098 & \textcolor{orange}{UN} & - \\ 
L\_92\_245& 13 & 0.173 & 0.030 & -16.9 &  5.1 & 0.00144 & 0.00105 & -0.00096 & 0.00085 & \ NV & 0.5008 \\ 
L\_92\_248& 10 & 0.412 & 0.047 & -12.8 &  3.3 & 0.00372 & 0.00377 & -0.00177 & 0.00175 & \textcolor{red}{V} & 0.0174 \\ 
L\_92\_249& 10 & 0.403 & 0.040 & -17.3 &  2.9 & 0.00331 & 0.00368 & -0.00229 & 0.00173 & \textcolor{red}{V} & 0.0005 \\ 
L\_92\_250& 13 & 0.055 & 0.030 & -46.0 & 19.1 & -0.00002 & 0.00107 & -0.00055 & 0.00092 & \ NV & 0.2647 \\ 
L\_93\_317& 12 & 0.230 & 0.031 & -31.0 &  3.9 & 0.00108 & 0.00095 & -0.00203 & 0.00092 & \ NV & 0.6470 \\ 
L\_93\_333& 16 & 0.167 & 0.026 & -26.4 &  4.6 & 0.00101 & 0.00262 & -0.00133 & 0.00102 & \ NV & 0.3953 \\ 
L\_93\_424& 12 & 0.253 & 0.030 & -35.7 &  3.5 & 0.00081 & 0.00096 & -0.00240 & 0.00107 & \ NV & 0.5936 \\ 
B\_0211+1051\_37& 9 & 0.761 & 0.040 & -6.1 &  1.5 & 0.00743 & 0.00118 & -0.00162 & 0.00116 & \ NV & 0.9640 \\ 
B\_0211+1051\_18& 8 & 0.857 & 0.044 &  1.4 &  1.5 & 0.00856 & 0.00135 & 0.00042 & 0.00146 & \ NV & 0.7017 \\ 
L\_94\_242& 12 & 0.530 & 0.033 &  8.9 &  1.8 & 0.00505 & 0.00129 & 0.00161 & 0.00095 & \ NV & 0.9261 \\ 
L\_94\_251& 11 & 0.425 & 0.033 & 12.0 &  2.2 & 0.00388 & 0.00096 & 0.00173 & 0.00099 & \ NV & 0.8108 \\ 
B\_0259+0747\_30& 9 & 1.153 & 0.046 & -43.4 &  1.1 & 0.00065 & 0.00311 & -0.01151 & 0.00148 & \ NV & 0.1567 \\ 
B\_0259+0747\_22& 8 & 0.826 & 0.051 & -68.0 &  1.8 & -0.00594 & 0.00485 & -0.00575 & 0.00154 & \textcolor{red}{V} & 0.0017 \\ 
B\_0303+4716\_121& 5 & 1.013 & 0.054 & -75.8 &  1.5 & -0.00892 & 0.00370 & -0.00481 & 0.00101 & \ NV & 0.3281 \\ 
H\_GSC02355& 10 & 5.998 & 0.042 & -27.2 &  0.2 & 0.03484 & 0.00247 & -0.04883 & 0.00217 & \ NV & 0.1163 \\ 
L\_95\_330& 12 & 0.962 & 0.034 & -5.5 &  1.0 & 0.00944 & 0.00111 & -0.00184 & 0.00089 & \ NV & 0.7842 \\ 
L\_95\_275& 6 & 0.752 & 0.044 & 46.2 &  1.7 & -0.00032 & 0.00121 & 0.00751 & 0.00104 & \ NV & 0.8362 \\ 
L\_95\_276& 4 & 0.204 & 0.051 & 44.8 &  7.5 & 0.00001 & 0.00234 & 0.00204 & 0.00416 & \textcolor{orange}{UN} & - \\ 
H\_HD283807& 7 & 0.066 & 0.040 & 59.5 & 22.6 & -0.00032 & 0.00116 & 0.00058 & 0.00084 & \ NV & 0.2401 \\ 
L\_96\_235& 3 & 0.150 & 0.060 & -19.9 & 12.7 & 0.00115 & 0.00009 & -0.00096 & 0.00066 & \textcolor{orange}{UN} & - \\ 
L\_97\_345& 3 & 1.411 & 0.064 & -0.5 &  1.3 & 0.01411 & 0.00106 & -0.00024 & 0.00020 & \textcolor{orange}{UN} & - \\ 
L\_97\_351& 1 & 1.013 & 0.094 & -35.1 &  2.7 & 0.00343 & 0.00114 & -0.00953 & 0.00091 & \textcolor{orange}{UN} & - \\ 
H\_HD255017& 2 & 0.237 & 0.071 & 29.8 &  9.2 & 0.00120 & 0.00031 & 0.00205 & 0.00073 & \textcolor{orange}{UN} & - \\ 
L\_98\_653& 1 & 0.439 & 0.112 & 75.4 &  7.6 & -0.00384 & 0.00116 & 0.00214 & 0.00098 & \textcolor{orange}{UN} & - \\ 
L\_98\_685& 1 & 0.201 & 0.114 & -9.5 & 20.6 & 0.00190 & 0.00116 & -0.00066 & 0.00095 & \textcolor{orange}{UN} & - \\ 
H\_HD57702& 1 & 0.332 & 0.115 &  1.1 & 10.8 & 0.00332 & 0.00115 & 0.00013 & 0.00095 & \textcolor{orange}{UN} & - \\ 
L\_RU\_152D& 1 & 0.297 & 0.113 &  2.8 & 12.1 & 0.00295 & 0.00113 & 0.00029 & 0.00092 & \textcolor{orange}{UN} & - \\ 
L\_PG0918+029D& 1 & 0.372 & 0.226 & 89.1 & 22.8 & -0.00372 & 0.00226 & 0.00011 & 0.00231 & \textcolor{orange}{UN} & - \\ 
Z\_HD81418& 2 & 0.140 & 0.086 & 68.4 & 23.1 & -0.00102 & 0.00028 & 0.00096 & 0.00134 & \textcolor{orange}{UN} & - \\ 
Z\_HD85471& 6 & 0.141 & 0.047 & 35.4 & 10.2 & 0.00046 & 0.00129 & 0.00134 & 0.00141 & \ NV & 0.7495 \\ 
Z\_HD86321& 1 & 0.132 & 0.110 & -86.7 & 61.4 & -0.00131 & 0.00110 & -0.00015 & 0.00089 & \textcolor{orange}{UN} & - \\ 
Z\_HD87582& 5 & 0.147 & 0.054 & 21.6 & 11.6 & 0.00107 & 0.00132 & 0.00101 & 0.00094 & \ NV & 0.9888 \\ 
L\_PG1047+003& 1 & 0.209 & 0.136 & 48.2 & 25.4 & -0.00023 & 0.00152 & 0.00207 & 0.00136 & \textcolor{orange}{UN} & - \\ 
L\_PG1047+003B& 1 & 0.159 & 0.155 &  3.7 & 61.4 & 0.00157 & 0.00155 & 0.00020 & 0.00139 & \textcolor{orange}{UN} & - \\ 
L\_PG1047+003C& 1 & 0.190 & 0.145 & 75.0 & 61.4 & -0.00164 & 0.00148 & 0.00095 & 0.00133 & \textcolor{orange}{UN} & - \\ 
L\_G163\_50& 2 & 0.146 & 0.123 & 44.8 & 61.4 & 0.00001 & 0.00193 & 0.00146 & 0.00132 & \textcolor{orange}{UN} & - \\ 
L\_G163\_51& 3 & 0.192 & 0.080 & 28.8 & 13.7 & 0.00103 & 0.00146 & 0.00162 & 0.00007 & \textcolor{orange}{UN} & - \\ 
Z\_HD96589& 5 & 0.193 & 0.051 & 56.9 &  8.0 & -0.00078 & 0.00134 & 0.00177 & 0.00118 & \ NV & 0.9093 \\ 
Z\_HD97853& 6 & 0.159 & 0.043 & 42.7 &  8.3 & 0.00013 & 0.00070 & 0.00158 & 0.00074 & \ NV & 0.0839 \\ 
Z\_BD+35.2256& 9 & 0.146 & 0.039 & 32.4 &  8.1 & 0.00062 & 0.00242 & 0.00133 & 0.00088 & \ NV & 0.4036 \\ 
Z\_BD+29.2198& 10 & 0.192 & 0.033 & 48.8 &  5.1 & -0.00025 & 0.00143 & 0.00190 & 0.00087 & \ NV & 0.4883 \\ 
Z\_BD+31.2314& 8 & 0.122 & 0.040 & 51.3 & 10.3 & -0.00026 & 0.00108 & 0.00119 & 0.00104 & \ NV & 0.3385 \\ 
Z\_BD+25.2439& 7 & 0.131 & 0.042 & 44.8 &  9.8 & 0.00001 & 0.00131 & 0.00131 & 0.00071 & \ NV & 0.7856 \\ 
Z\_BD+18.2549& 7 & 0.170 & 0.042 & 48.5 &  7.5 & -0.00021 & 0.00121 & 0.00168 & 0.00091 & \ NV & 0.4239 \\ 
Z\_BD+32.2217& 8 & 0.104 & 0.040 & 51.7 & 12.3 & -0.00024 & 0.00077 & 0.00101 & 0.00090 & \ NV & 0.1073 \\ 
Z\_BD+22.2446& 7 & 0.159 & 0.042 & 43.9 &  8.0 & 0.00006 & 0.00120 & 0.00159 & 0.00097 & \ NV & 0.2292 \\ 
Z\_BD+40.2546& 11 & 0.164 & 0.032 & 47.3 &  5.8 & -0.00013 & 0.00107 & 0.00164 & 0.00097 & \ NV & 0.5166 \\ 
Z\_BD+30.2290& 8 & 0.203 & 0.039 & 43.7 &  5.7 & 0.00009 & 0.00097 & 0.00202 & 0.00094 & \ NV & 0.6699 \\ 
L\_104\_334& 8 & 0.187 & 0.042 & 45.9 &  6.6 & -0.00006 & 0.00150 & 0.00187 & 0.00161 & \ NV & 0.0743 \\ 
Z\_BD+39.2611& 1 & 0.055 & 0.112 & -88.3 & 61.4 & -0.00055 & 0.00112 & -0.00003 & 0.00090 & \textcolor{orange}{UN} & - \\ 
Z\_BD+16.2491& 8 & 0.212 & 0.039 & 51.1 &  5.4 & -0.00045 & 0.00151 & 0.00207 & 0.00085 & \ NV & 0.5325 \\ 
Z\_HD116513& 13 & 0.167 & 0.030 & 66.3 &  5.3 & -0.00113 & 0.00128 & 0.00123 & 0.00091 & \ NV & 0.7017 \\ 
L\_PG1323-085C& 2 & 0.275 & 0.097 & 64.0 & 11.1 & -0.00169 & 0.00062 & 0.00217 & 0.00098 & \textcolor{orange}{UN} & - \\ 
L\_PG1323-085B& 6 & 0.173 & 0.047 & 35.8 &  8.3 & 0.00054 & 0.00102 & 0.00164 & 0.00060 & \ NV & 0.0854 \\ 
L\_PG1323-085D& 6 & 0.142 & 0.045 & 59.0 &  9.6 & -0.00067 & 0.00065 & 0.00126 & 0.00039 & \textcolor{red}{V} & 0.0049 \\ 
Z\_BD+31.2505& 11 & 0.179 & 0.033 & 52.6 &  5.4 & -0.00047 & 0.00145 & 0.00173 & 0.00084 & \ NV & 0.9933 \\ 
Z\_BD+35.2465& 4 & 0.190 & 0.053 & 56.7 &  8.5 & -0.00075 & 0.00057 & 0.00174 & 0.00080 & \textcolor{orange}{UN} & - \\ 
Z\_BD+30.2431& 7 & 0.208 & 0.043 & 51.6 &  6.1 & -0.00048 & 0.00126 & 0.00203 & 0.00100 & \ NV & 0.7369 \\ 
Z\_HD120010& 2 & 0.171 & 0.071 & 57.5 & 13.4 & -0.00073 & 0.00003 & 0.00155 & 0.00010 & \textcolor{orange}{UN} & - \\ 
Z\_HD121859& 13 & 0.226 & 0.031 & 71.5 &  3.9 & -0.00181 & 0.00097 & 0.00136 & 0.00078 & \ NV & 0.1764 \\ 
L\_106\_700& 11 & 0.494 & 0.033 & 72.4 &  1.9 & -0.00404 & 0.00090 & 0.00285 & 0.00067 & \ NV & 0.1658 \\ 
Z\_HD138733& 15 & 0.068 & 0.030 & 89.5 & 14.7 & -0.00068 & 0.00070 & 0.00001 & 0.00091 & \textcolor{red}{V} & 0.0129 \\ 
L\_107\_599& 8 & 1.031 & 0.045 & 65.3 &  1.3 & -0.00670 & 0.00123 & 0.00784 & 0.00102 & \ NV & 0.6894 \\ 
L\_107\_602& 10 & 0.963 & 0.040 & 60.1 &  1.2 & -0.00483 & 0.00374 & 0.00833 & 0.00154 & \ NV & 0.8954 \\ 
L\_PG1633+099B& 13 & 0.532 & 0.032 & 66.1 &  1.8 & -0.00357 & 0.00140 & 0.00394 & 0.00097 & \ NV & 0.8427 \\ 
L\_PG1633+099D& 14 & 0.523 & 0.038 & 66.6 &  2.1 & -0.00358 & 0.00199 & 0.00381 & 0.00126 & \ NV & 0.8657 \\ 
Z\_HD153752& 16 & 0.107 & 0.027 & 61.4 &  7.6 & -0.00058 & 0.00102 & 0.00090 & 0.00046 & \textcolor{red}{V} & 0.0222 \\ 
B\_1725+1152\_11& 9 & 0.938 & 0.041 & 74.1 &  1.3 & -0.00797 & 0.00169 & 0.00495 & 0.00087 & \ NV & 0.8484 \\ 
B\_1725+1152\_24& 11 & 0.610 & 0.036 & 65.2 &  1.7 & -0.00396 & 0.00195 & 0.00464 & 0.00079 & \ NV & 0.3462 \\ 
B\_1725+1152\_35& 15 & 0.681 & 0.037 & 65.2 &  1.6 & -0.00442 & 0.00183 & 0.00518 & 0.00111 & \ NV & 0.9634 \\ 
B\_1725+1152\_113& 12 & 0.942 & 0.034 & 72.2 &  1.0 & -0.00767 & 0.00220 & 0.00547 & 0.00076 & \ NV & 0.4971 \\ 
L\_109\_71& 3 & 1.443 & 0.065 & 85.2 &  1.3 & -0.01422 & 0.00032 & 0.00243 & 0.00058 & \textcolor{orange}{UN} & - \\ 
L\_109\_381& 15 & 1.470 & 0.030 & 80.7 &  0.6 & -0.01392 & 0.00127 & 0.00471 & 0.00089 & \ NV & 0.3730 \\ 
B\_1751+0939\_376& 14 & 0.795 & 0.033 & 83.1 &  1.2 & -0.00772 & 0.00296 & 0.00189 & 0.00163 & \ NV & 0.6864 \\ 
B\_1751+0939\_129& 13 & 0.887 & 0.040 & 79.3 &  1.3 & -0.00826 & 0.00203 & 0.00323 & 0.00145 & \ NV & 0.4327 \\ 
B\_1751+0939\_204& 11 & 0.147 & 0.045 & -19.3 &  9.4 & 0.00115 & 0.00418 & -0.00092 & 0.00141 & \textcolor{red}{V} & 0.0147 \\ 
L\_110\_229& 19 & 2.408 & 0.029 & 19.2 &  0.3 & 0.01887 & 0.00132 & 0.01495 & 0.00176 & \ NV & 0.3421 \\ 
L\_110\_233& 14 & 2.620 & 0.033 & 12.0 &  0.4 & 0.02395 & 0.00167 & 0.01062 & 0.00070 & \ NV & 0.1857 \\ 
L\_111\_1965& 17 & 1.216 & 0.028 & 74.2 &  0.7 & -0.01035 & 0.00094 & 0.00637 & 0.00099 & \ NV & 0.7499 \\ 
L\_111\_1969& 19 & 1.229 & 0.028 & 75.9 &  0.6 & -0.01082 & 0.00110 & 0.00583 & 0.00082 & \textcolor{red}{V} & 0.0219 \\ 
H\_HD344776& 16 & 6.101 & 0.034 & 25.3 &  0.2 & 0.03880 & 0.00095 & 0.04709 & 0.00074 & \textcolor{red}{V} & 0.0006 \\ 
B\_1959+6508\_179& 10 & 1.087 & 0.035 & -55.6 &  0.9 & -0.00392 & 0.00127 & -0.01014 & 0.00087 & \ NV & 0.3940 \\ 
B\_1959+6508\_73& 14 & 0.623 & 0.029 & -48.4 &  1.3 & -0.00073 & 0.00177 & -0.00619 & 0.00069 & \ NV & 0.7553 \\ 
B\_1959+6508\_108& 12 & 0.955 & 0.033 & -53.8 &  1.0 & -0.00288 & 0.00325 & -0.00911 & 0.00189 & \textcolor{red}{V} & 0.0004 \\ 
B\_1959+6508\_104& 14 & 0.953 & 0.029 & -46.6 &  0.9 & -0.00054 & 0.00181 & -0.00951 & 0.00116 & \ NV & 0.8586 \\ 
B\_1959+6508\_38& 13 & 1.587 & 0.030 & -51.1 &  0.5 & -0.00333 & 0.00118 & -0.01552 & 0.00083 & \ NV & 0.4246 \\ 
B\_2015-0137\_102& 18 & 1.097 & 0.029 & -82.0 &  0.7 & -0.01054 & 0.00218 & -0.00303 & 0.00158 & \ NV & 0.3969 \\ 
B\_2022+7611\_1& 14 & 0.651 & 0.031 & -9.2 &  1.4 & 0.00618 & 0.00098 & -0.00205 & 0.00095 & \ NV & 0.8763 \\ 
B\_2042+7508\_28& 13 & 1.088 & 0.030 & 34.9 &  0.8 & 0.00374 & 0.00102 & 0.01022 & 0.00109 & \ NV & 0.2030 \\ 
L\_112\_805& 18 & 0.196 & 0.030 & 78.6 &  4.5 & -0.00180 & 0.00154 & 0.00076 & 0.00195 & \ NV & 0.1500 \\ 
L\_112\_822& 16 & 0.349 & 0.029 & 81.7 &  2.4 & -0.00334 & 0.00147 & 0.00100 & 0.00124 & \ NV & 0.1203 \\ 
B\_2042+7508\_17& 11 & 0.995 & 0.039 & 44.2 &  1.1 & 0.00028 & 0.00390 & 0.00995 & 0.00200 & \textcolor{red}{V} & $1.4\times10^{-5}$ \\ 
L\_113\_339& 8 & 0.198 & 0.040 & -76.4 &  5.9 & -0.00176 & 0.00130 & -0.00090 & 0.00096 & \ NV & 0.5852 \\ 
L\_113\_241& 13 & 0.249 & 0.038 & -38.0 &  4.4 & 0.00060 & 0.00208 & -0.00242 & 0.00183 & \ NV & 0.1112 \\ 
B\_2202+4216\_25& 14 & 1.063 & 0.030 & 64.5 &  0.8 & -0.00670 & 0.00195 & 0.00826 & 0.00086 & \textcolor{red}{V} & 0.0348 \\ 
B\_2202+4216\_129& 18 & 0.359 & 0.032 & 71.1 &  2.5 & -0.00283 & 0.00195 & 0.00220 & 0.00126 & \ NV & 0.6288 \\ 
B\_2202+4216\_239& 13 & 1.116 & 0.039 & -70.2 &  1.0 & -0.00860 & 0.00215 & -0.00711 & 0.00178 & \textcolor{red}{V} & 0.0206 \\ 
L\_PG2213-006A& 13 & 0.379 & 0.040 & -22.6 &  3.1 & 0.00267 & 0.00409 & -0.00269 & 0.00208 & \textcolor{red}{V} & 0.0181 \\ 
H\_BD+62.2078& 17 & 7.471 & 0.036 & 72.7 &  0.1 & -0.06156 & 0.00126 & 0.04233 & 0.00102 & \textcolor{red}{V} & 0.0365 \\ 
B\_2253+1608\_23& 11 & 0.224 & 0.039 & -48.4 &  5.1 & -0.00026 & 0.00100 & -0.00222 & 0.00110 & \ NV & 0.1920 \\ 
B\_2340+8015\_34& 11 & 1.229 & 0.037 & 37.9 &  0.9 & 0.00301 & 0.00225 & 0.01192 & 0.00158 & \ NV & 0.1264 \\ 
B\_2340+8015\_109& 4 & 1.273 & 0.057 & 18.7 &  1.3 & 0.01011 & 0.00051 & 0.00773 & 0.00105 & \textcolor{orange}{UN} & - \\ 
B\_2340+8015\_99& 15 & 1.147 & 0.029 & 13.3 &  0.7 & 0.01025 & 0.00285 & 0.00513 & 0.00194 & \textcolor{red}{V} & 0.0006 \\ 
L\_115\_420& 15 & 0.199 & 0.028 & -73.2 &  4.2 & -0.00165 & 0.00102 & -0.00110 & 0.00104 & \ NV & 0.7984 \\ 
L\_PG2349+002& 6 & 0.798 & 0.059 & -26.0 &  2.1 & 0.00490 & 0.00541 & -0.00629 & 0.00479 & \textcolor{red}{V} & $1.7\times10^{-9}$ \\ 
\hline
\label{tab:res}
\end{longtable}
}
\twocolumn
\clearpage

\section{Monitoring data plots} \label{ap:b}

\begin{figure*}
  \centering
  \includegraphics[width=0.95\textwidth]{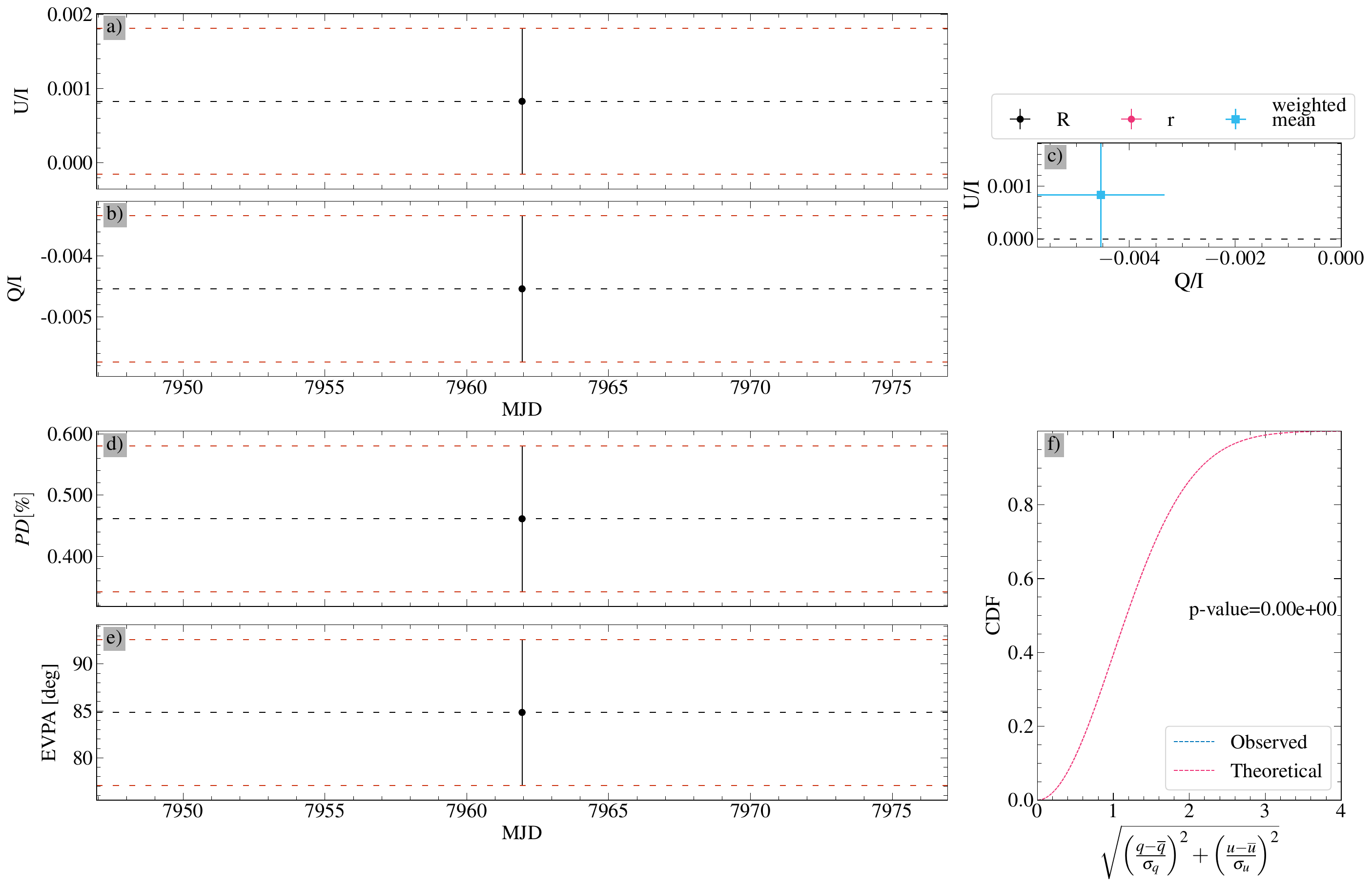}
  \caption{Evolution of polarization parameters of B\_0017+8135\_79, which is found to be variable. (a, b) - Evolution of the relative Stokes parameters. The dashed black line shows the weighted average, the red dashed lines show the corresponding 1$\sigma$ uncertainty. (c) - Distribution of measurements on the relative Stokes parameters plane. (d, e) - Evolution of the polarization degree and the electric vector position angle. The dashed black line shows the weighted average, the red dashed lines show the corresponding 1$\sigma$ uncertainty. (f) - EDF of measured polarization in both bands together with expected CDF of polarization measurements for a constant source with similar uncertainties.}
  \label{fig:B_0017+8135_79}
\end{figure*}

\begin{figure*}
  \centering
  \includegraphics[width=0.95\textwidth]{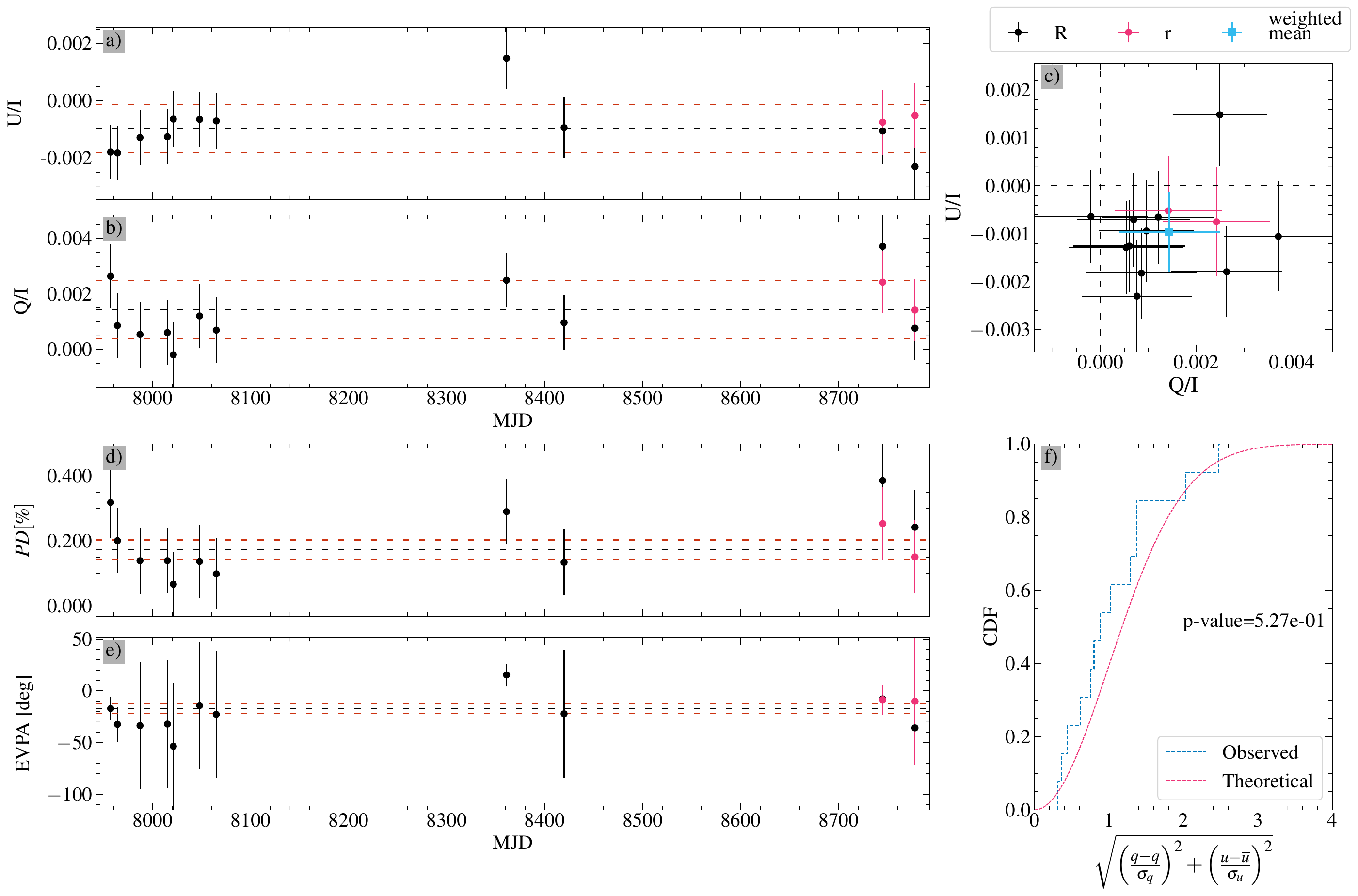}
  \caption{Same as Fig.~\ref{fig:B_0017+8135_82} for L\_92\_245, which is found to be stable.}
  \label{fig:L_92_245}
\end{figure*}

\clearpage

\begin{figure*}
  \centering
  \includegraphics[width=0.95\textwidth]{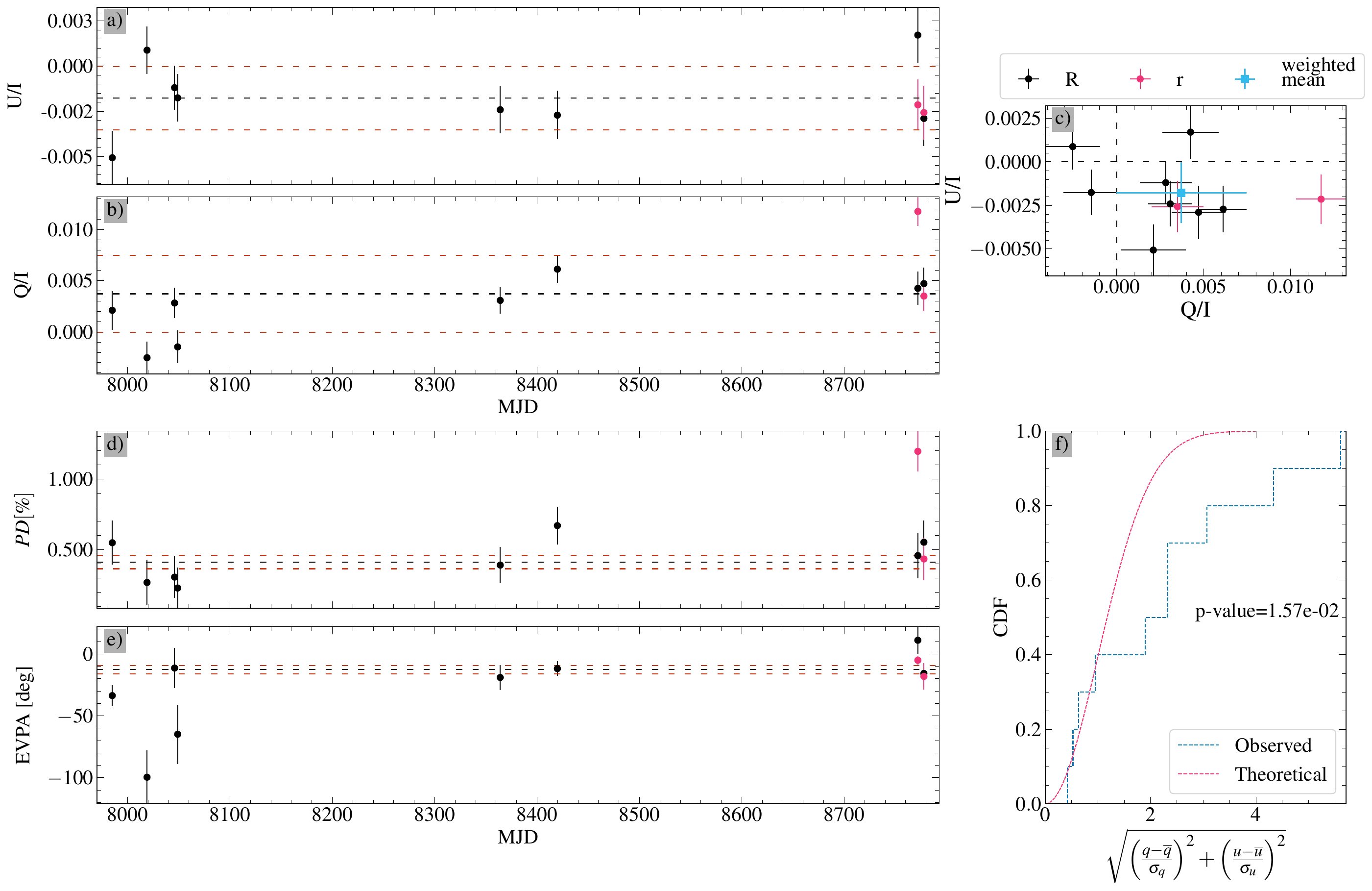}
  \caption{Same as Fig.~\ref{fig:B_0017+8135_82} for L\_92\_248, which is found to be variable. }
  \label{fig:L_92_248}
\end{figure*}

\begin{figure*}
  \centering
  \includegraphics[width=0.95\textwidth]{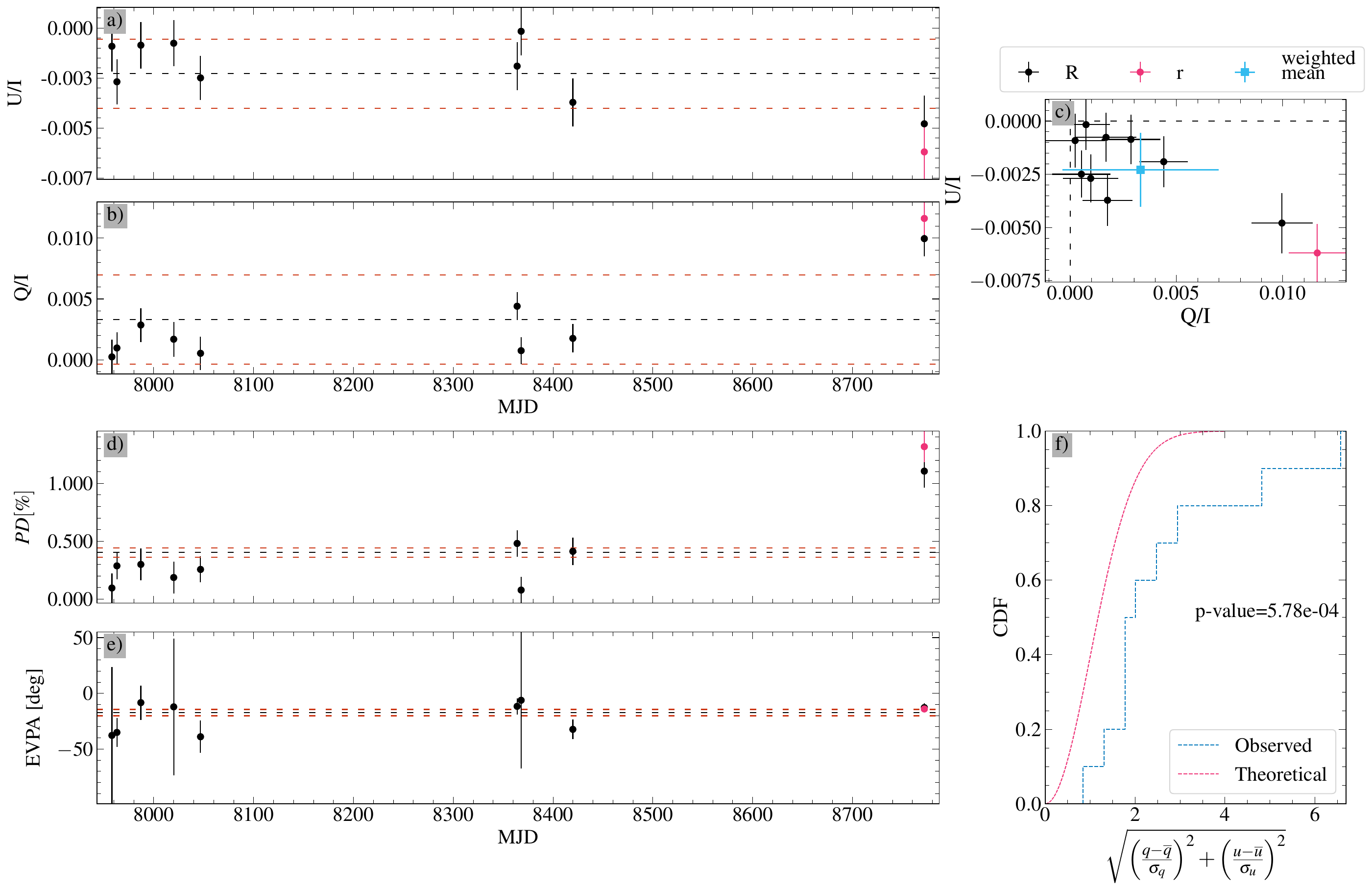}
  \caption{Same as Fig.~\ref{fig:B_0017+8135_82} for L\_92\_249, which is found to be variable. }
  \label{fig:L_92_249}
\end{figure*}

\clearpage

\begin{figure*}
  \centering
  \includegraphics[width=0.95\textwidth]{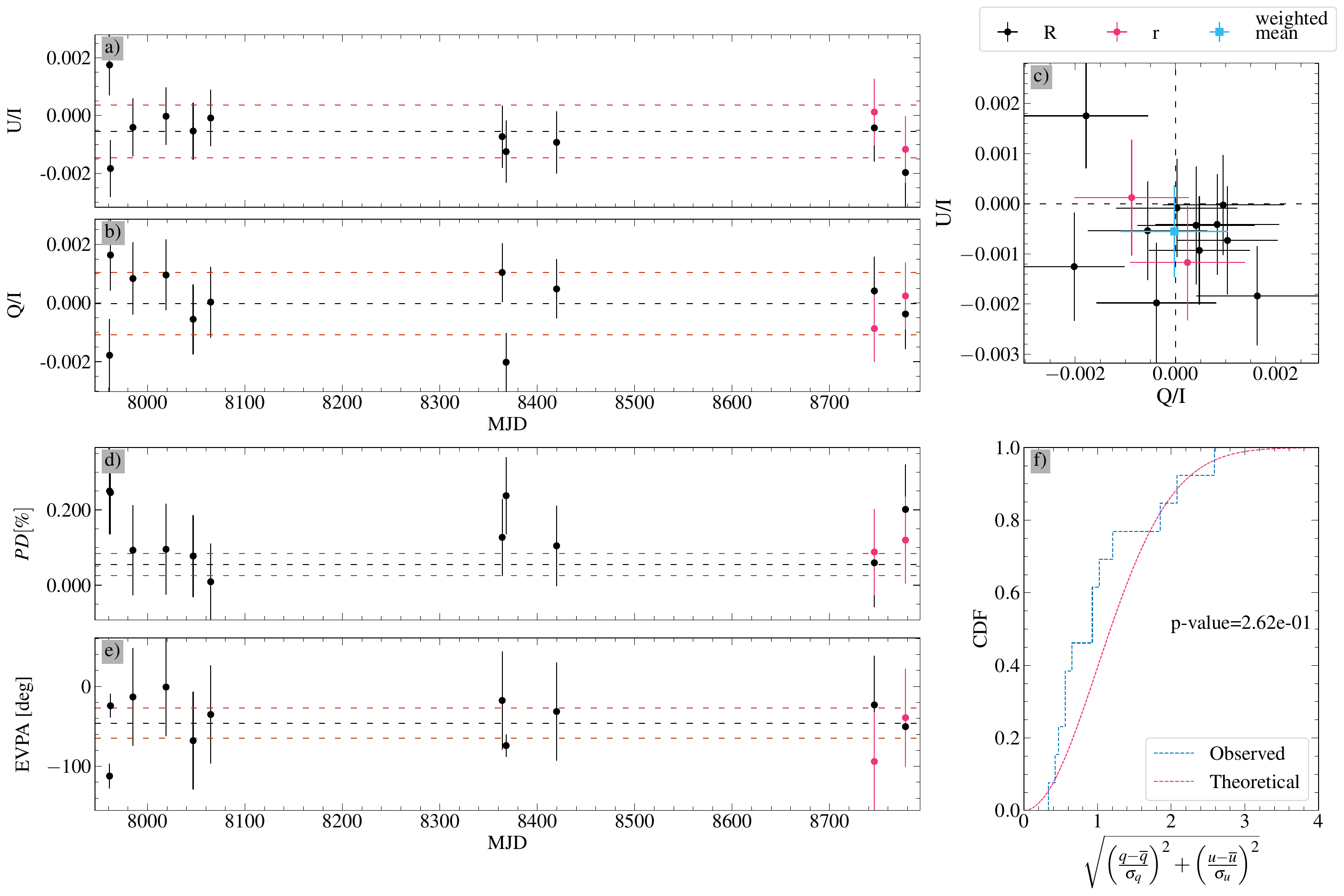}
  \caption{Same as Fig.~\ref{fig:B_0017+8135_82} for L\_92\_250, which is found to be stable. }
  \label{fig:L_92_250}
\end{figure*}

\begin{figure*}
  \centering
  \includegraphics[width=0.95\textwidth]{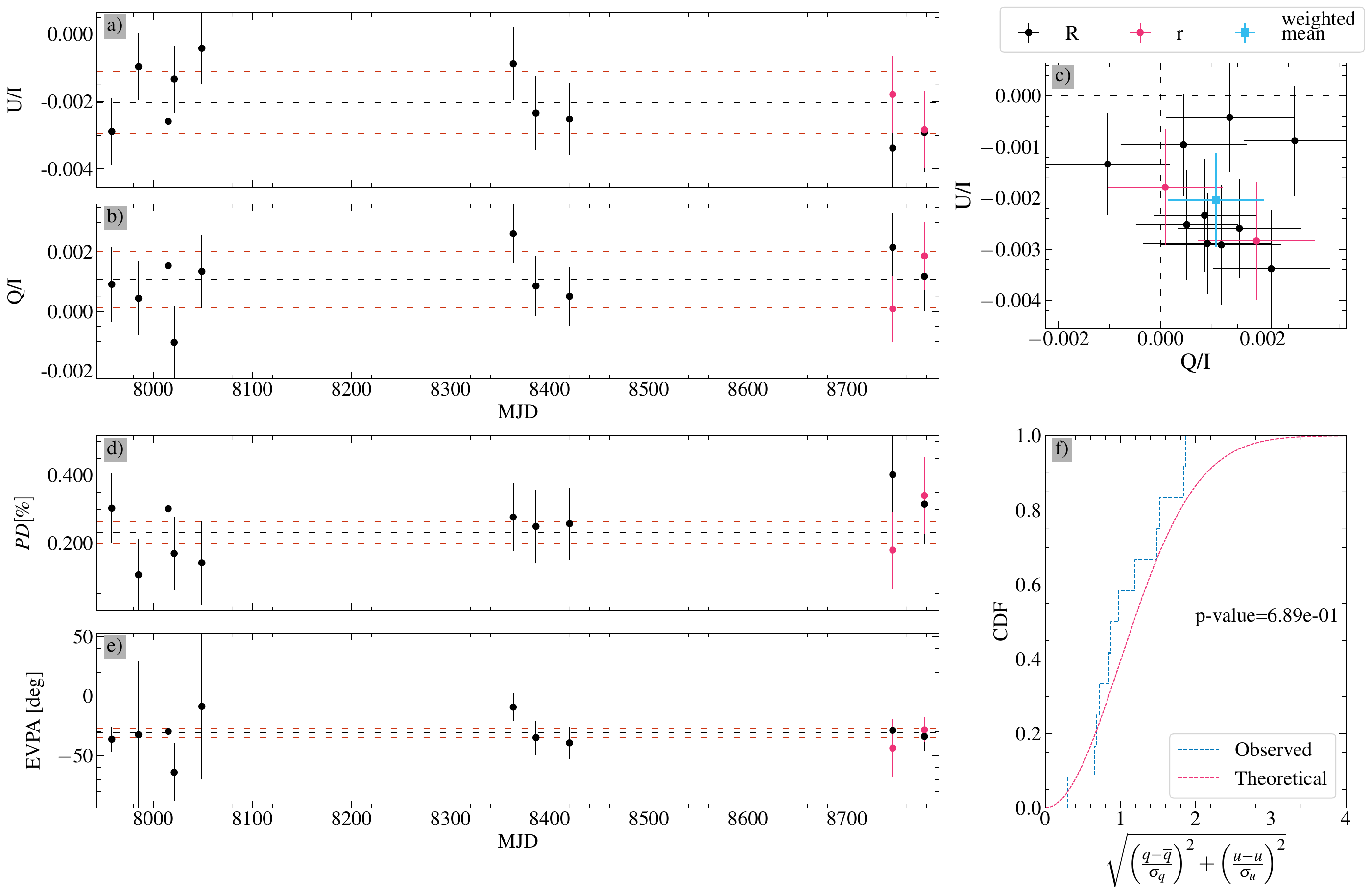}
  \caption{Same as Fig.~\ref{fig:B_0017+8135_82} for L\_93\_317, which is found to be stable. }
  \label{fig:L_93_317}
\end{figure*}

\clearpage

\begin{figure*}
  \centering
  \includegraphics[width=0.95\textwidth]{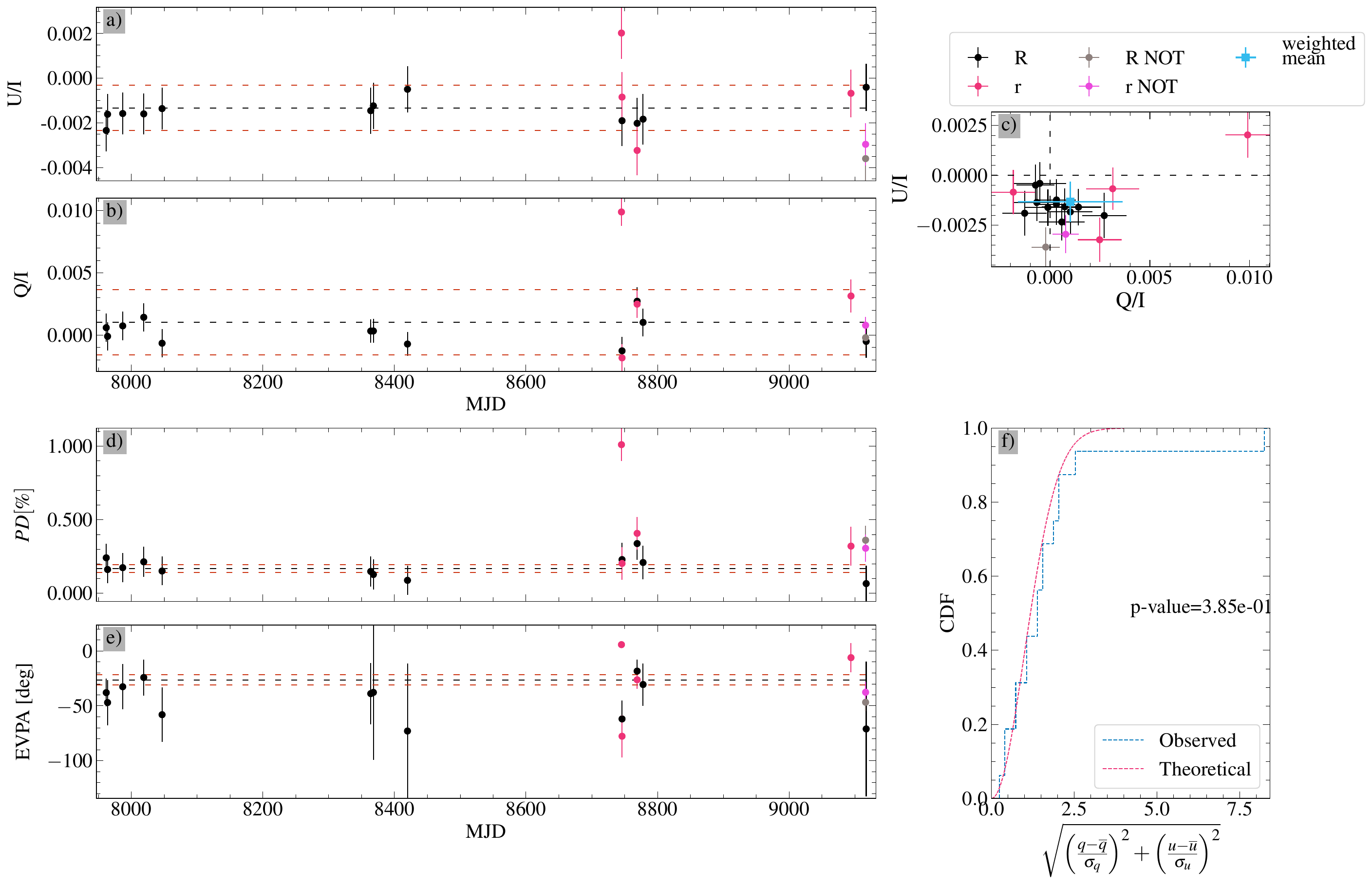}
  \caption{Same as Fig.~\ref{fig:B_0017+8135_82} for L\_93\_333, which is found to be stable. }
  \label{fig:L_93_333}
\end{figure*}

\begin{figure*}
  \centering
  \includegraphics[width=0.95\textwidth]{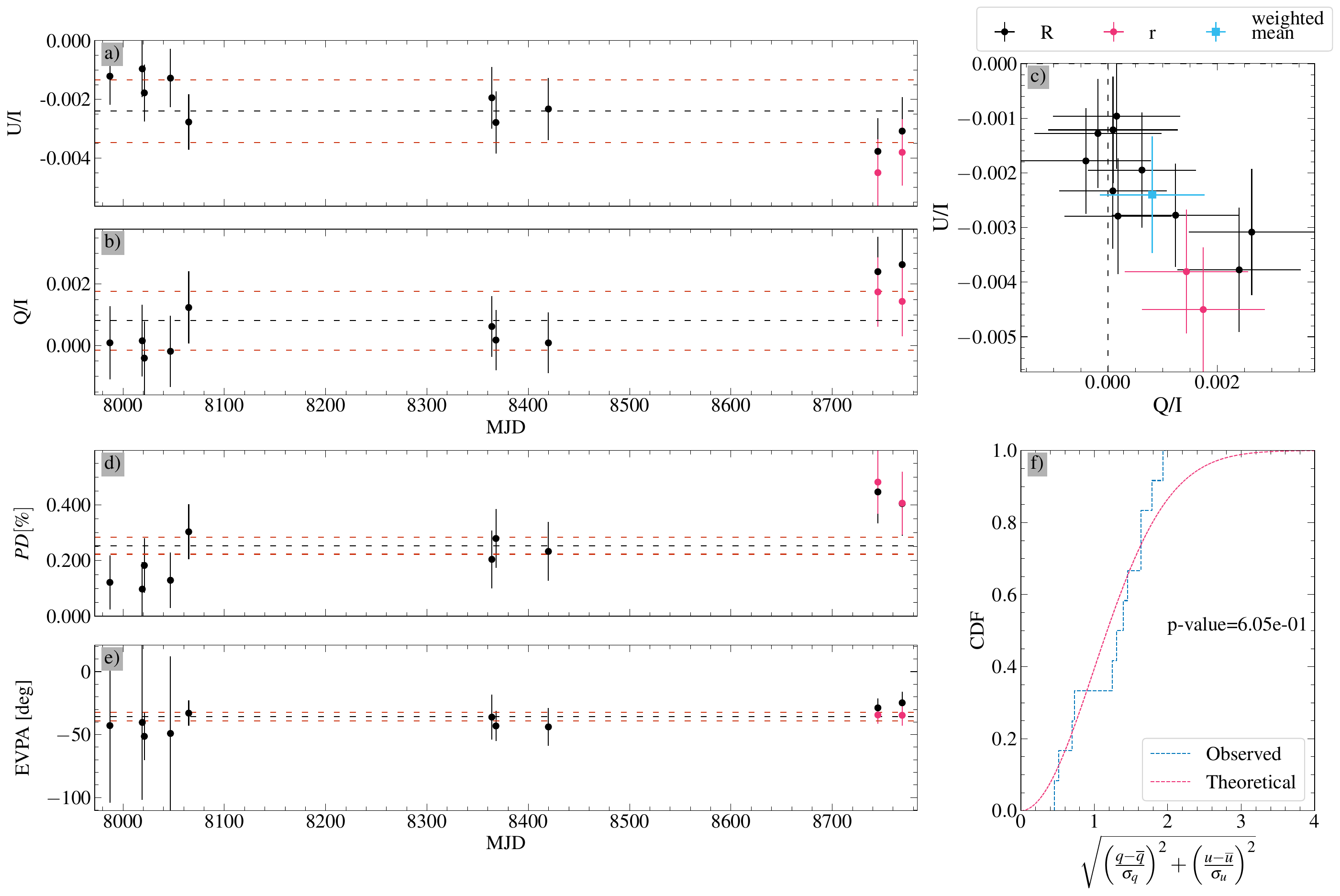}
  \caption{Same as Fig.~\ref{fig:B_0017+8135_82} for L\_93\_424, which is found to be stable. }
  \label{fig:L_93_424}
\end{figure*}

\clearpage

\begin{figure*}
  \centering
  \includegraphics[width=0.95\textwidth]{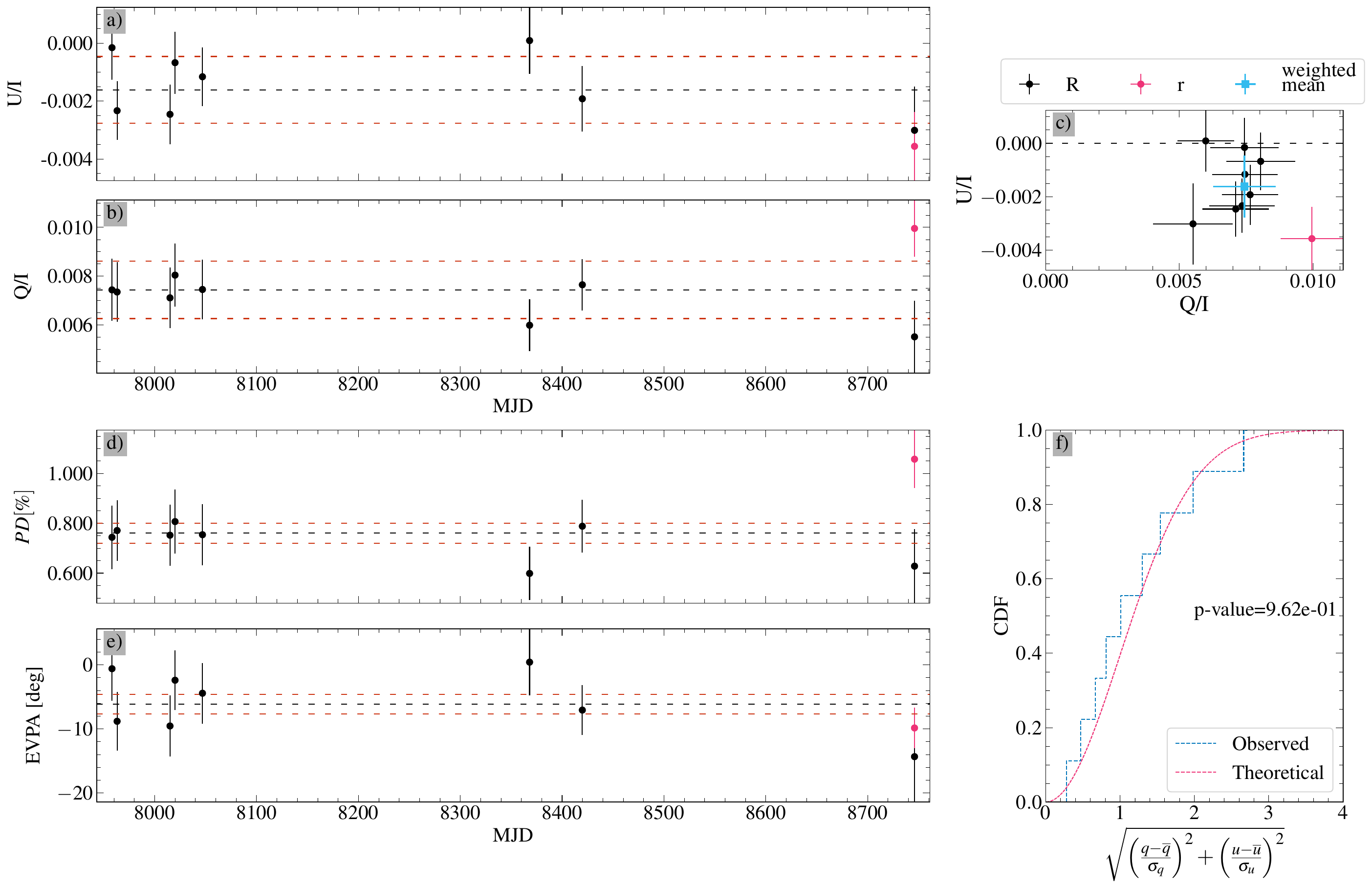}
  \caption{Same as Fig.~\ref{fig:B_0017+8135_82} for B\_0211+1051\_37, which is found to be stable. }
  \label{fig:B_0211+1051_37}
\end{figure*}

\begin{figure*}
  \centering
  \includegraphics[width=0.95\textwidth]{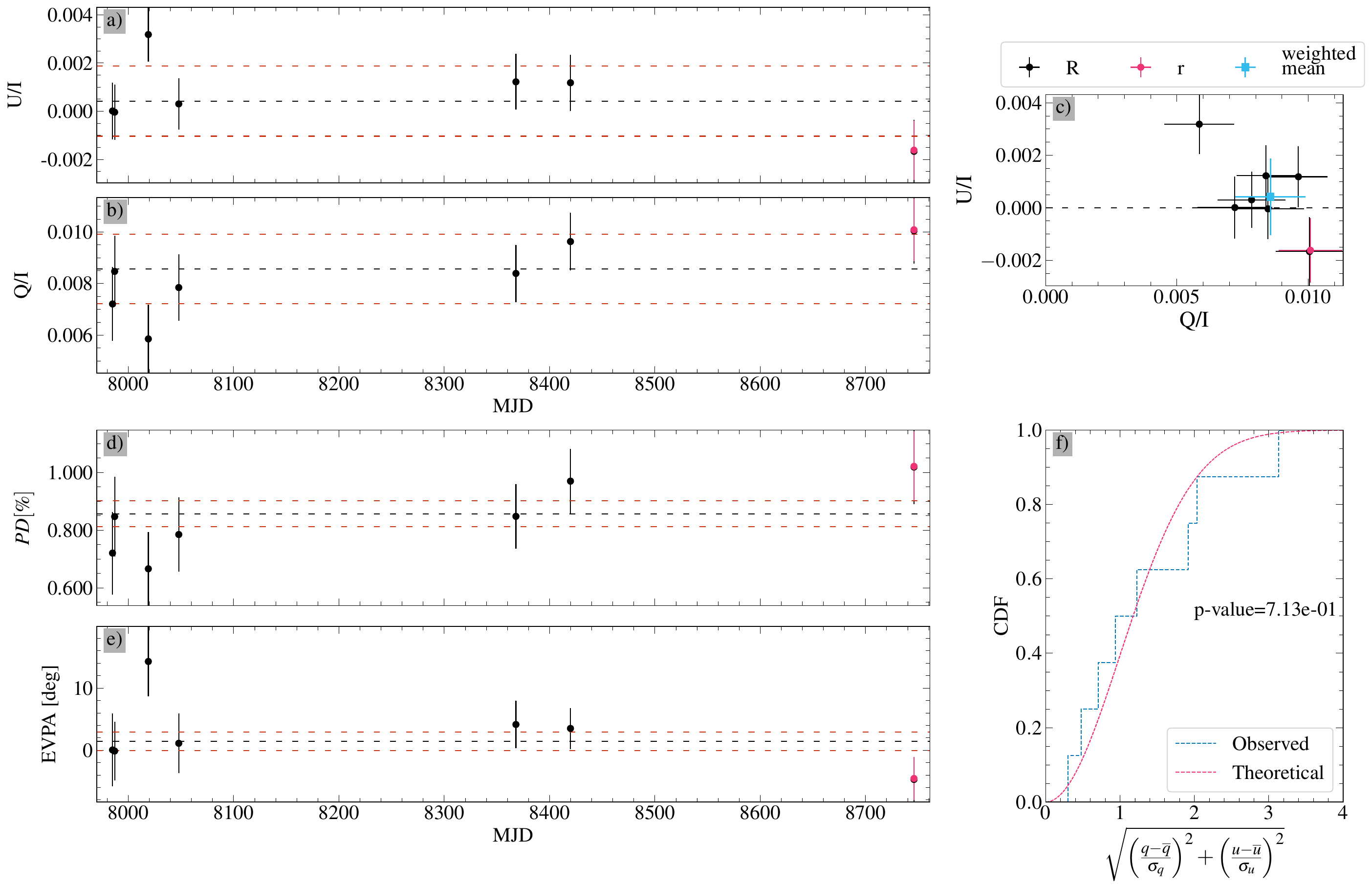}
  \caption{Same as Fig.~\ref{fig:B_0017+8135_82} for B\_0211+1051\_18, which is found to be stable. }
  \label{fig:B_0211+1051_18}
\end{figure*}

\clearpage

\begin{figure*}
  \centering
  \includegraphics[width=0.95\textwidth]{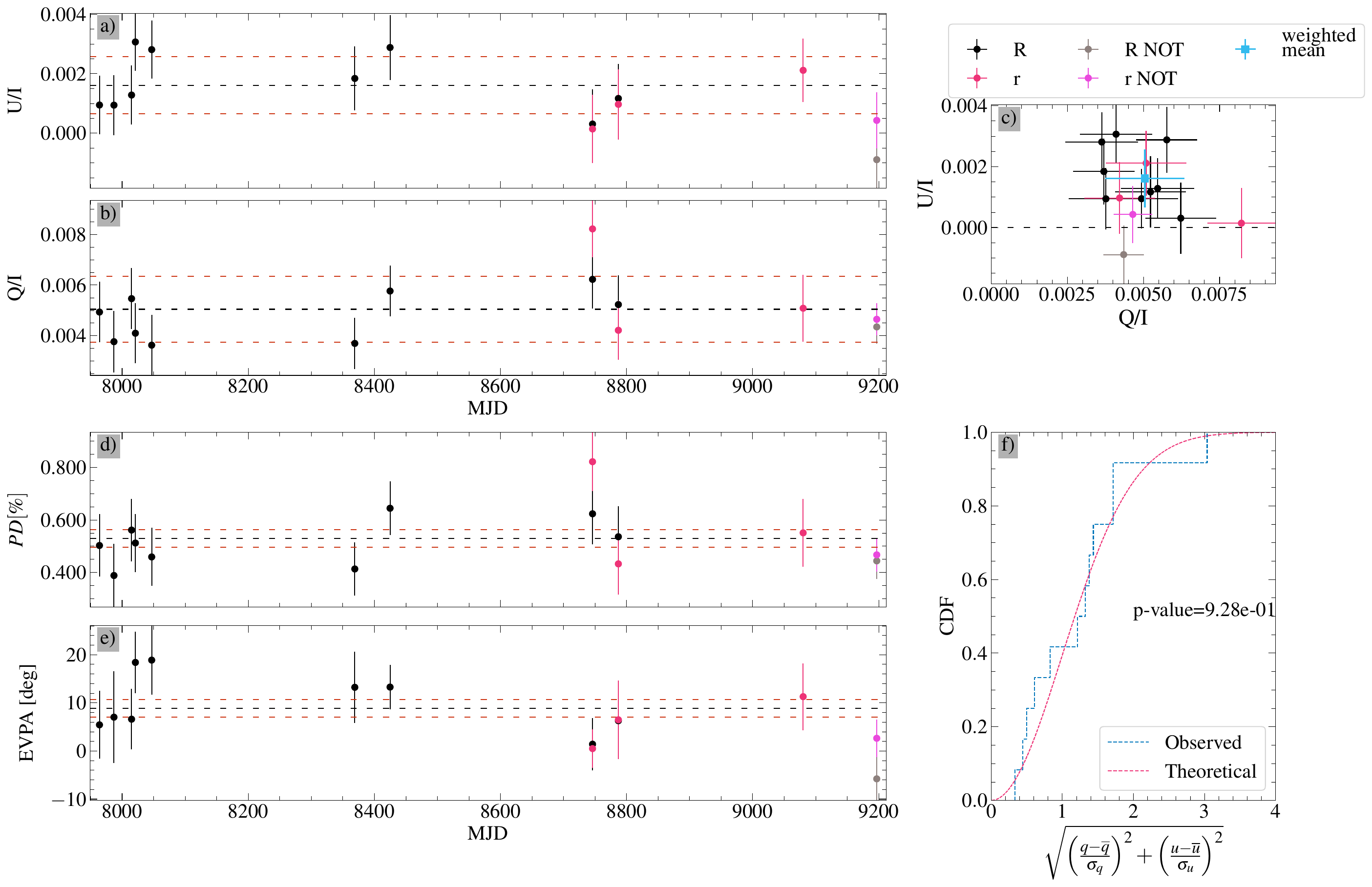}
  \caption{Same as Fig.~\ref{fig:B_0017+8135_82} for L\_94\_242, which is found to be stable. }
  \label{fig:L_94_242}
\end{figure*}

\begin{figure*}
  \centering
  \includegraphics[width=0.95\textwidth]{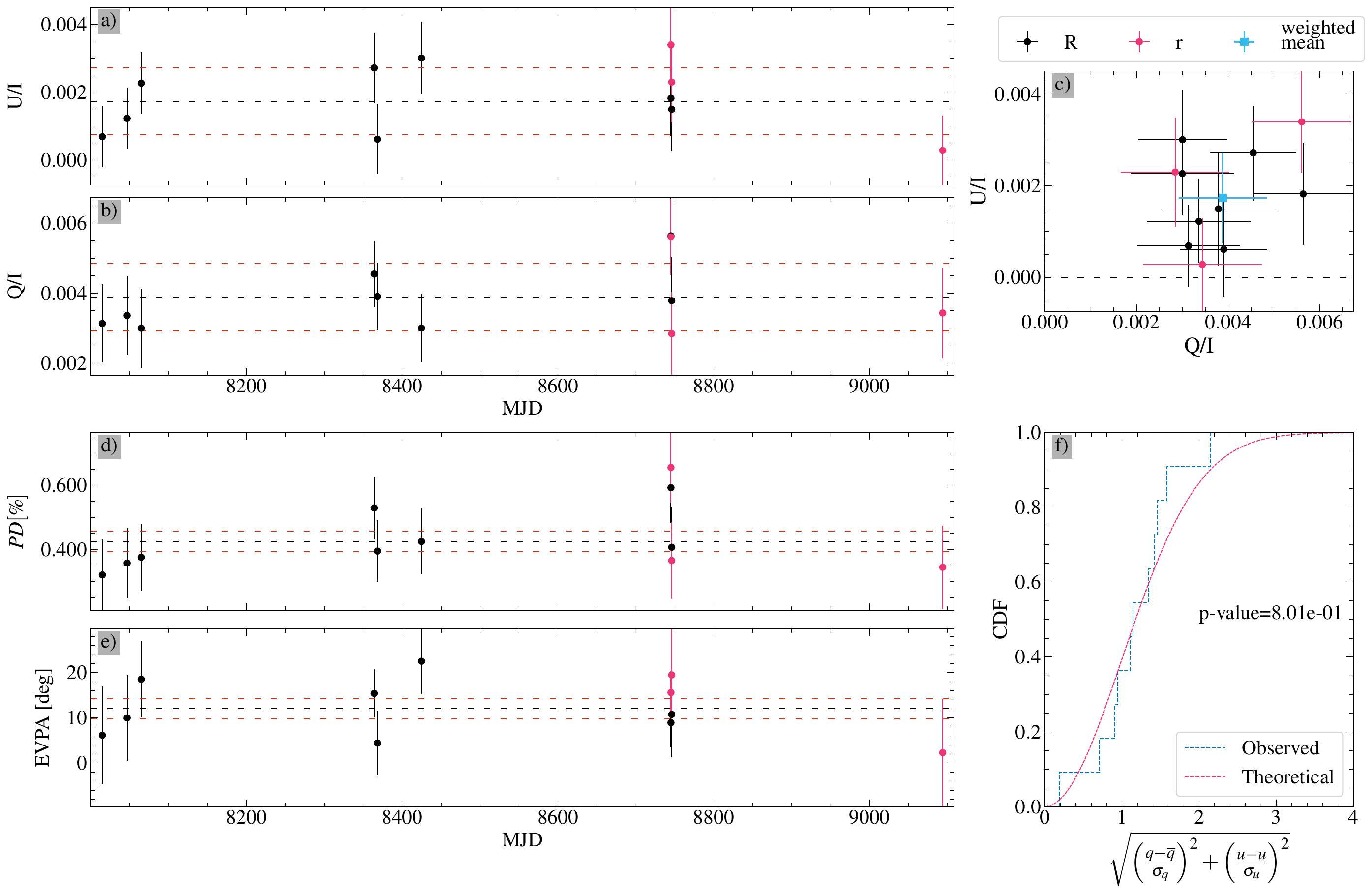}
  \caption{Same as Fig.~\ref{fig:B_0017+8135_82} for L\_94\_251, which is found to be stable. }
  \label{fig:L_94_251}
\end{figure*}

\clearpage

\begin{figure*}
  \centering
  \includegraphics[width=0.95\textwidth]{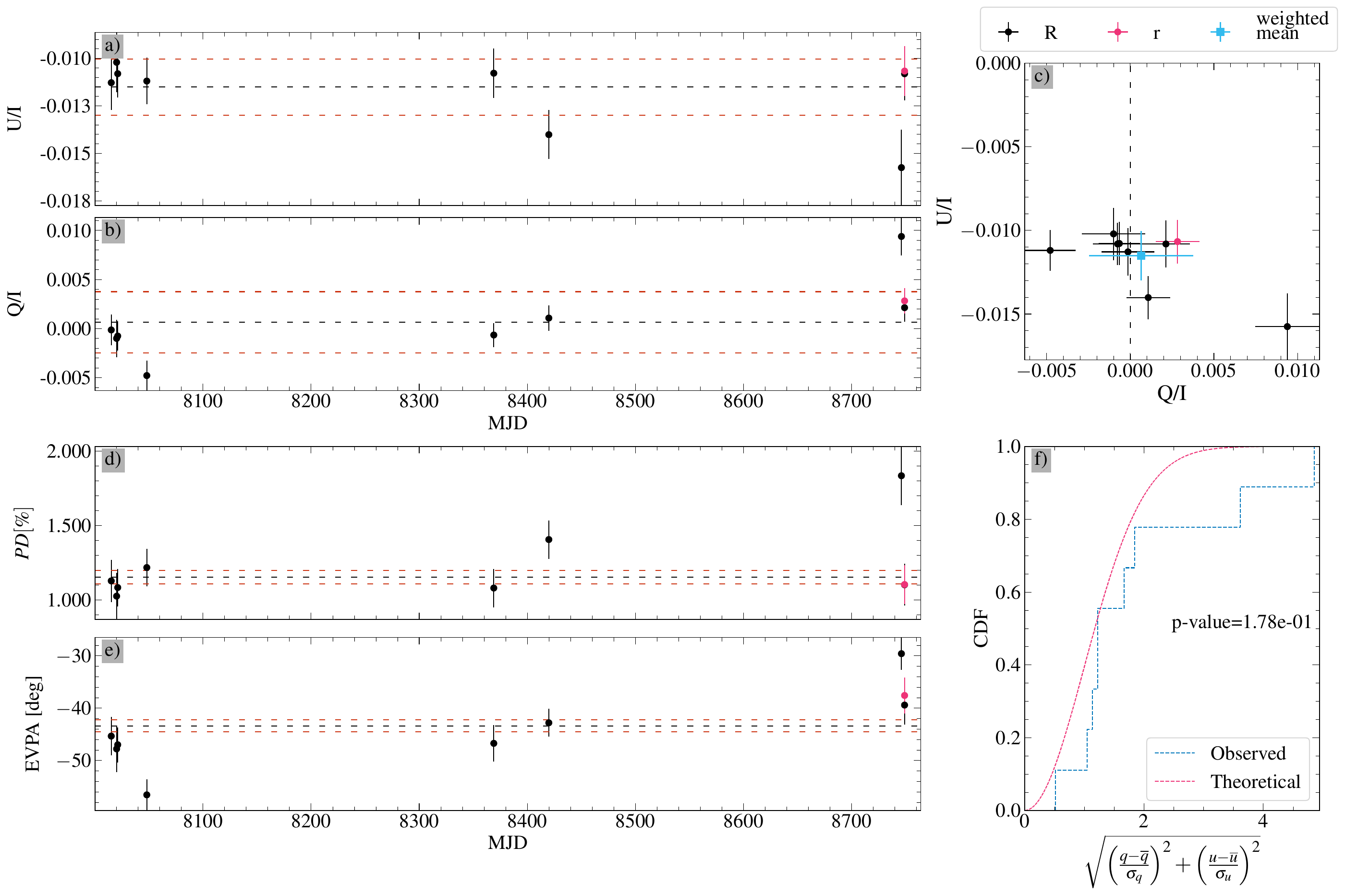}
  \caption{Same as Fig.~\ref{fig:B_0017+8135_82} for B\_0259+0747\_30, which is found to be stable. }
  \label{fig:B_0259+0747_30}
\end{figure*}

\begin{figure*}
  \centering
  \includegraphics[width=0.95\textwidth]{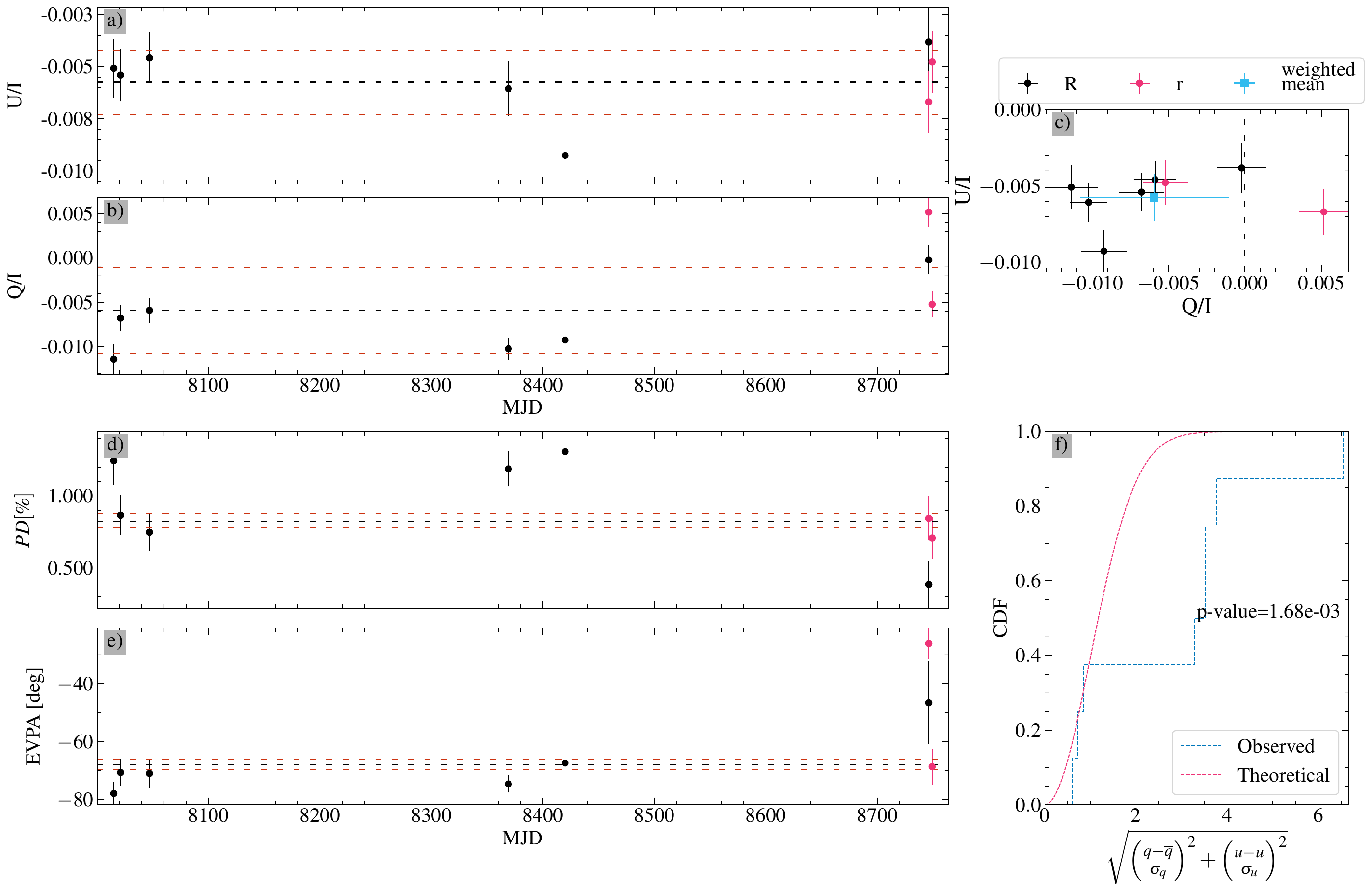}
  \caption{Same as Fig.~\ref{fig:B_0017+8135_82} for B\_0259+0747\_22, which is found to be variable. }
  \label{fig:B_0259+0747_22}
\end{figure*}

\clearpage

\begin{figure*}
  \centering
  \includegraphics[width=0.95\textwidth]{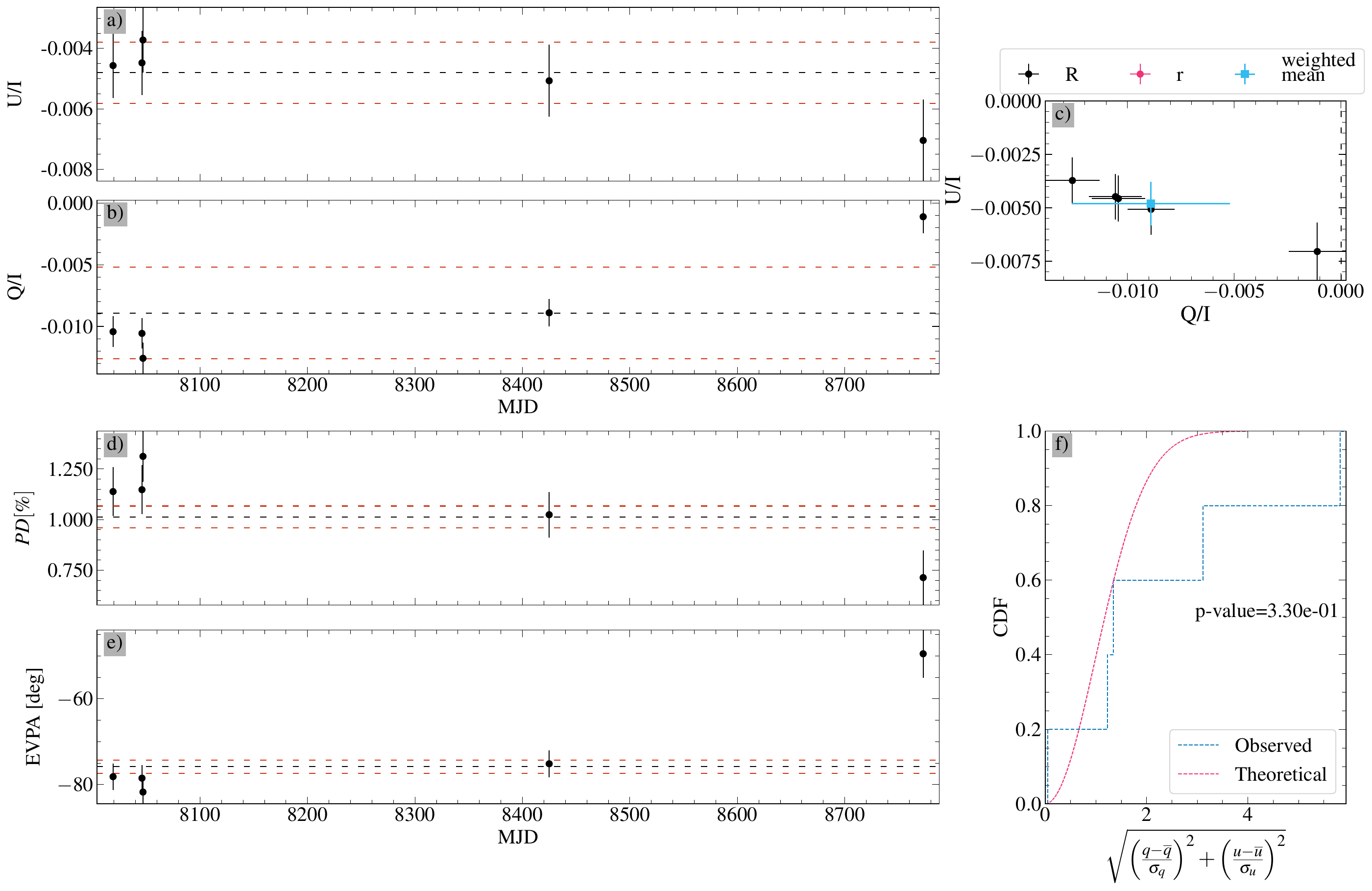}
  \caption{Same as Fig.~\ref{fig:B_0017+8135_82} for B\_0303+4716\_121, which is found to be stable. }
  \label{fig:B_0303+4716_121}
\end{figure*}

\begin{figure*}
  \centering
  \includegraphics[width=0.95\textwidth]{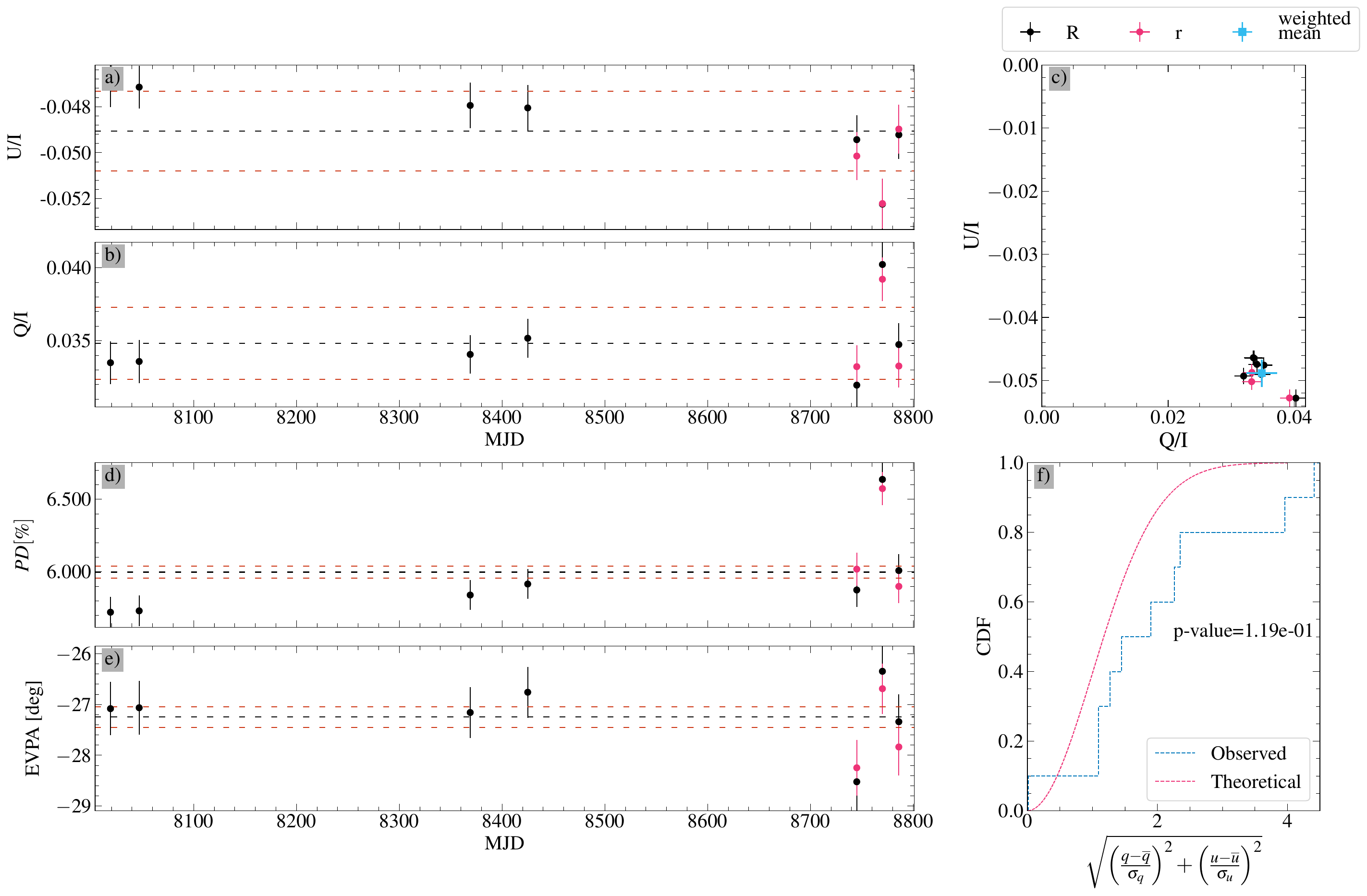}
  \caption{Same as Fig.~\ref{fig:B_0017+8135_82} for H\_GSC02355, which is found to be stable. }
  \label{fig:H_GSC02355}
\end{figure*}

\clearpage

\begin{figure*}
  \centering
  \includegraphics[width=0.95\textwidth]{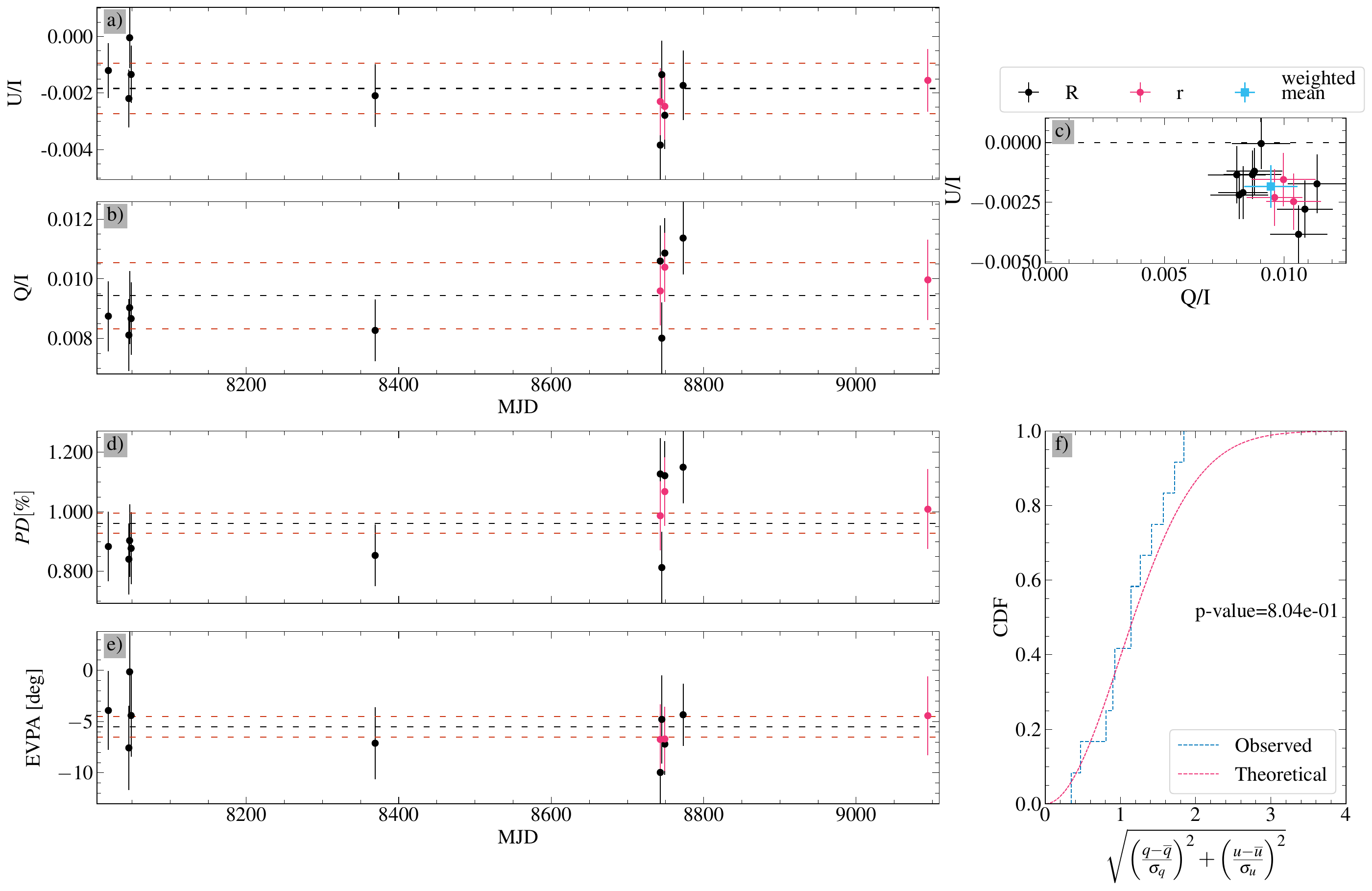}
  \caption{Same as Fig.~\ref{fig:B_0017+8135_82} for L\_95\_330, which is found to be stable. }
  \label{fig:L_95_330}
\end{figure*}

\begin{figure*}
  \centering
  \includegraphics[width=0.95\textwidth]{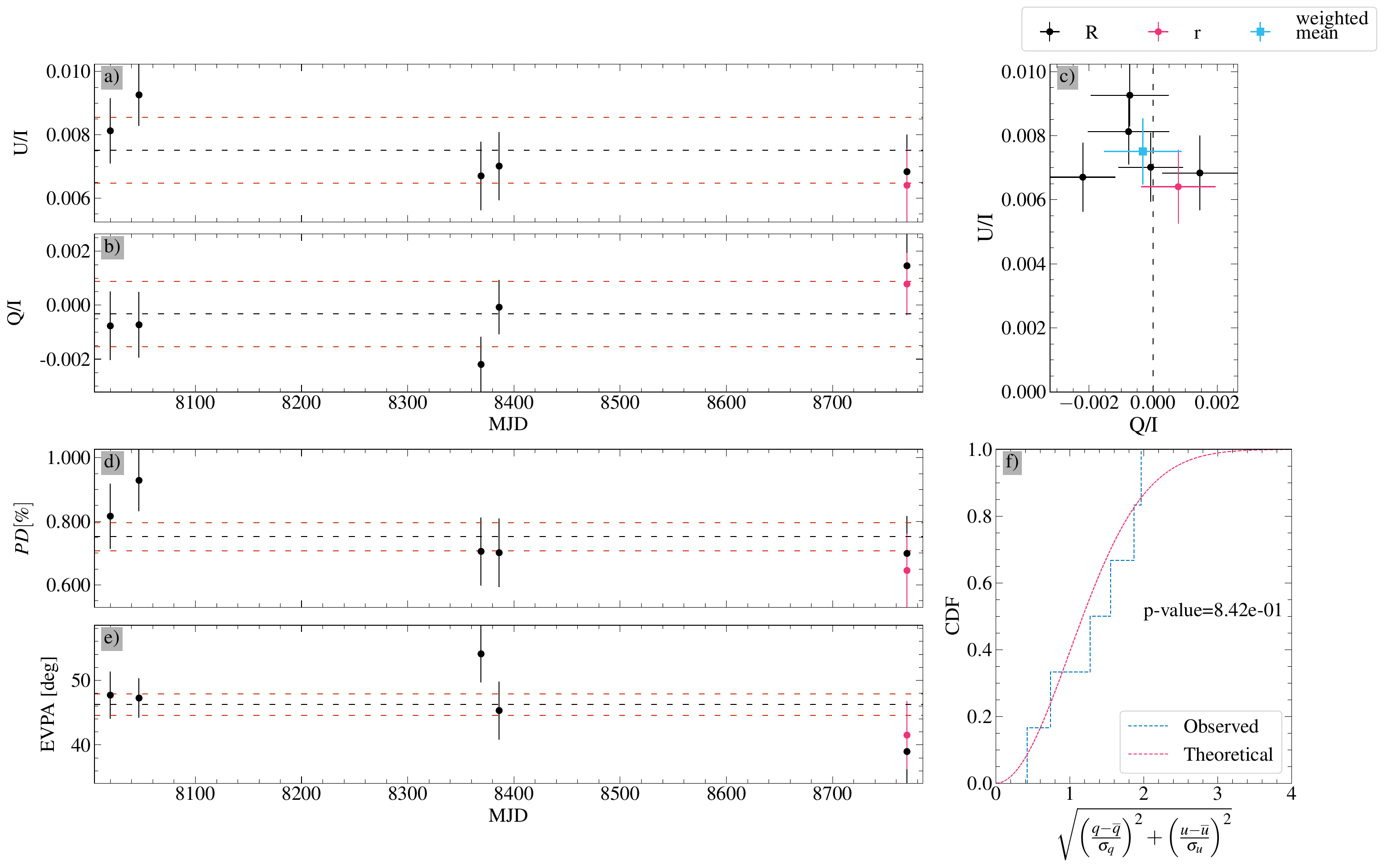}
  \caption{Same as Fig.~\ref{fig:B_0017+8135_82} for L\_95\_275, which is found to be stable. }
  \label{fig:L_95_275}
\end{figure*}

\clearpage

\begin{figure*}
  \centering
  \includegraphics[width=0.95\textwidth]{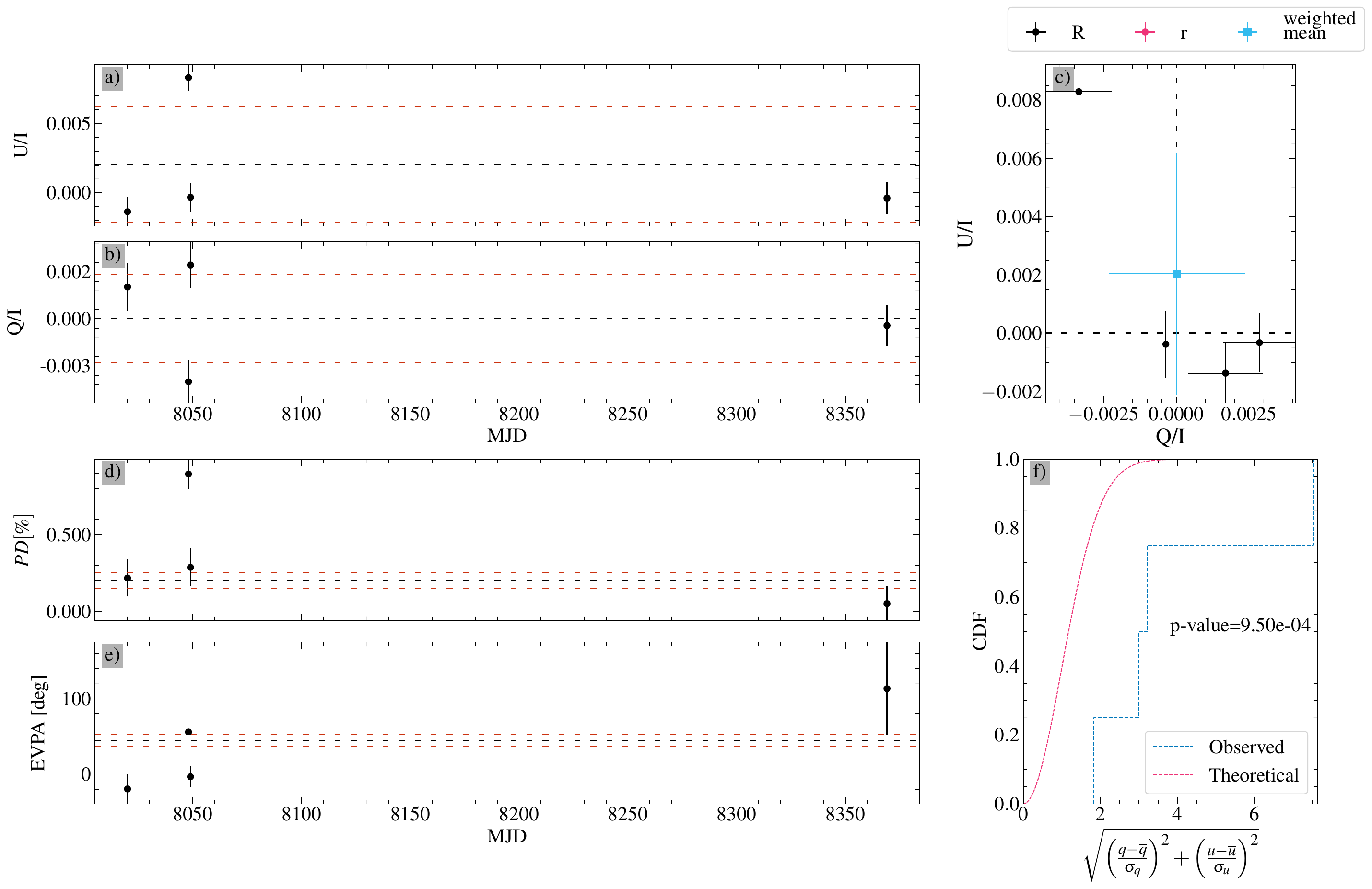}
  \caption{Same as Fig.~\ref{fig:B_0017+8135_82} for L\_95\_276, which has not enough measurements to judge it as varible of stable. }
  \label{fig:L_95_276}
\end{figure*}

\begin{figure*}
  \centering
  \includegraphics[width=0.95\textwidth]{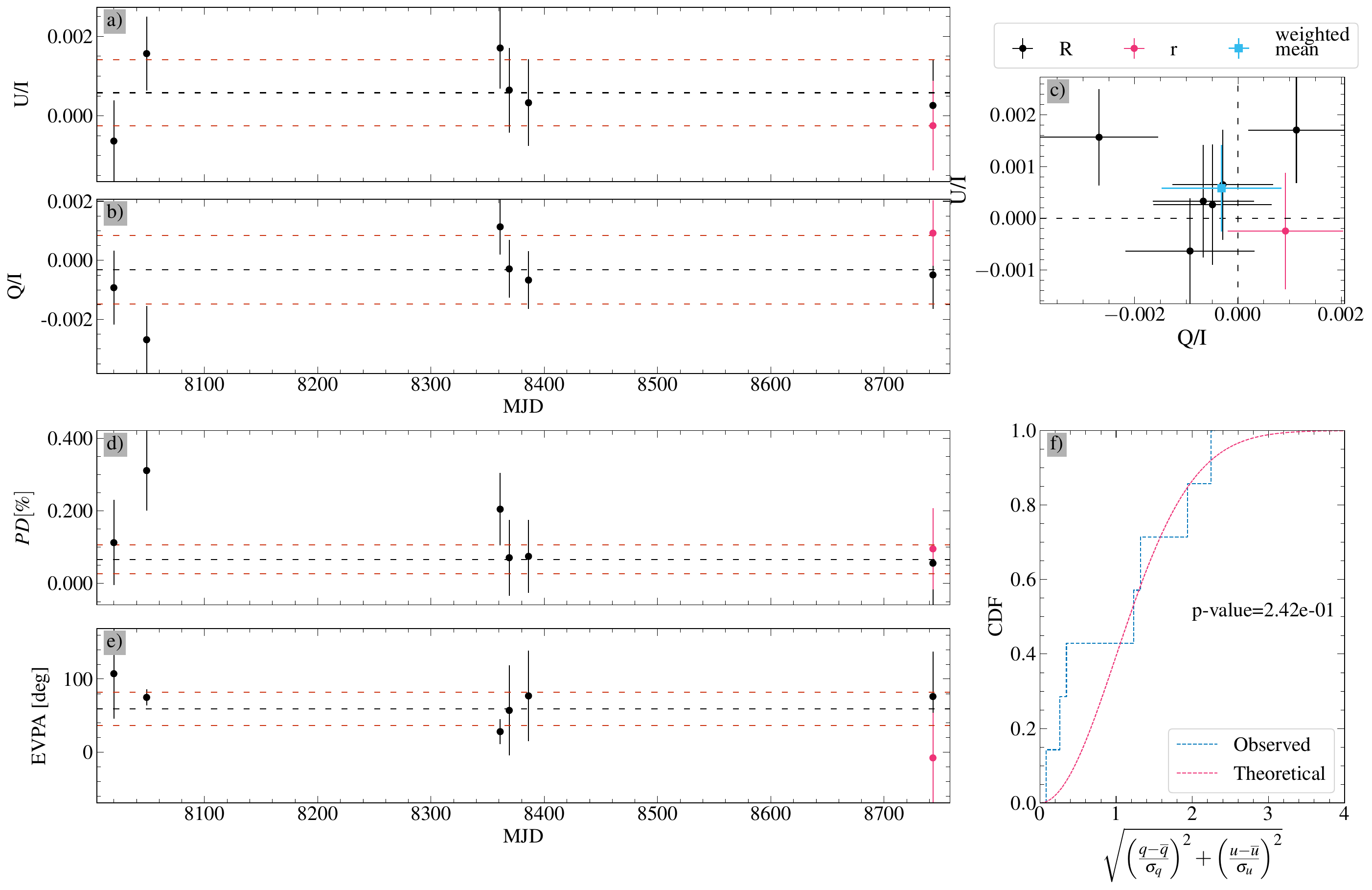}
  \caption{Same as Fig.~\ref{fig:B_0017+8135_82} for H\_HD283807, which is found to be stable. }
  \label{fig:H_HD283807}
\end{figure*}

\clearpage

\begin{figure*}
  \centering
  \includegraphics[width=0.95\textwidth]{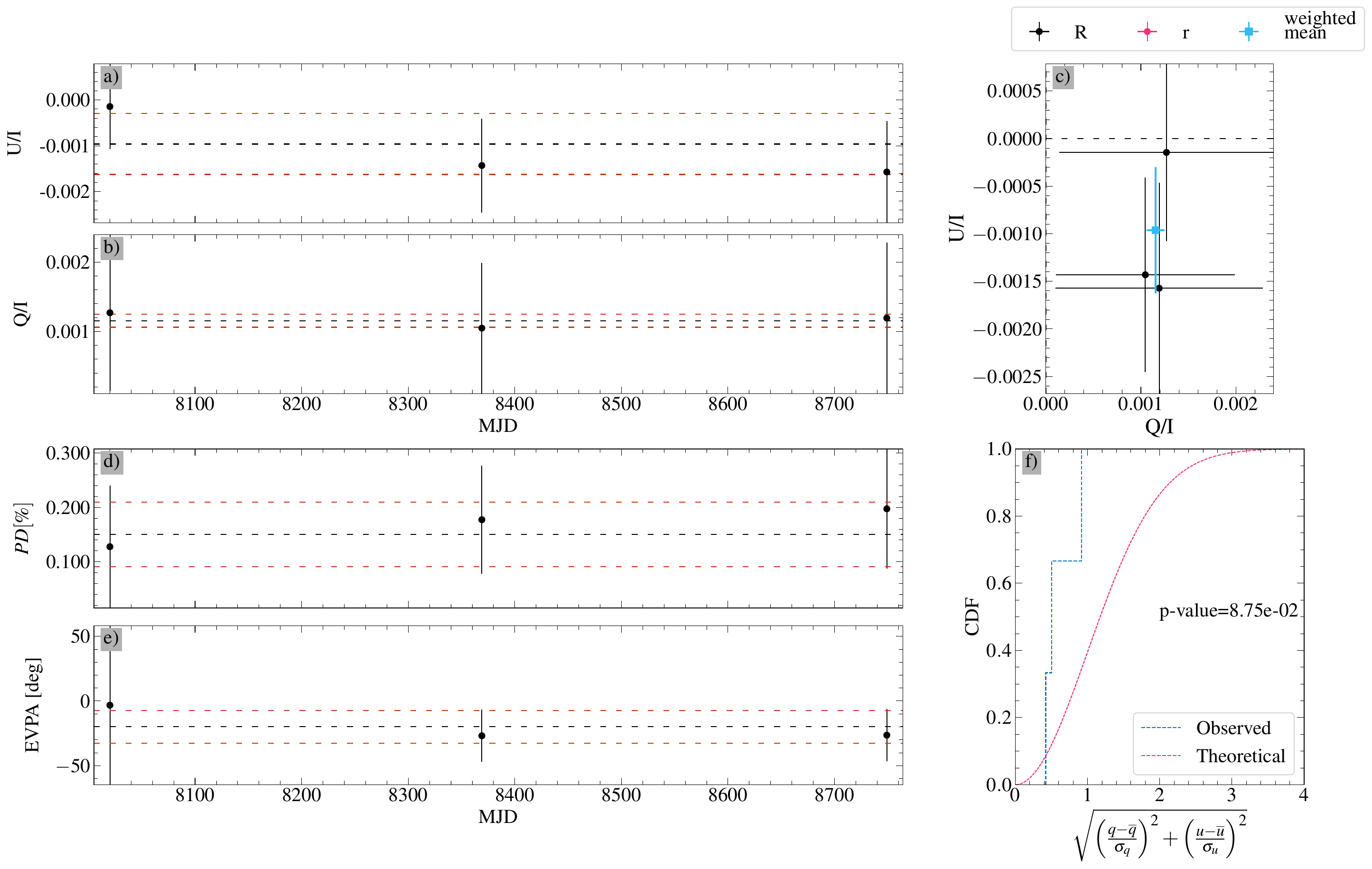}
  \caption{Same as Fig.~\ref{fig:B_0017+8135_82} for L\_96\_235, which has not enough measurements to judge it as varible of stable. }
  \label{fig:L_96_235}
\end{figure*}

\begin{figure*}
  \centering
  \includegraphics[width=0.95\textwidth]{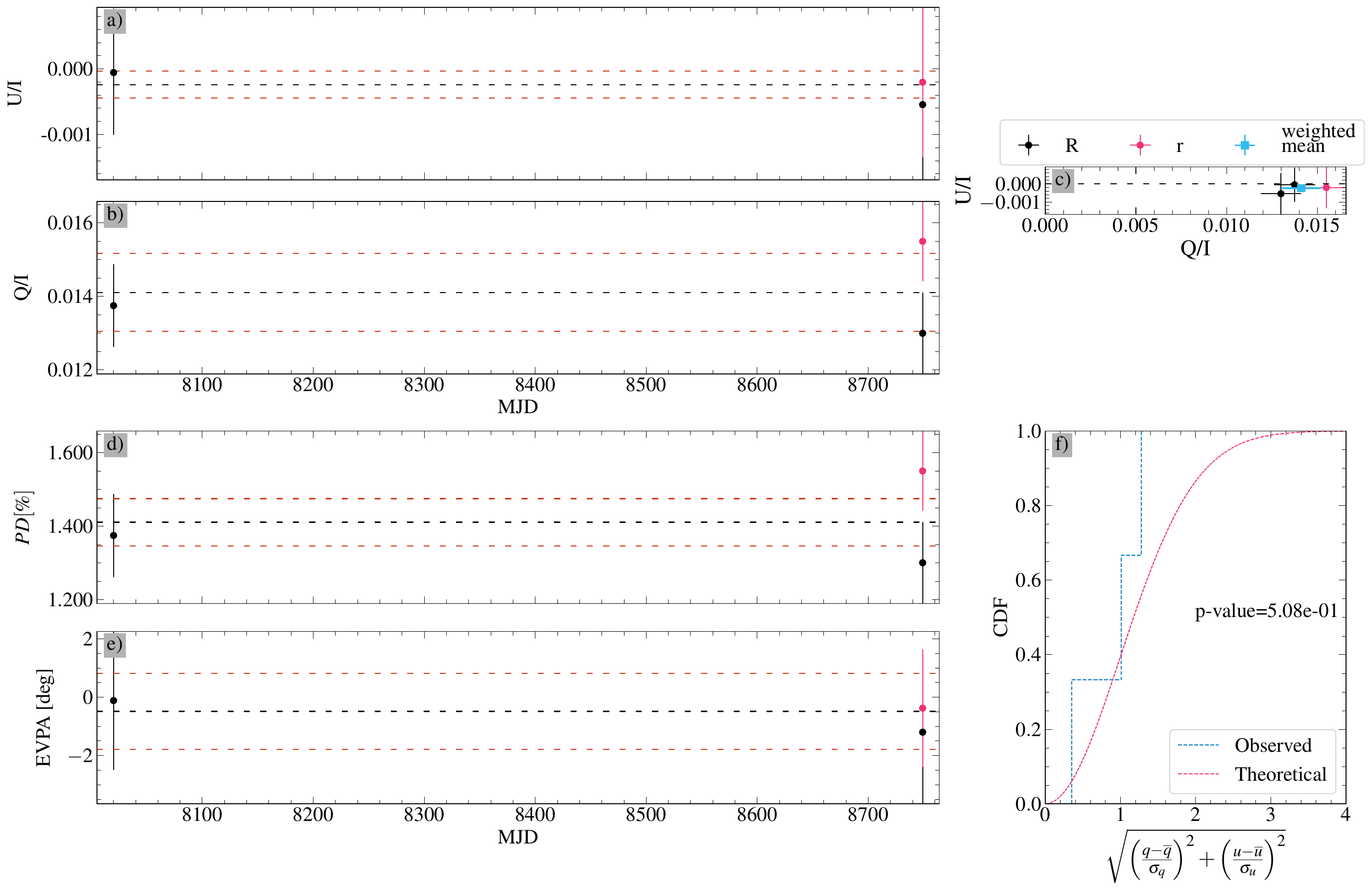}
  \caption{Same as Fig.~\ref{fig:B_0017+8135_82} for L\_97\_345, which has not enough measurements to judge it as varible of stable. }
  \label{fig:L_97_345}
\end{figure*}

\clearpage

\begin{figure*}
  \centering
  \includegraphics[width=0.95\textwidth]{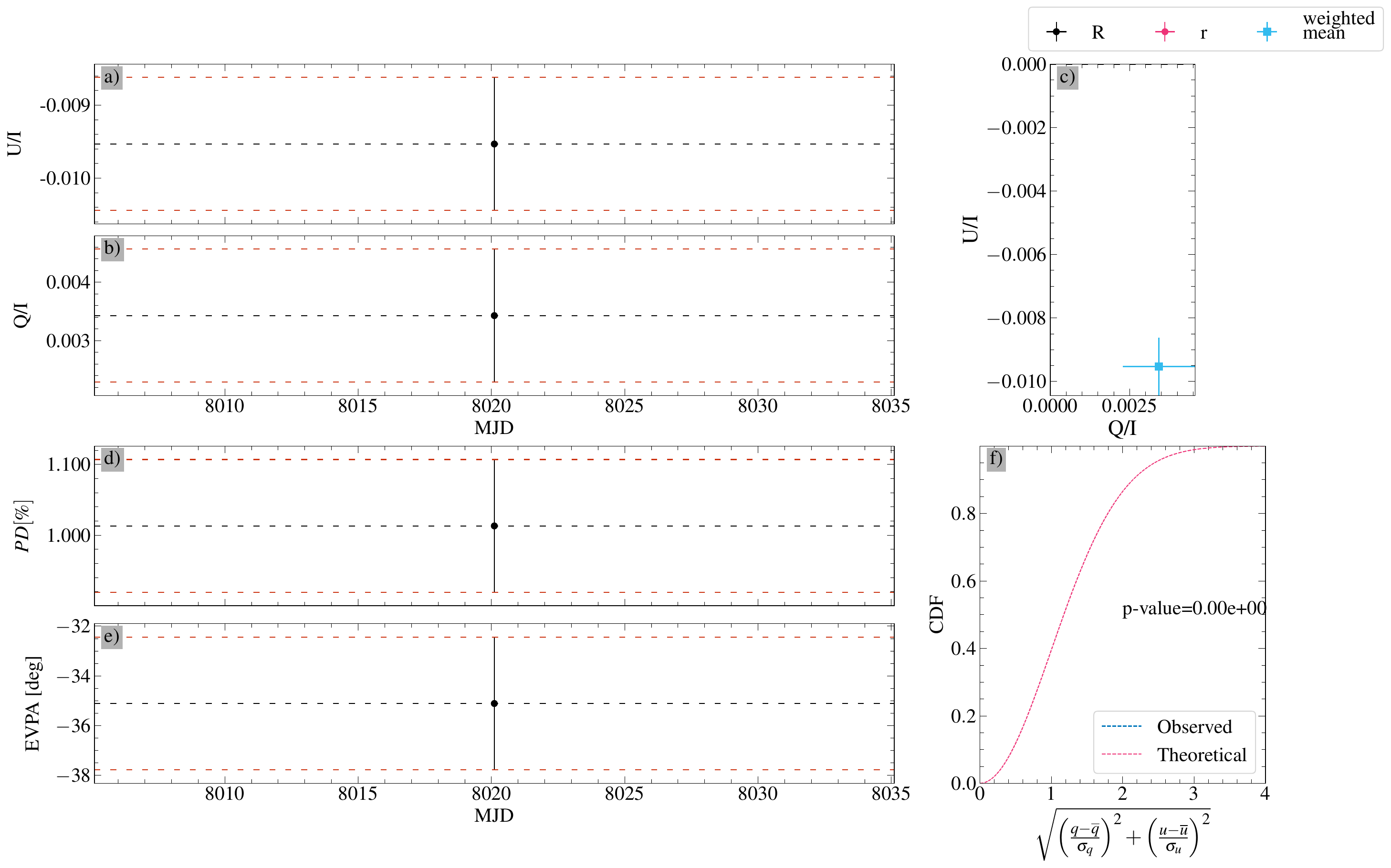}
  \caption{Same as Fig.~\ref{fig:B_0017+8135_82} for L\_97\_351, which has not enough measurements to judge it as varible of stable. }
  \label{fig:L_97_351}
\end{figure*}

\begin{figure*}
  \centering
  \includegraphics[width=0.95\textwidth]{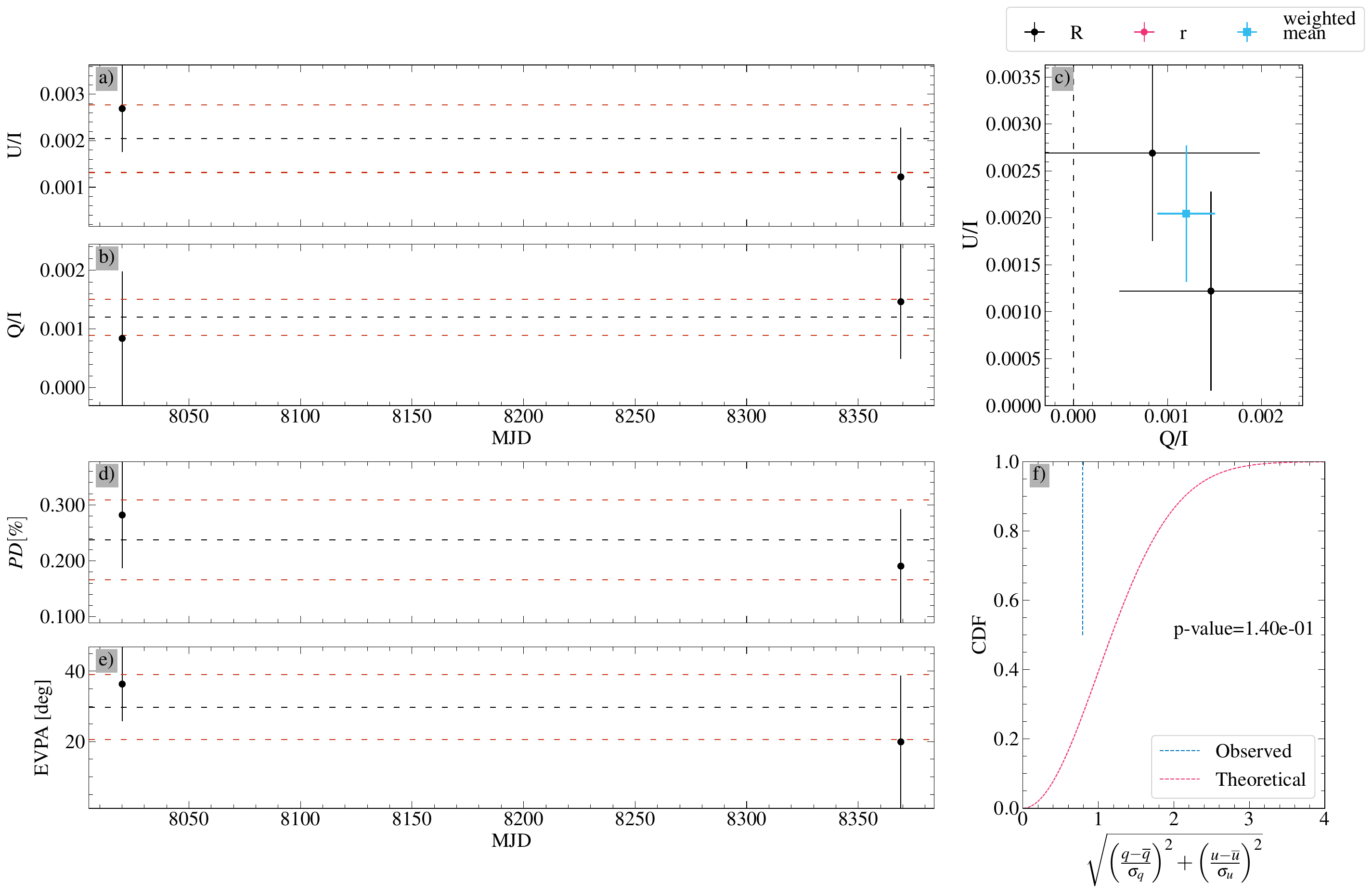}
  \caption{Same as Fig.~\ref{fig:B_0017+8135_82} for H\_HD255017, which has not enough measurements to judge it as varible of stable. }
  \label{fig:H_HD255017}
\end{figure*}

\clearpage

\begin{figure*}
  \centering
  \includegraphics[width=0.95\textwidth]{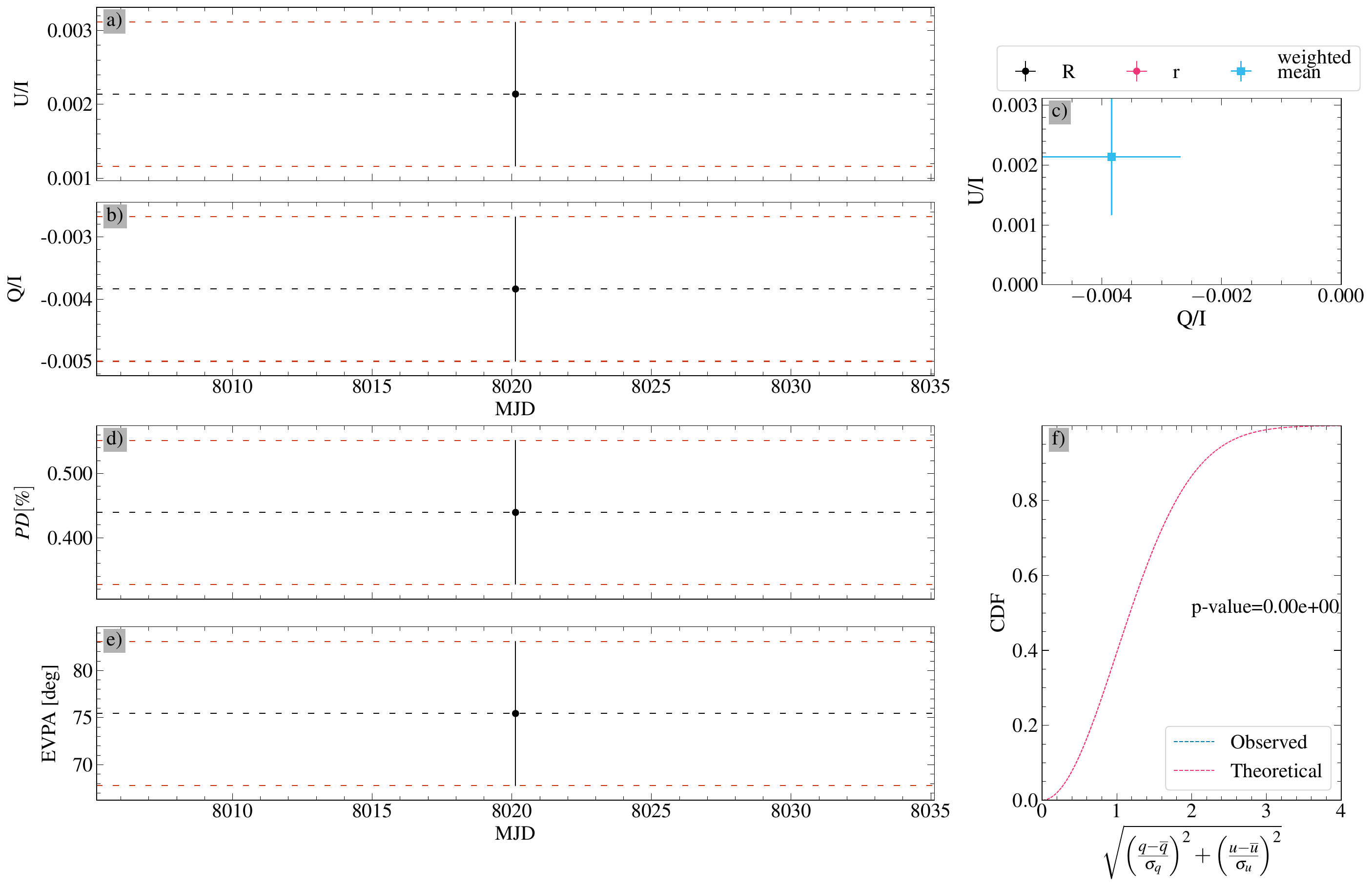}
  \caption{Same as Fig.~\ref{fig:B_0017+8135_82} for L\_98\_653, which has not enough measurements to judge it as varible of stable. }
  \label{fig:L_98_653}
\end{figure*}

\begin{figure*}
  \centering
  \includegraphics[width=0.95\textwidth]{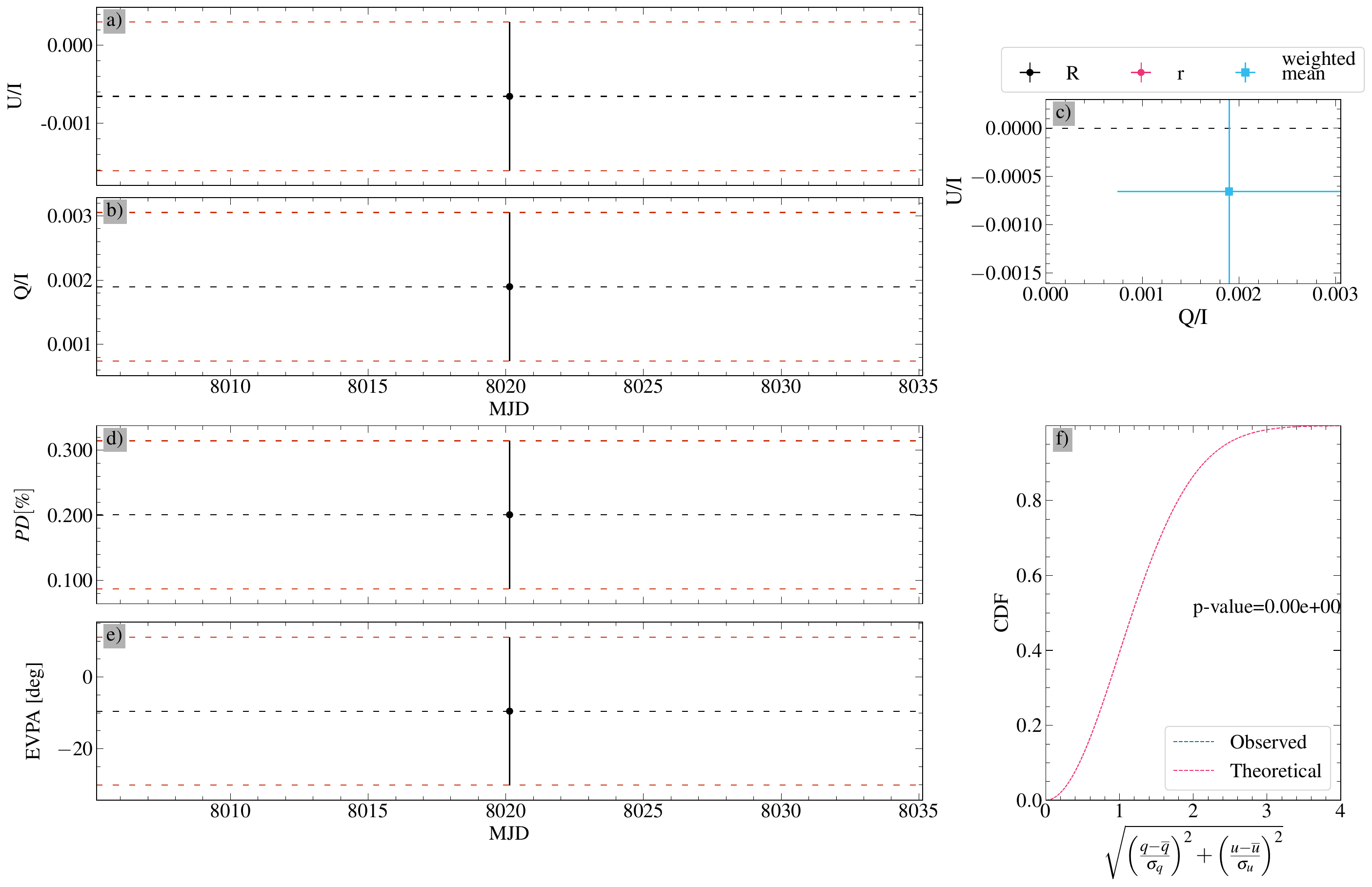}
  \caption{Same as Fig.~\ref{fig:B_0017+8135_82} for L\_98\_685, which has not enough measurements to judge it as varible of stable. }
  \label{fig:L_98_685}
\end{figure*}

\clearpage

\begin{figure*}
  \centering
  \includegraphics[width=0.95\textwidth]{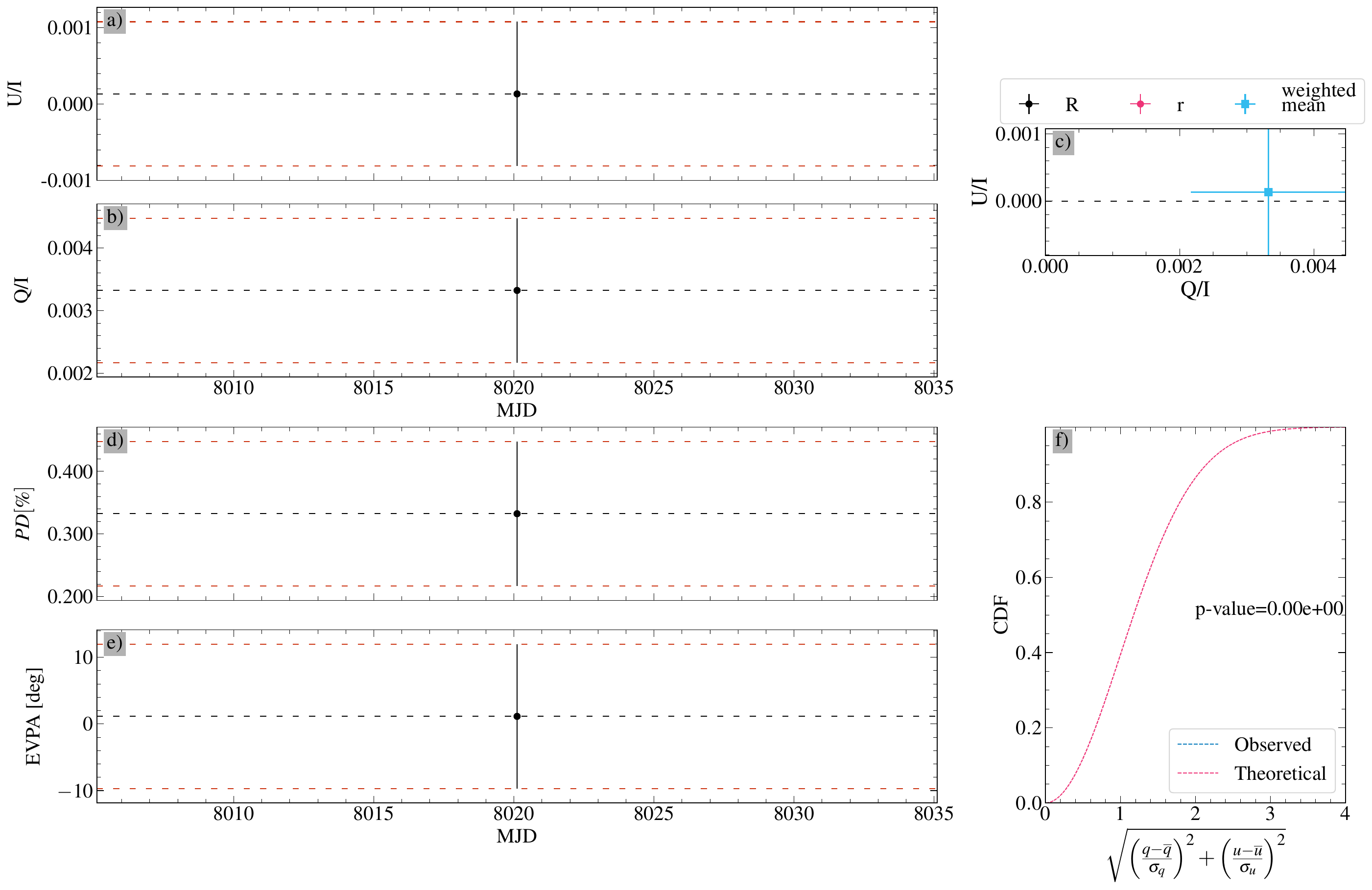}
  \caption{Same as Fig.~\ref{fig:B_0017+8135_82} for H\_HD57702, which has not enough measurements to judge it as varible of stable. }
  \label{fig:H_HD57702}
\end{figure*}

\begin{figure*}
  \centering
  \includegraphics[width=0.95\textwidth]{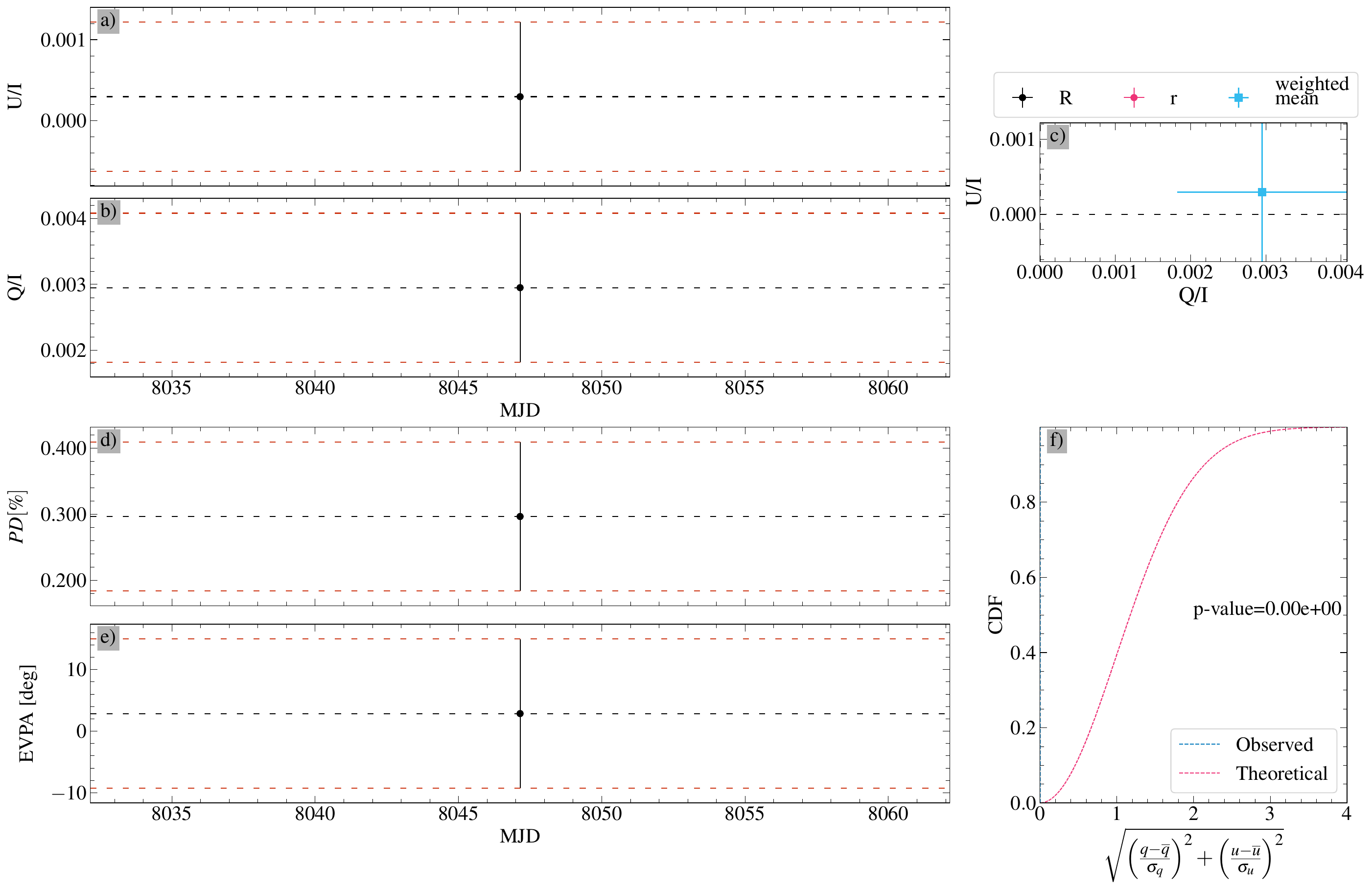}
  \caption{Same as Fig.~\ref{fig:B_0017+8135_82} for L\_RU\_152D, which has not enough measurements to judge it as varible of stable. }
  \label{fig:L_RU_152D}
\end{figure*}

\clearpage

\begin{figure*}
  \centering
  \includegraphics[width=0.95\textwidth]{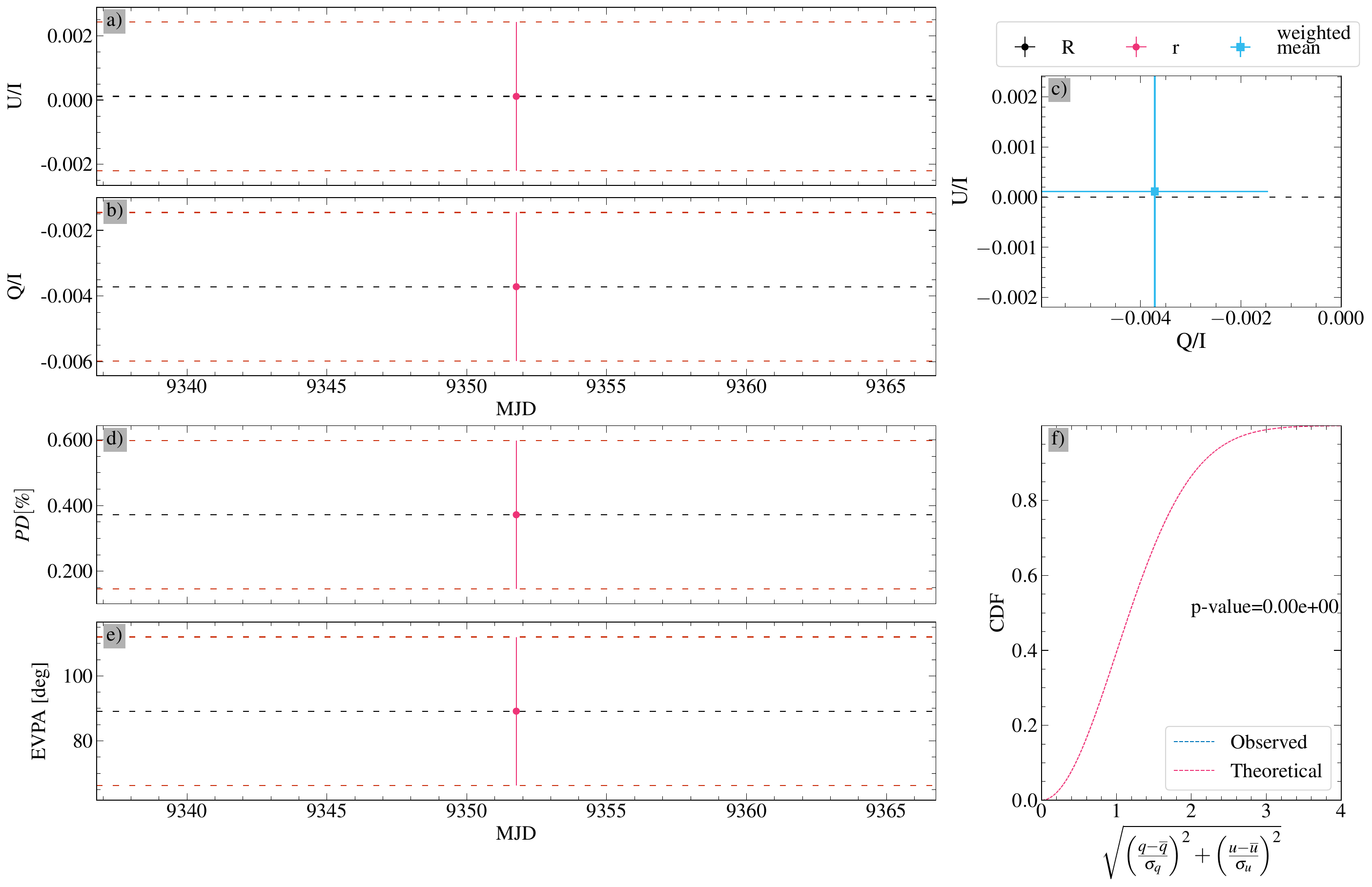}
  \caption{Same as Fig.~\ref{fig:B_0017+8135_82} for L\_PG0918+029D, which has not enough measurements to judge it as varible of stable. }
  \label{fig:L_PG0918+029D}
\end{figure*}

\begin{figure*}
  \centering
  \includegraphics[width=0.95\textwidth]{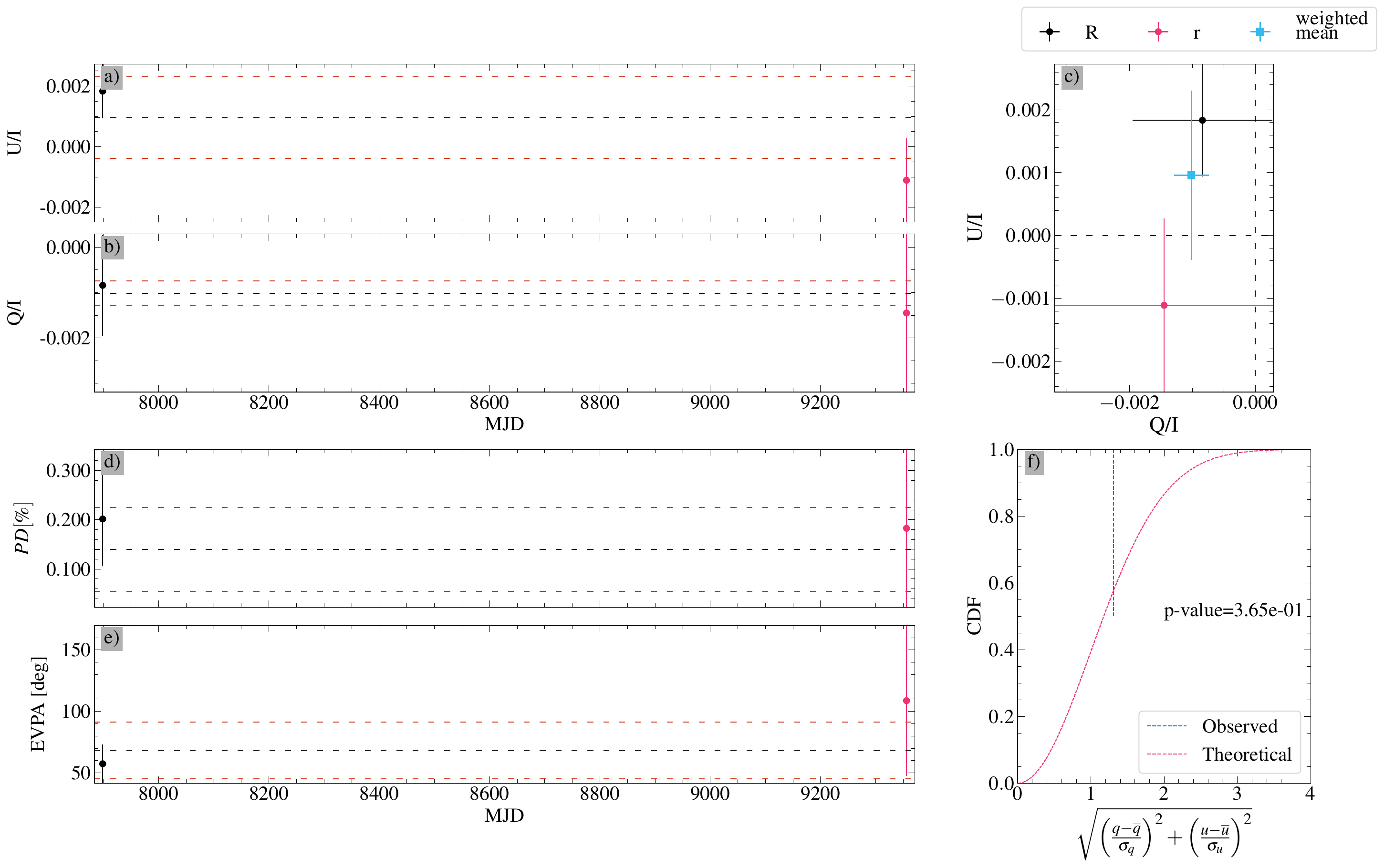}
  \caption{Same as Fig.~\ref{fig:B_0017+8135_82} for Z\_HD81418, which has not enough measurements to judge it as varible of stable. }
  \label{fig:Z_HD81418}
\end{figure*}

\clearpage

\begin{figure*}
  \centering
  \includegraphics[width=0.95\textwidth]{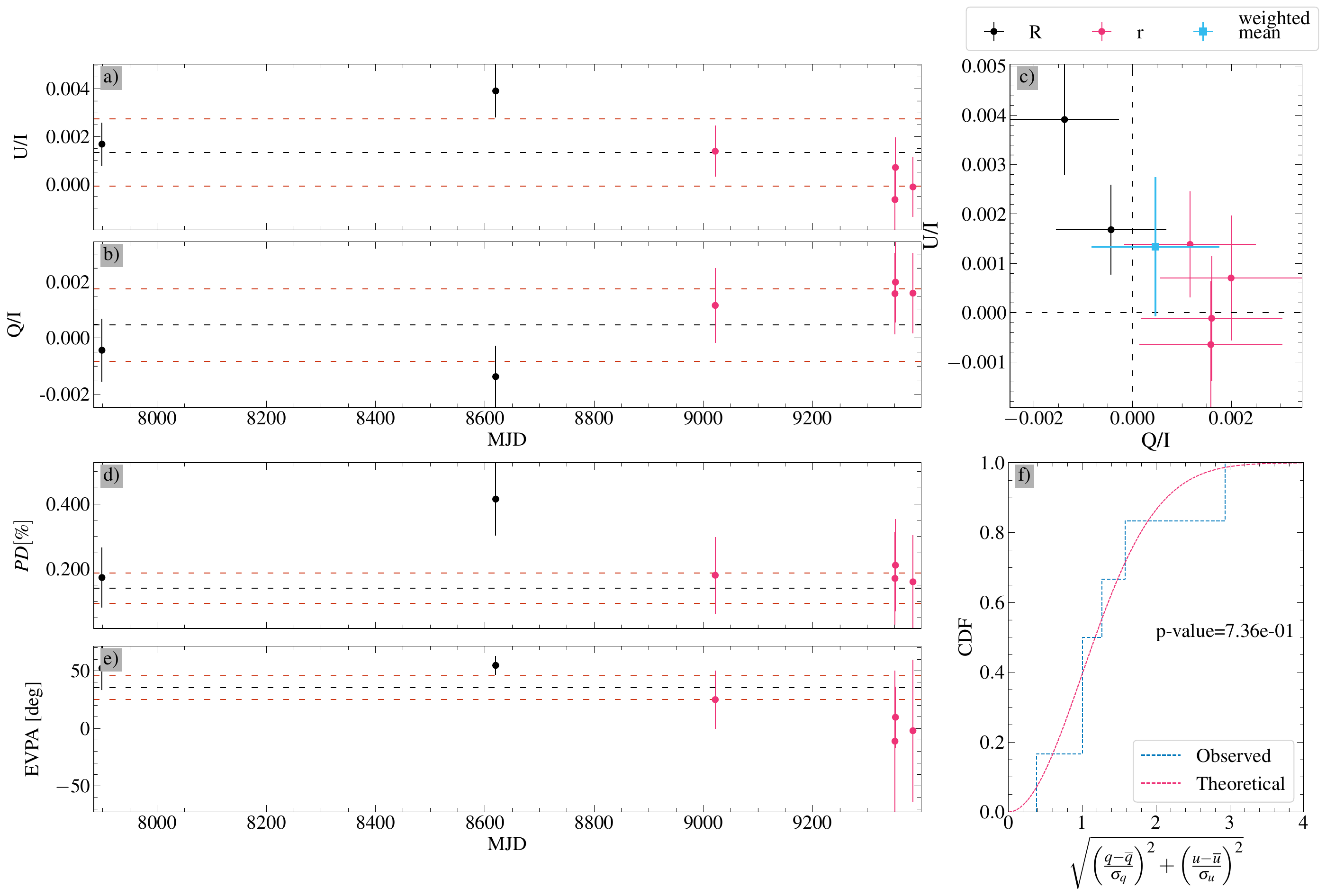}
  \caption{Same as Fig.~\ref{fig:B_0017+8135_82} for Z\_HD85471, which is found to be stable. }
  \label{fig:Z_HD85471}
\end{figure*}

\begin{figure*}
  \centering
  \includegraphics[width=0.95\textwidth]{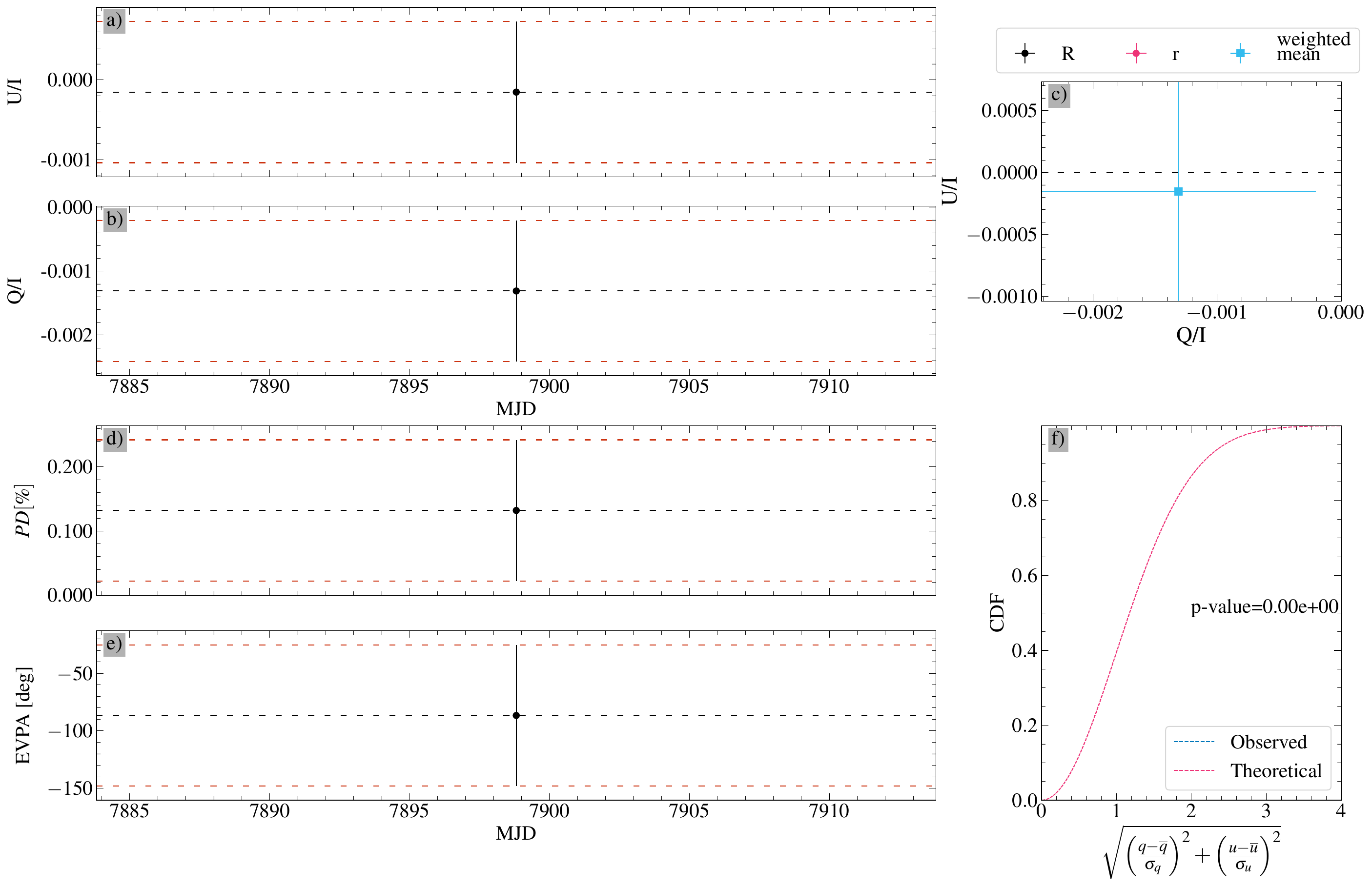}
  \caption{Same as Fig.~\ref{fig:B_0017+8135_82} for Z\_HD86321, which has not enough measurements to judge it as varible of stable. }
  \label{fig:Z_HD86321}
\end{figure*}

\clearpage

\begin{figure*}
  \centering
  \includegraphics[width=0.95\textwidth]{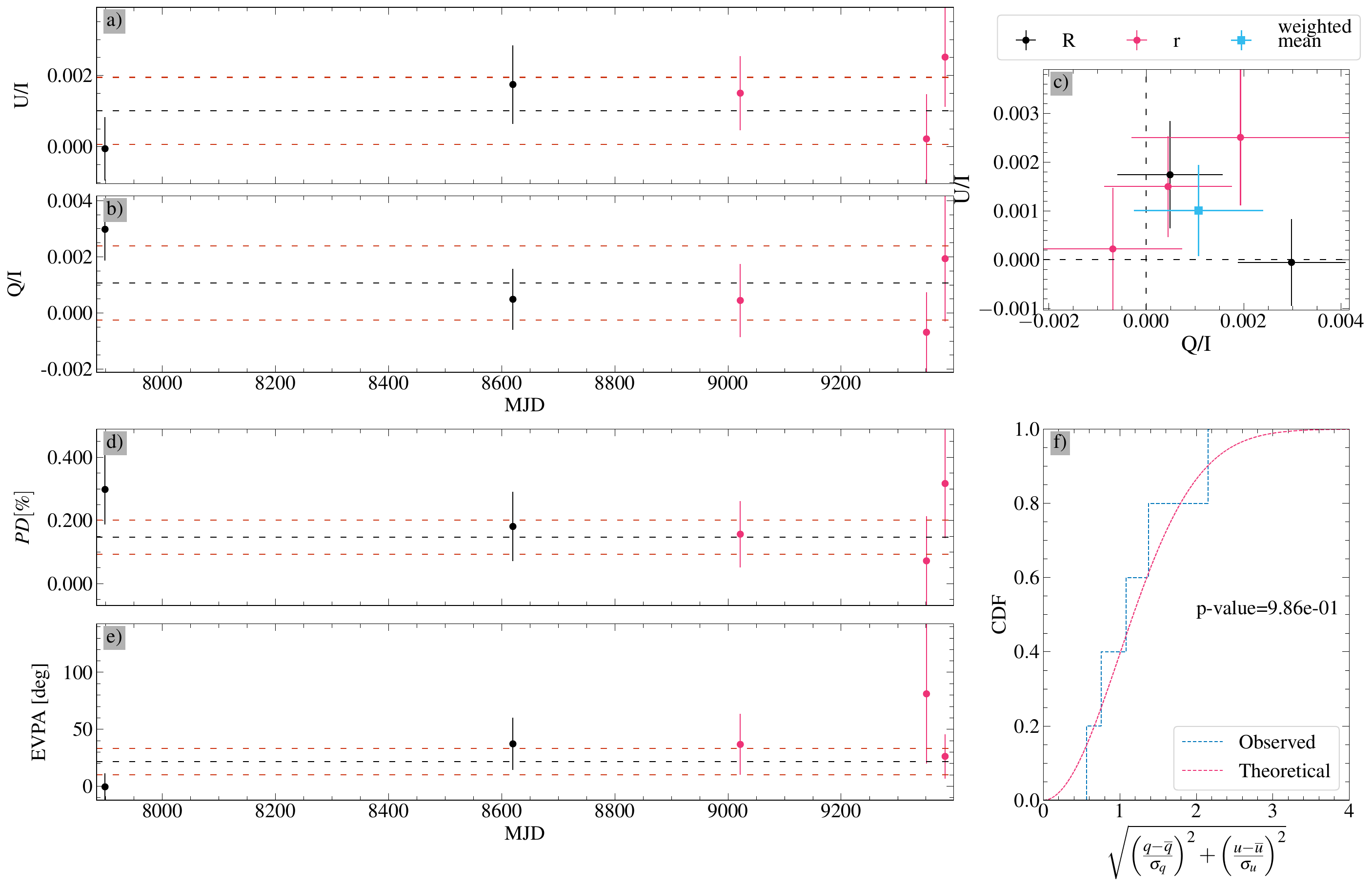}
  \caption{Same as Fig.~\ref{fig:B_0017+8135_82} for Z\_HD87582, which is found to be stable. }
  \label{fig:Z_HD87582}
\end{figure*}

\begin{figure*}
  \centering
  \includegraphics[width=0.95\textwidth]{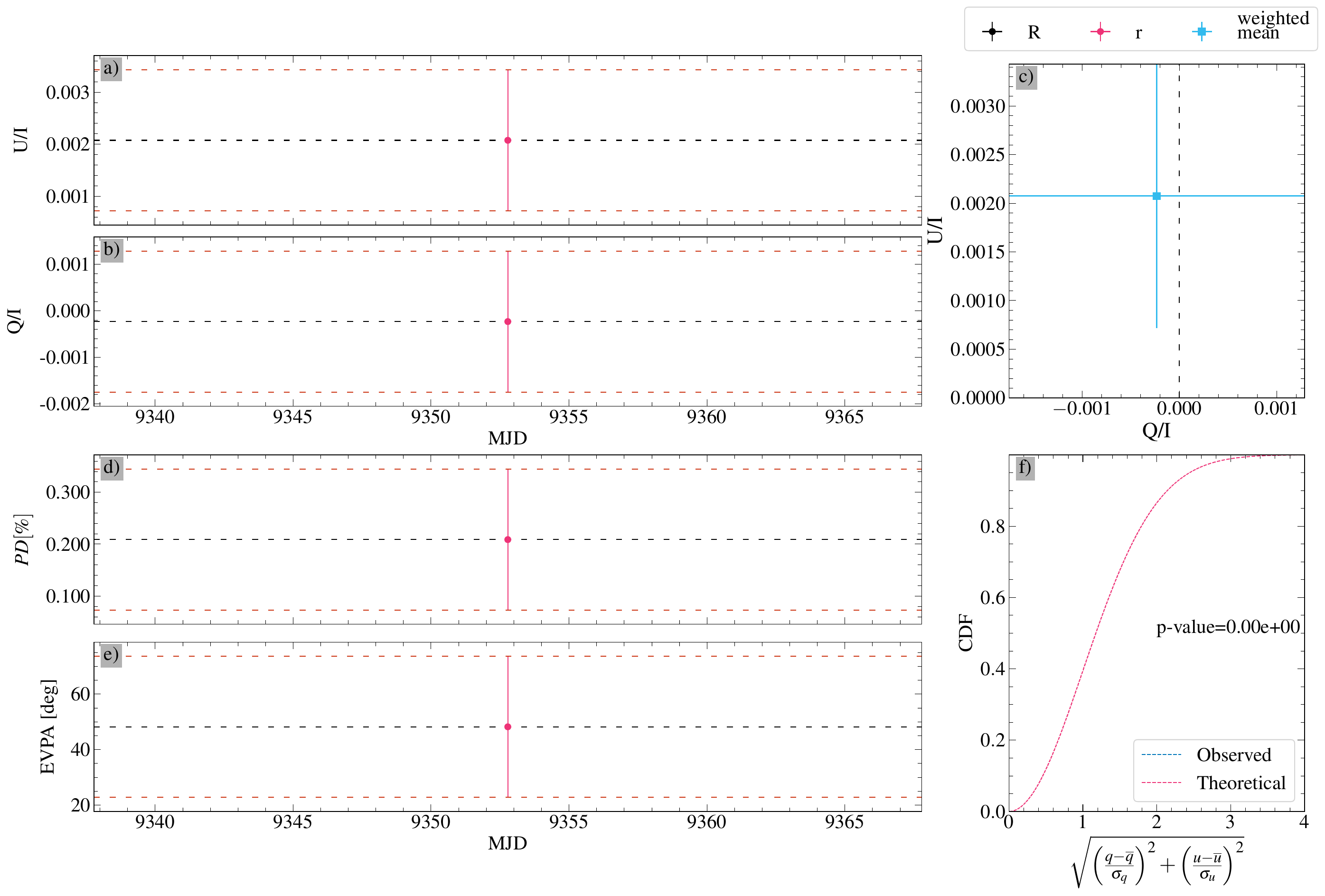}
  \caption{Same as Fig.~\ref{fig:B_0017+8135_82} for L\_PG1047+003, which has not enough measurements to judge it as varible of stable. }
  \label{fig:L_PG1047+003}
\end{figure*}

\clearpage

\begin{figure*}
  \centering
  \includegraphics[width=0.95\textwidth]{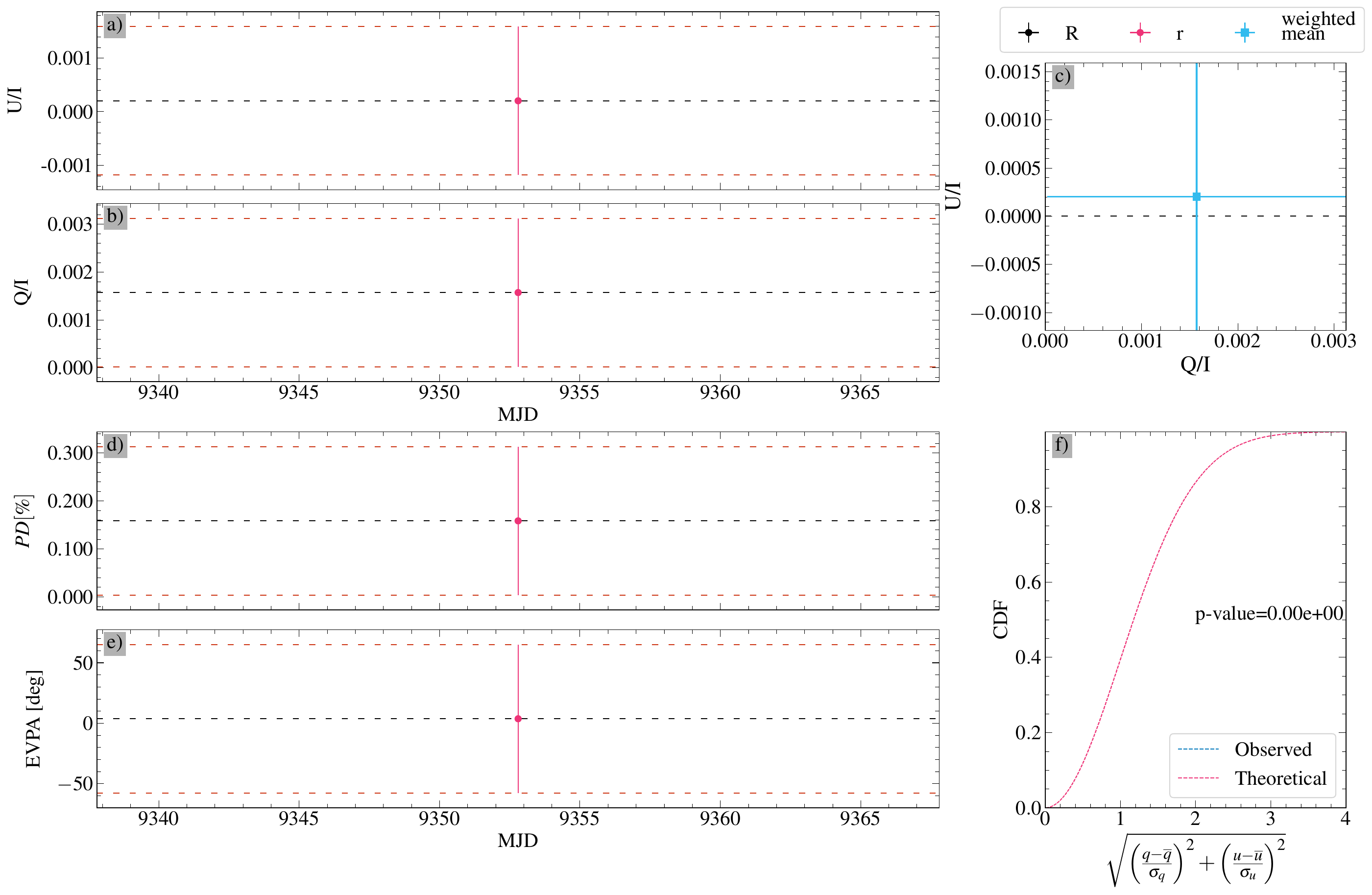}
  \caption{Same as Fig.~\ref{fig:B_0017+8135_82} for L\_PG1047+003B, which has not enough measurements to judge it as varible of stable. }
  \label{fig:L_PG1047+003B}
\end{figure*}

\begin{figure*}
  \centering
  \includegraphics[width=0.95\textwidth]{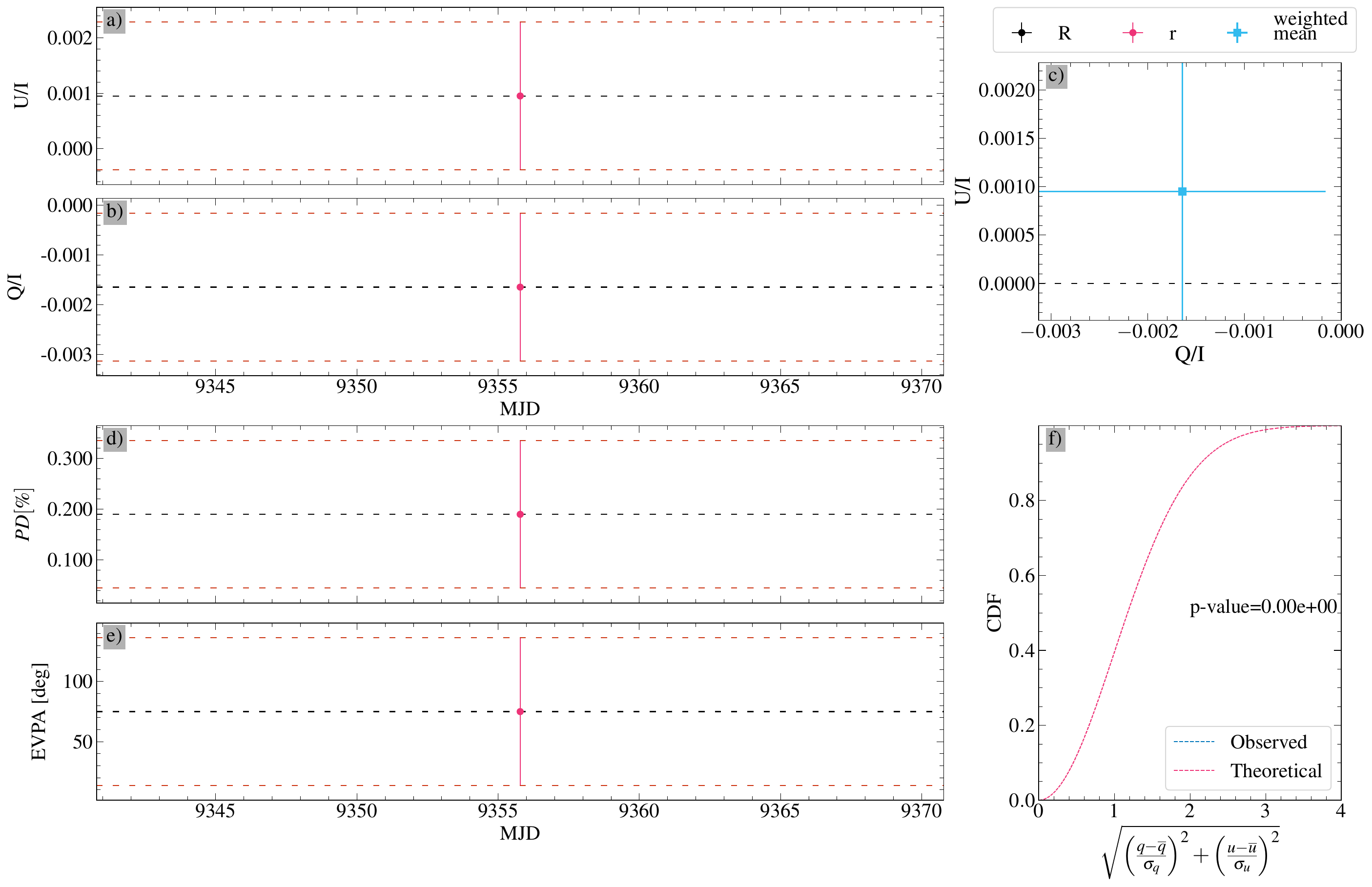}
  \caption{Same as Fig.~\ref{fig:B_0017+8135_82} for L\_PG1047+003C, which has not enough measurements to judge it as varible of stable. }
  \label{fig:L_PG1047+003C}
\end{figure*}

\clearpage

\begin{figure*}
  \centering
  \includegraphics[width=0.95\textwidth]{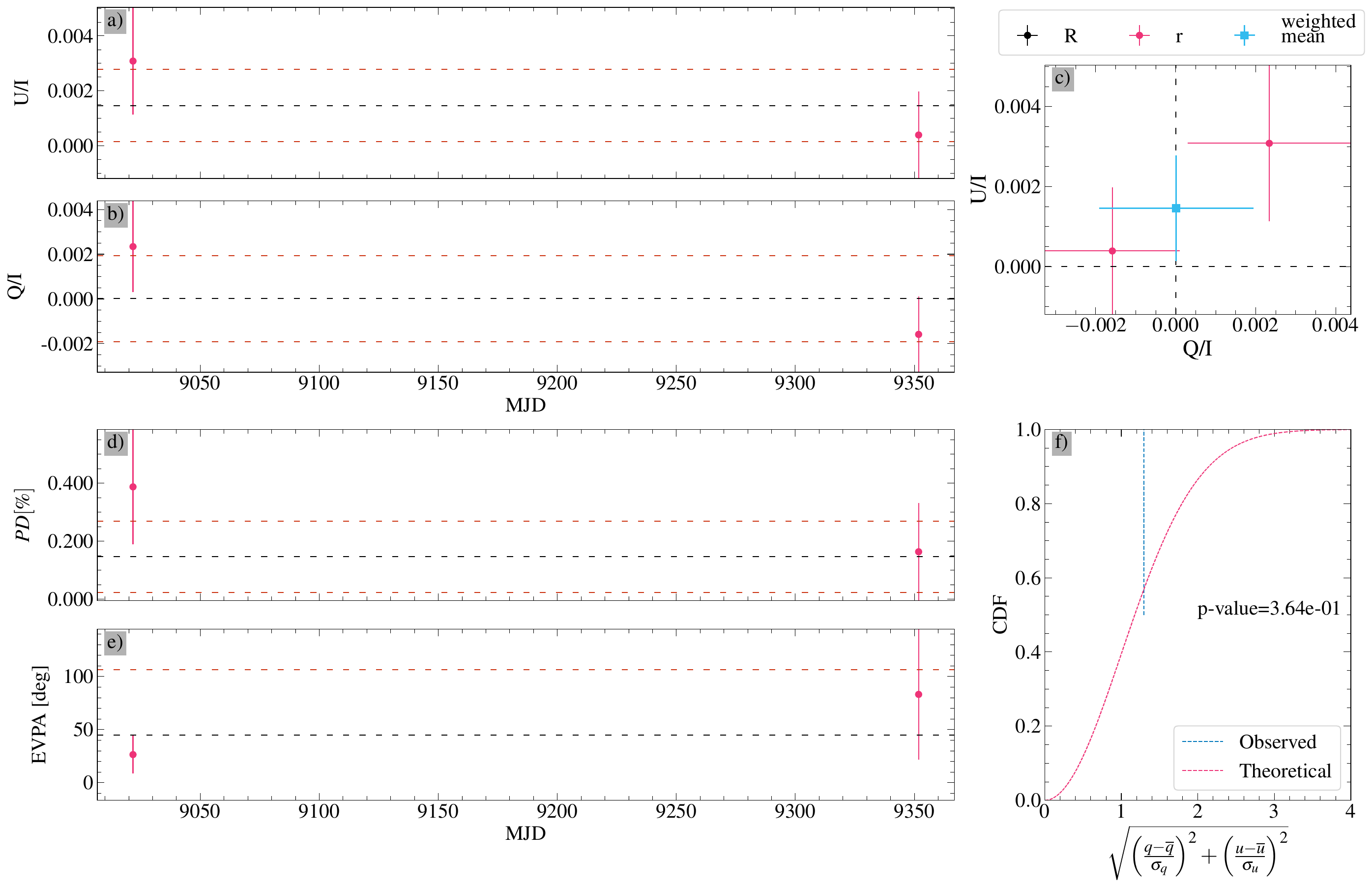}
  \caption{Same as Fig.~\ref{fig:B_0017+8135_82} for L\_G163\_50, which has not enough measurements to judge it as varible of stable. }
  \label{fig:L_G163_50}
\end{figure*}

\begin{figure*}
  \centering
  \includegraphics[width=0.95\textwidth]{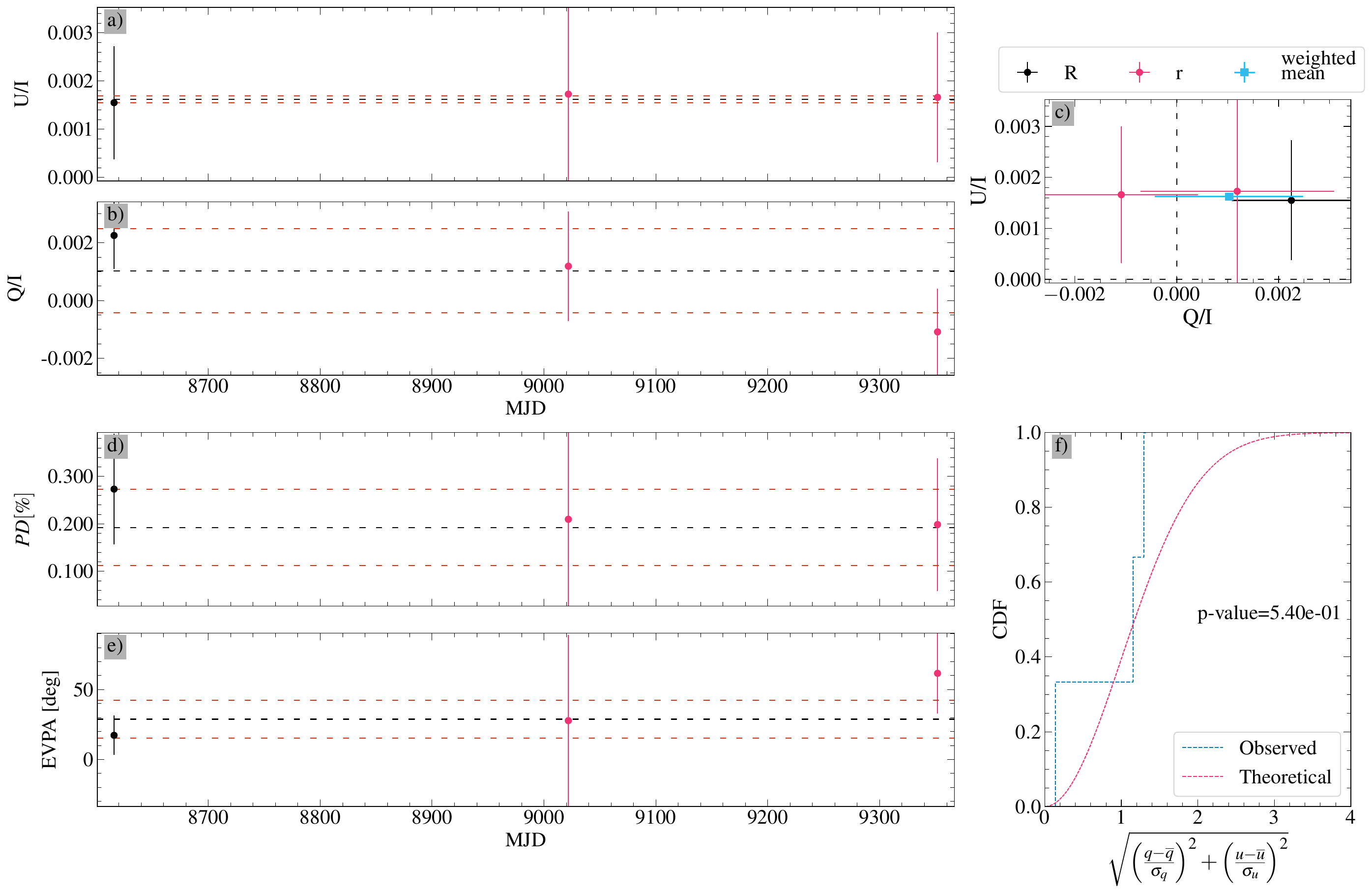}
  \caption{Same as Fig.~\ref{fig:B_0017+8135_82} for L\_G163\_51, which has not enough measurements to judge it as varible of stable. }
  \label{fig:L_G163_51}
\end{figure*}

\clearpage

\begin{figure*}
  \centering
  \includegraphics[width=0.95\textwidth]{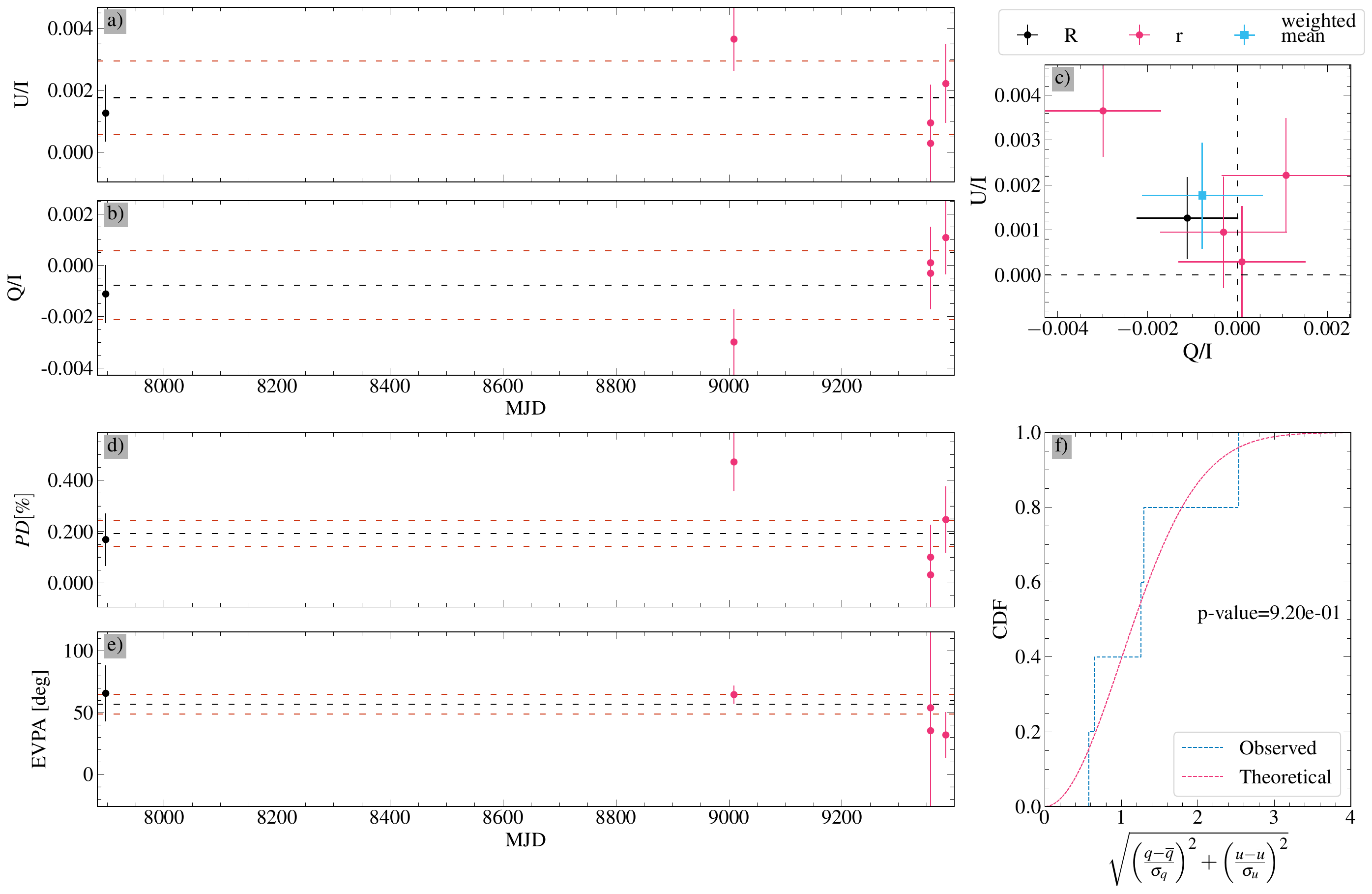}
  \caption{Same as Fig.~\ref{fig:B_0017+8135_82} for Z\_HD96589, which is found to be stable. }
  \label{fig:Z_HD96589}
\end{figure*}

\begin{figure*}
  \centering
  \includegraphics[width=0.95\textwidth]{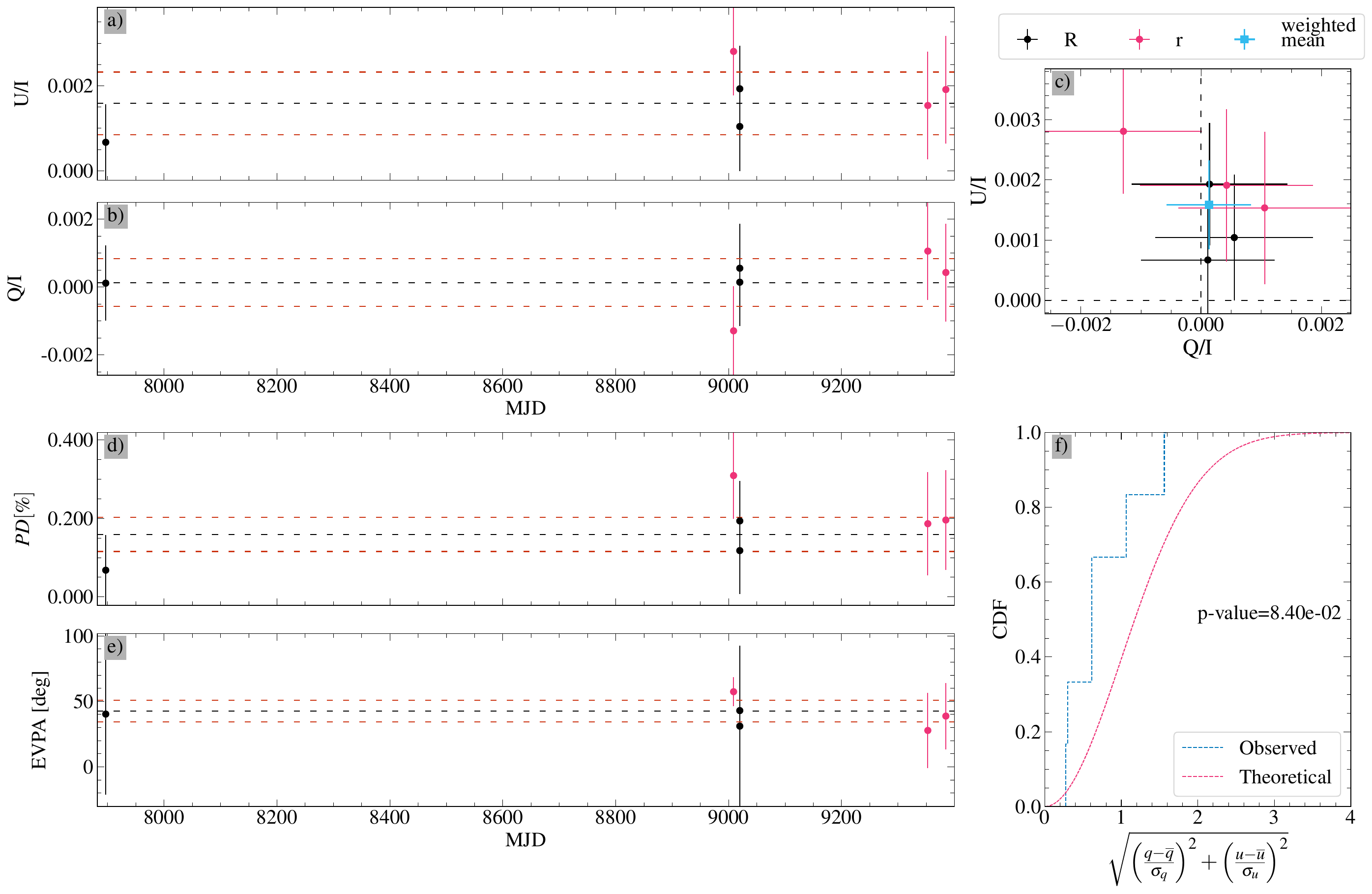}
  \caption{Same as Fig.~\ref{fig:B_0017+8135_82} for Z\_HD97853, which is found to be stable. }
  \label{fig:Z_HD97853}
\end{figure*}

\clearpage

\begin{figure*}
  \centering
  \includegraphics[width=0.95\textwidth]{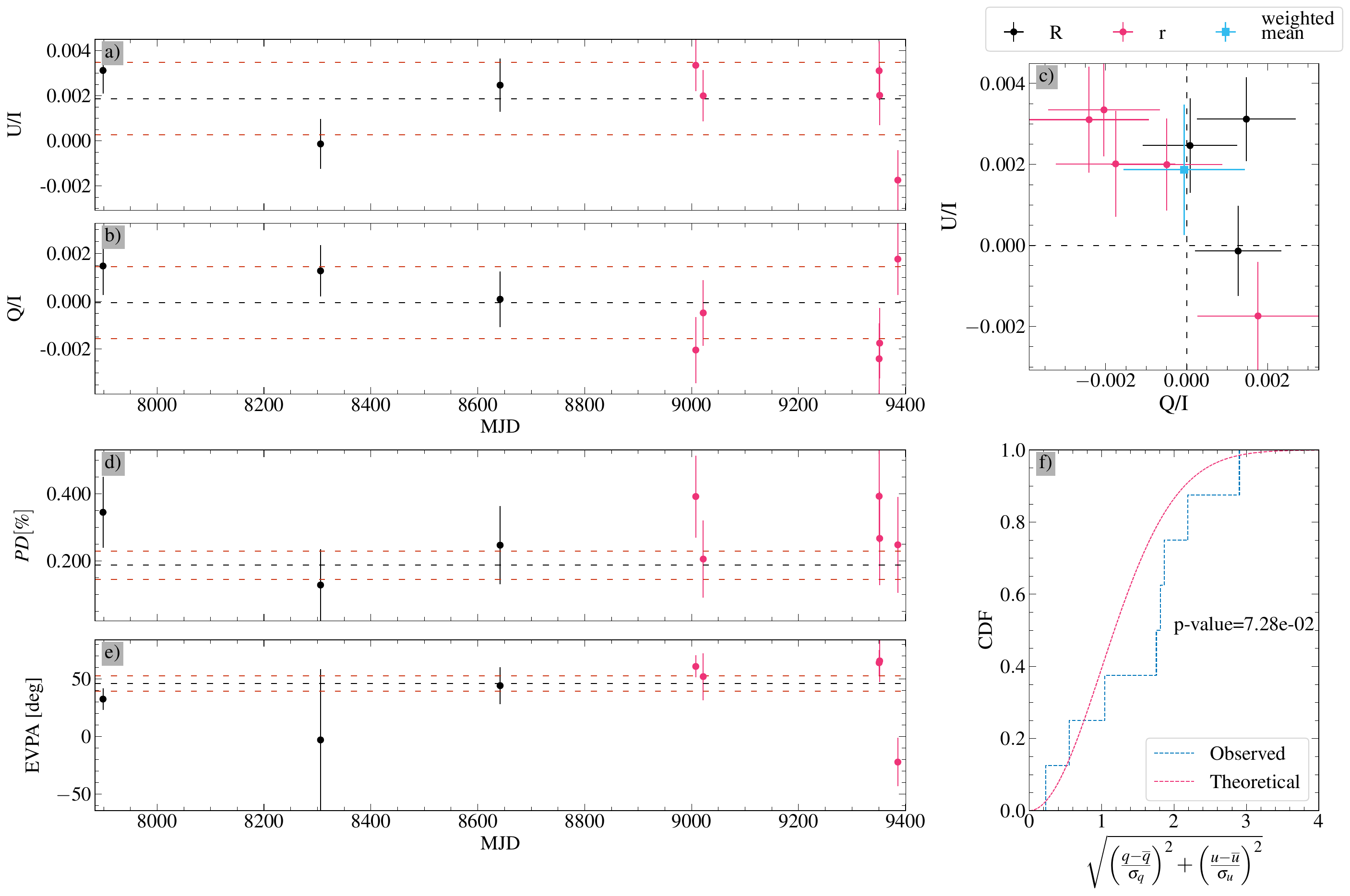}
  \caption{Same as Fig.~\ref{fig:B_0017+8135_82} for L\_104\_334, which is found to be stable. }
  \label{fig:L_104_334}
\end{figure*}

\begin{figure*}
  \centering
  \includegraphics[width=0.95\textwidth]{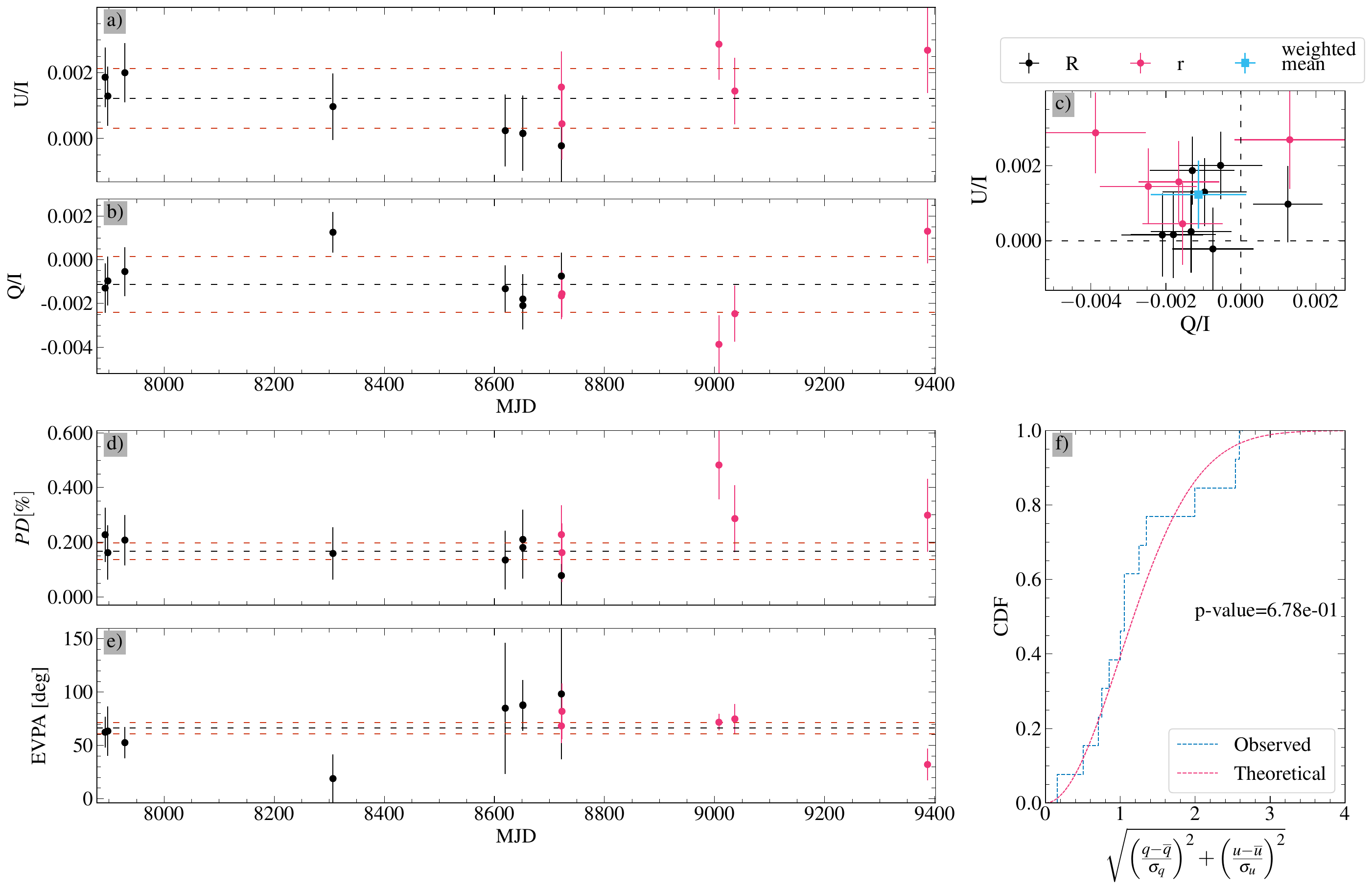}
  \caption{Same as Fig.~\ref{fig:B_0017+8135_82} for Z\_HD116513, which is found to be stable. }
  \label{fig:Z_HD116513}
\end{figure*}

\clearpage

\begin{figure*}
  \centering
  \includegraphics[width=0.95\textwidth]{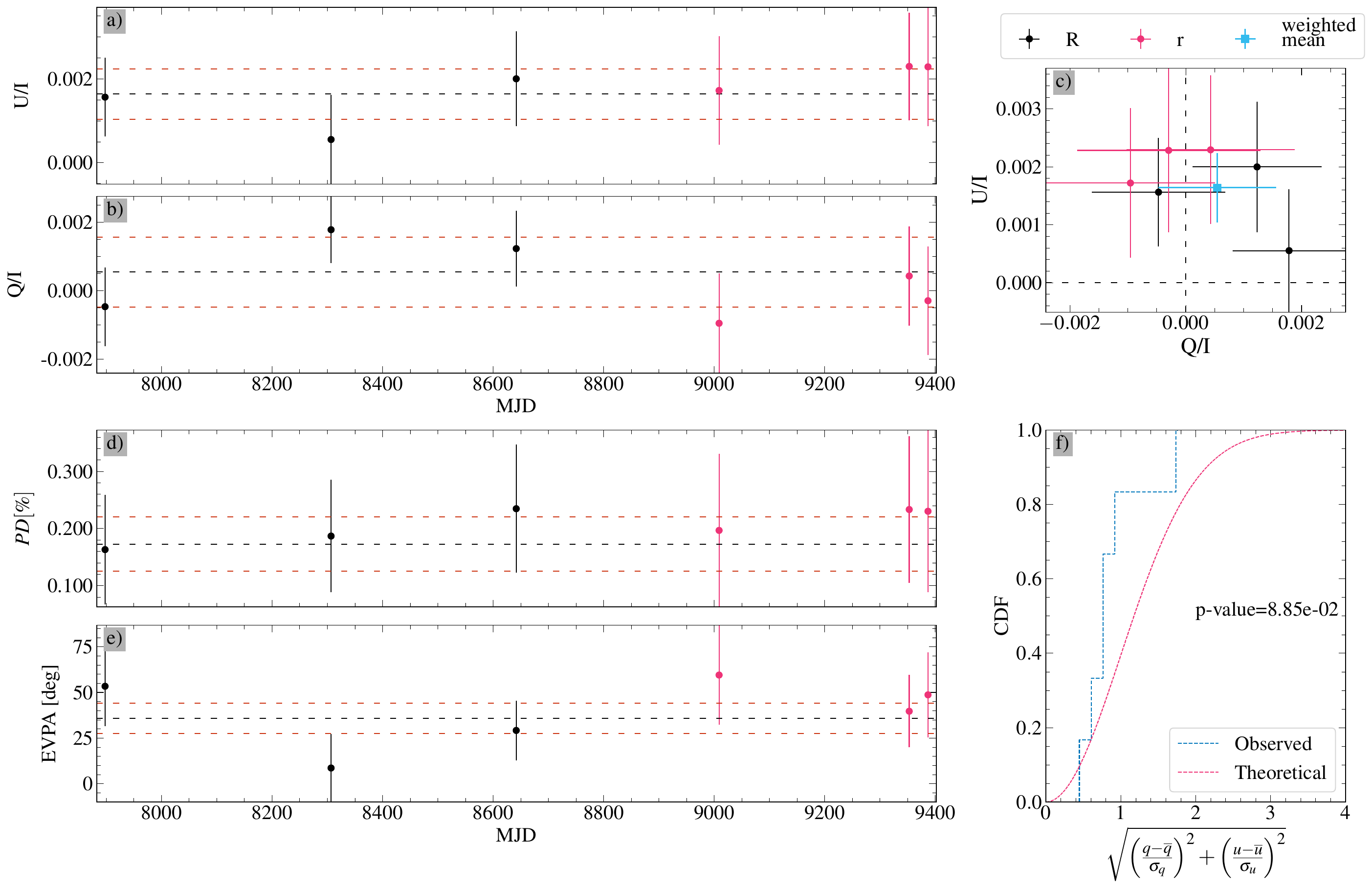}
  \caption{Same as Fig.~\ref{fig:B_0017+8135_82} for L\_PG1323$-$085B, which is found to be stable. }
  \label{fig:L_PG1323-085B}
\end{figure*}

\begin{figure*}
  \centering
  \includegraphics[width=0.95\textwidth]{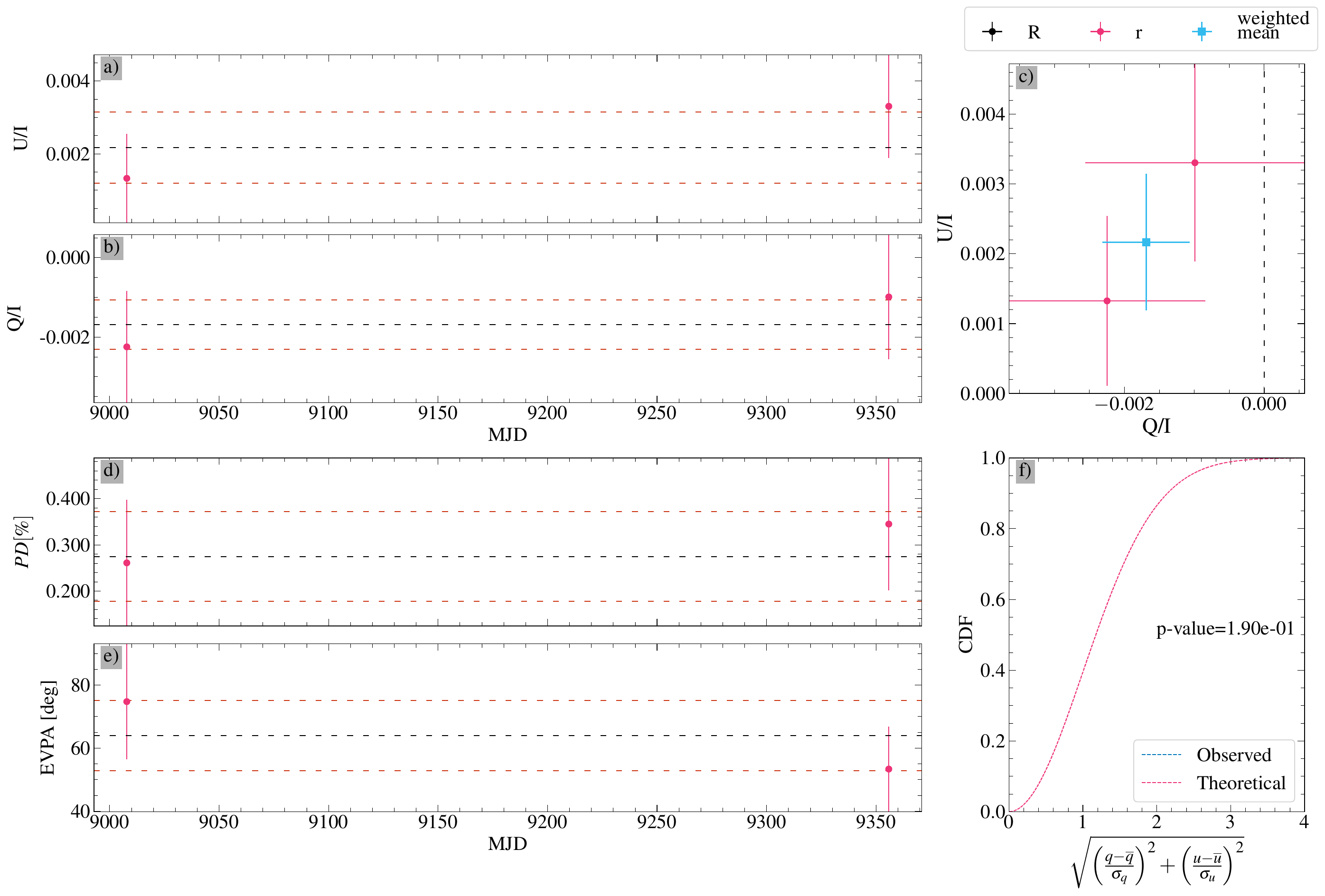}
  \caption{Same as Fig.~\ref{fig:B_0017+8135_82} for L\_PG1323$-$085C, which has not enough measurements to judge it as varible of stable. }
  \label{fig:L_PG1323-085C}
\end{figure*}

\clearpage

\begin{figure*}
  \centering
  \includegraphics[width=0.95\textwidth]{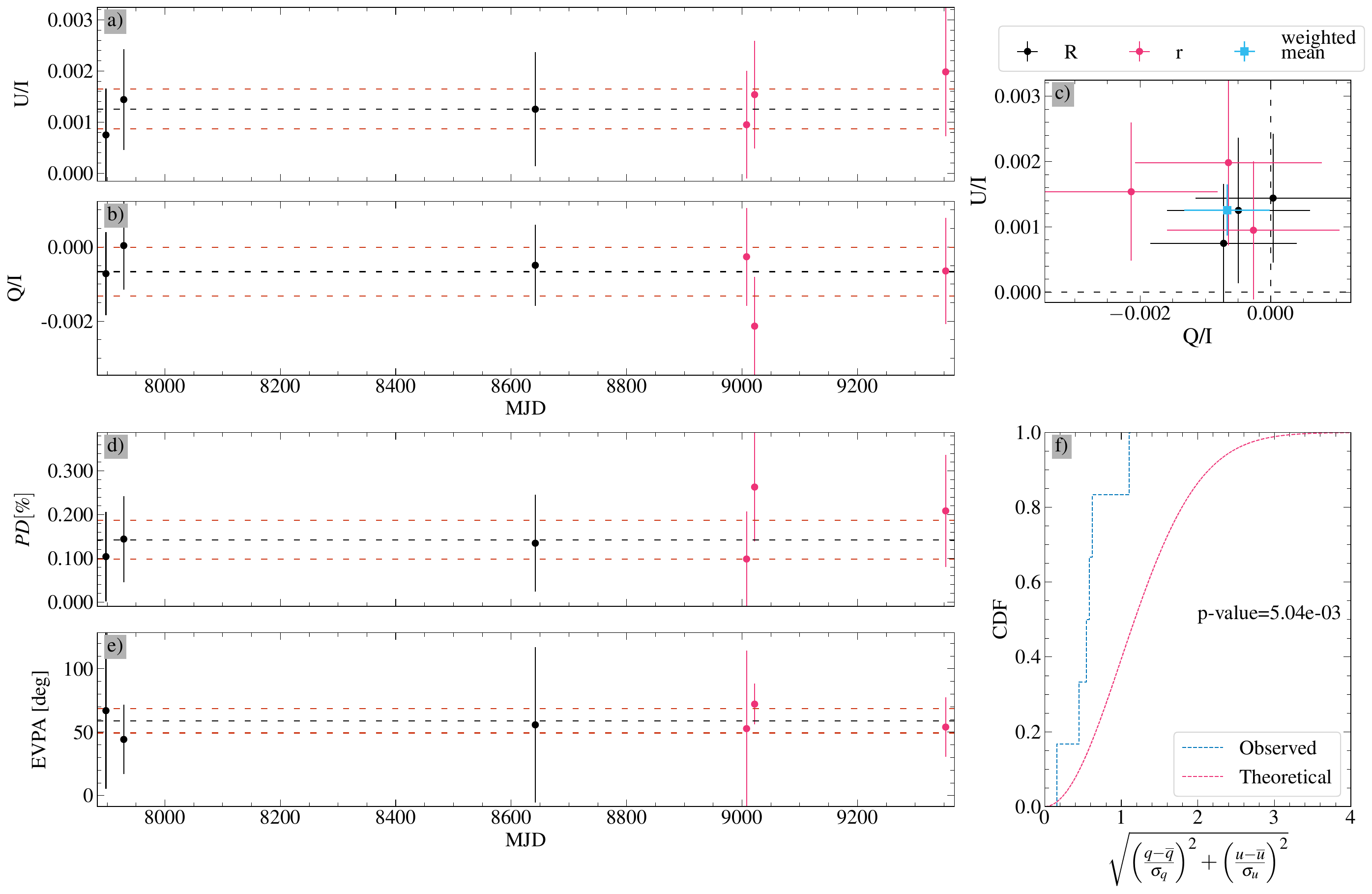}
  \caption{Same as Fig.~\ref{fig:B_0017+8135_82} for L\_PG1323$-$085D, which is found to be variable. }
  \label{fig:L_PG1323-085D}
\end{figure*}

\begin{figure*}
  \centering
  \includegraphics[width=0.95\textwidth]{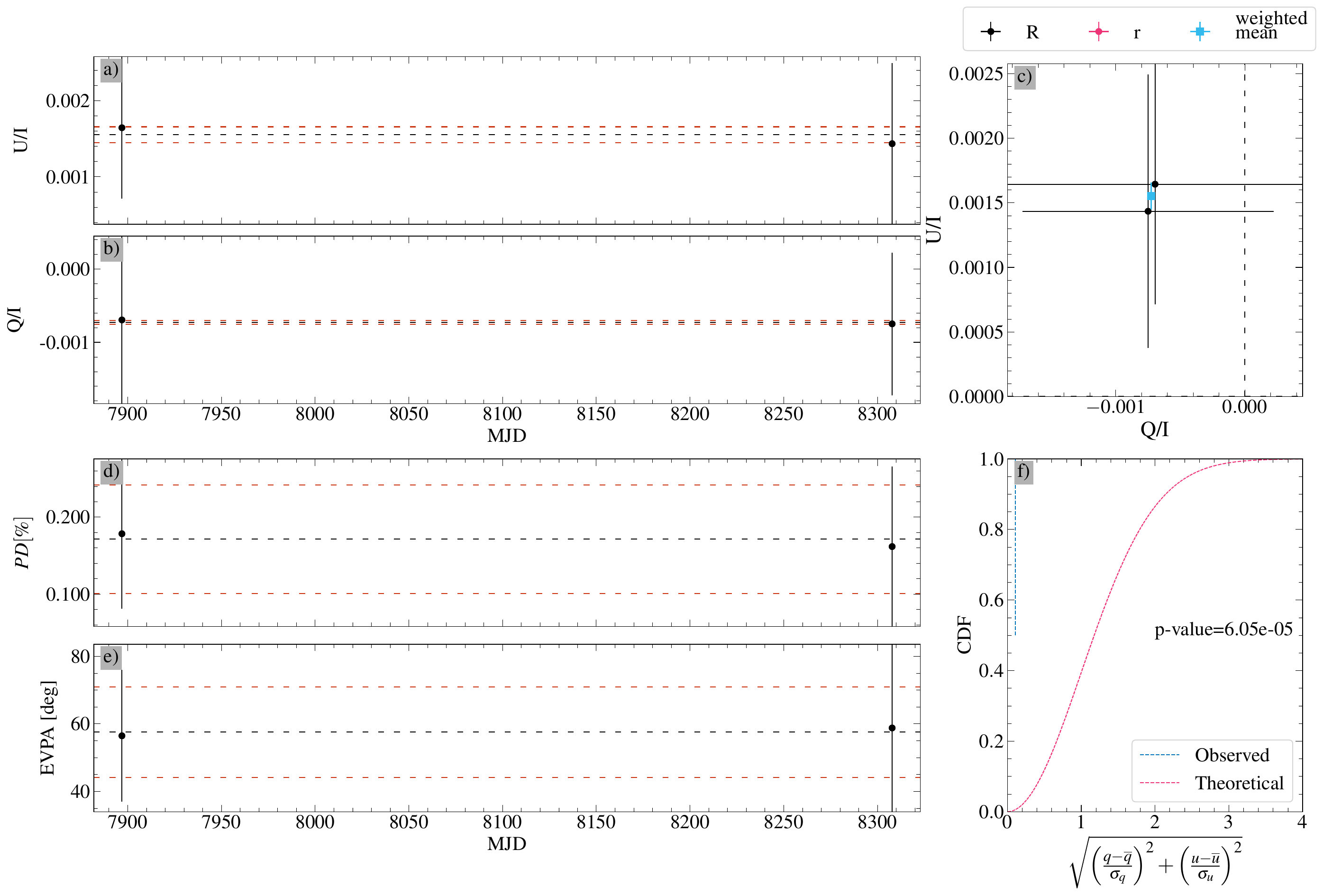}
  \caption{Same as Fig.~\ref{fig:B_0017+8135_82} for Z\_HD120010, which has not enough measurements to judge it as varible of stable. }
  \label{fig:Z_HD120010}
\end{figure*}

\clearpage

\begin{figure*}
  \centering
  \includegraphics[width=0.95\textwidth]{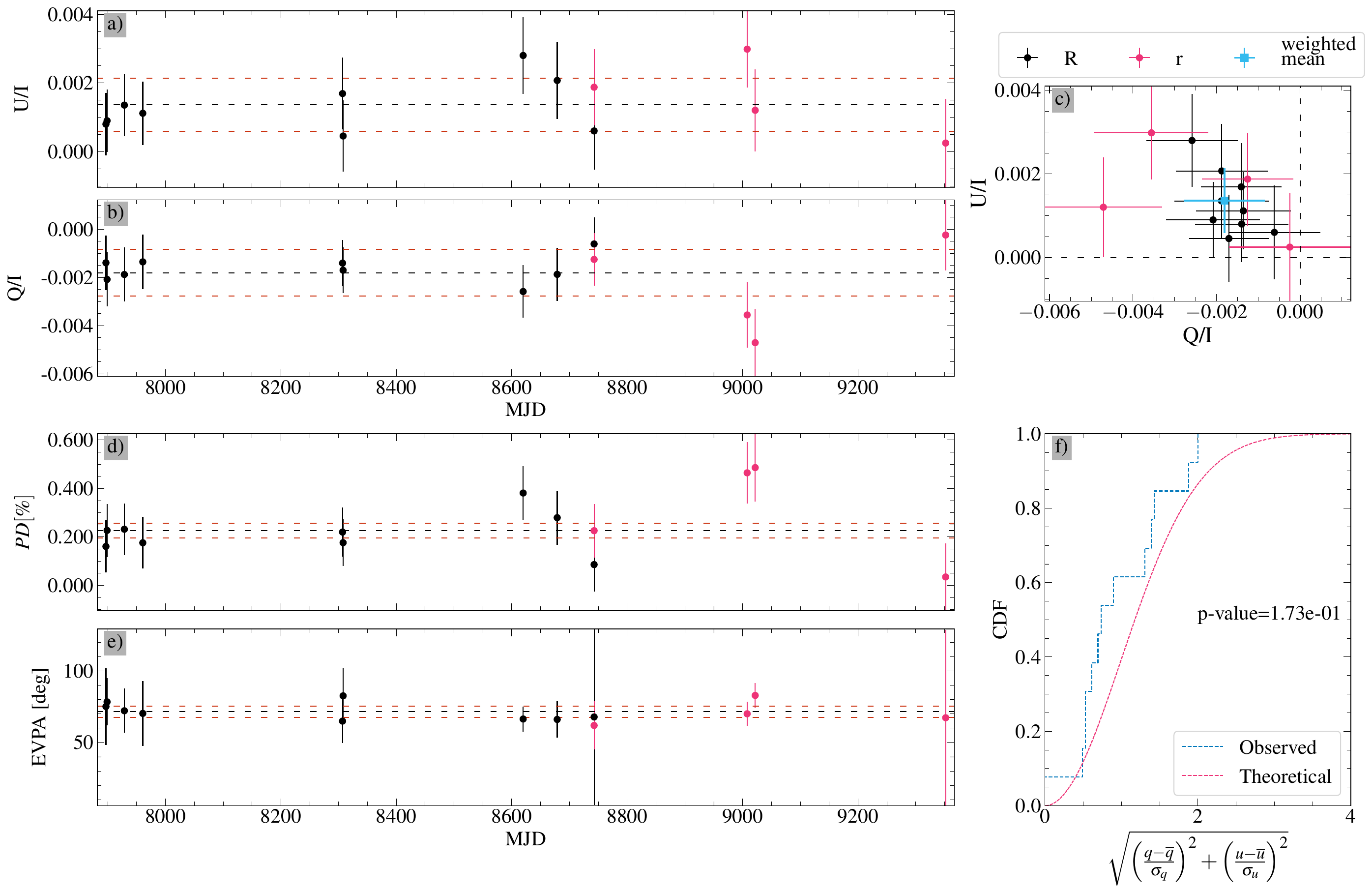}
  \caption{Same as Fig.~\ref{fig:B_0017+8135_82} for Z\_HD121859, which is found to be stable. }
  \label{fig:Z_HD121859}
\end{figure*}

\begin{figure*}
  \centering
  \includegraphics[width=0.95\textwidth]{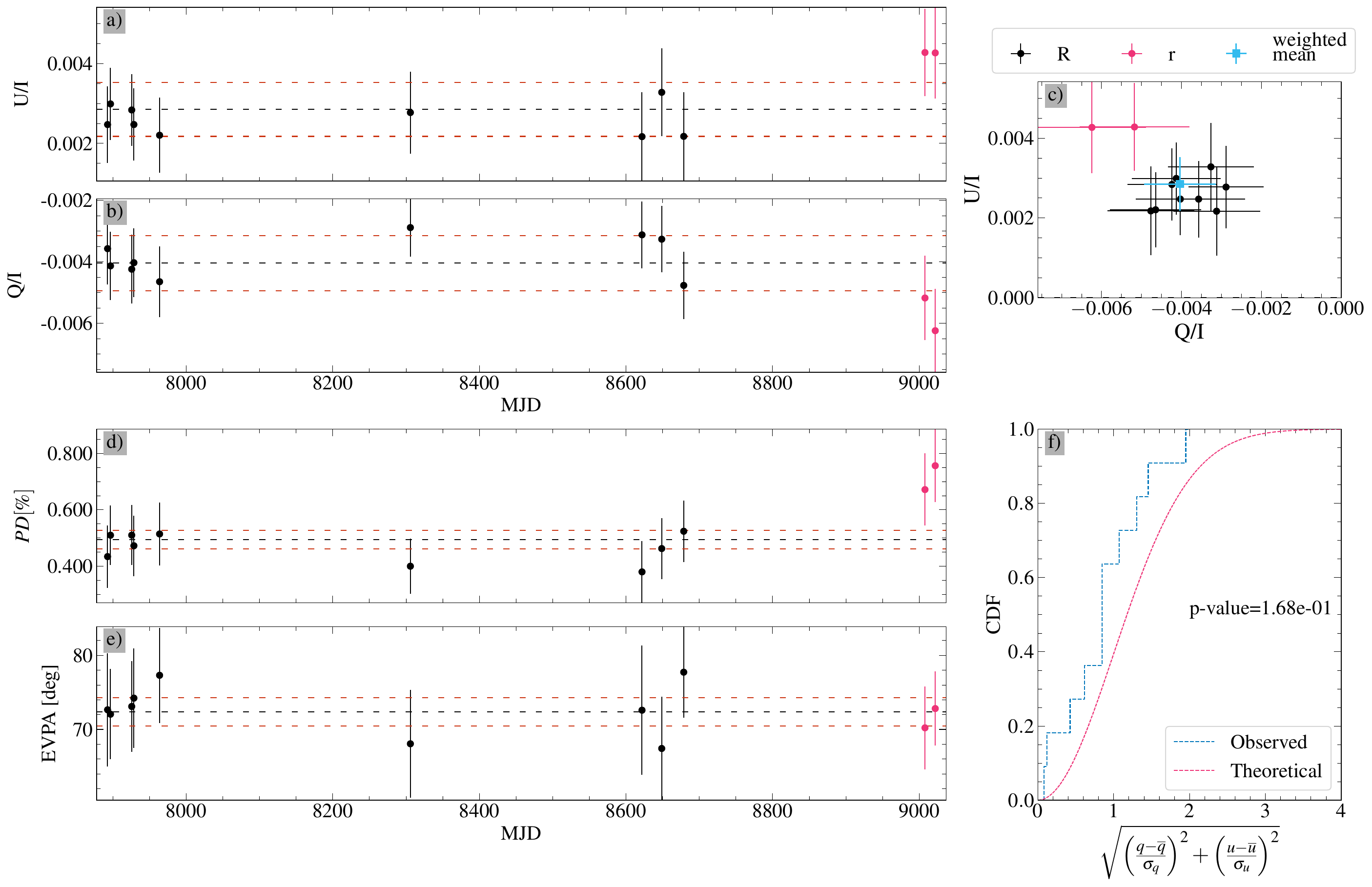}
  \caption{Same as Fig.~\ref{fig:B_0017+8135_82} for L\_106\_700, which is found to be stable. }
  \label{fig:L_106_700}
\end{figure*}

\clearpage

\begin{figure*}
  \centering
  \includegraphics[width=0.95\textwidth]{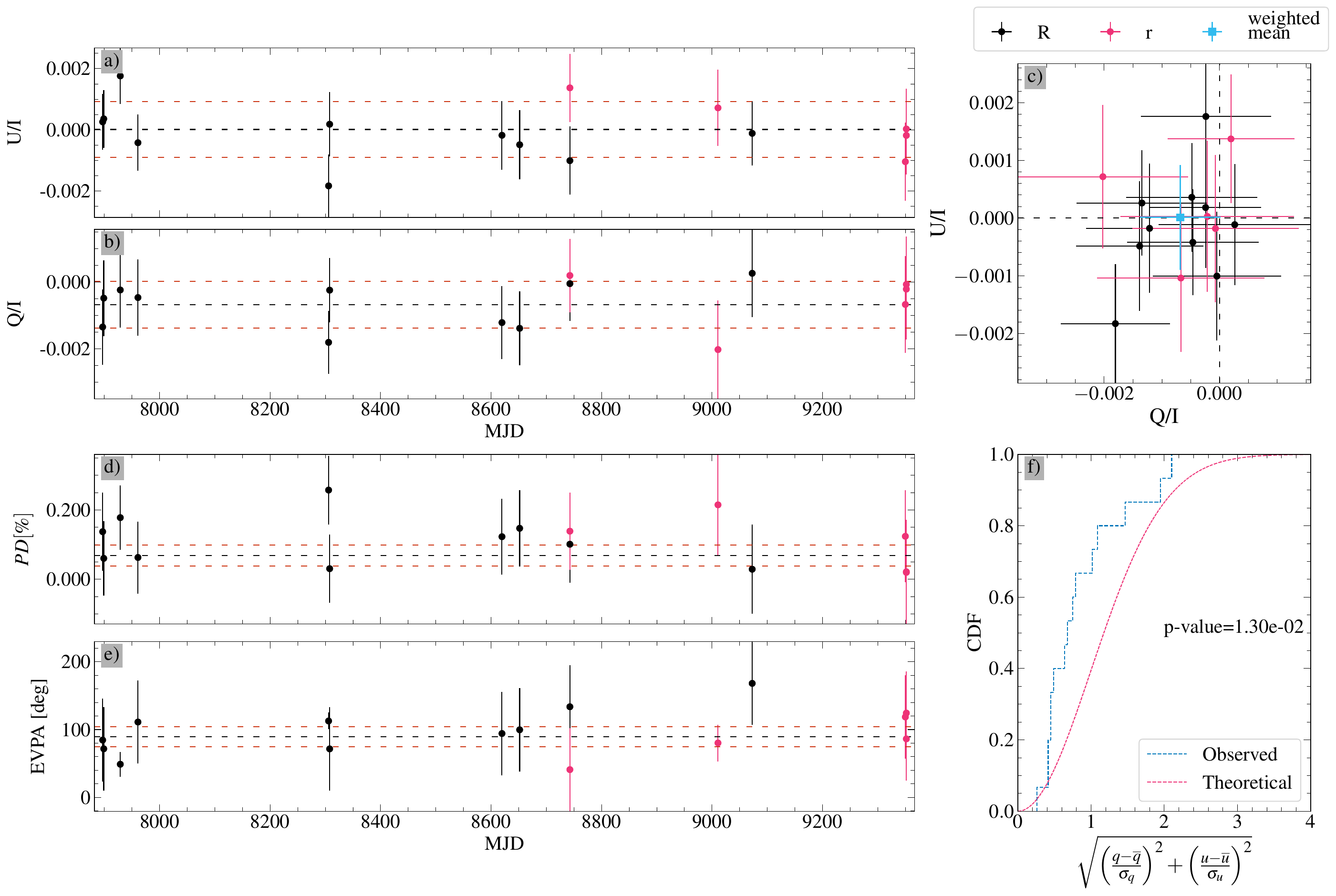}
  \caption{Same as Fig.~\ref{fig:B_0017+8135_82} for Z\_HD138733, which is found to be variable. }
  \label{fig:Z_HD138733}
\end{figure*}

\begin{figure*}
  \centering
  \includegraphics[width=0.95\textwidth]{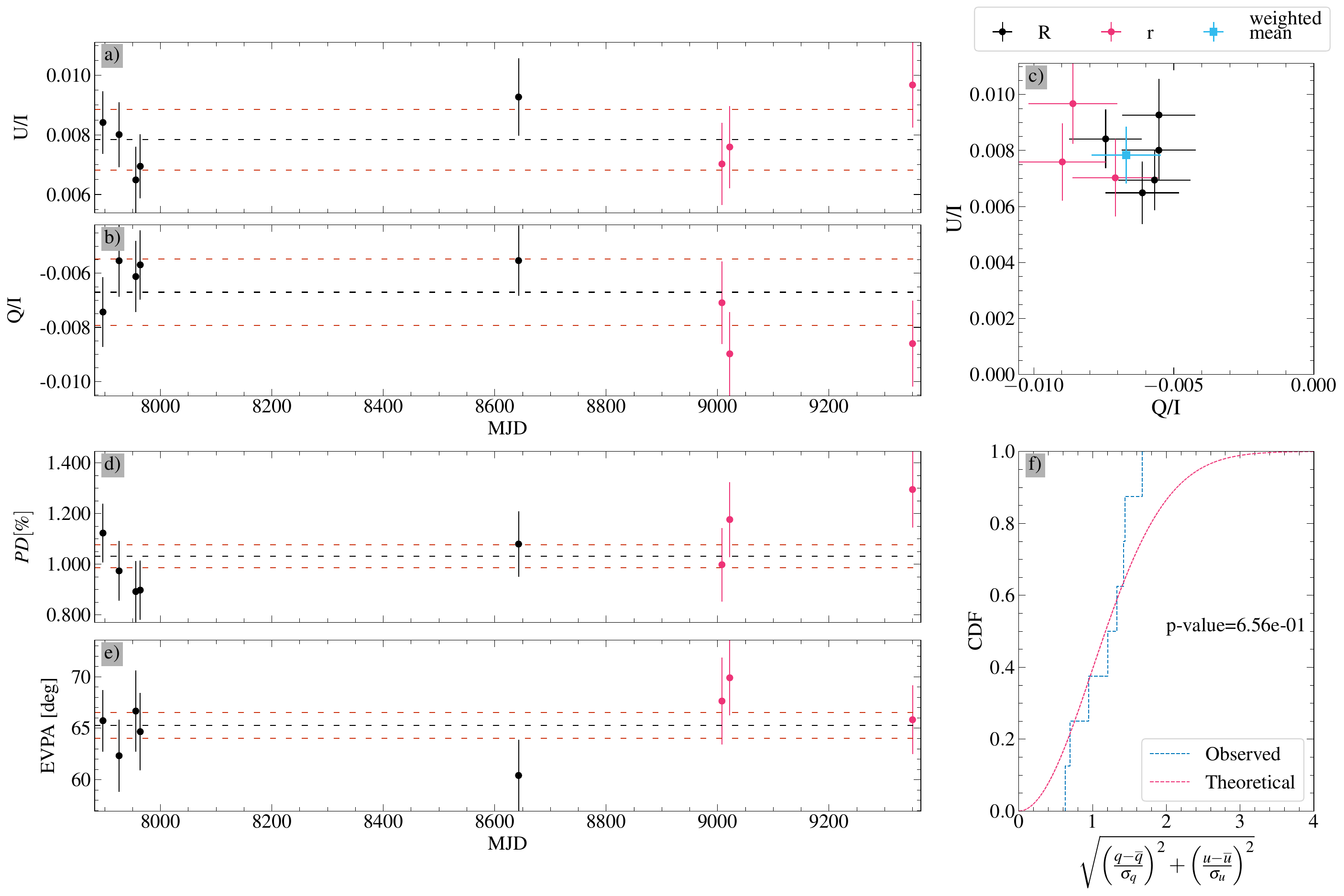}
  \caption{Same as Fig.~\ref{fig:B_0017+8135_82} for L\_107\_599, which is found to be stable. }
  \label{fig:L_107_599}
\end{figure*}

\clearpage

\begin{figure*}
  \centering
  \includegraphics[width=0.95\textwidth]{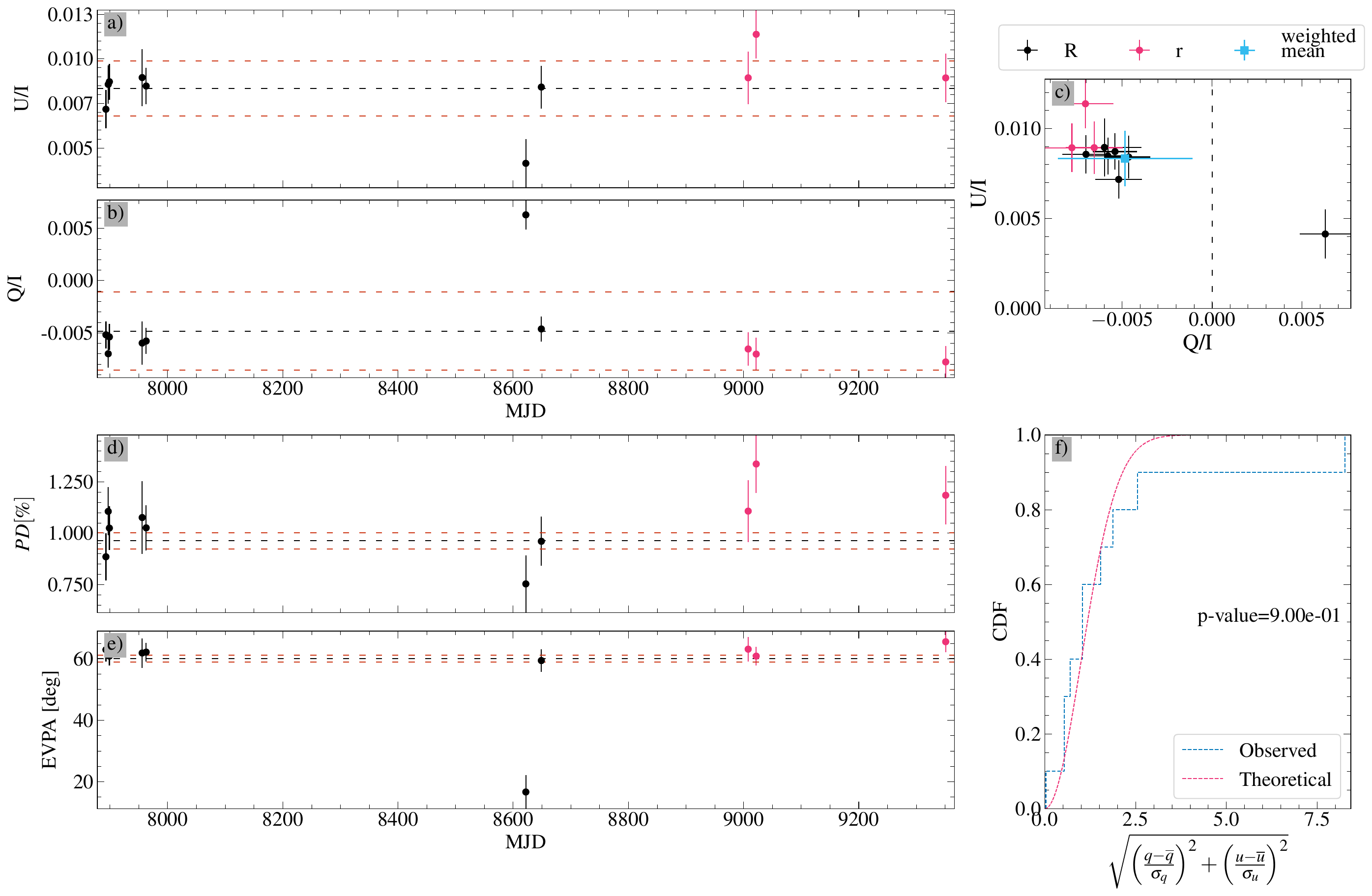}
  \caption{Same as Fig.~\ref{fig:B_0017+8135_82} for L\_107\_602, which is found to be stable. }
  \label{fig:L_107_602}
\end{figure*}

\begin{figure*}
  \centering
  \includegraphics[width=0.95\textwidth]{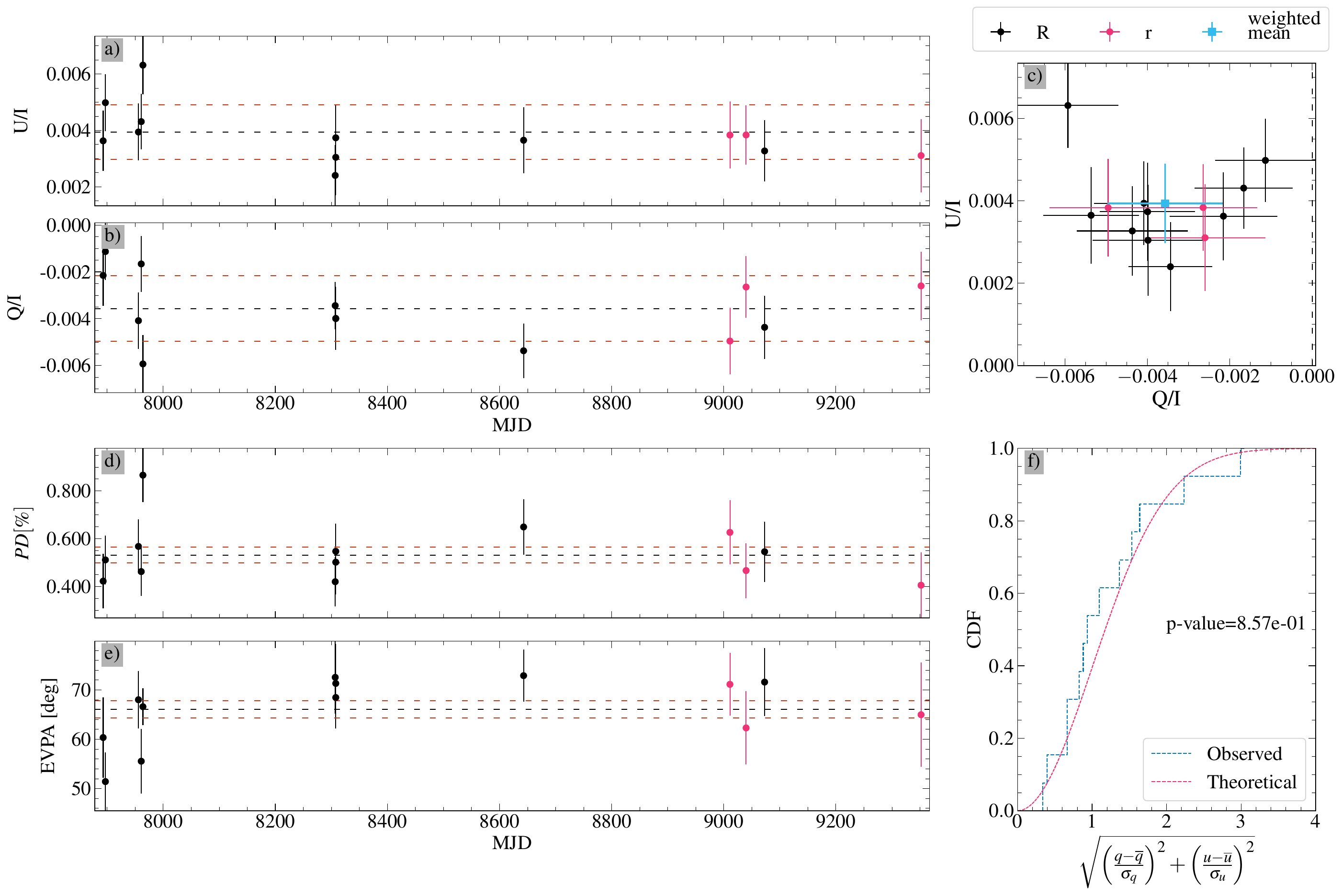}
  \caption{Same as Fig.~\ref{fig:B_0017+8135_82} for L\_PG1633+099B, which is found to be stable. }
  \label{fig:L_PG1633+099B}
\end{figure*}

\clearpage

\begin{figure*}
  \centering
  \includegraphics[width=0.95\textwidth]{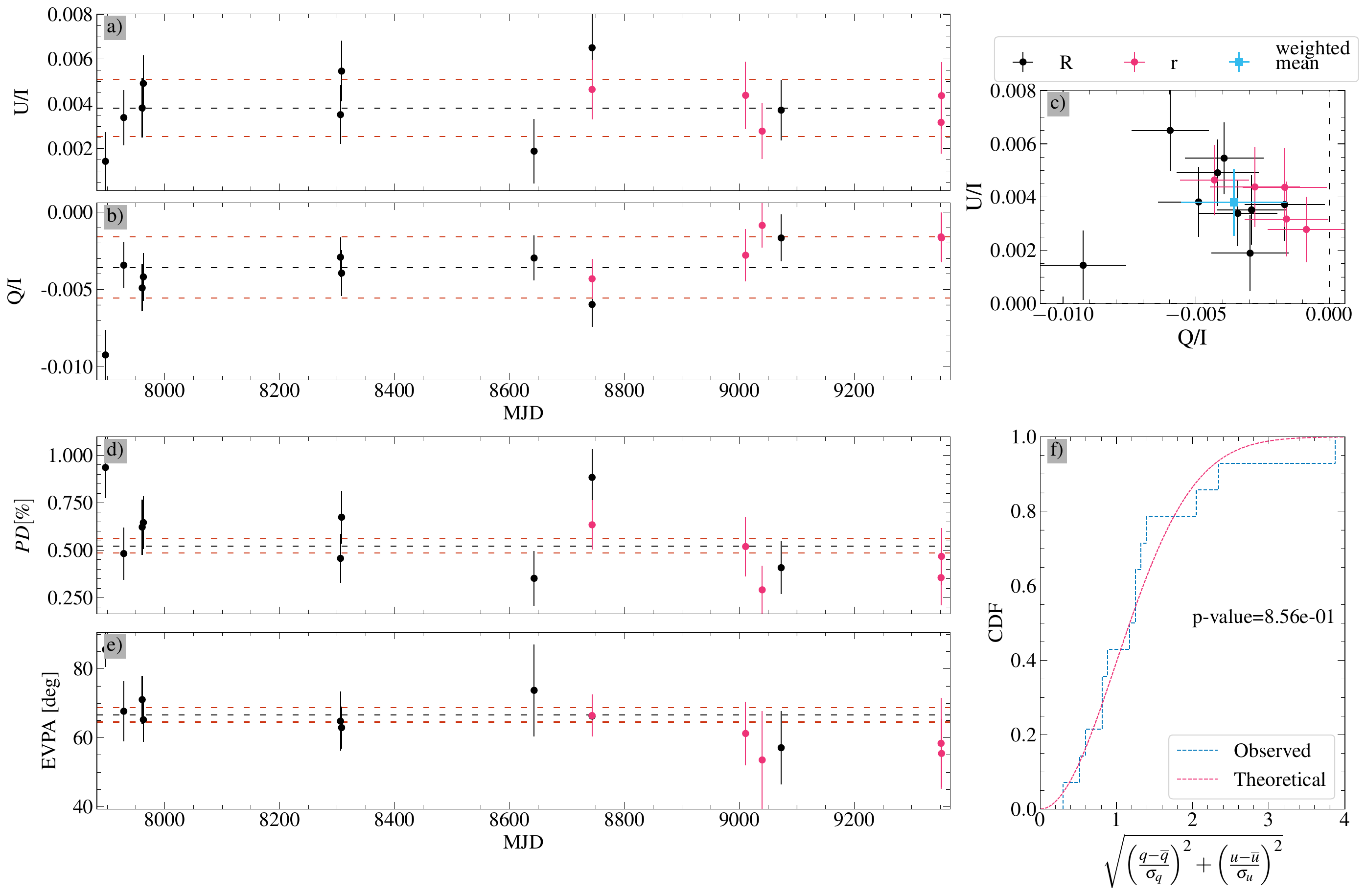}
  \caption{Same as Fig.~\ref{fig:B_0017+8135_82} for L\_PG1633+099D, which is found to be stable. }
  \label{fig:L_PG1633+099D}
\end{figure*}

\begin{figure*}
  \centering
  \includegraphics[width=0.95\textwidth]{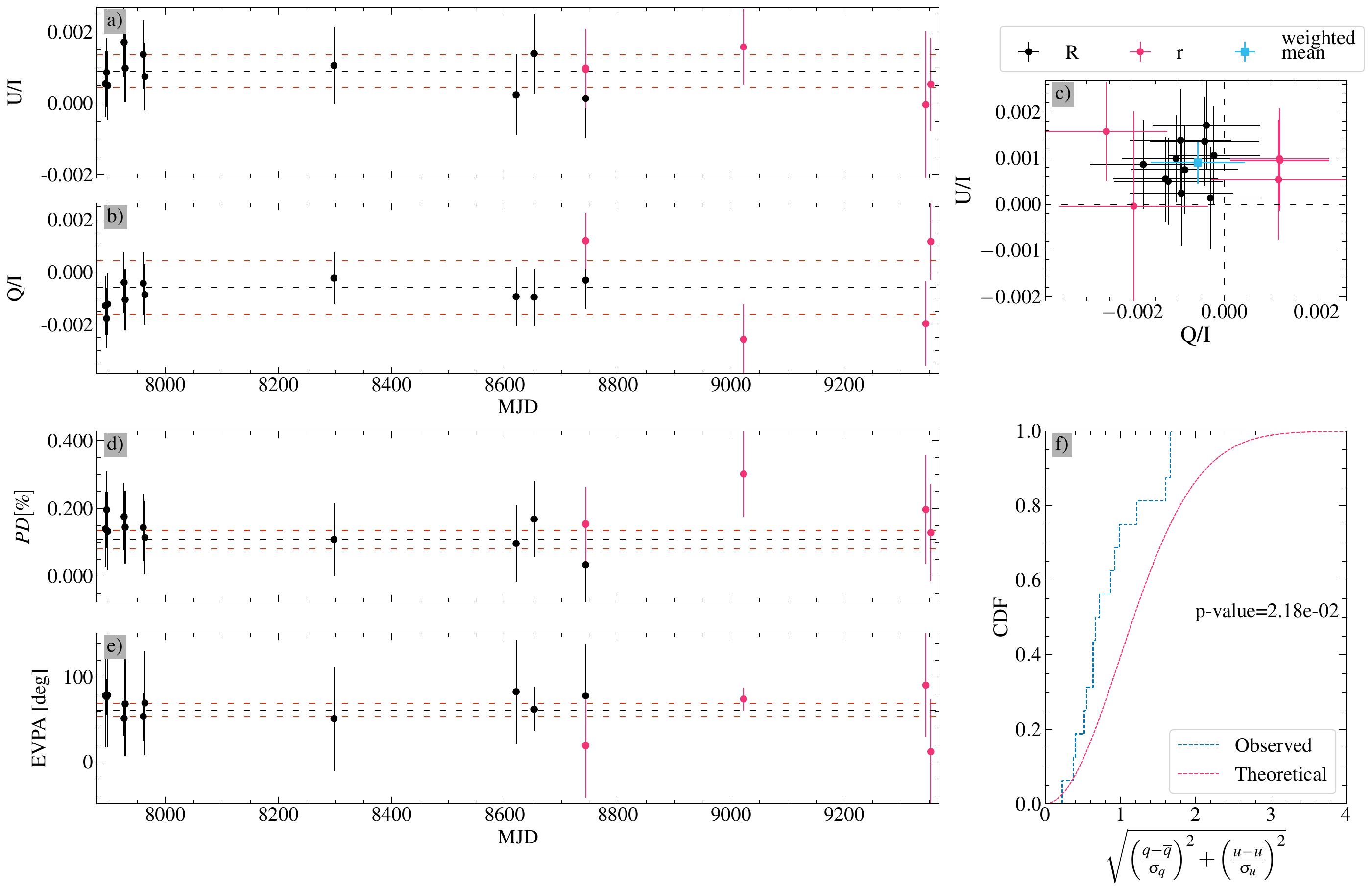}
  \caption{Same as Fig.~\ref{fig:B_0017+8135_82} for Z\_HD153752, which is found to be variable. }
  \label{fig:Z_HD153752}
\end{figure*}

\clearpage

\begin{figure*}
  \centering
  \includegraphics[width=0.95\textwidth]{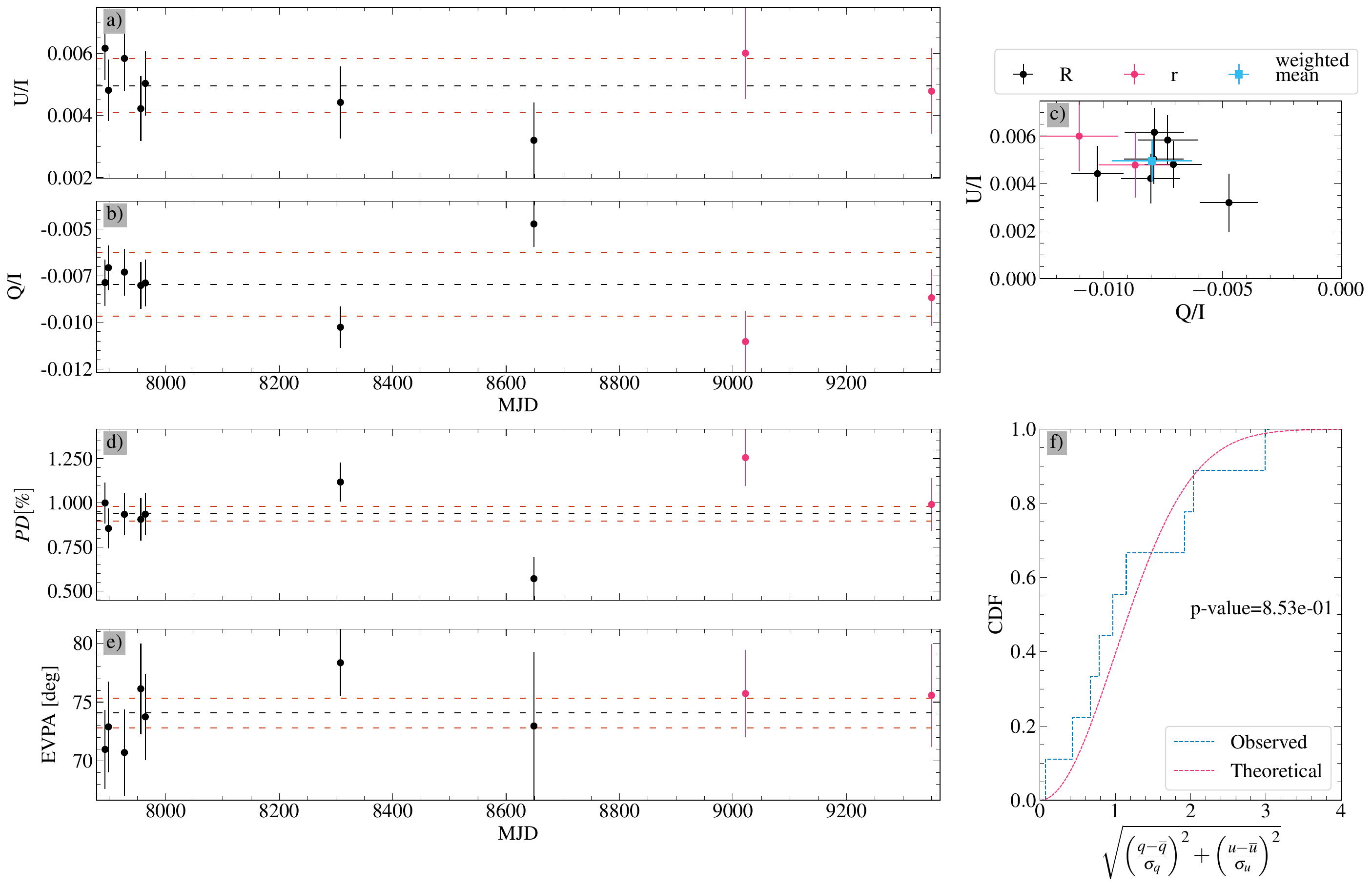}
  \caption{Same as Fig.~\ref{fig:B_0017+8135_82} for B\_1725+1152\_11, which is found to be stable. }
  \label{fig:B_1725+1152_11}
\end{figure*}

\begin{figure*}
  \centering
  \includegraphics[width=0.95\textwidth]{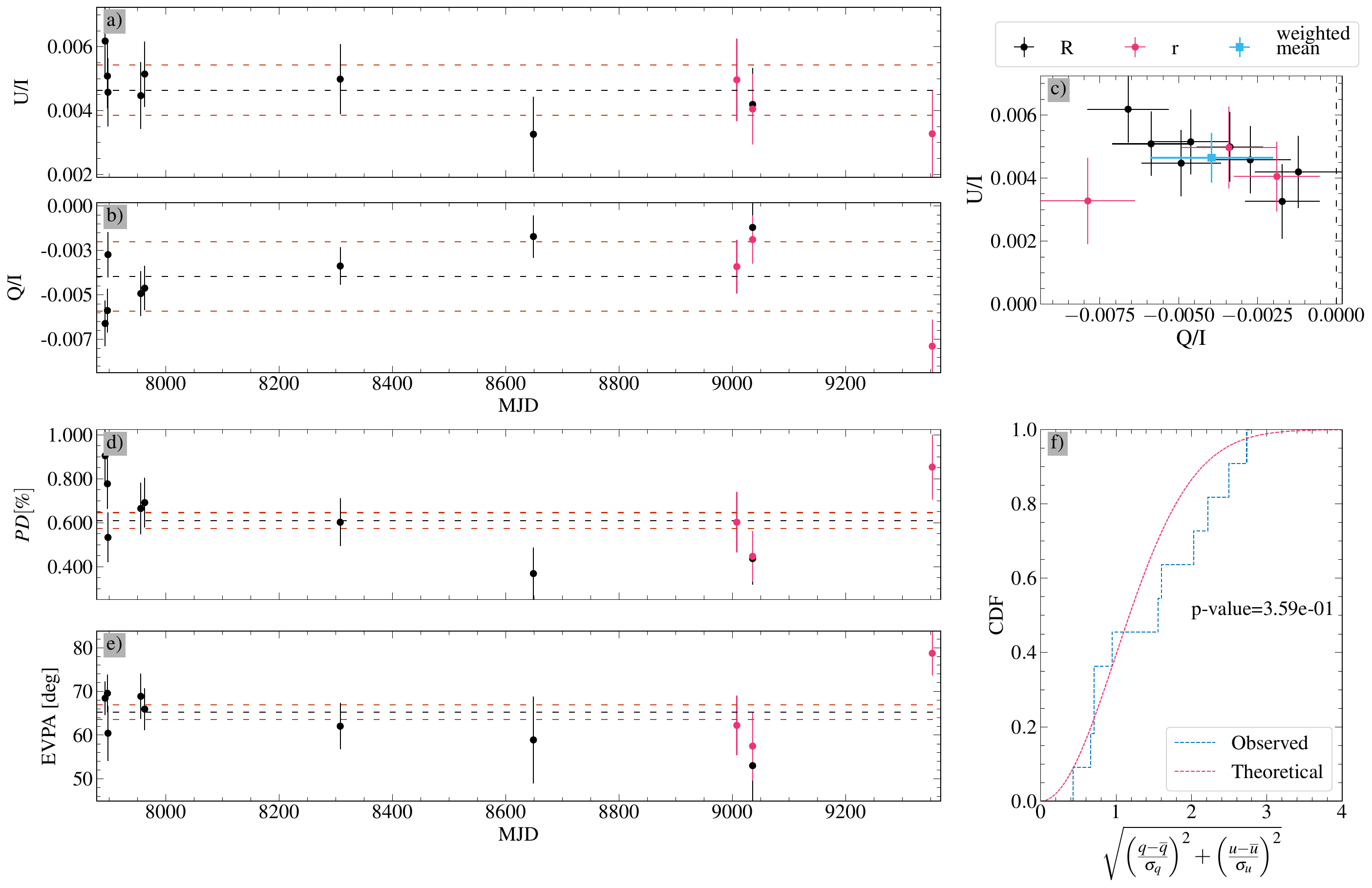}
  \caption{Same as Fig.~\ref{fig:B_0017+8135_82} for B\_1725+1152\_24, which is found to be stable. }
  \label{fig:B_1725+1152_24}
\end{figure*}

\clearpage

\begin{figure*}
  \centering
  \includegraphics[width=0.95\textwidth]{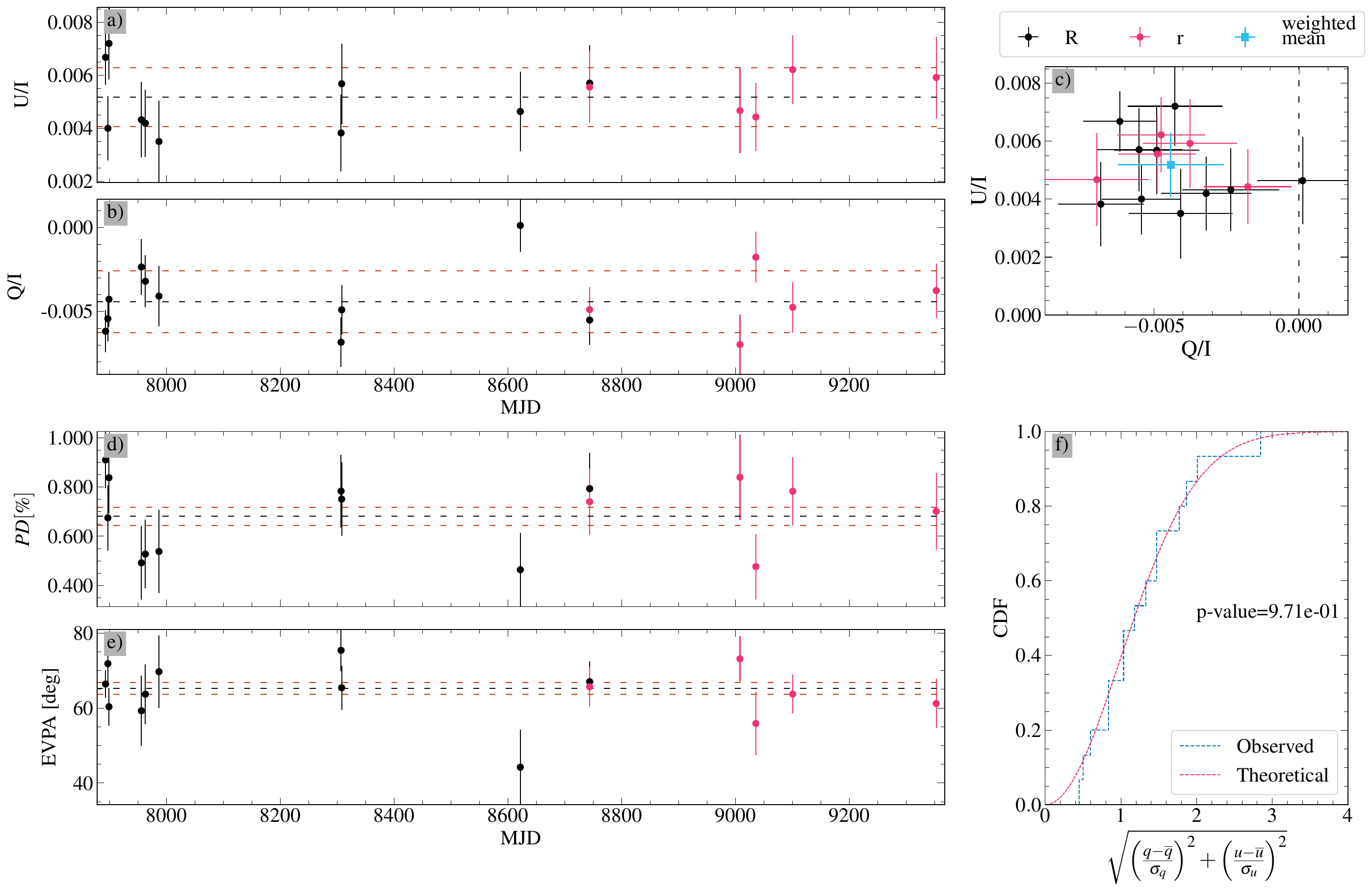}
  \caption{Same as Fig.~\ref{fig:B_0017+8135_82} for B\_1725+1152\_35, which is found to be stable. }
  \label{fig:B_1725+1152_35}
\end{figure*}

\begin{figure*}
  \centering
  \includegraphics[width=0.95\textwidth]{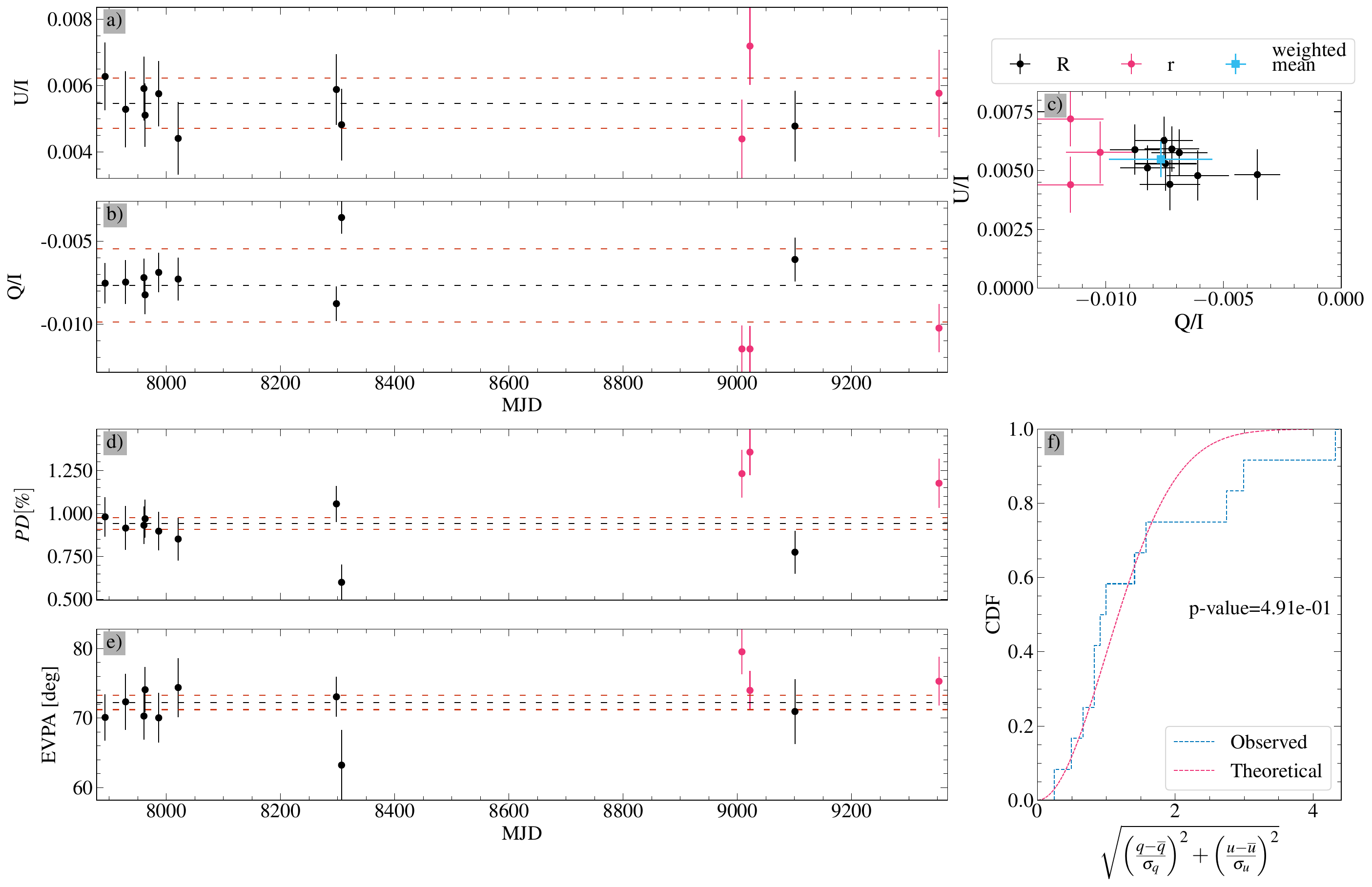}
  \caption{Same as Fig.~\ref{fig:B_0017+8135_82} for B\_1725+1152\_113, which is found to be stable. }
  \label{fig:B_1725+1152_113}
\end{figure*}

\clearpage

\begin{figure*}
  \centering
  \includegraphics[width=0.95\textwidth]{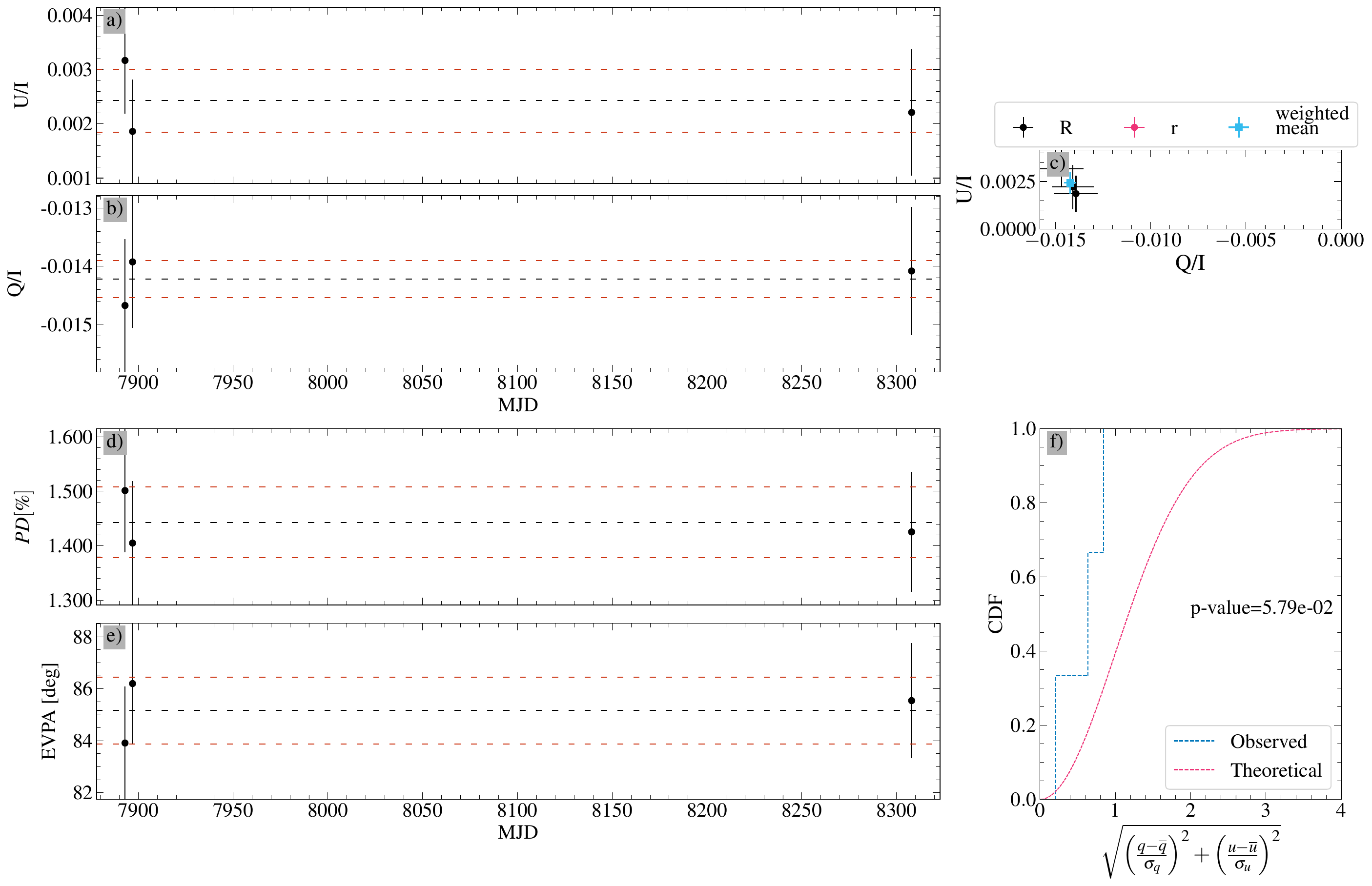}
  \caption{Same as Fig.~\ref{fig:B_0017+8135_82} for L\_109\_71, which has not enough measurements to judge it as varible of stable. }
  \label{fig:L_109_71}
\end{figure*}

\begin{figure*}
  \centering
  \includegraphics[width=0.95\textwidth]{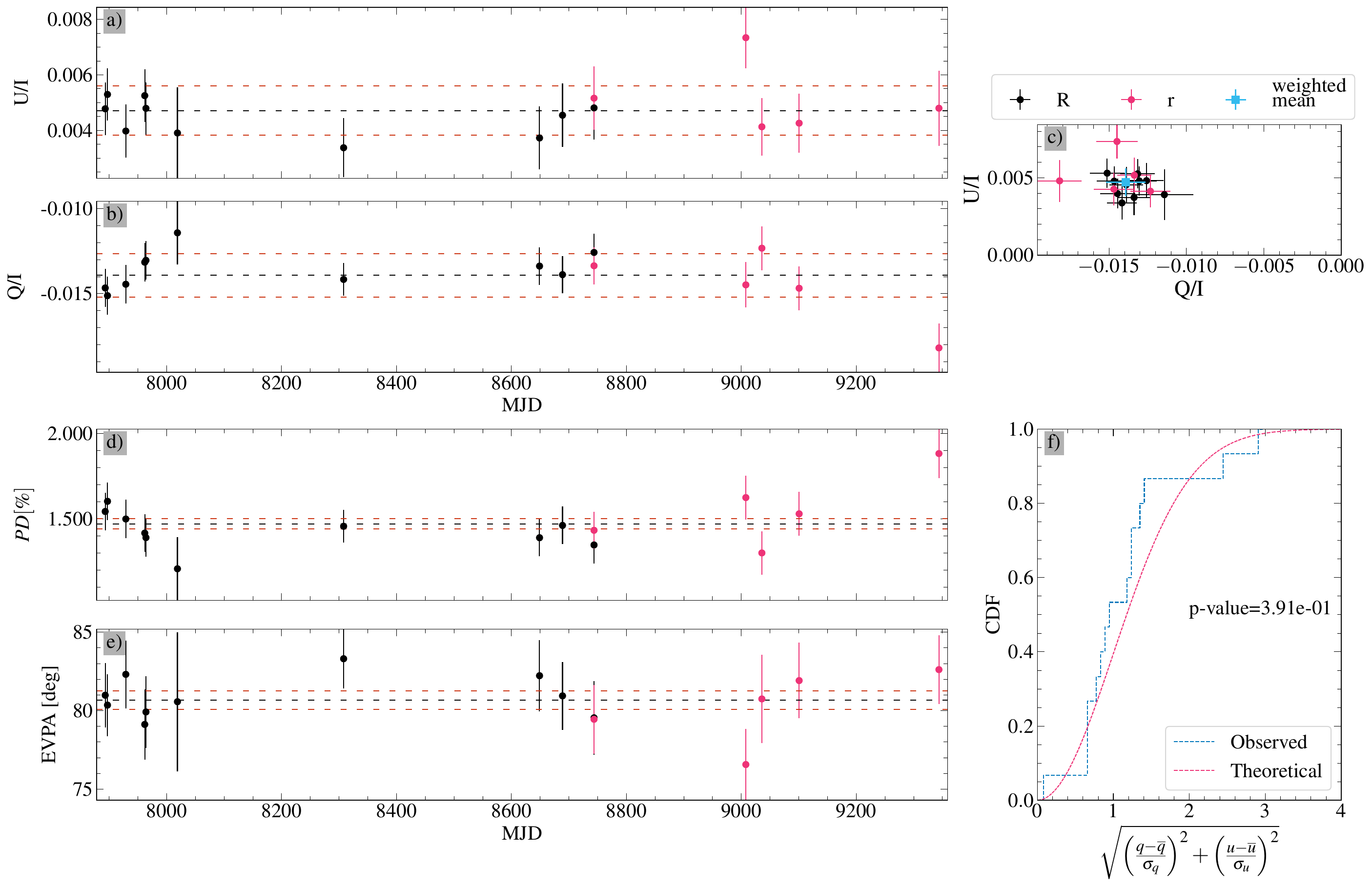}
  \caption{Same as Fig.~\ref{fig:B_0017+8135_82} for L\_109\_381, which is found to be stable. }
  \label{fig:L_109_381}
\end{figure*}

\clearpage

\begin{figure*}
  \centering
  \includegraphics[width=0.95\textwidth]{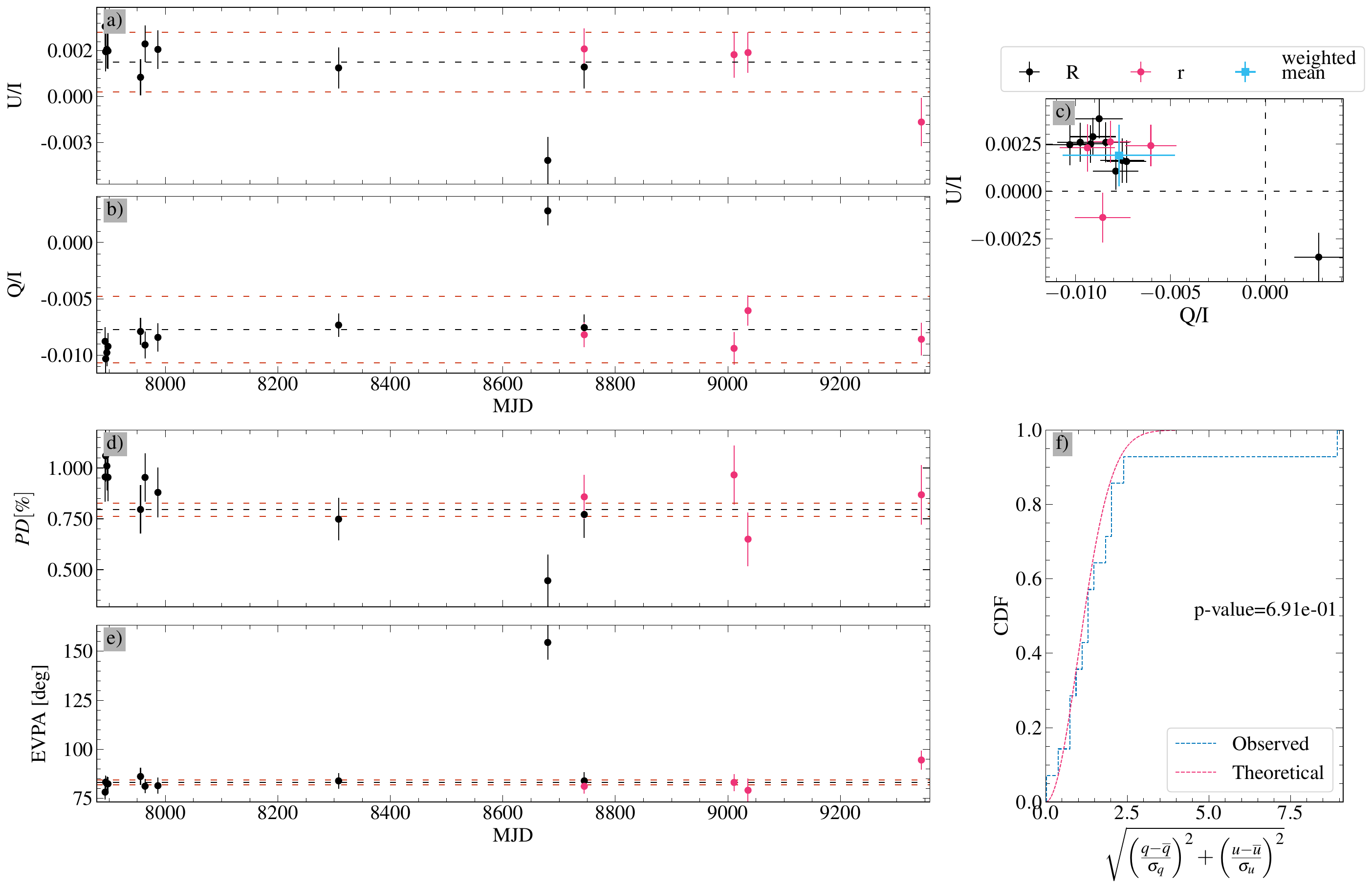}
  \caption{Same as Fig.~\ref{fig:B_0017+8135_82} for B\_1751+0939\_376, which is found to be stable. }
  \label{fig:B_1751+0939_376}
\end{figure*}

\begin{figure*}
  \centering
  \includegraphics[width=0.95\textwidth]{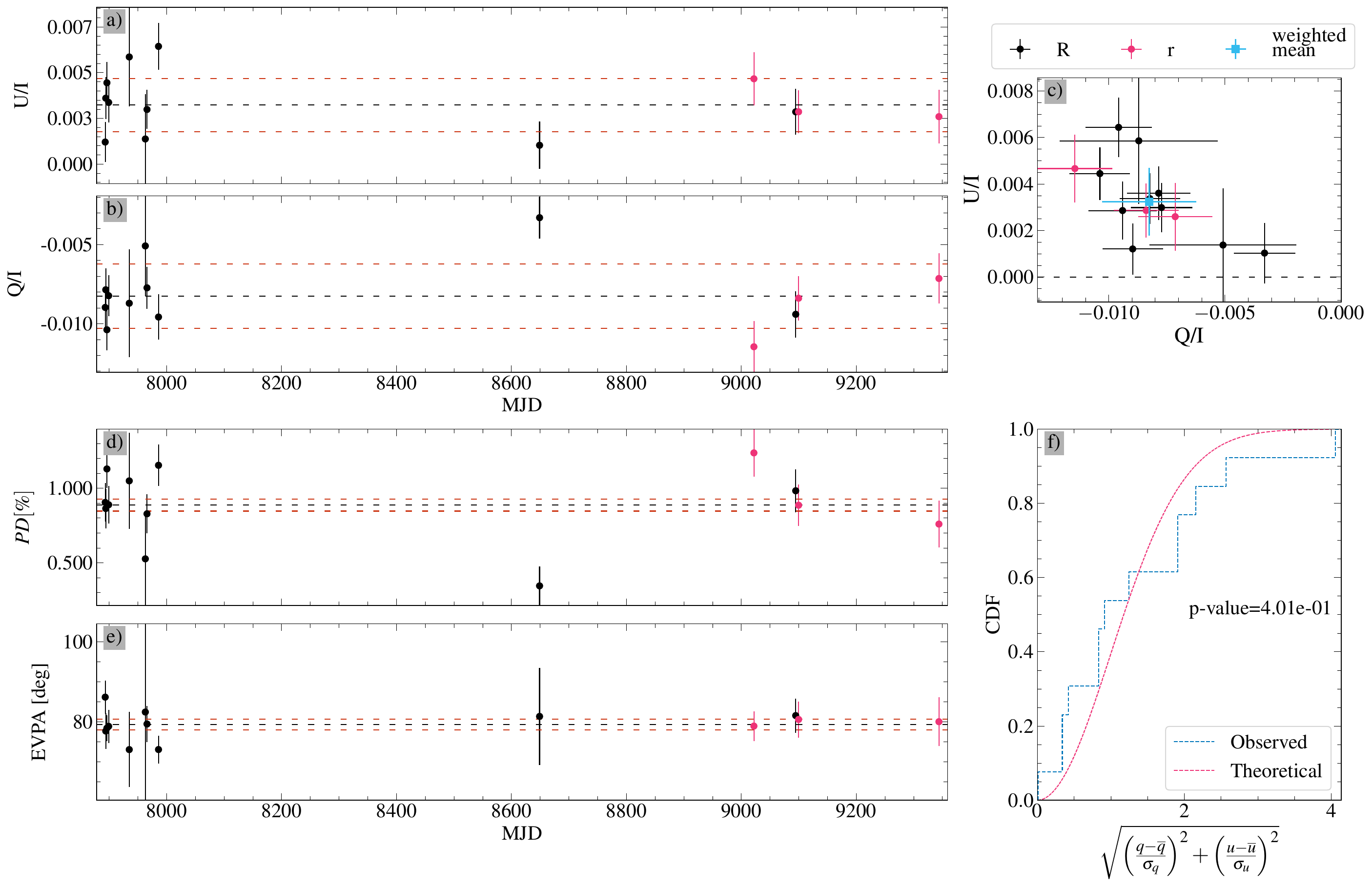}
  \caption{Same as Fig.~\ref{fig:B_0017+8135_82} for B\_1751+0939\_129, which is found to be stable. }
  \label{fig:B_1751+0939_129}
\end{figure*}

\clearpage

\begin{figure*}
  \centering
  \includegraphics[width=0.95\textwidth]{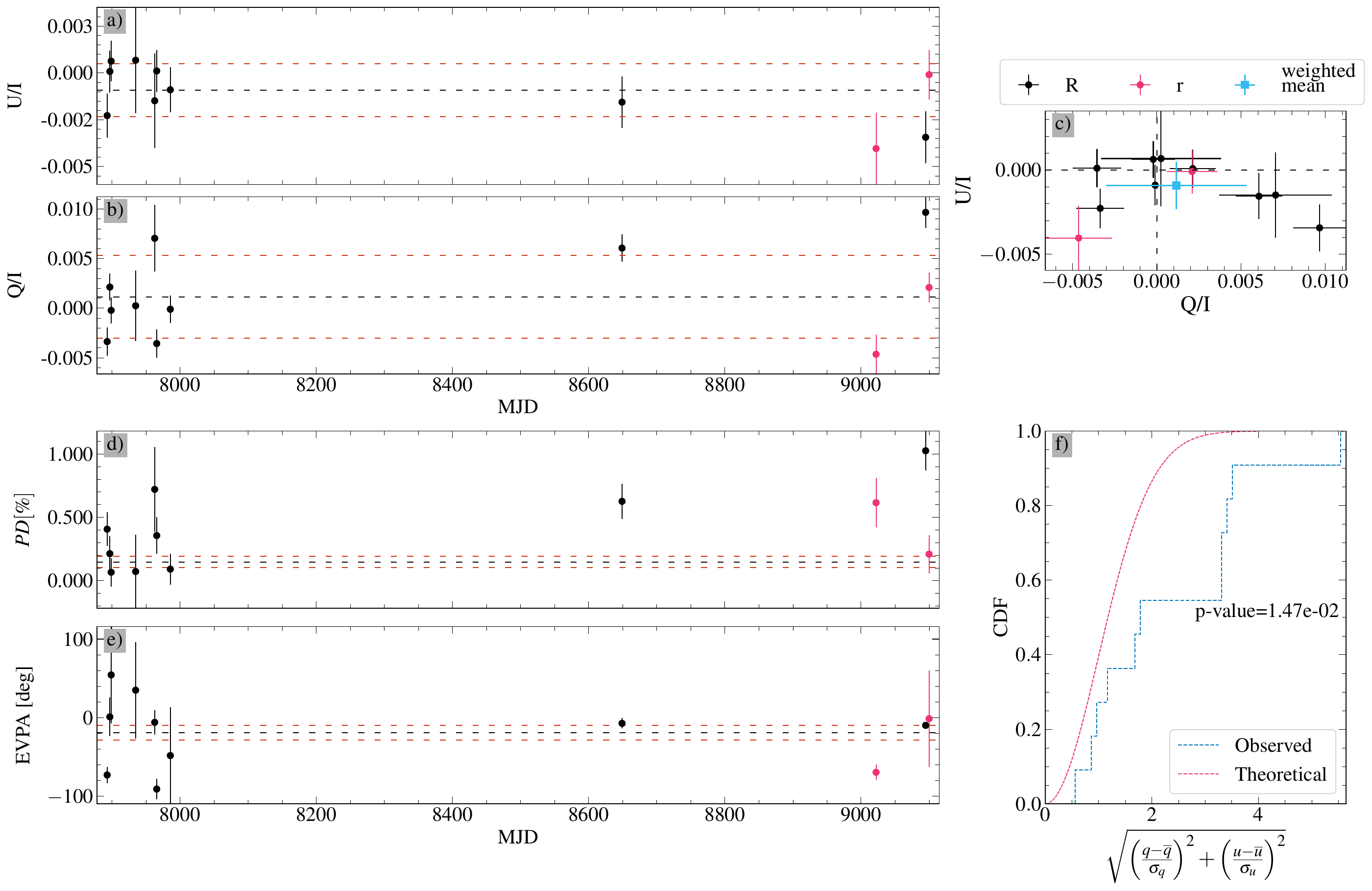}
  \caption{Same as Fig.~\ref{fig:B_0017+8135_82} for B\_1751+0939\_204, which is found to be variable. }
  \label{fig:B_1751+0939_204}
\end{figure*}

\begin{figure*}
  \centering
  \includegraphics[width=0.95\textwidth]{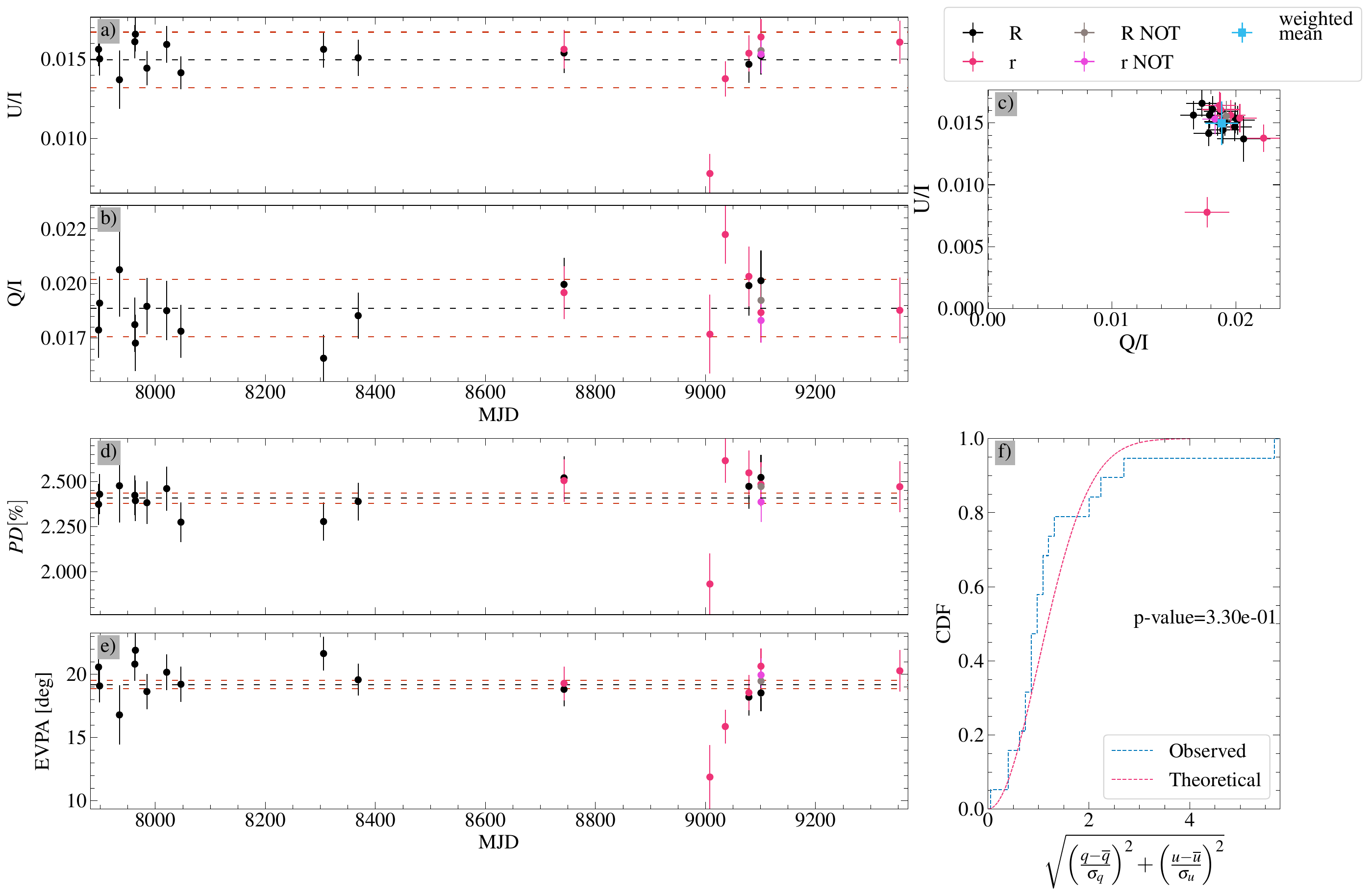}
  \caption{Same as Fig.~\ref{fig:B_0017+8135_82} for L\_110\_229, which is found to be stable. }
  \label{fig:L_110_229}
\end{figure*}

\clearpage

\begin{figure*}
  \centering
  \includegraphics[width=0.95\textwidth]{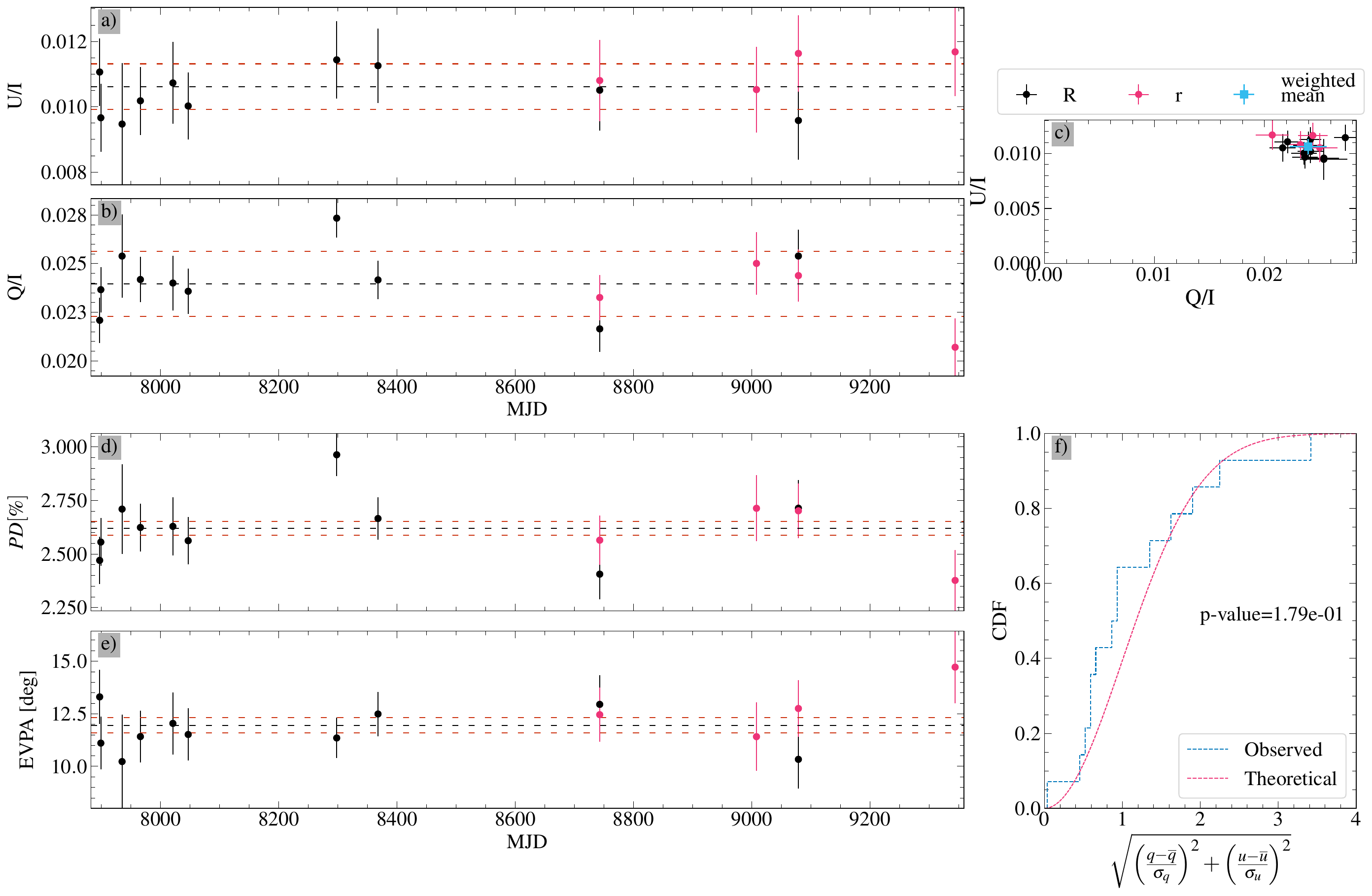}
  \caption{Same as Fig.~\ref{fig:B_0017+8135_82} for L\_110\_233, which is found to be stable. }
  \label{fig:L_110_233}
\end{figure*}

\begin{figure*}
  \centering
  \includegraphics[width=0.95\textwidth]{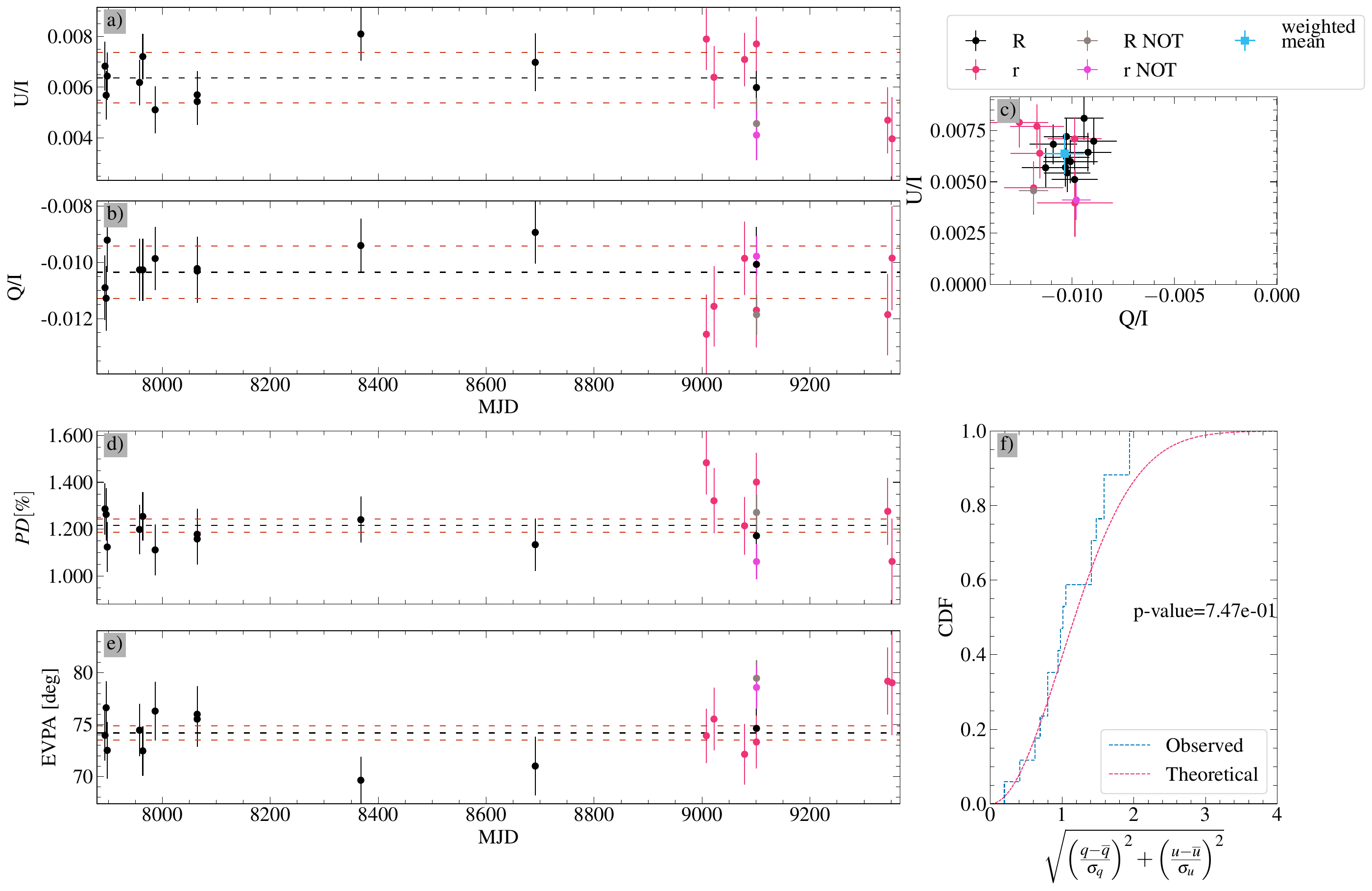}
  \caption{Same as Fig.~\ref{fig:B_0017+8135_82} for L\_111\_1965, which is found to be stable. }
  \label{fig:L_111_1965}
\end{figure*}

\clearpage

\begin{figure*}
  \centering
  \includegraphics[width=0.95\textwidth]{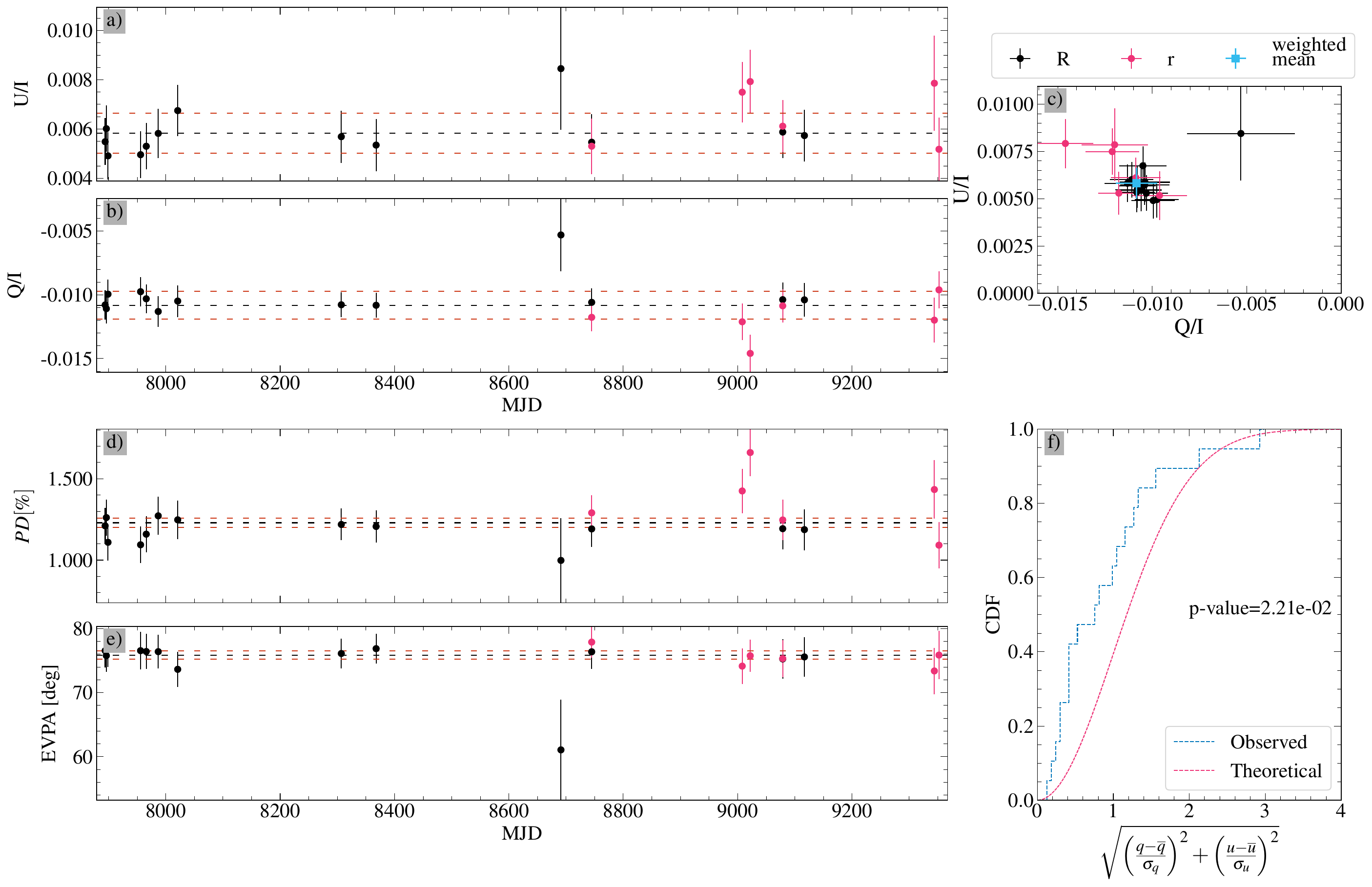}
  \caption{Same as Fig.~\ref{fig:B_0017+8135_82} for L\_111\_1969, which is found to be variable. }
  \label{fig:L_111_1969}
\end{figure*}

\begin{figure*}
  \centering
  \includegraphics[width=0.95\textwidth]{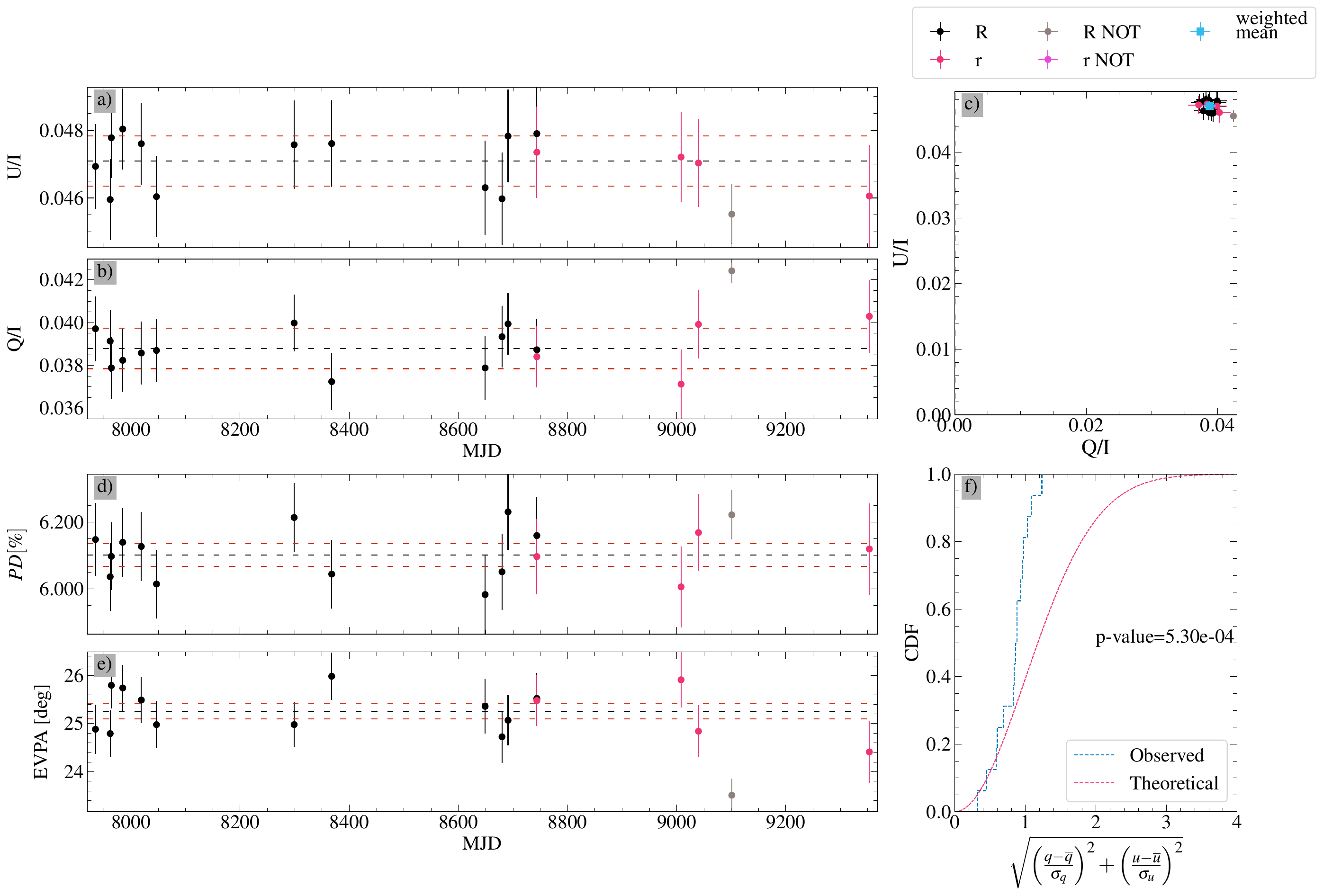}
  \caption{Same as Fig.~\ref{fig:B_0017+8135_82} for H\_HD344776, which is found to be variable. }
  \label{fig:H_HD344776}
\end{figure*}

\clearpage

\begin{figure*}
  \centering
  \includegraphics[width=0.95\textwidth]{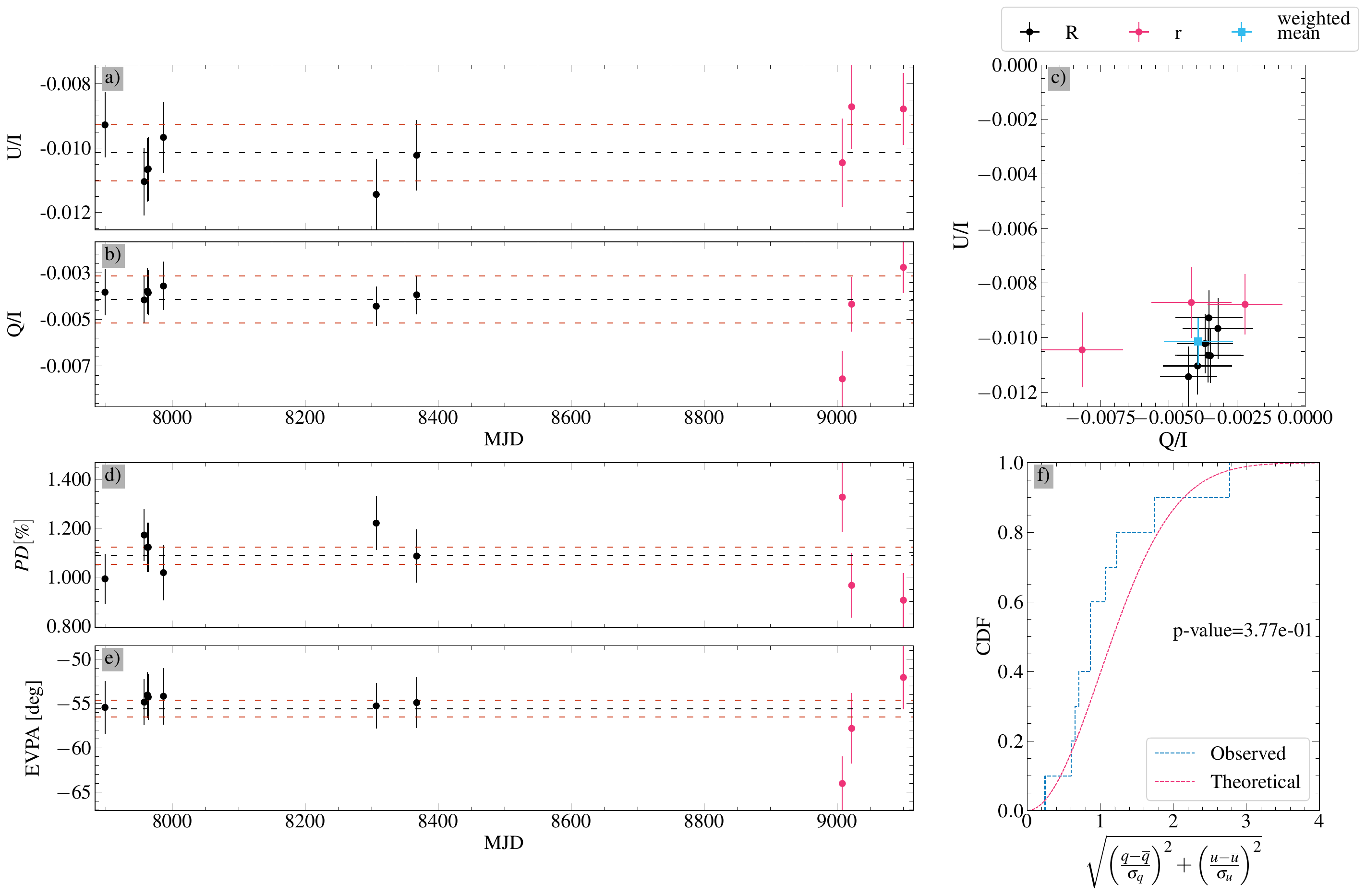}
  \caption{Same as Fig.~\ref{fig:B_0017+8135_82} for B\_1959+6508\_179, which is found to be stable. }
  \label{fig:B_1959+6508_179}
\end{figure*}

\begin{figure*}
  \centering
  \includegraphics[width=0.95\textwidth]{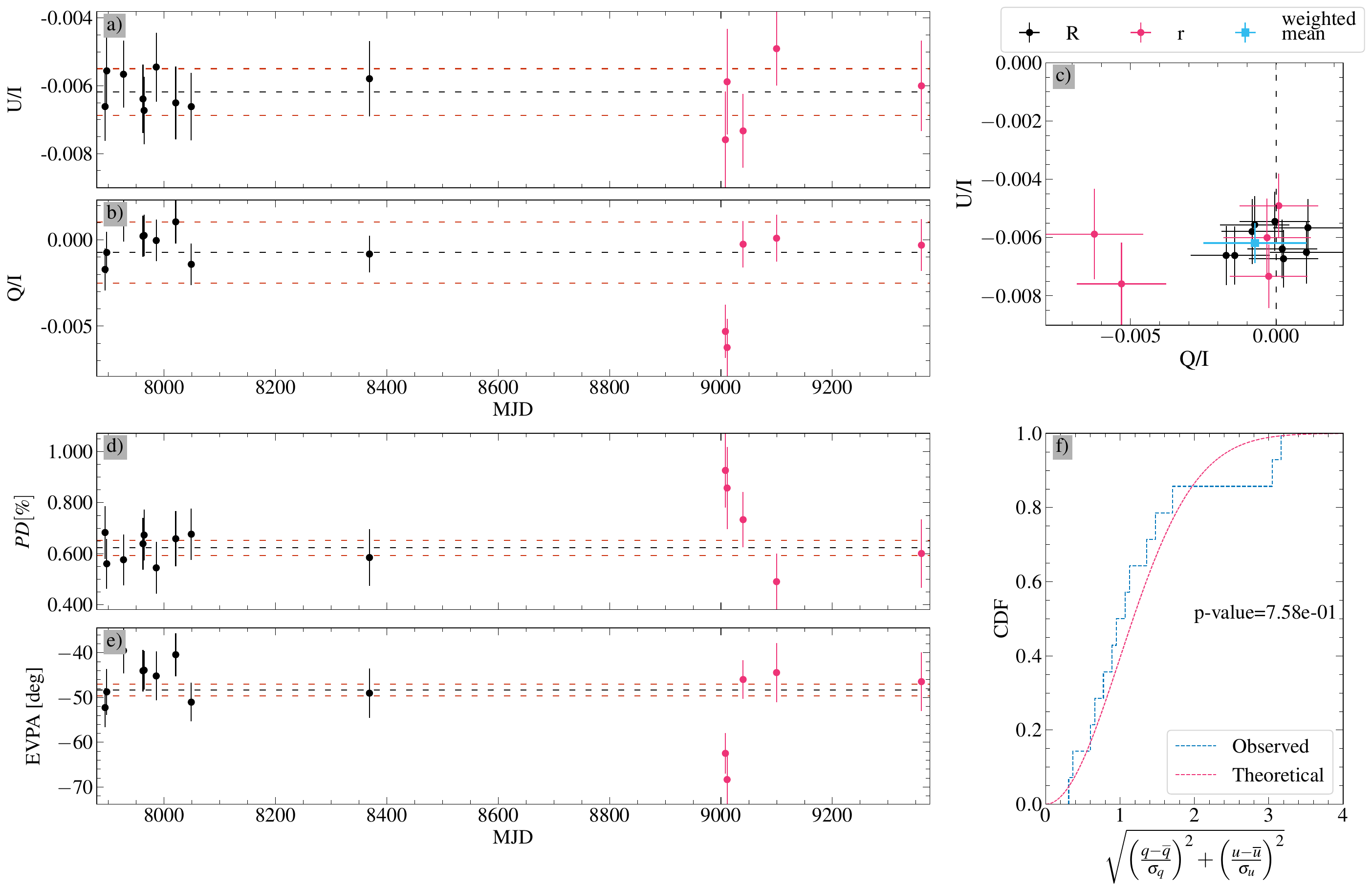}
  \caption{Same as Fig.~\ref{fig:B_0017+8135_82} for B\_1959+6508\_73, which is found to be stable. }
  \label{fig:B_1959+6508_73}
\end{figure*}

\clearpage

\begin{figure*}
  \centering
  \includegraphics[width=0.95\textwidth]{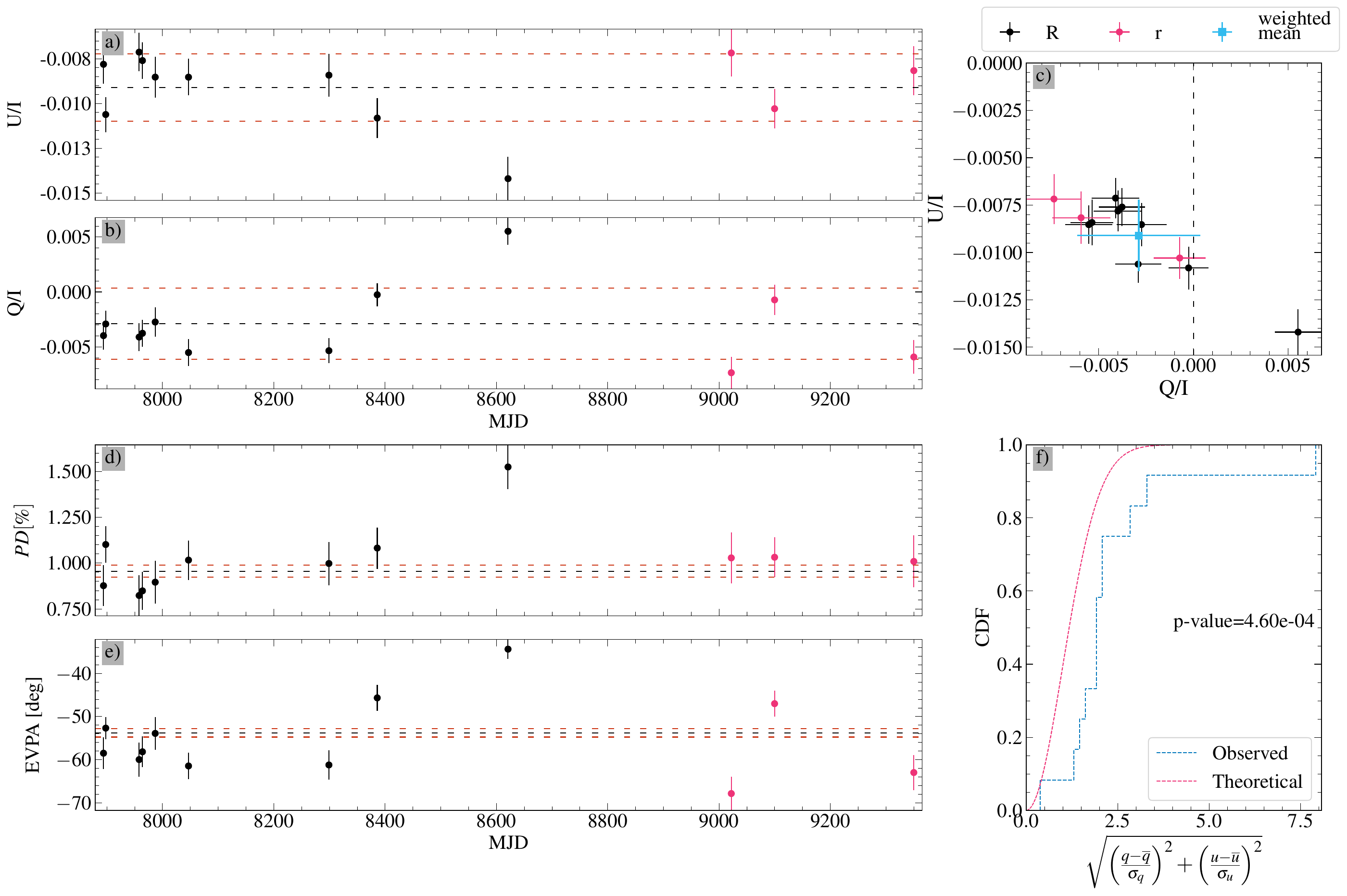}
  \caption{Same as Fig.~\ref{fig:B_0017+8135_82} for B\_1959+6508\_108, which is found to be variable. }
  \label{fig:B_1959+6508_108}
\end{figure*}

\begin{figure*}
  \centering
  \includegraphics[width=0.95\textwidth]{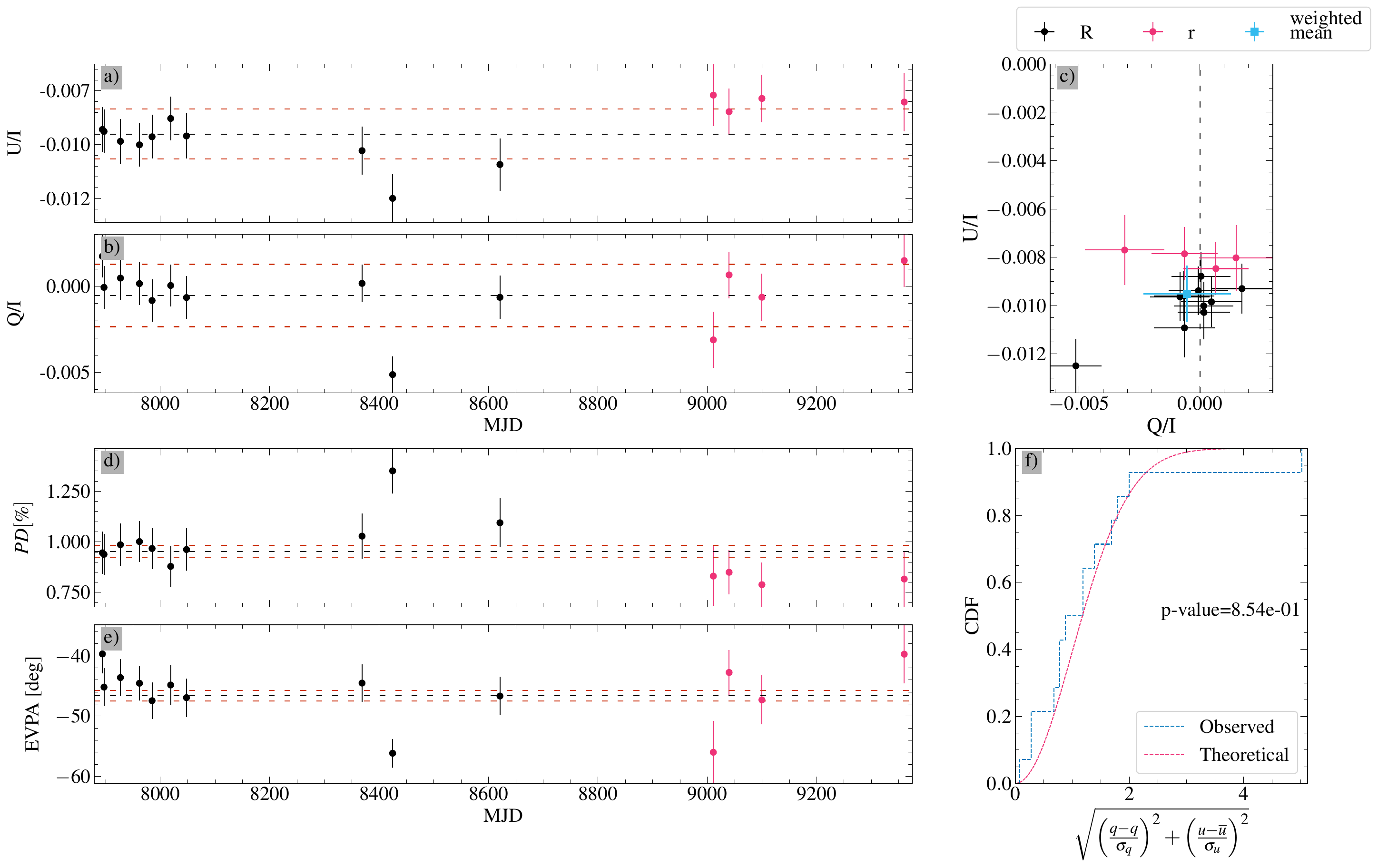}
  \caption{Same as Fig.~\ref{fig:B_0017+8135_82} for B\_1959+6508\_104, which is found to be stable. }
  \label{fig:B_1959+6508_104}
\end{figure*}

\clearpage

\begin{figure*}
  \centering
  \includegraphics[width=0.95\textwidth]{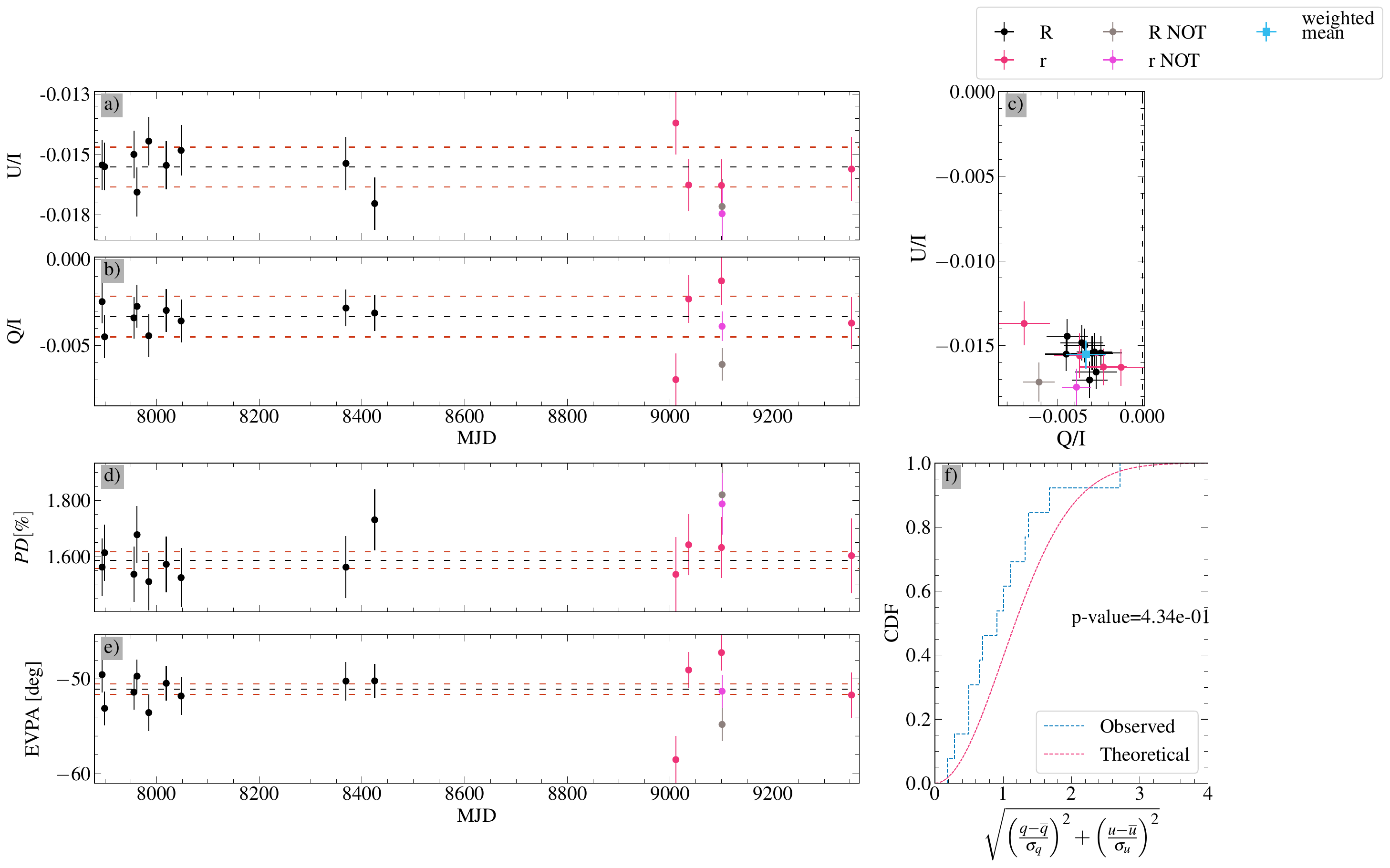}
  \caption{Same as Fig.~\ref{fig:B_0017+8135_82} for B\_1959+6508\_38, which is found to be stable. }
  \label{fig:B_1959+6508_38}
\end{figure*}

\begin{figure*}
  \centering
  \includegraphics[width=0.95\textwidth]{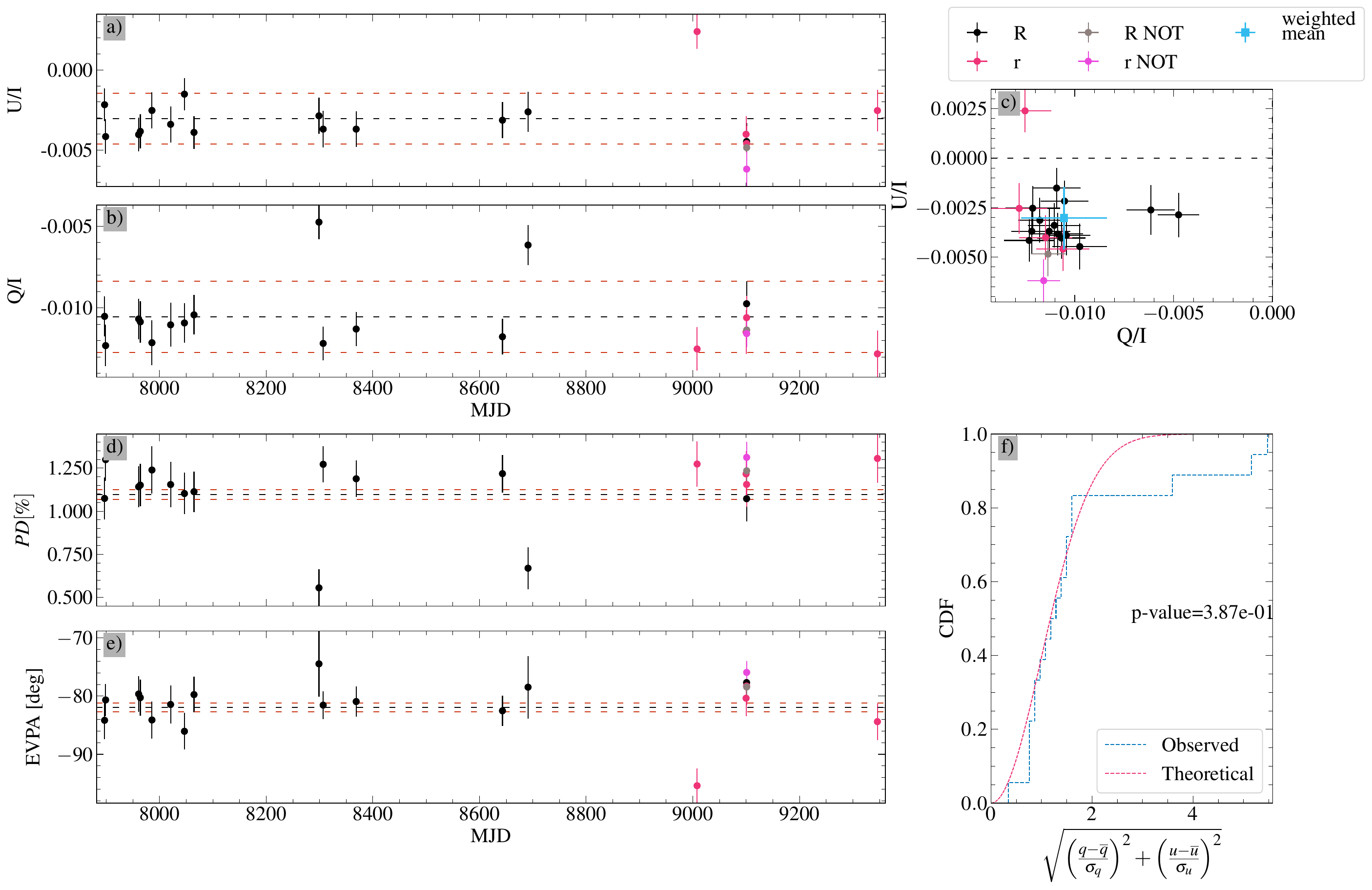}
  \caption{Same as Fig.~\ref{fig:B_0017+8135_82} for B\_2015$-$0137\_102, which is found to be stable. }
  \label{fig:B_2015-0137_102}
\end{figure*}

\clearpage

\begin{figure*}
  \centering
  \includegraphics[width=0.95\textwidth]{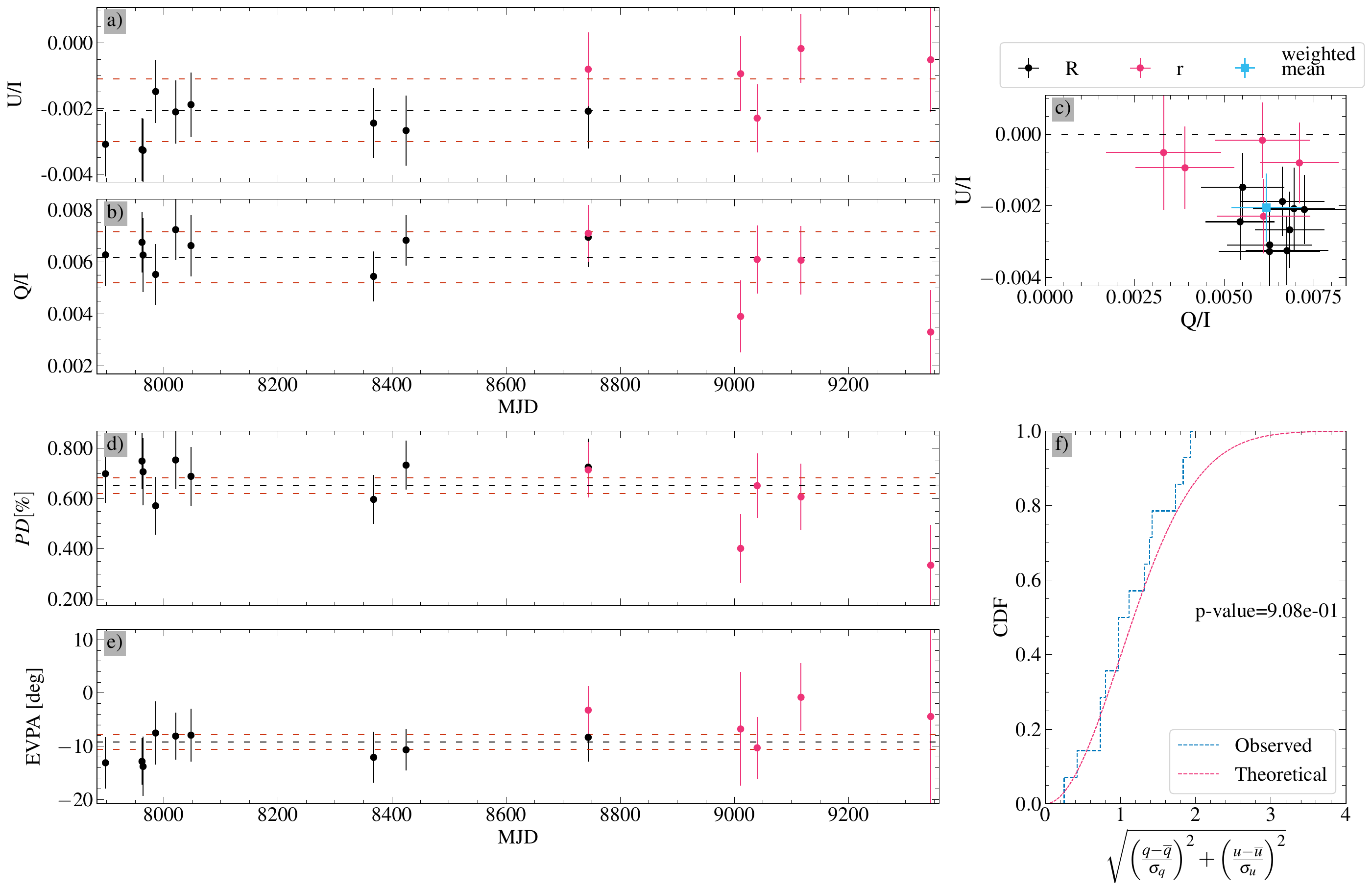}
  \caption{Same as Fig.~\ref{fig:B_0017+8135_82} for B\_2022+7611\_1, which is found to be stable. }
  \label{fig:B_2022+7611_1}
\end{figure*}

\begin{figure*}
  \centering
  \includegraphics[width=0.95\textwidth]{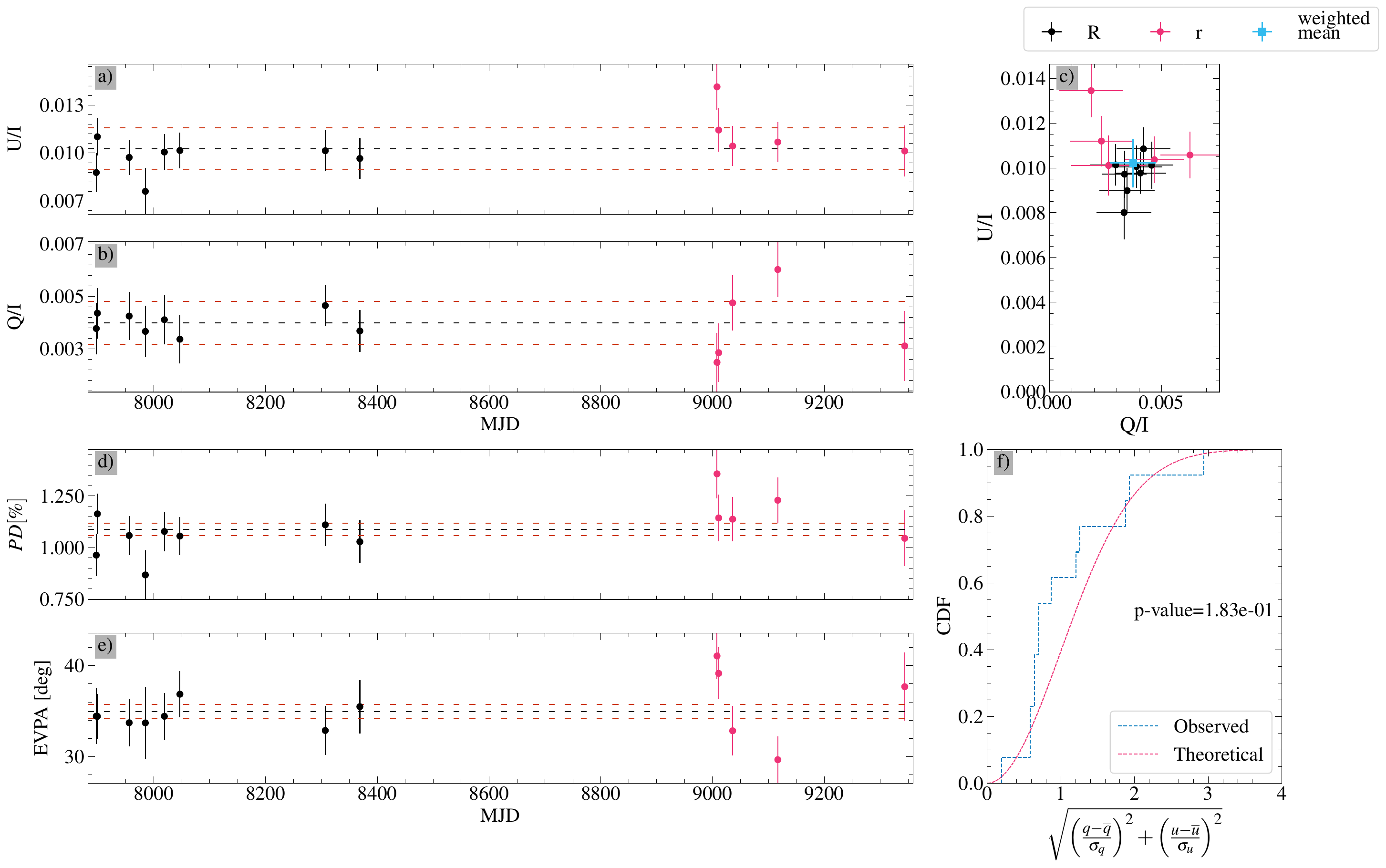}
  \caption{Same as Fig.~\ref{fig:B_0017+8135_82} for B\_2042+7508\_28, which is found to be stable. }
  \label{fig:B_2042+7508_28}
\end{figure*}

\clearpage

\begin{figure*}
  \centering
  \includegraphics[width=0.95\textwidth]{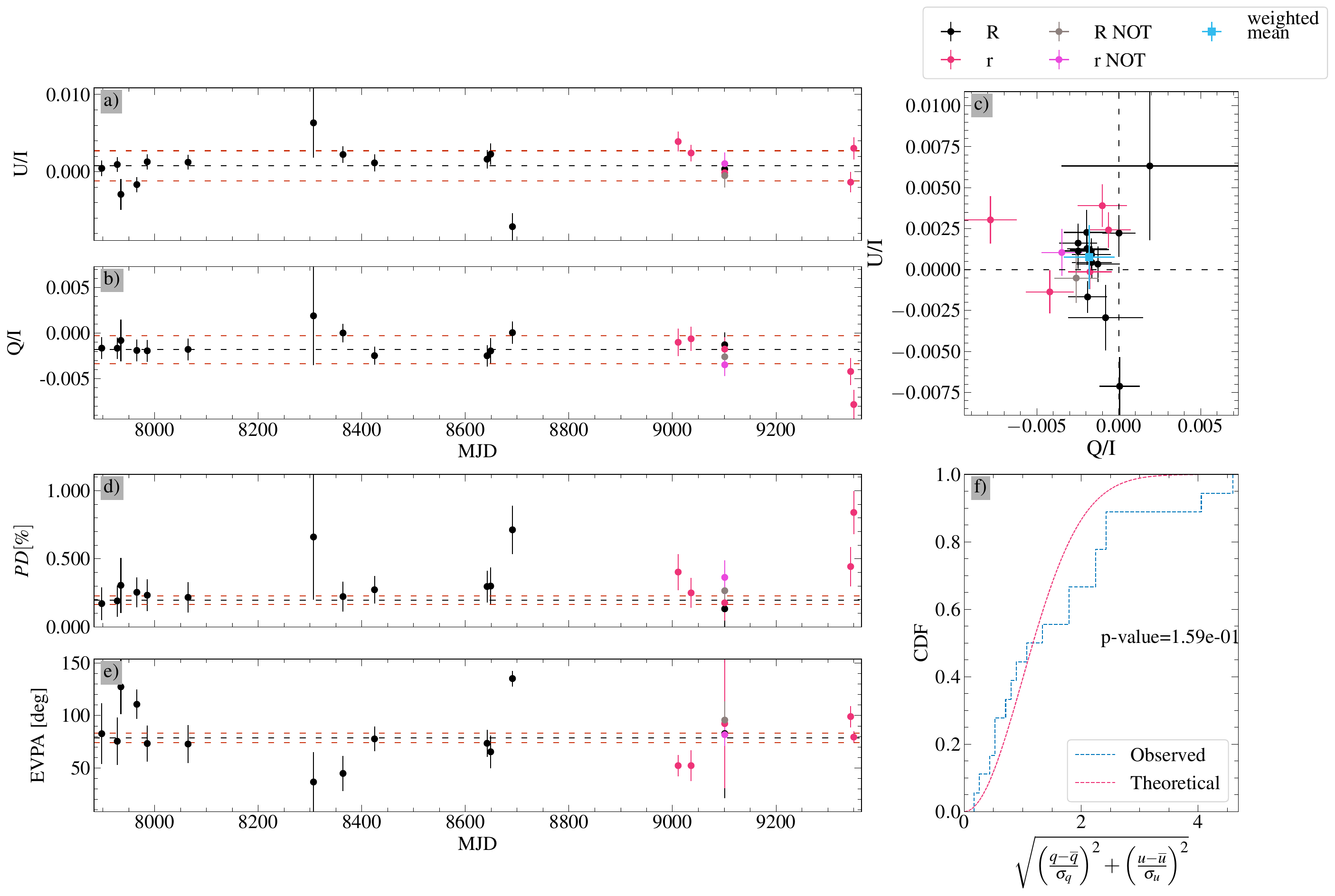}
  \caption{Same as Fig.~\ref{fig:B_0017+8135_82} for L\_112\_805, which is found to be stable. }
  \label{fig:L_112_805}
\end{figure*}

\begin{figure*}
  \centering
  \includegraphics[width=0.95\textwidth]{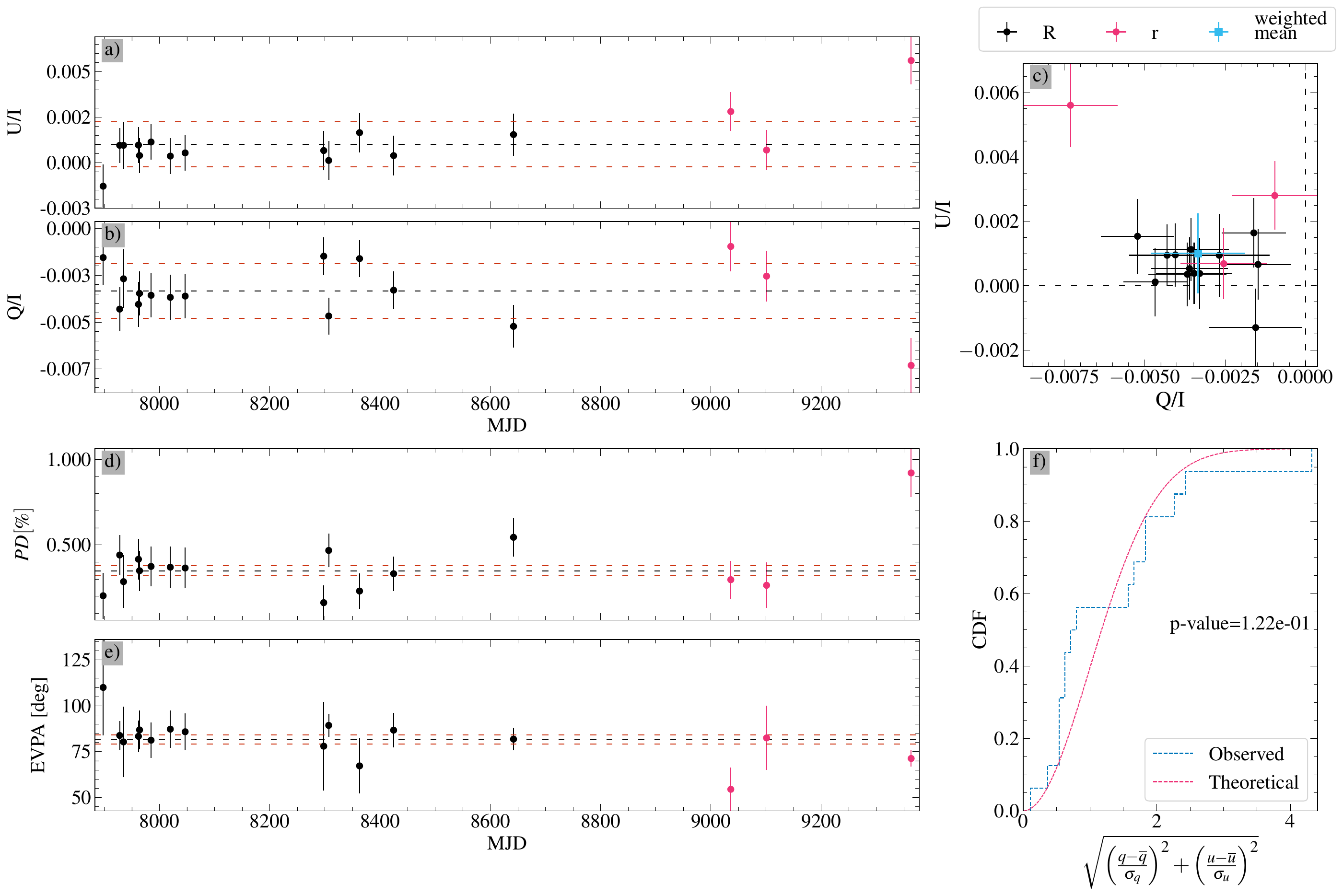}
  \caption{Same as Fig.~\ref{fig:B_0017+8135_82} for L\_112\_822, which is found to be stable. }
  \label{fig:L_112_822}
\end{figure*}

\clearpage

\begin{figure*}
  \centering
  \includegraphics[width=0.95\textwidth]{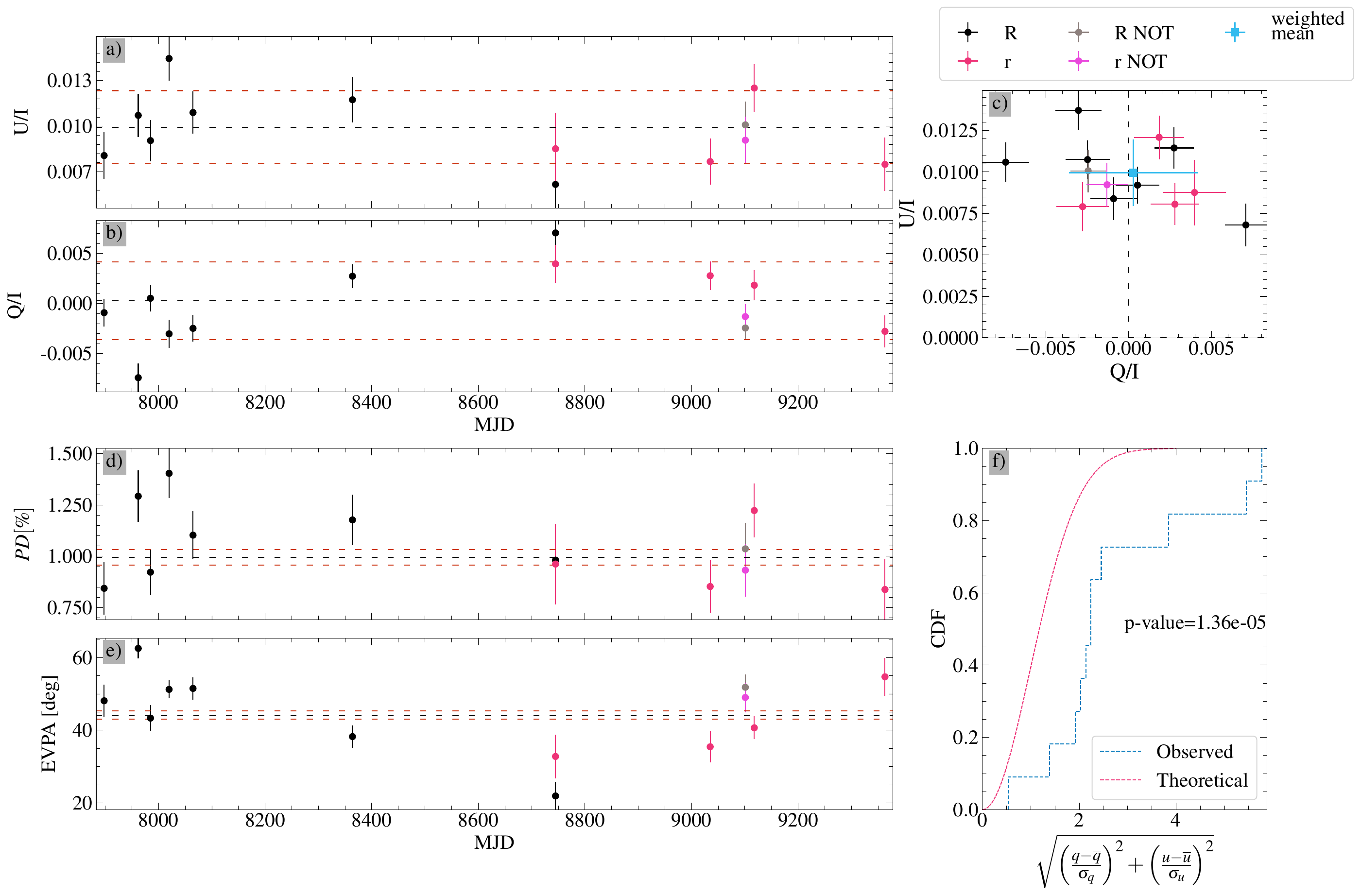}
  \caption{Same as Fig.~\ref{fig:B_0017+8135_82} for B\_2042+7508\_17, which is found to be variable. }
  \label{fig:B_2042+7508_17}
\end{figure*}

\begin{figure*}
  \centering
  \includegraphics[width=0.95\textwidth]{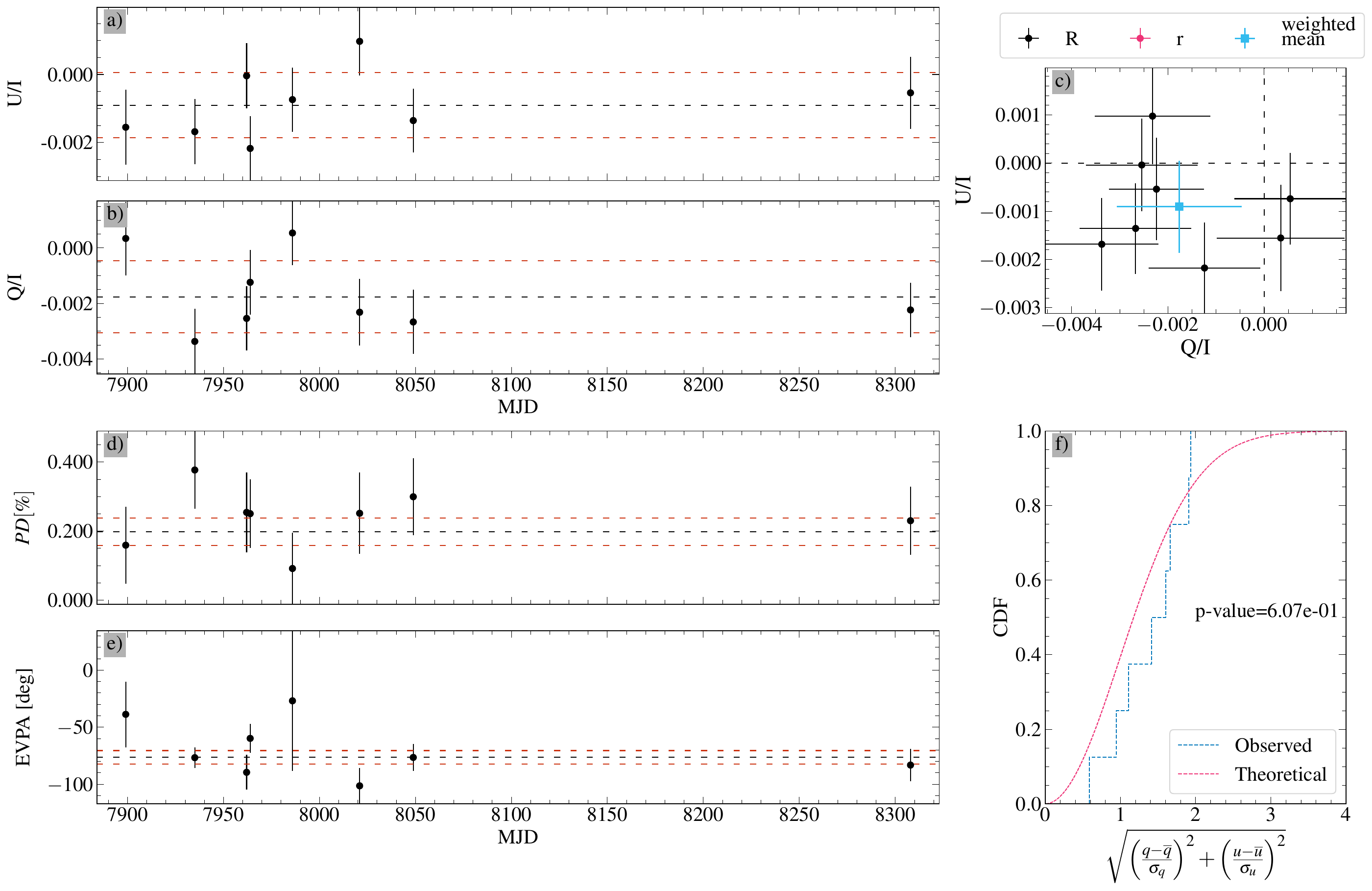}
  \caption{Same as Fig.~\ref{fig:B_0017+8135_82} for L\_113\_339, which is found to be stable. }
  \label{fig:L_113_339}
\end{figure*}

\clearpage

\begin{figure*}
  \centering
  \includegraphics[width=0.95\textwidth]{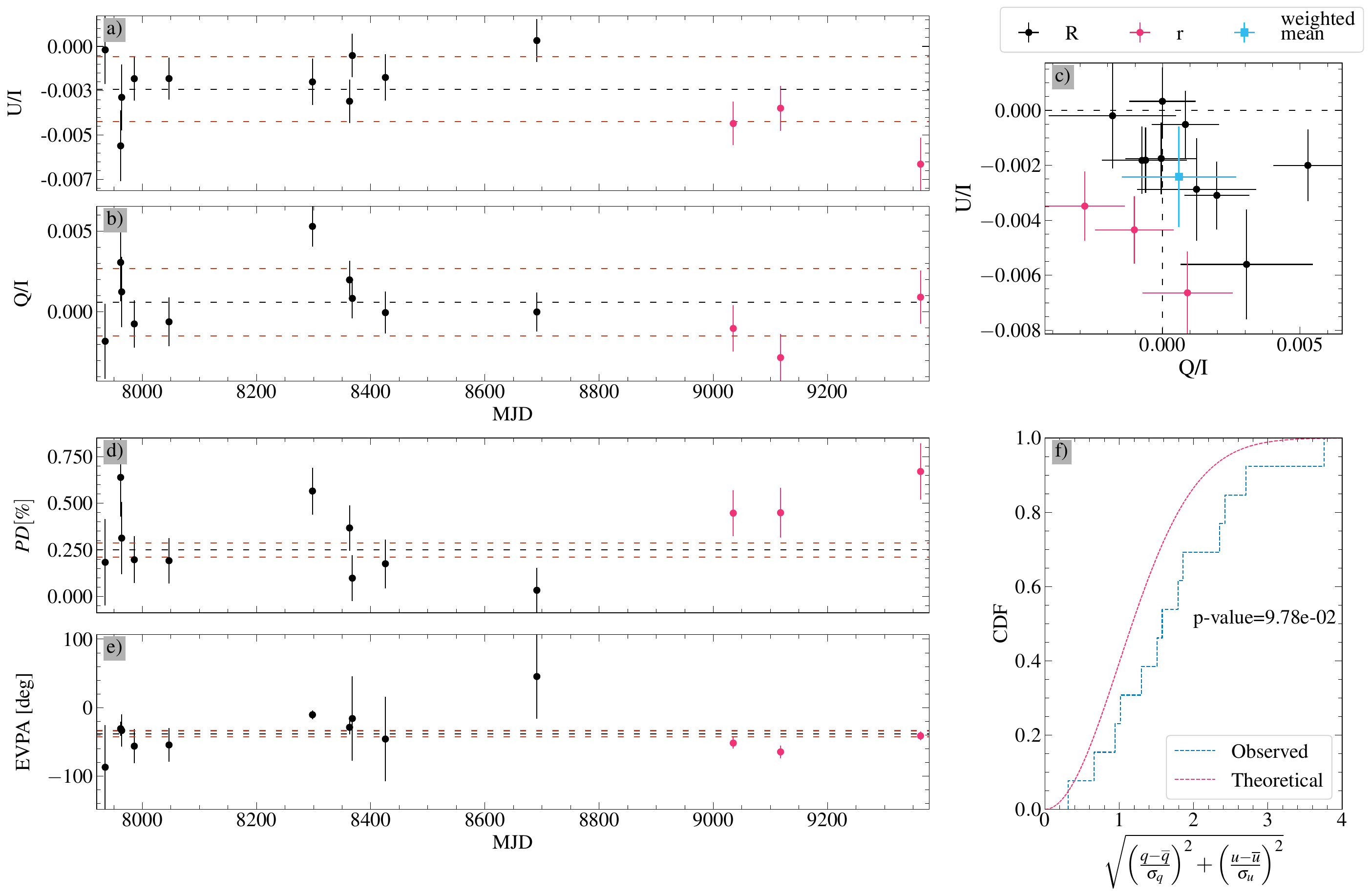}
  \caption{Same as Fig.~\ref{fig:B_0017+8135_82} for L\_113\_241, which is found to be stable. }
  \label{fig:L_113_241}
\end{figure*}

\begin{figure*}
  \centering
  \includegraphics[width=0.95\textwidth]{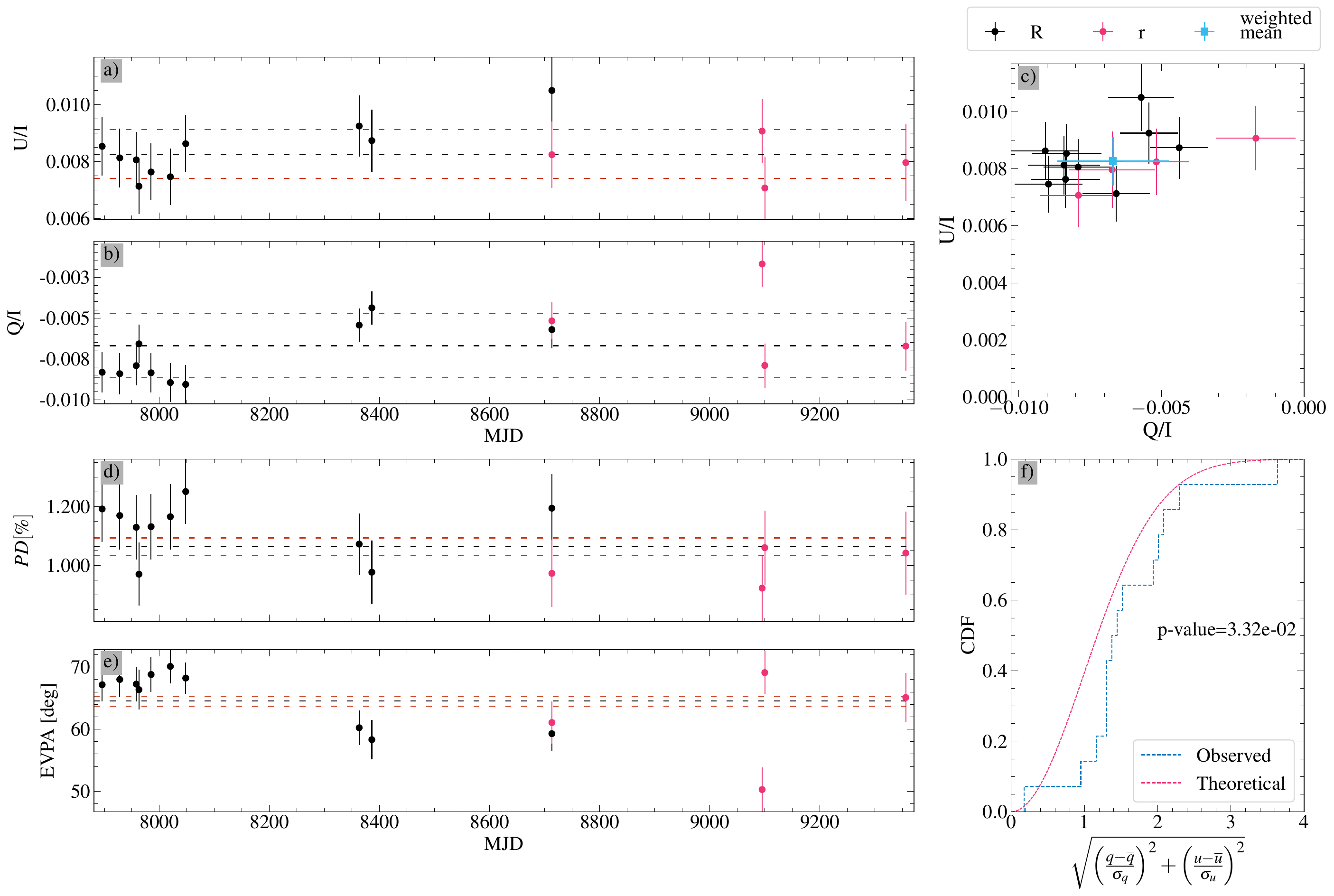}
  \caption{Same as Fig.~\ref{fig:B_0017+8135_82} for B\_2202+4216\_25, which is found to be variable. }
  \label{fig:B_2202+4216_25}
\end{figure*}

\clearpage

\begin{figure*}
  \centering
  \includegraphics[width=0.95\textwidth]{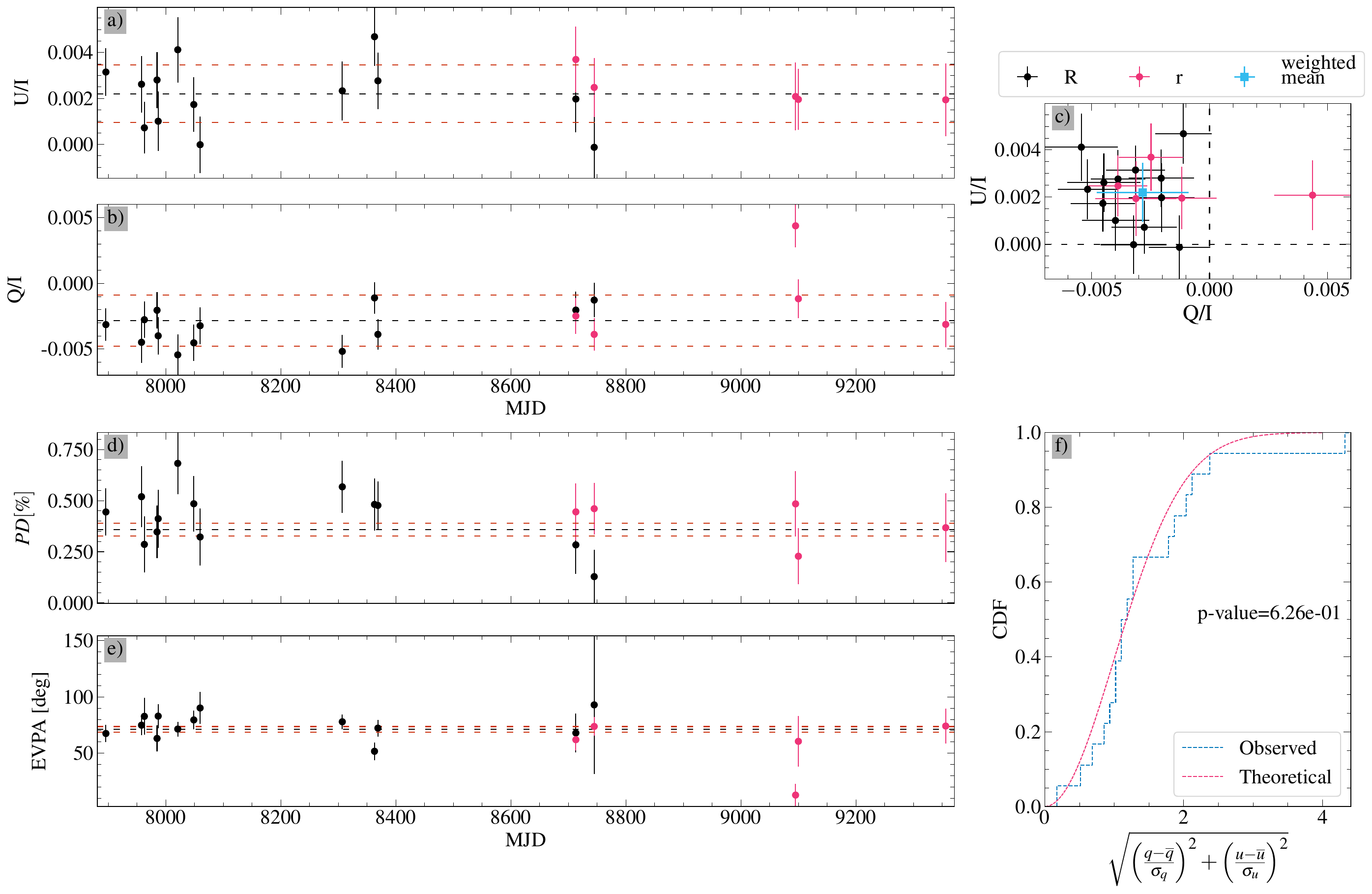}
  \caption{Same as Fig.~\ref{fig:B_0017+8135_82} for B\_2202+4216\_129, which is found to be stable. }
  \label{fig:B_2202+4216_129}
\end{figure*}

\begin{figure*}
  \centering
  \includegraphics[width=0.95\textwidth]{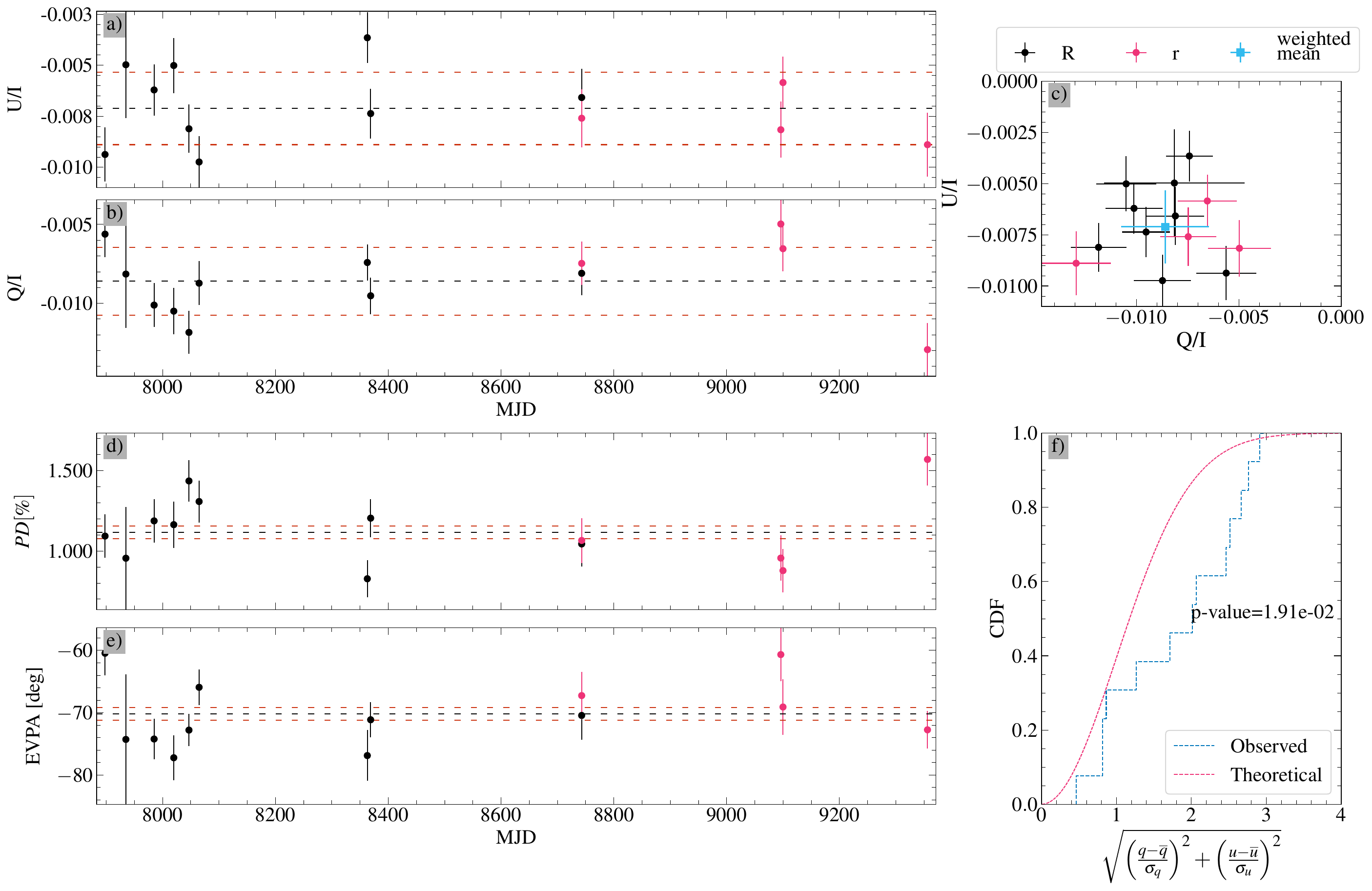}
  \caption{Same as Fig.~\ref{fig:B_0017+8135_82} for B\_2202+4216\_239, which is found to be variable. }
  \label{fig:B_2202+4216_239}
\end{figure*}

\clearpage

\begin{figure*}
  \centering
  \includegraphics[width=0.95\textwidth]{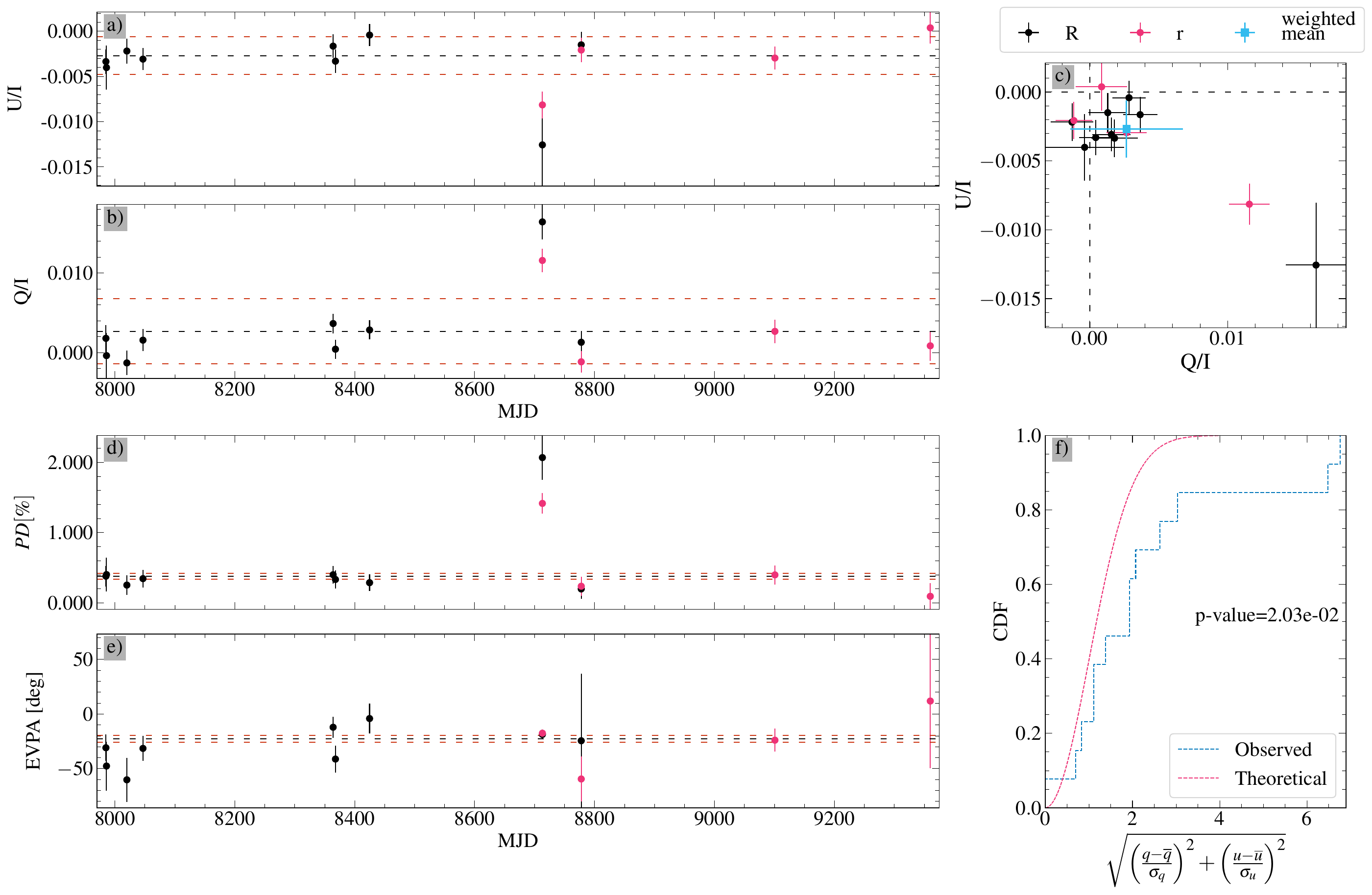}
  \caption{Same as Fig.~\ref{fig:B_0017+8135_82} for L\_PG2213$-$006A, which is found to be variable. }
  \label{fig:L_PG2213-006A}
\end{figure*}

\begin{figure*}
  \centering
  \includegraphics[width=0.95\textwidth]{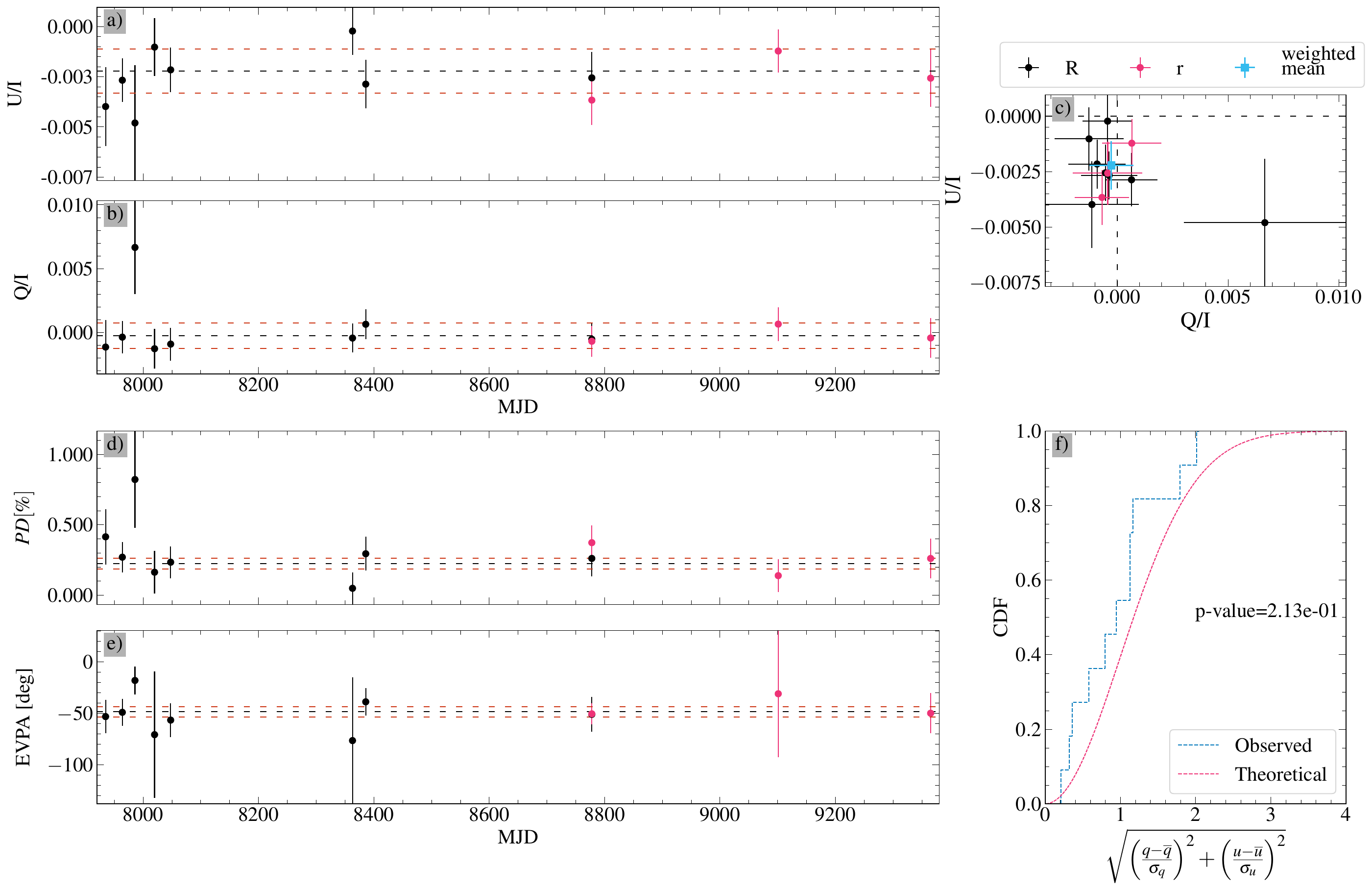}
  \caption{Same as Fig.~\ref{fig:B_0017+8135_82} for B\_2253+1608\_23, which is found to be stable. }
  \label{fig:B_2253+1608_23}
\end{figure*}

\clearpage

\begin{figure*}
  \centering
  \includegraphics[width=0.95\textwidth]{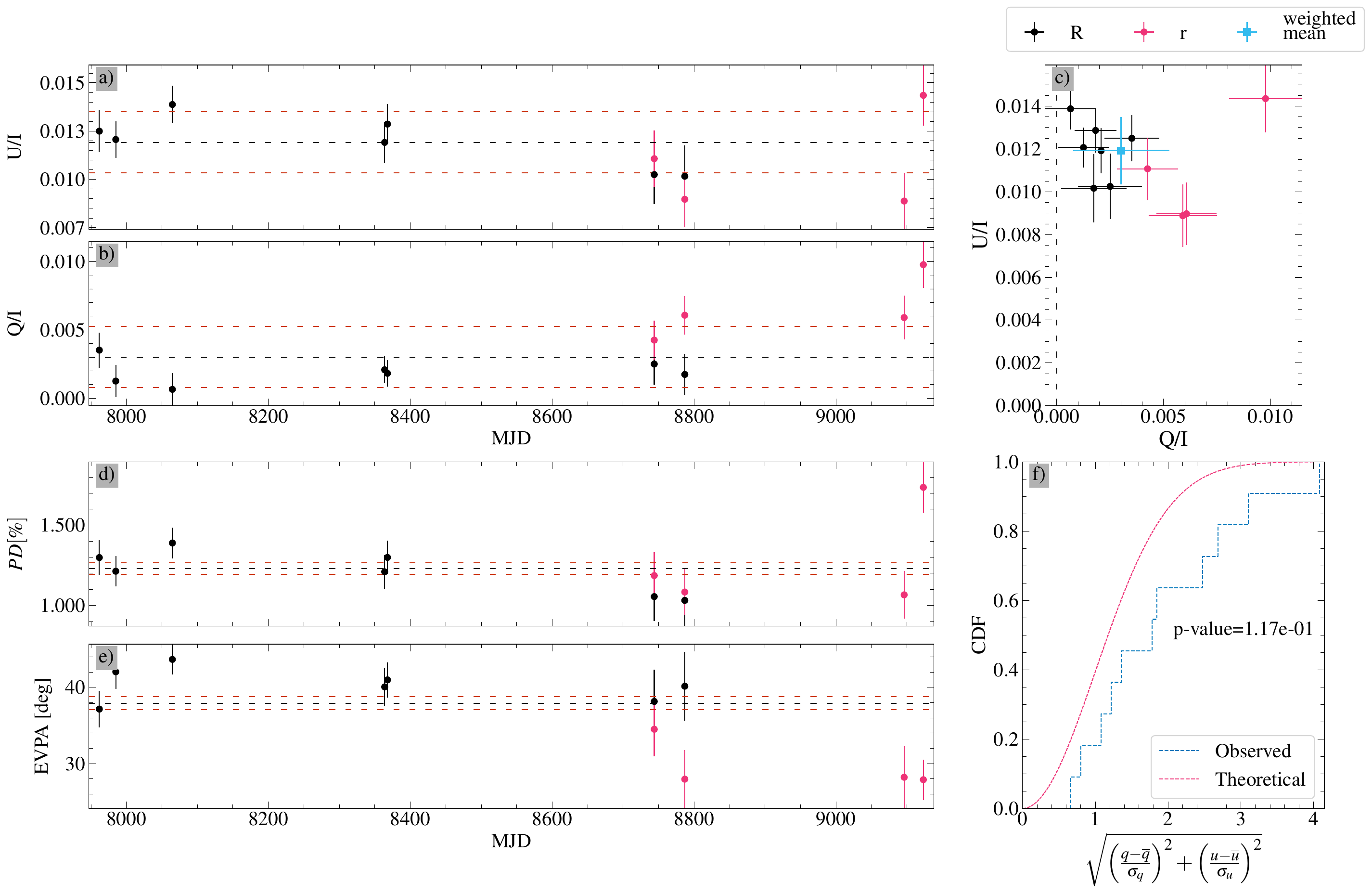}
  \caption{Same as Fig.~\ref{fig:B_0017+8135_82} for B\_2340+8015\_34, which is found to be stable. }
  \label{fig:B_2340+8015_34}
\end{figure*}

\begin{figure*}
  \centering
  \includegraphics[width=0.95\textwidth]{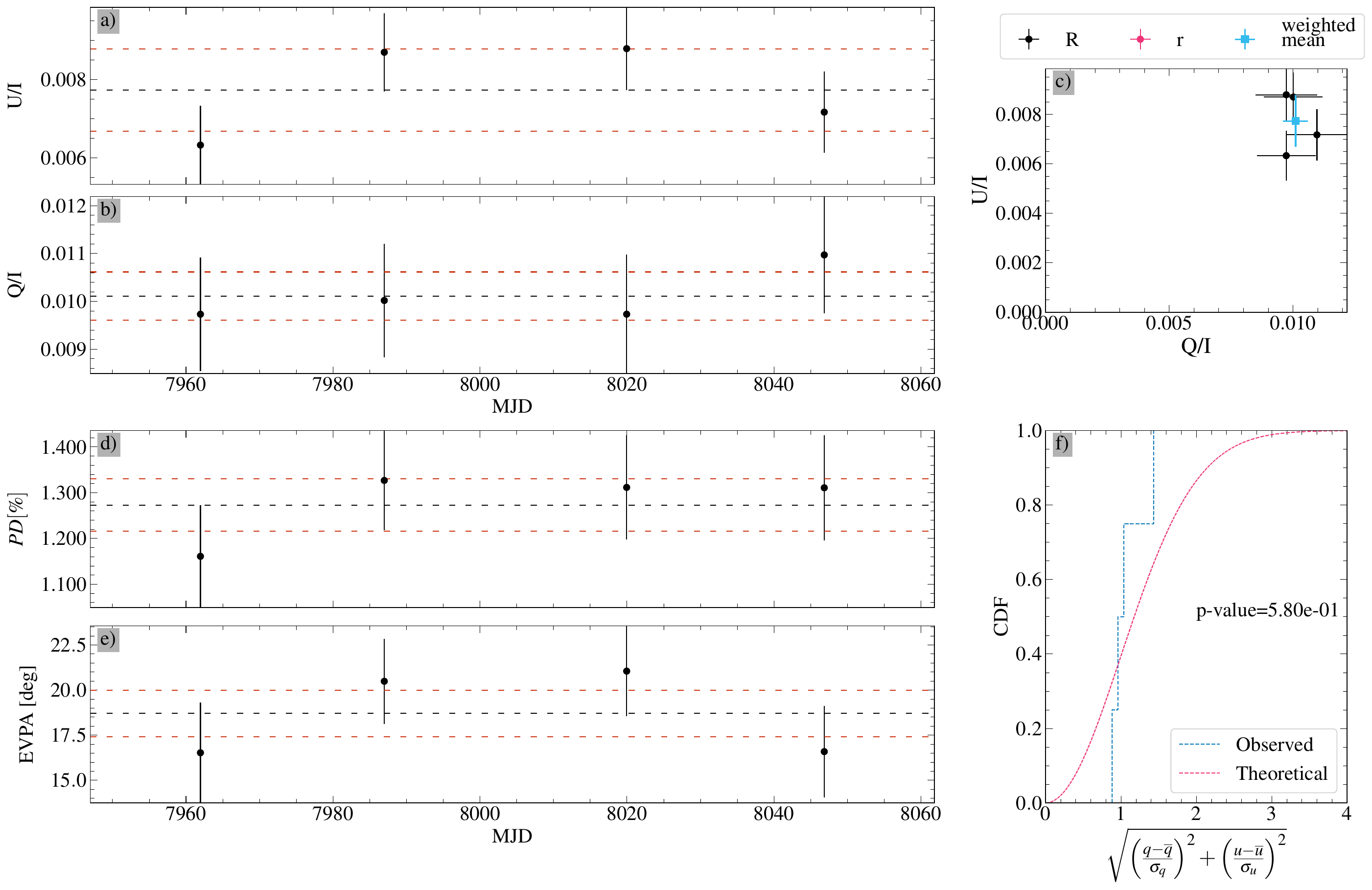}
  \caption{Same as Fig.~\ref{fig:B_0017+8135_82} for B\_2340+8015\_109, which has not enough measurements to judge it as varible of stable. }
  \label{fig:B_2340+8015_109}
\end{figure*}

\clearpage

\begin{figure*}
  \centering
  \includegraphics[width=0.95\textwidth]{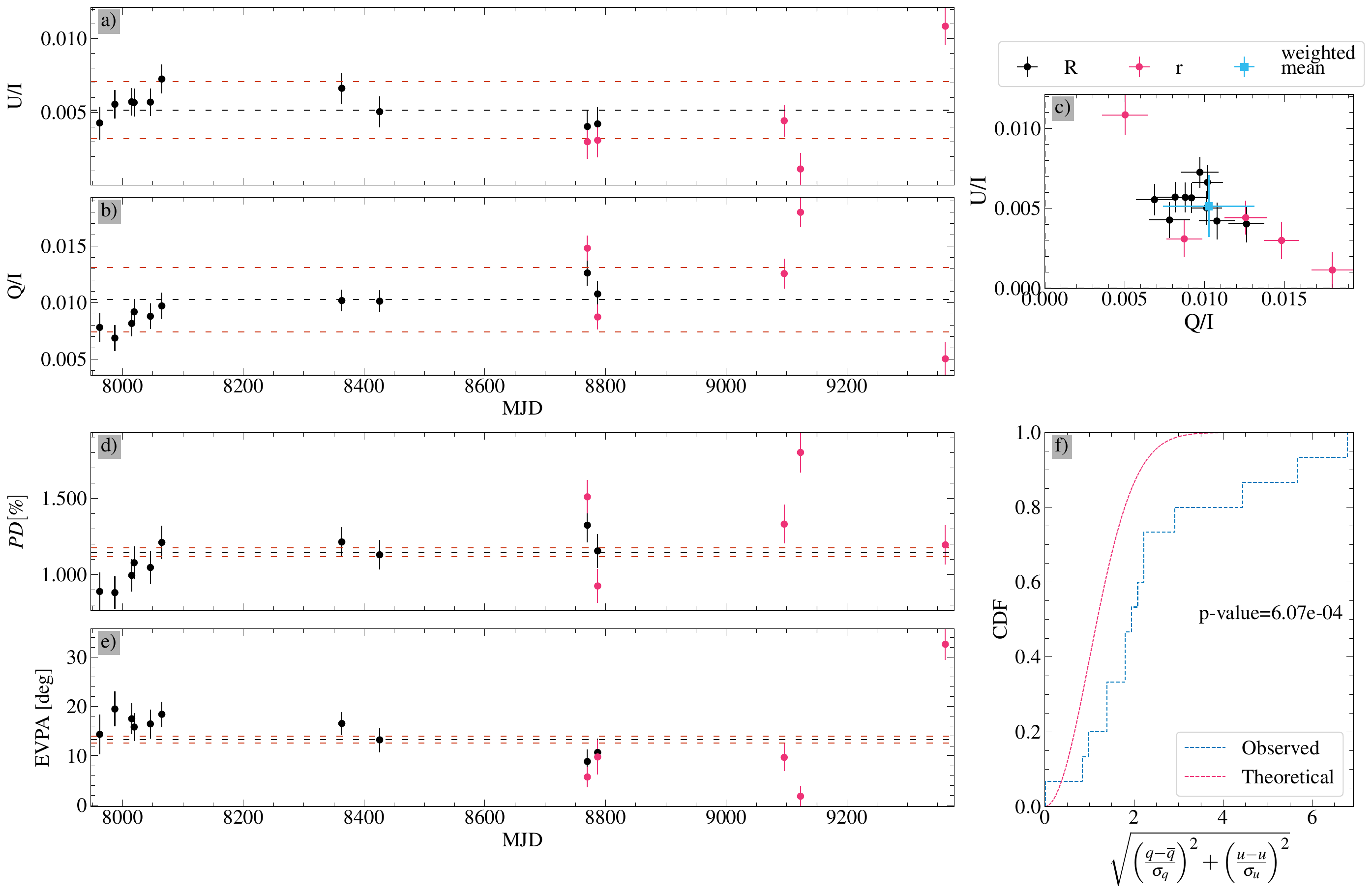}
  \caption{Same as Fig.~\ref{fig:B_0017+8135_82} for B\_2340+8015\_99, which is found to be variable. }
  \label{fig:B_2340+8015_99}
\end{figure*}

\begin{figure*}
  \centering
  \includegraphics[width=0.95\textwidth]{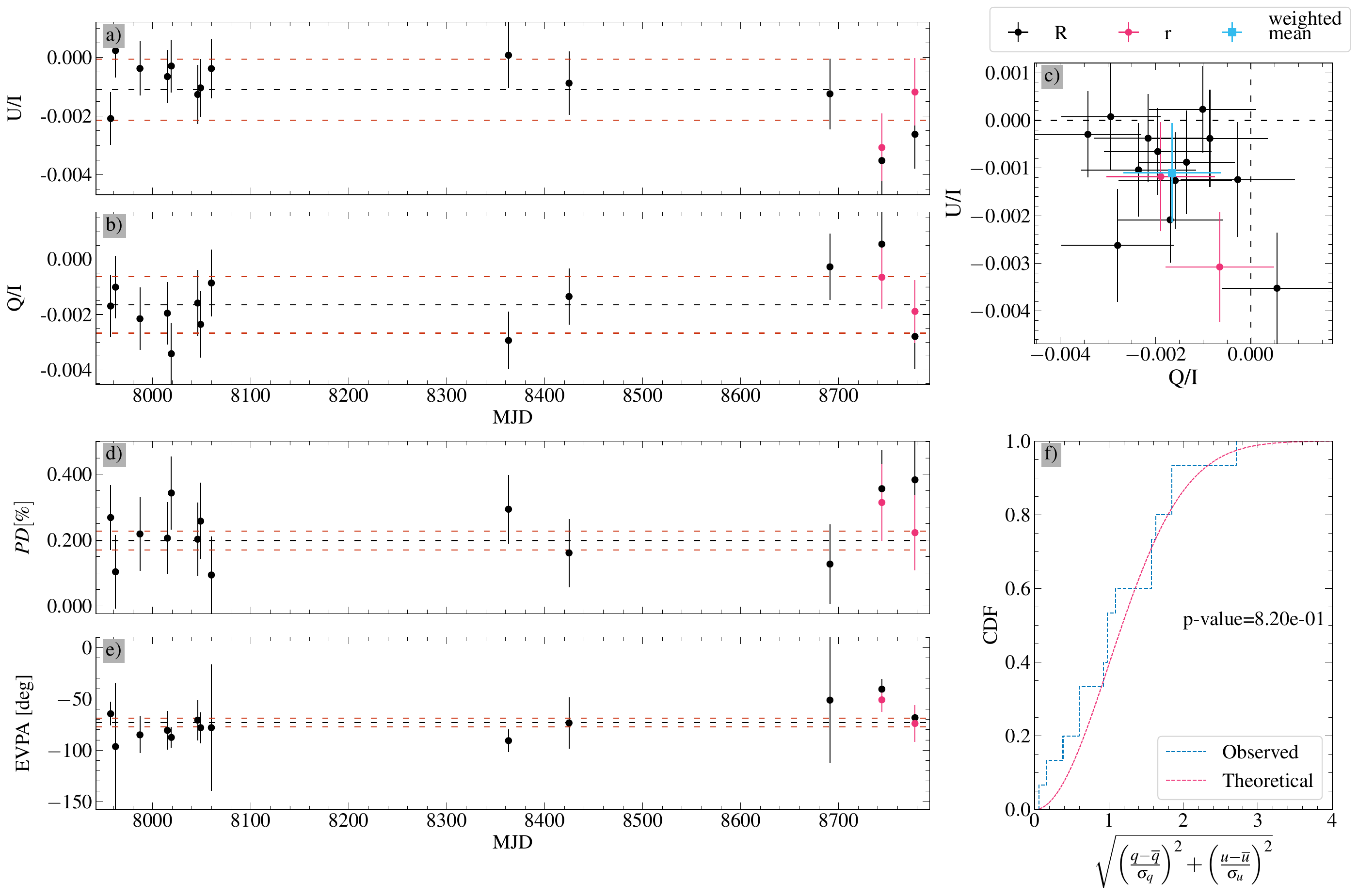}
  \caption{Same as Fig.~\ref{fig:B_0017+8135_82} for L\_115\_420, which is found to be stable. }
  \label{fig:L_115_420}
\end{figure*}

\clearpage

\begin{figure*}
  \centering
  \includegraphics[width=0.95\textwidth]{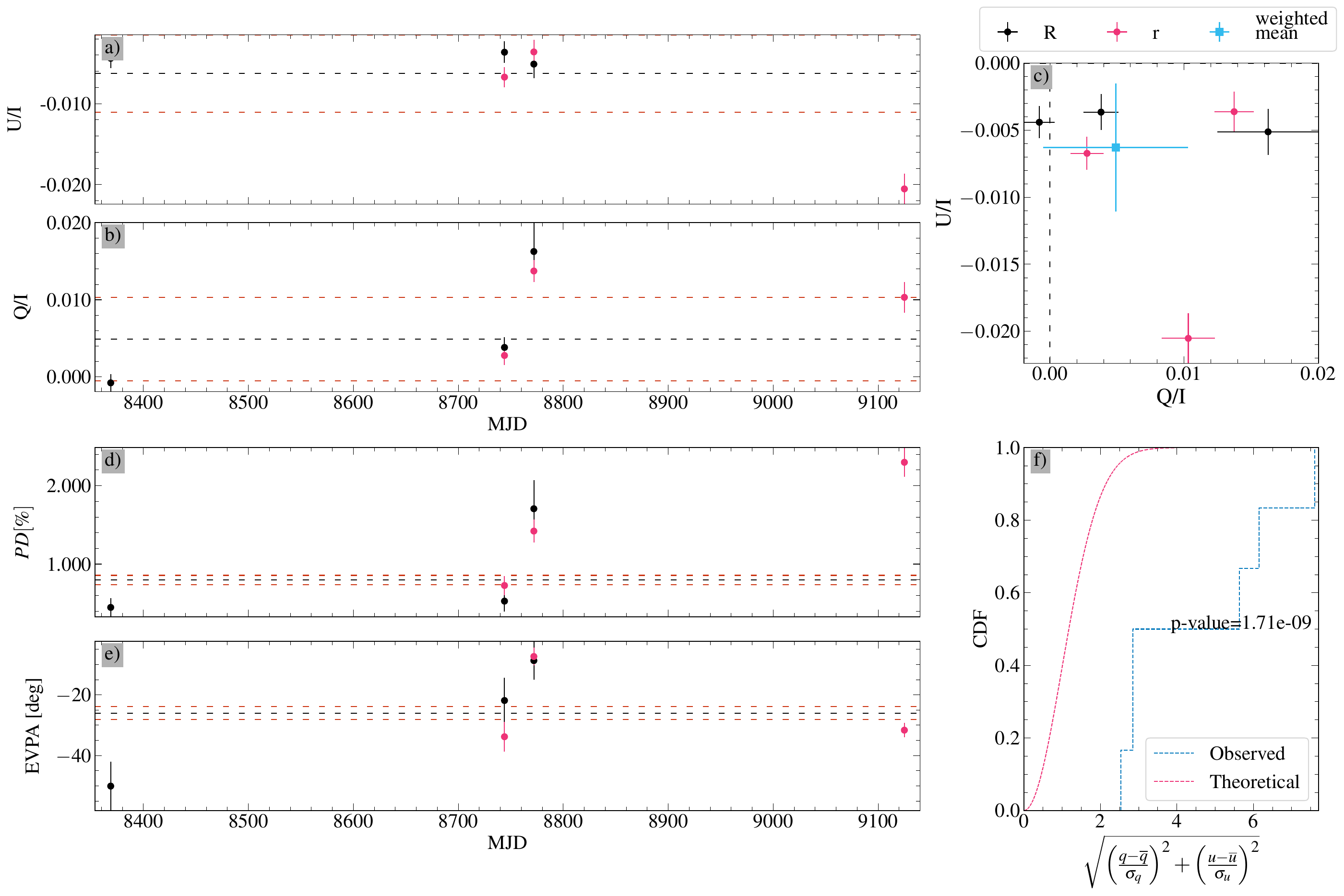}
  \caption{Same as Fig.~\ref{fig:B_0017+8135_82} for L\_PG2349+002, which is found to be variable. }
  \label{fig:L_PG2349+002}
\end{figure*}

\end{document}